\documentclass{bmcart}

\usepackage{amsthm,amsmath,amssymb}
\usepackage{graphicx}
\usepackage[utf8]{inputenc} 

\startlocaldefs
\endlocaldefs

\begin{document}

\begin{frontmatter}

\begin{fmbox}
\dochead{Review}


\title{Rydberg-Stark deceleration of atoms and molecules}


\author[
   addressref={aff1},                   
   corref={aff1},                       
   email={s.hogan@ucl.co.uk}   
]{\inits{SD}\fnm{Stephen D} \snm{Hogan}}


\address[id=aff1]{
  \orgname{Department of Physics and Astronomy, University College London}, 
  \street{Gower Street},                     %
  \postcode{WC1E 6BT}                                
  \city{London},                              
  \cny{UK}                                    
}

\begin{artnotes}
\end{artnotes}

\end{fmbox}


\begin{abstractbox}

\begin{abstract} 
The large electric dipole moments associated with highly excited Rydberg states of atoms and molecules make gas-phase samples in these states very well suited to deceleration and trapping using inhomogeneous electric fields. The methods of Rydberg-Stark deceleration with which this can be achieved are reviewed here. Using these techniques, the longitudinal motion of beams of atoms and molecules moving at speeds as high as 2500~m/s have been manipulated, with changes in kinetic energy of up to $|\Delta E_{\mathrm{kin}}|=1.3\times10^{-20}$~J ($|\Delta E_{\mathrm{kin}}|/e=80$~meV or $|\Delta E_{\mathrm{kin}}|/hc=650$~cm$^{-1}$) achieved, while decelerated and trapped samples with number densities of $10^6$--$10^7$~cm$^{-3}$ and translational temperatures of $\sim150$~mK have been prepared. Applications of these samples in areas of research at the interface between physics and physical chemistry are discussed.
\end{abstract}


\begin{keyword}
\kwd{Rydberg states of atoms and molecules}
\kwd{Stark effect}
\kwd{Stark deceleration}
\kwd{Cold atoms and molecules}
\end{keyword}


\end{abstractbox}
%

\end{frontmatter}





\section*{Introduction}

\subsection*{Rydberg states of atoms and molecules}
\begin{figure}
\begin{center}
\includegraphics[width=0.98\textwidth]{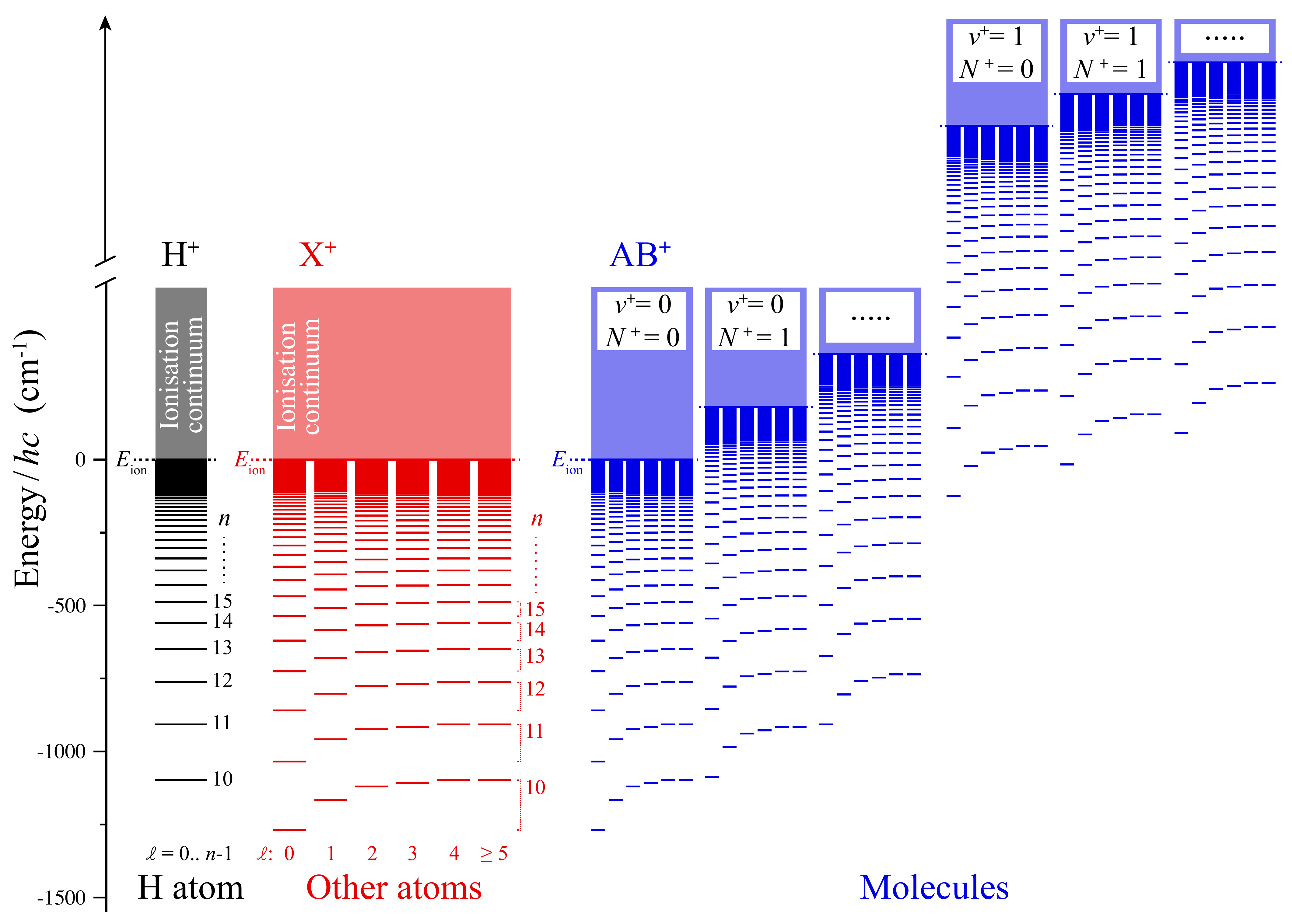}
\caption{\csentence{Rydberg states of atoms and molecules} Schematic diagram of series of Rydberg states in the hydrogen atom (H), in other non-hydrogenic atoms (X), and in molecules (AB).}
\label{fig:RydbergSeries}
\end{center}
\end{figure}

All atoms and molecules possess Rydberg states. These are excited electronic states of high principal quantum number, $n$, that form series converging to each quantum state (electronic, vibrational, rotational, spin-orbit or hyperfine) of the atomic or molecular ion core (see Fig.~\ref{fig:RydbergSeries}). To first order the energies of these states are given by the Rydberg formula~\cite{rydberg80a}
\begin{eqnarray}
E_{n\ell} = E_{\mathrm{ion}} - \frac{hc\,R_{M}}{(n-\delta_{\ell})^2},\label{eq:Ry}
\end{eqnarray}
where $E_{\mathrm{ion}}$ is the energy associated with the Rydberg series limit, $R_{M}=R_{\infty}\,\mu_{\mathrm{red}}/m_{\mathrm{e}}$ is the Rydberg constant corrected for the reduced mass, $\mu_{\mathrm{red}}=Mm_{\mathrm{e}}/(M+m_{\mathrm{e}})$, of the atom or molecule for which $M$ is the mass of the ion core and $m_{\mathrm{e}}$ is the electron mass, $\delta_{\ell}$ is a quantum defect which is dependent on the orbital angular momentum quantum number, $\ell$, of the Rydberg electron, and $h$ and $c$ are the Planck constant and speed of light in vacuum, respectively.

Rydberg states converging to the lowest ionisation limit of atoms decay via spontaneous emission. The corresponding fluorescence lifetimes of the shortest-lived low-$\ell$ states typically exceed $1~\mu$s for values of $n>30$. In molecules low-$\ell$ Rydberg states often predissociate on timescales $\ll1~\mu$s before spontaneous emission can occur~\cite{merkt97a,softley04a}. However, for molecules with a stable ion core, provided the Rydberg electron possesses sufficient orbital angular momentum (typically if $\ell\geq4$) predissociation cannot occur directly, and long-lived states with characteristic properties similar to those of Rydberg states in the H atom result. 

In non-hydrogenic atoms and in molecules, low-$\ell$ Rydberg states generally posses non-zero quantum defects, i.e. $\delta_{\ell}>0$. These arise because the Rydberg electron penetrates the non-spherically symmetric, incompletely screened, ion core to which it is bound. As a result these states, for which $\ell\lesssim4$, are more tightly bound than higher-$\ell$ states with the same values of $n$, for which $\delta_{\ell}\simeq0$ [see Eq.~(\ref{eq:Ry}) and Fig.~\ref{fig:RydbergSeries}]. Consequently, for each value of $n$ Rydberg states with $\ell\geq4$ can, to a good approximation, be considered degenerate in energy in the same way that all states in the H atom, or other hydrogenic atoms (e.g., D, T, He$^+$, Li$^{2+}$, $\overline{\textrm{H}}$ and Ps), are.

The $\ell$-degeneracy of hydrogenic Rydberg states leads to linear Stark energy shifts, $E_{\mathrm{Stark}}$, in an electric field, $\vec{F}$~\cite{pauli26a}. These linear Stark shifts can be expressed in the form
\begin{eqnarray}
E_{\mathrm{Stark}} = -\vec{\mu}_{\mathrm{elec}}\cdot\vec{F},\label{eq:EStark}
\end{eqnarray}
and are therefore a consequence of each state possessing an electric dipole moment, $\vec{\mu}_{\mathrm{elec}}$. For each value of $n$, the maximum induced electric dipole moment $\mu_{\mathrm{max}}\simeq(3/2)\,n^2 e\,a_0$, where $e$ is the electron charge and $a_0$ is the Bohr radius corrected for the reduced mass and the charge of the ion core~\cite{damburg83a}. These electric dipole moments exceed 1000~D for $n>16$, and make states with high values of $n$ particularly sensitive to electric fields~\cite{osterwalder99a}. Furthermore, in the presence of inhomogeneous fields forces, $\vec{f}$, can be exerted on atoms or molecules in such states where
\begin{eqnarray}
\vec{f} &=& -\nabla\,E_{\mathrm{Stark}}\nonumber\\
            &=& \phantom{-}\nabla(\vec{\mu}_{\mathrm{elec}}\cdot\vec{F}).
\end{eqnarray}
From the dependence of these forces on the relative orientation of the electric dipole moment and electric field vectors, states with dipole moments oriented parallel (antiparallel) to the field exhibit negative (positive) Stark energy shifts. Therefore in inhomogeneous fields atoms or molecules in these states are forced toward higher-field (lower-field) regions. Because of this states with positive Stark shifts are often known as low-field-seeking states, while states with negative Stark shifts are known as high-field-seeking states. Forces of the kind described by Eq. (3), experienced by atoms or molecules possessing non-zero electric dipole moments in inhomogeneous electric fields, have played central roles in experiments involving focussing~\cite{bennewitz55a}, state-selection~\cite{gordon55a,brooks76a,stolte82a,parker89a}, and ultimately multistage deceleration and trapping of polar ground state molecules~\cite{bethlem99a,bethlem00a,vandemeerakker12a}, and are analogous to those exploited in the experiments of Gerlach and Stern involving silver atoms with non-zero magnetic dipole moments in inhomogeneous magnetic fields~\cite{gerlach21a,gerlach22a,gerlach22b}. 

In addition to the large electric dipole moments that result from $\ell$-mixing of hydrogenic Rydberg states in electric fields, the resulting Rydberg-Stark states also exhibit  fluorescence lifetimes that are significantly longer than those of pure low-$\ell$ states typically prepared by laser photoexcitation. The fluorescence lifetimes of the Stark states with the largest electric dipole moments which can be prepared by two-photon excitation are on the order of 100~$\mu$s for values of $n>30$ and scale with $n^{3}$ -- $n^{5}$~\cite{gallagher94a}.

Over that last 10 years experimental techniques have been developed to exploit these long lifetimes and large electric dipole moments to accelerate, decelerate, transport and trap gas-phase samples of atoms and molecules in hydrogenic Rydberg states using inhomogeneous electric fields. The ubiquity of hydrogenic Rydberg states in atoms and molecules means that these techniques, which are reviewed here, can be applied to a wide range of species including those that cannot be easily laser-cooled using current technologies~\cite{shuman10a,zhelyazkova14a}, homonuclear diatomic molecules which do not possess significant electric or magnetic dipole moments in their ground states and therefore cannot be readily decelerated using other methods, e.g., multistage Stark~\cite{bethlem99a,vandemeerakker12a} or Zeeman~\cite{vanhaecke07a,hogan07a,narevicius08a} deceleration, and exotic species such as the positronium atom with short-lived ground states that are prone to decay by annihilation~\cite{alramadhan94a,vallery03a}. In this respect these Rydberg-Stark deceleration techniques represent one of several direct approaches to the preparation of chemically important cold molecules~\cite{bell09a,carr09a}, others include multistage Stark deceleration~\cite{bethlem99a,vandemeerakker12a}, multistage Zeeman deceleration~\cite{vanhaecke07a,hogan07a,narevicius08a}, optical Stark deceleration~\cite{fulton04a}, buffer gas cooling~\cite{messer84a,doyle95a}, and sympathetic cooling of molecular ions~\cite{willitsch08a,staanum08a}

\begin{figure}
\begin{center}
\includegraphics[width=0.55\textwidth]{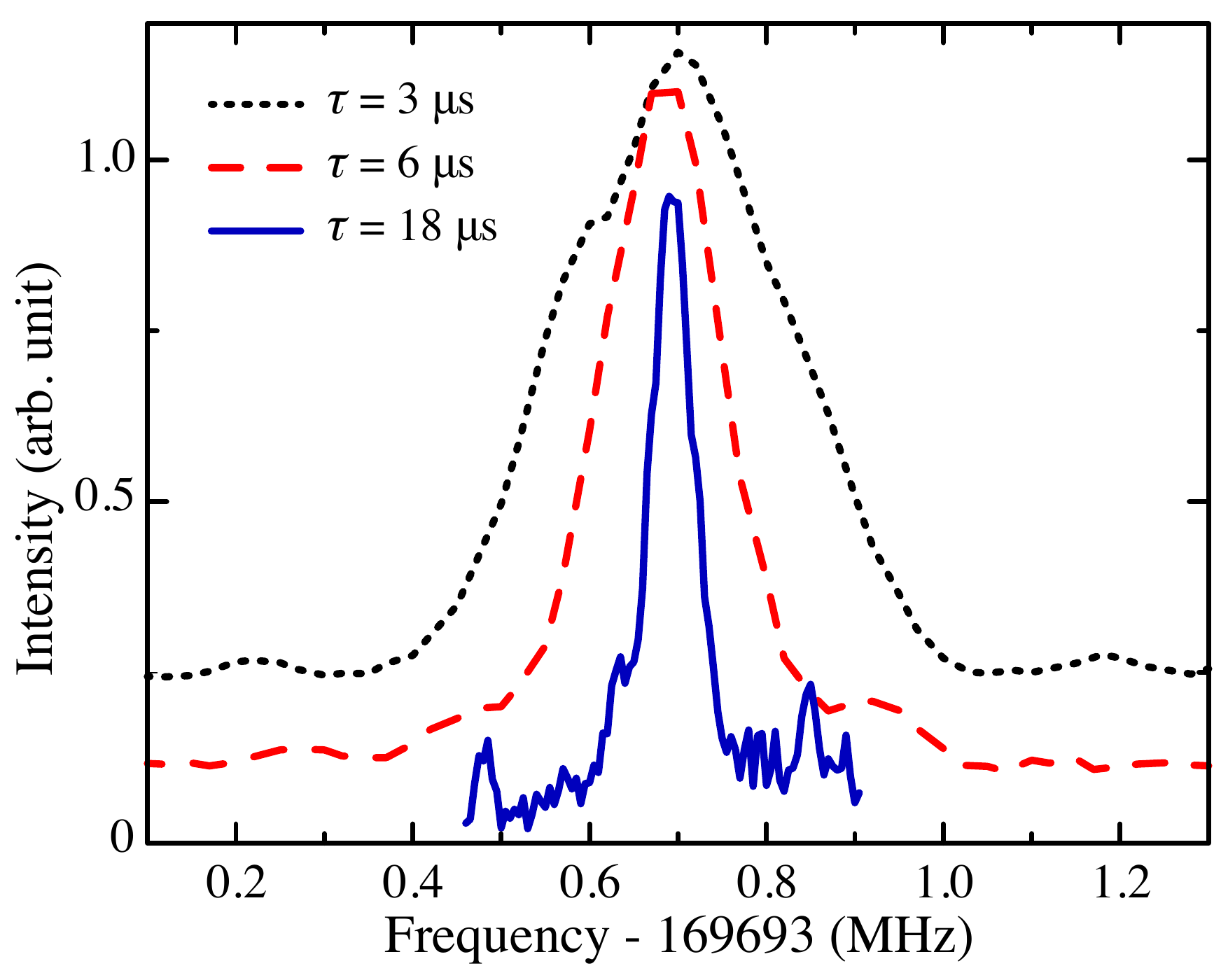}
\caption{\csentence{High-resolution millimeter-wave spectra of Kr} Millimeter-wave spectra of the $77\mathrm{d}[3/2](J'=1)\rightarrow93\mathrm{p}[3/2](J=1)$ transition in Kr recorded following interaction times between the atoms and the millimeter-wave field of $\tau=3$, 6 and 18~$\mu$s as indicated. The corresponding transition line-widths are 350, 180 and 60~kHz, respectively. From Ref.~\cite{osterwalder99a} with permission.}
\label{fig:mmwKr}
\end{center}
\end{figure}

In addition to the use of inhomogeneous electric fields, alternative approaches to trapping Rydberg atoms have also been developed. These include magnetic trapping Rb atoms in high-angular-momentum Rydberg states prepared by collisional $m$-mixing of low-$\ell$ Rydberg states with values of $n$ close to 130~\cite{choi06a}. The investigation of approaches to magnetic trapping atoms in such high-$\ell$ states is of relevance to the production and confinement of anti-hydrogen~\cite{amoretti02a}. In addition, confinement of Rb Rydberg atoms in optical lattices has also been achieved~\cite{anderson11a}.

\subsection*{Applications of decelerated beams of Rydberg atoms and molecules}

Highly excited Rydberg states of atoms and molecules play important roles in many areas at the interface between physics and physical chemistry. Transitions between individual Rydberg states of carbon with values of $n$ exceeding 1000 have been observed in absorption spectra following recombination in the interstellar medium~\cite{stepkin07a}. In the upper atmosphere of the Earth atoms and molecules in excited states including Rydberg states, are also expected to play important roles in the decay processes and reactivity of atmospheric plasmas~\cite{wayne93a}. Rydberg-Stark deceleration and electric trapping of Rydberg atoms and molecules using the techniques reviewed here opens up opportunities to experimentally study these decay processes in controlled laboratory environments on timescales exceeding 1~ms, which, prior to the development of these methods, was not possible. 

High-resolution laser, millimeter-wave and microwave spectroscopy of atomic and molecular Rydberg states is of importance in studies of the role of nuclear spins in photoionisation~\cite{merkt02a}, spectroscopic studies of the interactions of Rydberg atoms and molecules with surfaces~\cite{hogan12a,thiele14a}, and in the precise determination of ionisation and dissociation energies~\cite{osterwalder04a,liu09b,liu10a,sprecher10a}. In many of these experiments the achievable frequency resolution is not limited by the bandwidth of the radiation sources used, but instead by the interaction times of the atomic or molecular samples with the radiation field. Examples of this effect of interaction-time broadening in vacuum-ultraviolet--millimeter-wave double-resonance spectra of the $77\mathrm{d}[3/2](J'=1)\rightarrow93\mathrm{p}[3/2](J=1)$ transition in Kr can be seen in Fig.~\ref{fig:mmwKr}~\cite{osterwalder99a}. These spectra were recorded following preparation of the $77\mathrm{d}[3/2](J'=1)$ state by single-photon excitation from the ground state, after which the excited atoms interacted with a narrow-bandwidth millimeter-wave field for a period of time, $\tau$, before the population in the $93\mathrm{p}[3/2](J=1)$ state was detected by selective pulsed electric-field ionisation. The effect of interaction-time broadening can be clearly seen as $\tau$ is increased from $3~\mu$s to $\tau=18~\mu$s and the measured line-width decreases from $350$~kHz to $60$~kHz. The resolution in these experiments is not limited by the bandwidth of the millimeter-wave source but instead by the interaction time between the atoms and the radiation field. Further improvements in resolution in these experiments require longer interaction times. In precision spectroscopic studies of ground state atoms or molecules this is often achieved using the Ramsey method of separated oscillatory fields~\cite{ramsey50a}. However, the sensitivity of high Rydberg states to stray or inhomogeneous electric fields makes it challenging to achieve sufficient control over these fields in an extended volume to exploit these methods. For this reason the most appropriate approach to increasing interaction times is to exploit decelerated beams.

The development of methods for preparing quantum-state-selected velocity-controlled beams of atoms and molecules which possess electric or magnetic dipole moments in their ground or low-lying metastable states has given rise to opportunities to perform low-energy scattering experiments with collision energy resolution on the order of $E_{\mathrm{kin}}/hc = 0.01$~cm$^{-1}$ (see, e.g.,~\cite{henson12a,shagam13a}). In a similar vein, a range of scattering studies, involving atoms or molecules in high Rydberg states, are expected to benefit from the opportunities to prepare cold, velocity-tuneable beams using Rydberg-Stark deceleration. These range from studies of the interactions of Rydberg atoms and molecules with surfaces, to investigations of energy transfer in collisions between samples of Rydberg atoms or molecules and ground states species. 

The interactions of atoms and molecules in high Rydberg states with surfaces are of importance in several areas of research, including, e.g., cavity-quantum-electrodynamics at vacuum--solid-state interfaces~\cite{hogan12a,hinds97a}, experiments involving the photoexcitation of Rydberg states of samples confined in miniature vapor cells~\cite{kubler10a}, and studies of charge transfer~\cite{softley04a,gray88a}. At distances of $<10~\mu$m from conducting surfaces, the interaction of a Rydberg atom or molecule with its image-dipole in the surface contributes to state-changing and attractive forces toward the surface~\cite{hinds97a,anderson88a,sandoghdar92a}. Investigations of these processes are of importance in developing accurate models for charge-transfer (ionisation) into the surfaces~\cite{gray88a,hill00a,lloyd05a}. Several studies of these surface-ionisation processes have been carried out using beams of Rydberg atoms and molecules. These have included experiments with beams of K and Xe atoms~\cite{gray88a,hill00a} and H$_2$ molecules~\cite{lloyd05a}, and with beams of H atoms prepared in Stark states with large electric dipole moments which permitted investigations of the role that the orientation of the electric dipole moment with respect to the surface had on the ionisation dynamics~\cite{so11a}. Recent experimental studies of resonant charge-transfer from H Rydberg atoms at Cu(100) surfaces have highlighted a dependence of the ionization process on the velocity of the incoming Rydberg atoms~\cite{gibbard15a}. This would suggest that it will be of interest in future studies to exploit velocity-controlled or decelerated beams to obtain precise control over the positions and momenta of the atoms or molecules. 

\begin{figure}
\begin{center}
\includegraphics[width=0.9\textwidth]{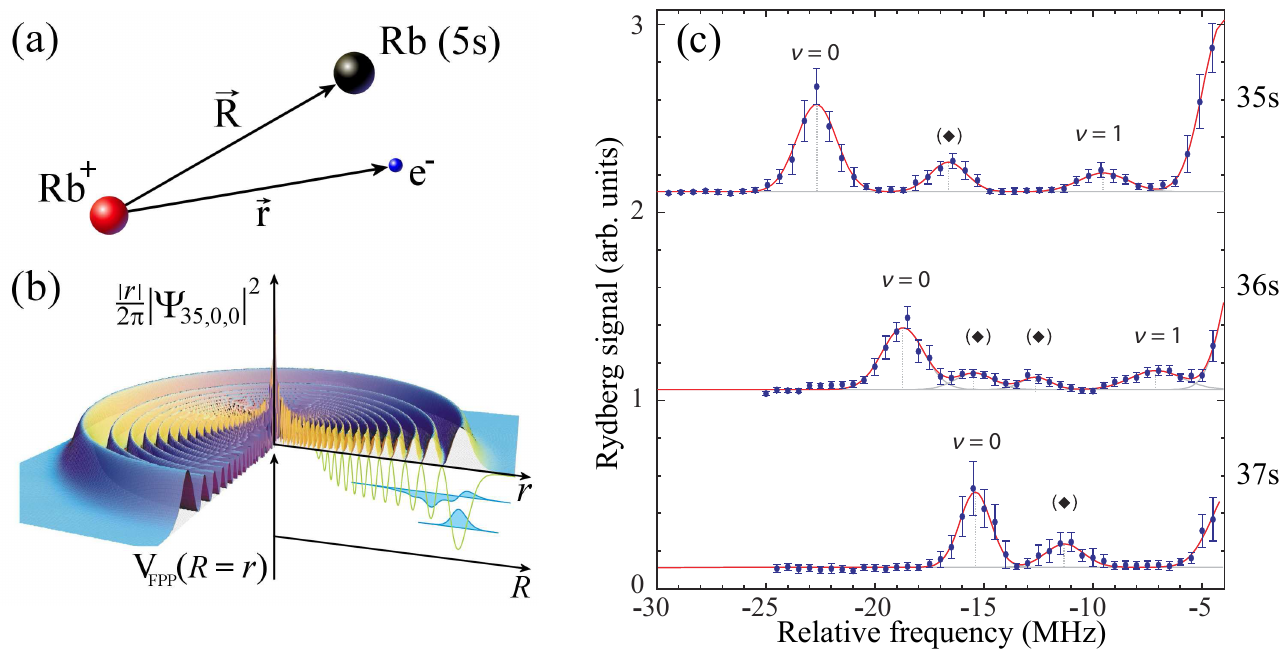}
\caption{\csentence{Long-range Rydberg molecules.} (a) and (b) the scattering process involving an electron in a diffuse Rydberg orbital and a ground state Rb atom which gives rise to sets of bound molecular eigenstates. (c) Experimentally measured transitions to the lowest vibrational states of long-range Rb$_2$ Rydberg molecules in the vicinity of the 37s (bottom), 36s (middle) and 35s (top) single-atom Rydberg states. From Ref.~\cite{bendkowsky09a} with permission.}
\label{fig:longrange_molecules}
\end{center}
\end{figure}

The high polarizability and large electric dipole moments of high Rydberg states give rise to strong van der Waals, dipole-dipole and higher order electric multipole interactions~\cite{gallagher08a,comparat10a}. These have been exploited in laser cooled samples of ultracold atoms to blockade photoexcitation~\cite{anderson98a,mourachko98a,vogt06a}, study cooperative effects~\cite{pritchard10a}, and prepare multiparticle entangled states~\cite{wilk10a,isenhower10a}. However, these properties, which are a result of the large spatial extent of Rydberg electron wavefunctions, also play important roles in the interactions of samples in Rydberg states with ground state atoms or molecules~\cite{smith78a}. Perhaps the most spectacular consequences of these interactions are seen in long-range Rydberg molecules~\cite{green00a,bendkowsky09a}. As depicted in Fig.~\ref{fig:longrange_molecules}(a) and (b) these molecules arise as a result of the scattering of a slow electron in a diffuse Rydberg orbital from a ground state atom or molecule. These bound molecular states have so far only been observed upon photoassociation in samples of laser cooled atoms [see Fig.~\ref{fig:longrange_molecules}(c)], however, it can be expected that they should also play a role in very-low--energy atomic or molecular scattering experiments in which one of the collision partners is prepared in a high Rydberg state. The development of methods such as those reviewed here for preparing a wider range of cold, decelerated beams of atoms and molecules in Rydberg states has the potential to open up exciting opportunities for studies of this unique gas-phase chemistry at long-range. 

\begin{figure}
\begin{center}
\includegraphics[width=0.7\textwidth]{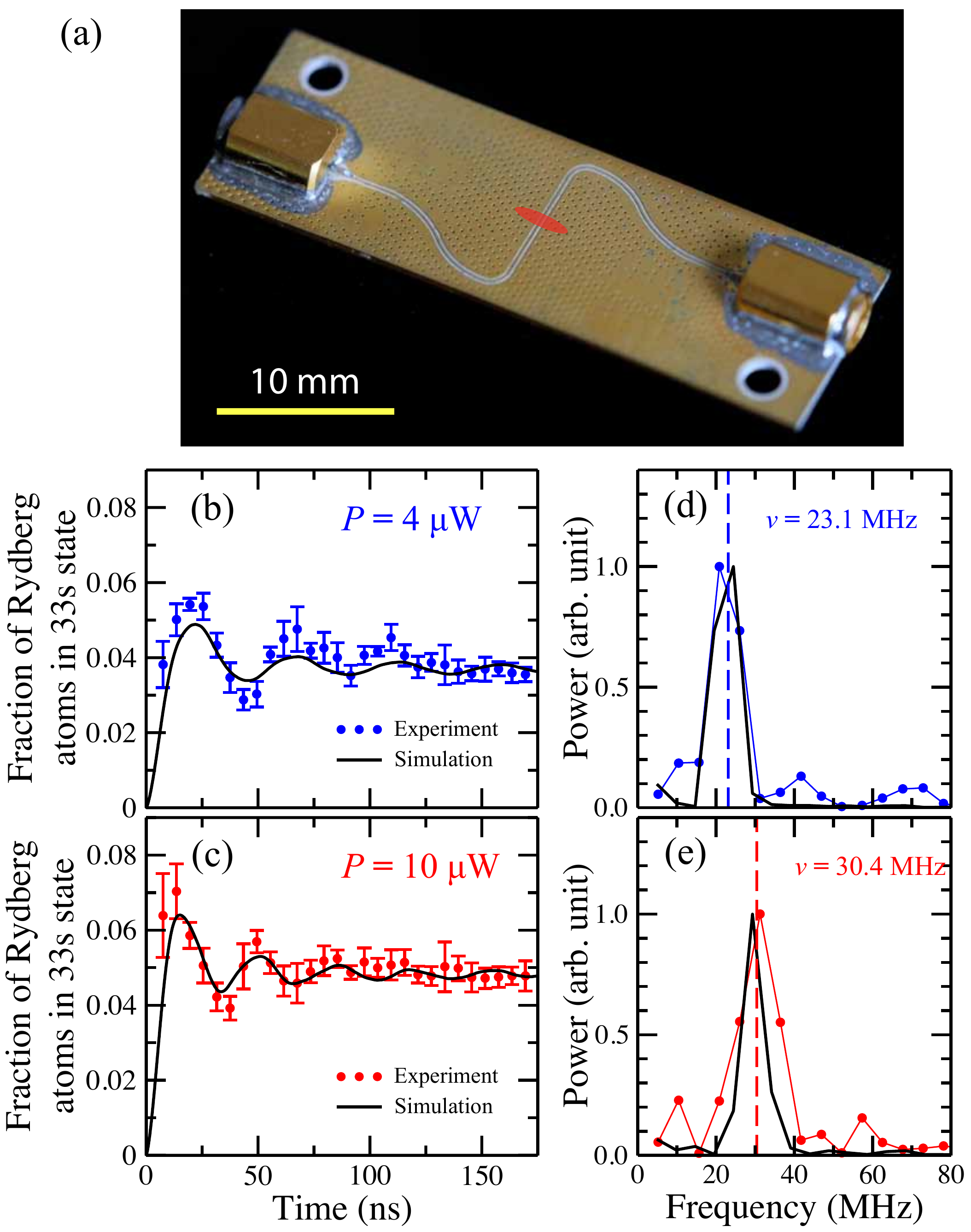}
\caption{\csentence{Coupling Rydberg atoms to microwave circuits.} (a) Photograph of a coplanar microwave waveguide with the position of a sample of Rydberg atoms, when probed by a pulsed microwave field propagating along the transmission line, indicated schematically by the red shaded region. (b-c) Rabi oscillations observed in an ensemble of Rydberg atoms coupled to the microwave field surrounding the waveguide for microwave powers of $4~\mu$W and $10~\mu$W at the source. (d-e) Fourier transforms of (b) and (c). From Ref.~\cite{hogan12a}.}
\label{fig:microwaveRy}
\end{center}
\end{figure}

Highly excited Rydberg states of atoms and molecules are very sensitive to resonant electromagnetic fields at microwave or millimeter-wave frequencies. This is a consequence of (1) the fact that the energy differences between states for which $\Delta n=1$ scale with $n^{-3}$ and correspond to transition frequencies $<400$~GHz for $n\geq25$; and (2) the large electric dipole transition moments for these $\Delta n=1$ transitions which scale with $n^2$ and approach $\sim150\,e a_0$ for $n=25$. These properties, combined with their long lifetimes, have seen Rydberg atoms play an important role in studies of microwave cavity quantum electrodynamics (QED) since the field was established~\cite{walther06a}. 

Recent implementations of microwave cavity QED in two-dimensional surface-based superconducting microwave circuits~\cite{wallraff04a} have led to a new role for Rydberg atoms in hybrid cavity QED experiments~\cite{rabl06a}. In these hybrid gas-phase--solid-state systems the Rydberg atoms are considered as long-coherence-time quantum bits (qubits) which will be coupled via two-dimensional chip-based superconducting microwave resonators to solid-state devices. These hybrid quantum systems take advantage of the long coherence times offered by gas-phase atoms as qubits, the strong coupling that can be achieved between Rydberg atoms and microwave resonators because of the their large electric dipole transition moments, and the scalability offered by micro-fabricated superconducting circuits to open new avenues of study in cavity QED at vacuum--solid-state interfaces, and potential applications in quantum information processing. 

Several approaches have been pursued in these experiments, including the realisation of atom-chips containing microwave circuitry with which cold samples of Rb have been prepared and then photoexcited to Rydberg states~\cite{carter12a,hermann14a}, and the preparation of beams of Rydberg atoms which propagate above the surfaces containing the microwave circuits (see Fig.~\ref{fig:microwaveRy})~\cite{hogan12a}. The latter approach has several advantages. The chip-based circuits can be located in a cryogenic environment where direct laser access is not required, Rydberg state photoexcitation can be carried out in a region of the apparatus which is spatially separated from the microwave circuits permitting finer control over the initial-state preparation process, and atoms or molecules can be selected for use in the experiments to ensure that effects of adsorption on the cryogenic surfaces is minimised~\cite{thiele14a}. An example of the coherent coupling of beams of helium Rydberg atoms to pulsed microwave fields surrounding a co-planar waveguide are displayed in Fig.~\ref{fig:microwaveRy}(b) and~(c)~\cite{hogan12a}. The dependence of the observed Rabi frequency for microwave transitions between Rydberg states on the microwave power can be clearly seen in the Fourier transforms of the experimental and calculated data [Fig.~\ref{fig:microwaveRy}(d) and~(e)]. In these experiments the dephasing of the Rabi oscillations was dominated by the spatial spread of the Rydberg atom beam in the inhomogeneous stray electric fields above the surface of the waveguide and the motion of the atoms. This work has in part provided motivation for the development of the chip-based guides, decelerators and traps for beams of Rydberg atoms and molecules reviewed here.

\begin{figure}
\begin{center}
\includegraphics[width=0.8\textwidth]{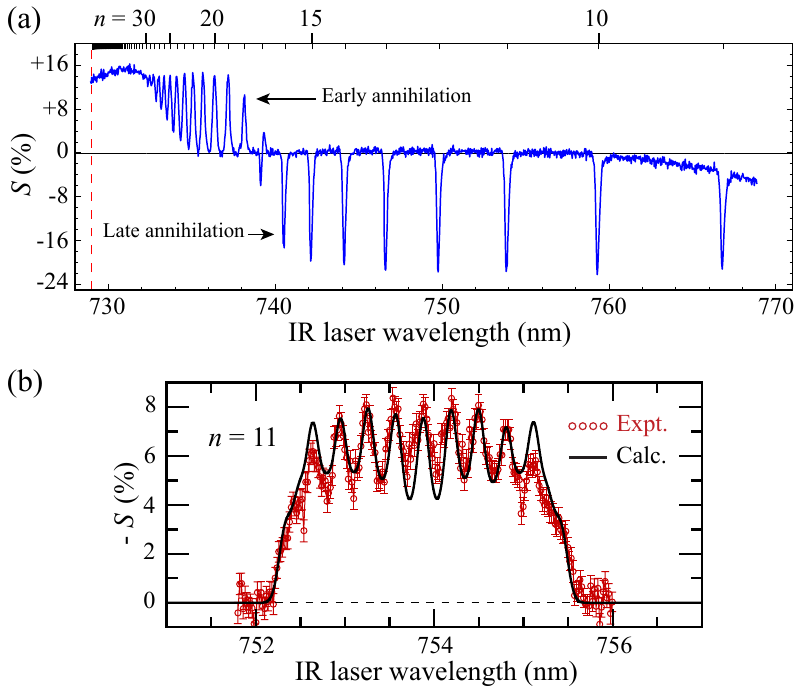}
\caption{\csentence{Spectra of Rydberg states of positronium.} (a) Two-colour two-photon excitation spectrum of Rydberg states of Ps with values of $n$ from 9 up to the ionisation limit. States with values of $n>17$ are detected when they annihilate after ionisation at a wire grid close to the photoexcitation region (early annihilation). States with values of $n<17$ pass through this grid and are detected when they annihilate at the walls of the vacuum chamber (late annihilation). (b) When photoexcitation is carried out in the presence of an electric field of 1.9~kV/cm individual Rydberg-Stark states can be selectively prepared. From Ref.~\cite{wall15a}.}
\label{fig:specPs}
\end{center}
\end{figure}

Experimental techniques with which the translational motion of Rydberg atoms and molecules can be manipulated are of particular interest in experiments with antihydrogen ($\overline{\mathrm{H}}$) and positronium (Ps). Antihydrogen atoms formed by recombination of antiprotons and positrons~\cite{amoretti02a}, or by positron transfer in collisions between antiprotons and Rydberg positronium atoms~\cite{humberston87a,charlton90a,storry04a,kellerbauer08a} are produced in high Rydberg states. At present $\overline{\mathrm{H}}$ atoms that eventually decay to their ground state can be magnetically trapped~\cite{alpha10a,alpha11a}, however, no attempts have been made to confine the atoms while in the Rydberg states to improve the ultimate trapping efficiency after decay. In this respect, the implementation of methods to electrically trap Rydberg atoms in a wide rage of Stark states in the presence of strong background magnetic fields is of particular interest. In addition, the AEgIS experiment, currently under development at CERN, is designed to exploit inhomogeneous electric fields to accelerate Rydberg $\overline{\mathrm{H}}$ atoms to produce beams with tuneable velocities for antimatter gravity and spectroscopy experiments~\cite{kellerbauer08a}. 

The Ps atom, the bound state of an electron and a positron, is another unique system which is of interest in tests of the effect of the Earth's gravitational field of particles composed of antimatter,~\cite{mills02a,cassidy14a} and precision spectroscopy of fundamental importance~\cite{karshenboim05a}. The longer-lived triplet ground state of Ps has an annihilation lifetime of $142$~ns~\cite{vallery03a}, while the singlet state lives for $125$~ps~\cite{alramadhan94a}. However, when excited to high Rydberg states, as in the data in Fig.~\ref{fig:specPs}, the spatial overlap of the electron and positron wavefunctions is reduced with the result that for all excited triplet levels other than the $2\,^3$S$_1$ level the rate of fluorescence to the ground state is greater than the direct annihilation rate. Furthermore because of its reduced mass of exactly $0.5\,m_{\mathrm{e}}$, where $m_{\mathrm{e}}$ is the electron ($\equiv$ positron) mass, the fluorescence lifetimes of Rydberg states of Ps~\cite{wall15a,ziock90a,cassidy12a,hogan13b} are twice as long as those of states with the same values of $n$ in the H atom. For this reason Rydberg states of Ps are well suited for precision spectroscopic studies, and tests of antimatter gravity. The challenge associated with carrying out precision spectroscopy or gravity measurements with Rydberg Ps is the high speeds ($\sim10^5$~m/s) with which the samples, produced by implantation of pulsed positron beams into room temperature porous silica targets, move~\cite{cassidy10a}. However, these speeds correspond to kinetic energies $E_{\mathrm{kin}}/e \simeq 50$~meV ($E_{\mathrm{kin}}/hc \simeq 460$~cm$^{-1}$) that lie well within the $\sim80$~meV ($\equiv650$~cm$^{-1}$) changes in kinetic energy that have been achieved in Rydberg-Stark deceleration of fast beams of He Rydberg atoms using inhomogeneous electric fields~\cite{lancuba14a}. As a result Rydberg-Stark deceleration represents a viable route to the preparation of slowly moving, or electrically trapped, Ps atoms.

\section*{Rydberg states in electric fields}

In the presence of an external electric field $\vec{F}=(0,0,F)$, the Hamilton, $H$, of an atom with a single excited Rydberg electron can be expressed as
\begin{eqnarray}
H &=& H_0 + H_{\mathrm{S}}\nonumber\\
    &=& H_0 + eFz,\label{eq:HStark}
\end{eqnarray}
where $H_0$ is the Hamiltonian in the absence of the field, $H_{\mathrm{S}}$ is the Stark Hamiltonian representing the interaction with the field, $e$ is the electron charge, and $z$ represents the position in cartesian coordinates.

\begin{figure}
\begin{center}
\includegraphics[width=0.9\textwidth]{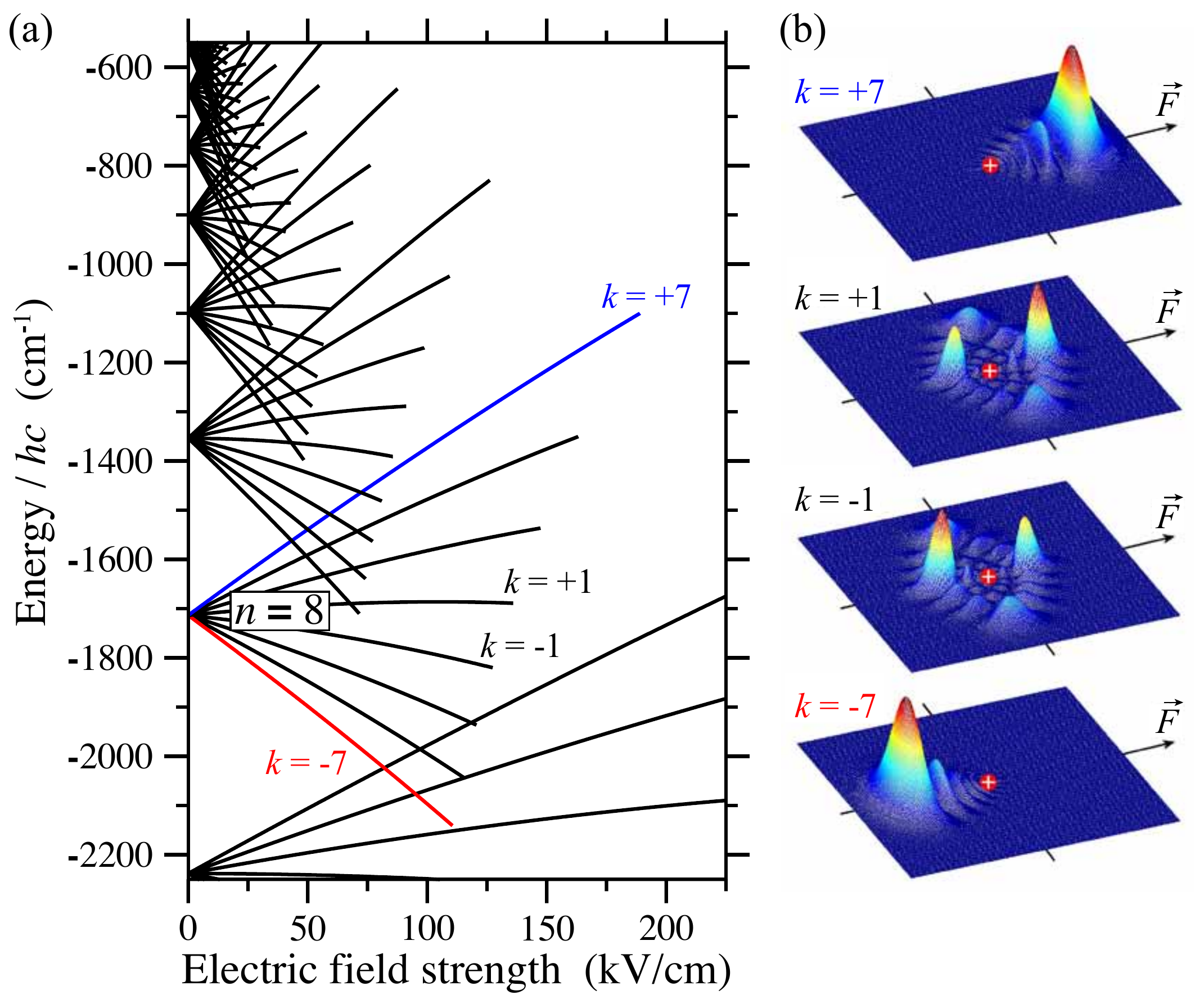}
\caption{\csentence{The Stark effect in Rydberg states of the H atom}. (a) Dependence of the energies of $m=0$ Stark states with values of $n$ from~7 to~14 on the strength of the electric field. (b) Electron probability density in a plane containing the electric field axis for $n=8$ Stark states with $k=-7, -1, +1$ and~$+7$. After Ref.~\cite{kleppner83a}.}
\label{fig:starkshifts}
\end{center}
\end{figure}

\subsection*{The Stark effect in Rydberg states of the hydrogen atom}

In hydrogenic Rydberg states the Schr\"odinger equation associated with the Hamiltonian $H$ can be solved in parabolic coordinates~\cite{bethe57a,englefield72a}. The wavefunctions obtained are characterised by four quantum numbers: $n$, $m$ the azimuthal quantum number, and two parabolic quantum numbers $n_1$ and $n_2$. These quantum numbers satisfy the condition that $n = n_1 + n_2 +|m|+1$. The energies of the eigenstates, generally referred to as Stark states, in the presence of the field can be expressed to second order as~\cite{damburg83a}
\begin{eqnarray}
E_{n\,n_1 n_2 m\,} &=& E_{\mathrm{ion}} -\frac{R_M\,hc}{n^2} + \frac{3}{2}n(n_1-n_2)\,ea_{0}F +\nonumber\\
			     & & \hspace*{0.4cm}-\frac{1}{16}n^4[17\,n^2 - 3(n_1-n_2)^2 - 9\,m^2 + 19]\frac{e^2a_{0}^2}{E_{\mathrm{h}}}F^{\,2} + \dots\label{eq:starkparabolic}
\end{eqnarray}
where $E_{\mathrm{h}}=2hc\,R_{M}$ and $a_{0}$ are the Hartree energy and the Bohr radius corrected for the reduced mass and the charge of the ion core, respectively. Often the difference between the two parabolic quantum numbers is denoted by the index $k$, such that $k=n_1-n_2$. For each value of $m$, the allowed values of $k$ range from $-(n-|m|-1)$ to $+(n-|m|-1)$ in intervals of 2. The resulting electric field dependence of the $m=0$ Stark states of the H atom with values of $n$ ranging from~7 to~14 are displayed in Fig.~\ref{fig:starkshifts}(a). From this energy level diagram it can be seen that in weak electric fields the first order, linear Stark shift dominates with the quadratic and higher order terms gradually increasing in significance as the field strength increases. This effect is most visible in this figure for the states with values of $k$ close to zero in this figure.

Comparison of the first order term in electric field strength in Eq.~(\ref{eq:starkparabolic}) with Eq.~(\ref{eq:EStark}) indicates that to each Stark state an electric dipole moment of $\vec{\mu}_{n\,k}=(0,0,-(3/2)\,nk\,e\,a_0)$ can be attributed. These electric dipole moments are a consequence of the spatial distribution of electron charge about the ion core in each eigenstate, as is evident from the electron probability density distributions displayed in Fig.~\ref{fig:starkshifts}(b) for the $n=8$ Stark states with $k=-7, -1, +1$ and $+7$~\cite{kleppner83a}. The noticeable spatial separation of the positive charge of the ion core, from the distribution of negative charge associated with the electron, seen for the $k=+7$ Stark state in Fig.~\ref{fig:starkshifts}(b) indicates clearly that the electric dipole moment (214~D) of this state is oriented antiparallel to $\vec{F}$ giving rise to the positive Stark energy shift of this state in Fig.~\ref{fig:starkshifts}(a). In the same way the negative Stark shift of the $k=-7$ state results from the orientation of its electric dipole moment parallel to $\vec{F}$. As can be seen in Fig.~\ref{fig:starkshifts}(b) the states with low values of $|k|$ in the middle of the Stark manifold have approximately equal distributions of electron charge on either side of the ion core and therefore small electric dipole moments (31~D) and weak linear Stark shifts. Typically Rydberg-Stark deceleration experiments have been performed following photoexcitation of outer Stark states with values of $n$ between 15 and 60, and electric dipole moments between 500~D and 13500~D.

\subsection*{Electric field ionisation of hydrogenic Rydberg states}

In manipulating the translational motion of atoms or molecules in hydrogenic Rydberg-Stark states using inhomogeneous electric fields, the maximal fields that can be employed are limited by the ionisation electric fields of the states in which the samples are prepared. The Stark contribution, $H_{\mathrm{S}}$, to the Hamiltonian in Eq.~(\ref{eq:HStark}) gives rise to a saddle point in the potential experienced by the Rydberg electron. If the energy of this saddle point lies below the energy of the excited Rydberg state, electric field ionisation will occur. This ionisation field depends strongly on the value of $n$, and also on the value of $k$. The outermost state with a negative Stark energy shift [e.g., the $k=-7$ state in Fig.~\ref{fig:starkshifts}(a)] typically ionises in fields equal to or larger than the classical ionisation field, $F_{\mathrm{class}}$, for which the energy of the Stark saddle point coincides with the energy of the Stark state in the field~\cite{gallagher94a}
\begin{eqnarray}
F_{\mathrm{class}} = \frac{F_0}{9n^4},\label{eq:calssion}
\end{eqnarray}
where $F_0=2hc\,R_{\mathrm{M}}/(ea_0)$, with $R_{\mathrm{M}}$ and $a_0$ adjusted to account for the reduced mass of the system and the charge of the ion core. More strictly the rate at which the electron tunnels through the barrier associated with the Stark saddle point must be considered in a complete description of the ionisation process. For a state $|n\,n_1n_2m\rangle$ the ionisation rate in an electric field $F$ is~\cite{damburg83a}
\begin{eqnarray}
\Gamma_{n\,n_1\,n_2\,m} &=& \frac{E_{\mathrm{h}}}{\hbar}\frac{(4C)^{2n_2+m+1}}{n^3\,n_2!\,(n_2+m)!}\nonumber\\ 
&& \hspace{-1.6cm} \times\exp\left[-\frac{2}{3}C - \frac{1}{4}n^3\,\frac{e\,a_0\,F}{E_{\mathrm{h}}}\,\left(34n_2^2+34n_2m\phantom{\frac{53}{3}}\right.\right.\nonumber\\
&& \hspace{+1cm}\left.\left. +46n_2+7m^2+23m+\frac{53}{3} \right) \right]
\label{eq:ionrate}
\end{eqnarray}
where
\begin{eqnarray}
C &=& \frac{1}{e\,a_0\sqrt{E_{\mathrm{h}}}}\frac{(-2E_{n\,n_1\,n_2\,m})^{3/2}}{F}
\end{eqnarray}
and $E_{n\,n_1\,n_2\,m}$ is the energy of the state, with respect to the field-free ionisation limit, in the presence of the electric field. With this in mind, the classical ionisation field, Eq.~(\ref{eq:calssion}), corresponds to the field in which the ionisation rate of the $k=-(n-1)$ state is $\sim10^8$~s$^{-1}$. Typically electric field switching times on the order of 10~ns are achieved in experiments in which pulsed electric field ionisation is employed for the detection of Rydberg atoms or molecules. To ensure complete ionisation, fields that lead to ionisation rates on the order of $10^8$~s$^{-1}$ are therefore required. Because the Rydberg electron has a non-zero probability of being located on the side of the ion core opposite to the Stark saddle point for states with values of $k>-(n-1)$, higher fields are required to achieve equivalent ionisation rates for these states. The result of this is that the fields for which similar ionisation rates occur for the $k=+(n-1)$ states is approximately $2F_{\mathrm{class}}$. Calculating the ionisation rate using Eq.~(\ref{eq:ionrate}), and accounting for energy shifts up to fourth-order in $F$, leads to ionisation rates of $10^8$~s$^{-1}$ for the $n=30$, $k=-29$ and $k=+29$ states of 740~V/cm and 1750~V/cm, respectively. 

\begin{figure}
\begin{center}
\includegraphics[width=0.95\textwidth]{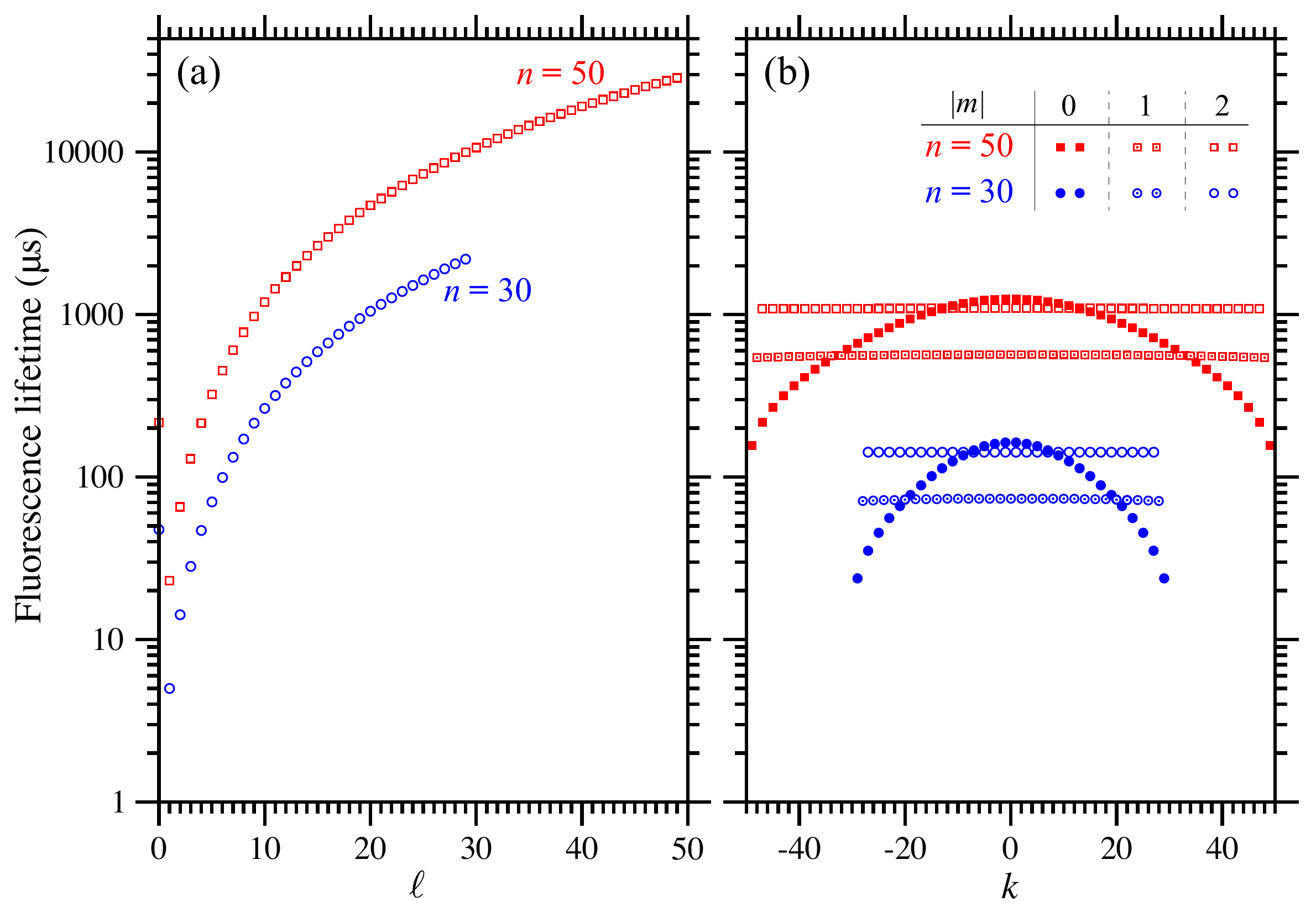}
\caption{\csentence{Fluorescence lifetimes of Rydberg states of the H atom}. (a) Fluorescence lifetimes of field-free states of the H atom with $n=30$ and $50$, for each allowed value of $\ell$. (b) Fluorescence lifetimes of $|m|=0, 1$ and $2$ Rydberg-Stark states with $n=30$ and $50$, and each allowed value of~$k$.}
\label{fig:Hlifetimes}
\end{center}
\end{figure}

\subsection*{Fluorescence lifetimes}

While an atom or molecule in a highly excited Rydberg state is energetically far from equilibrium, Rydberg-Stark states with sufficiently long fluorescence lifetimes to permit deceleration to zero mean velocity in the laboratory-fixed frame-of-reference and electric trapping can generally be photoexcited from a ground state, or low-lying intermediate state. The fluorescence rate, $\Gamma_{n\ell}$, of an excited Rydberg state, $|n\,\ell\rangle$, is given by the sum of the Einstein $A$ coefficients associated with all allowed decay pathways to energetically lower-lying states, $|n'\,\ell'\rangle$,~\cite{bethe57a}, i.e.,
\begin{eqnarray}
\Gamma_{n\ell} &=& \sum_{n'\ell'} A_{n'\ell',n\ell},
\end{eqnarray}
where~\cite{gallagher94a,foot05a}
\begin{eqnarray}
A_{n'\ell',n\ell} &=& \frac{\phantom{}2e^2\omega_{n'\ell',n\ell}^3}{3\epsilon_0\,h c^3}\frac{\ell_{\mathrm{max}}}{2\ell+1} |\langle n'\ell'|r|n\ell\rangle|^2,\label{eq:Acoeff}
\end{eqnarray}
with $\omega_{n'\ell',n\ell}=2\pi\nu_{n'\ell',n\ell}$ the angular frequency corresponding to the energy difference between the states, $\epsilon_0$ the vacuum permittivity, and $\ell_{\mathrm{max}} = \mathrm{max}(\ell, \ell')$. The fluorescence lifetime, $\tau_{n\ell}$, of the excited state is then $\tau_{n\ell}=\left(\Gamma_{n\ell}\right)^{-1}$. Because of the dependence of the Einstein $A$ coefficient on the cube of the transition frequency, the lifetimes of $n$p Rydberg states are typically dominated by decay to the ground state. For $|n\ell\rangle$ Rydberg states with $\ell\neq1$, decay via a single-photon electric-dipole transition to the 1s cannot occur and, as can be seen in Fig.~\ref{fig:Hlifetimes}(a), longer lifetimes result. 

The mixed-$\ell$ character of Rydberg-Stark states results in rates of fluorescence which lie between the fluorescence rates of the short-lived low-$\ell$ states and the longest-lived `circular' $\ell=n-1$ states. These fluorescence rates also have a strong dependence on the value of $|m|$ since Stark states with higher values of $|m|$ do not exhibit short-lived low-$\ell$ character because of the requirement that $\ell\geq|m|$. The fluorescence lifetime of each $|nn_1n_2m\rangle$ Rydberg-Stark state is determined by summing over the decay rates, $\Gamma_{n\ell}$, associated with all allowed decay pathways from the $|n\ell\rangle$ states into which it can be transformed weighted by the transformation coefficients~\cite{hiskes64a}. These transformation coefficients can be expressed in terms of Wigner-3J symbols as~\cite{gallagher94a},
\begin{eqnarray}
\langle nn_1n_2m|n\ell m\rangle &=& (-1)^{[(1-n+m+n_1-n_2)/2] + \ell}\sqrt{2\ell+1}\nonumber\\
						& &\hspace*{1.5cm} \times \left(\begin{array}{ccc}
						\frac{n-1}{2} & \frac{n-1}{2} & \ell\\
						\frac{m+n_1-n_2}{2} & \frac{m-n_1+n_2}{2} & -m
						\end{array}\right),
\end{eqnarray}
such that
\begin{eqnarray}
|nn_1n_2m\rangle &=& \sum_{\ell} |n\ell m\rangle\langle n\ell m|nn_1n_2m\rangle.
\end{eqnarray}
Consequently the fluorescence rate, $\Gamma_{nn_1n_2m}$, of each individual Rydberg-Stark state is
\begin{eqnarray}
\Gamma_{nn_1n_2m} = \sum_{n\ell} |\langle nn_1n_2m|n\ell m\rangle|^2\,\Gamma_{n\ell}
\end{eqnarray}
and the fluorescence lifetime, $\tau_{nn_1n_2m}=\left(\Gamma_{nn_1n_2m}\right)^{-1}$. The fluorescence lifetimes of all $n=30$ and $n=50$ Rydberg-Stark states of the hydrogen atom for which $|m|=0,1$ and $2$ are displayed in Fig.~\ref{fig:Hlifetimes}(b). From the data in this figure it can be seen that the fluorescence lifetimes of the $m=0$ Stark states exhibit a significant dependence on the value of $k=n_1-n_2$, while this is not the case for states with higher values of $|m|$. The fluorescence lifetimes of $|m|=2$ Stark states with values of $n>30$, which are typically prepared experimentally, exceed 140~$\mu$s.

\subsection*{The Stark effect in non-hydrogenic atoms and molecules}

\begin{figure}
\begin{center}
\includegraphics[width=0.95\textwidth]{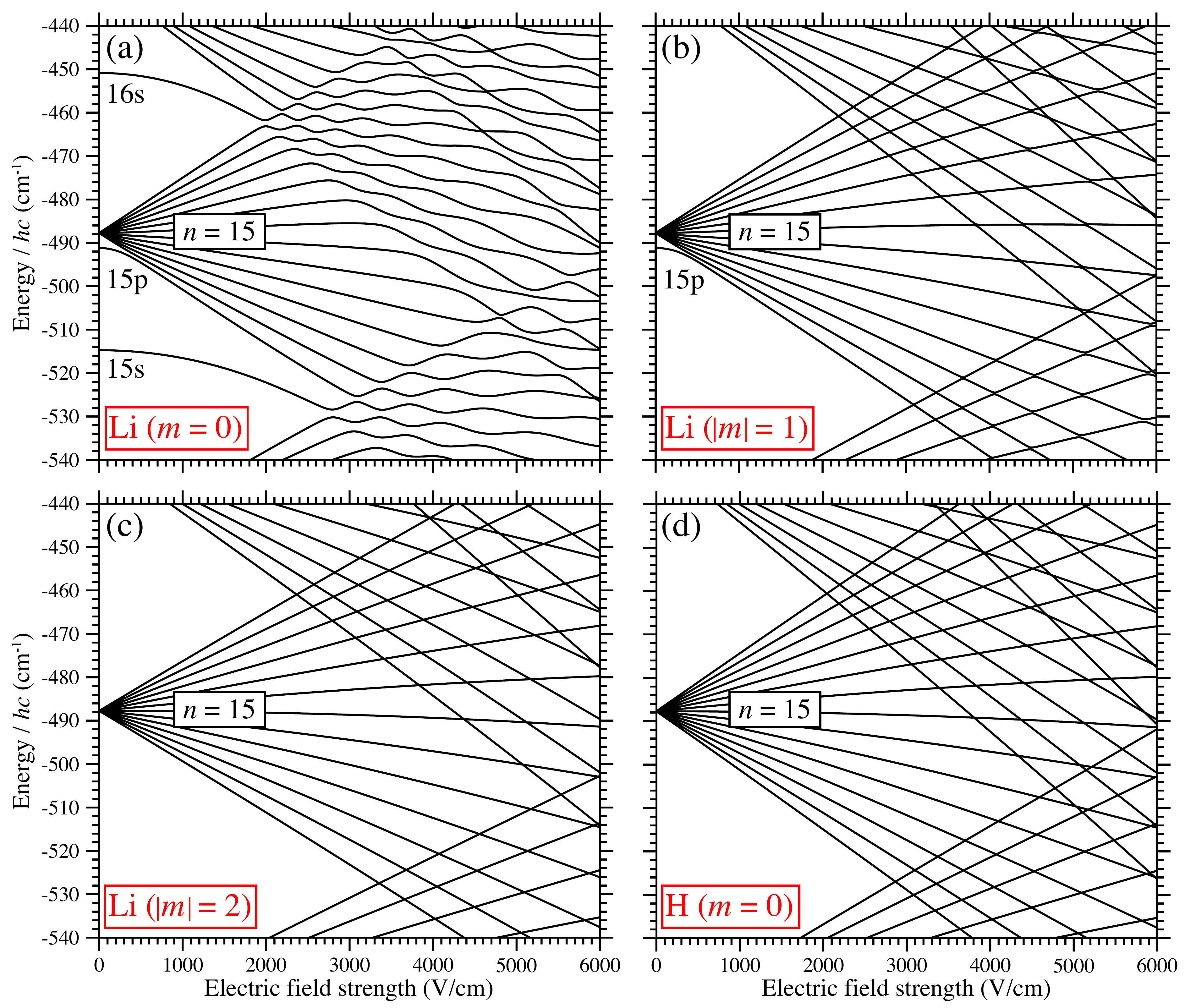}
\caption{\csentence{The Stark effect in non-hydrogenic atoms.} Energy level diagrams depicting the Stark effect in $n=15$ Rydberg states of Li with (a) $m=0$, (b) $|m|=1$, and (c) $|m|=2$, and (d) $n=15$ Rydberg states of the H atom with $m=0$.}
\label{fig:LiStark}
\end{center}
\end{figure}

In non-hydrogenic atoms and in molecules, the non-spherical symmetry of the ion core causes core-penetrating low-$\ell$ Rydberg states to be more strongly bound than the high-$\ell$ `hydrogenic' states. This effect is accounted for in the Rydberg formula in Eq.~(\ref{eq:Ry}) by the introduction of non-zero quantum defects, $\delta_{\ell}$, for these low-$\ell$ states. In general the values of the quantum defects are most significant for states with $\ell\leq4$, while for higher-$\ell$ states $\delta_{\ell}\simeq0$. In the presence of electric fields, non-hydrogenic low-$\ell$ Rydberg states exhibit quadratic Stark shifts in weak fields and give rise to large avoided crossings in Stark maps in higher fields. These effects can be seen clearly for the case of Li in Fig.~\ref{fig:LiStark}(a). The quantum defects of the s-, p- and d-states close to $n=15$ used in these calculations were $\delta_{\mathrm{s}}=0.399$, $\delta_{\mathrm{p}}=0.053$ and $\delta_{\mathrm{d}}=0.002$~\cite{zimmerman79a}. Because of these non-zero quantum defects, the 15s and 15p states are shifted to lower energies than the higher-$\ell$ states. These states then exhibit quadratic Stark shifts in fields below $\approx1000$~V/cm and give rise to the large avoided crossings in fields beyond the Inglis-Teller limit, $F_{\mathrm{IT}}=F_0/(3n^5)$, where states with values of $n$ which differ by $1$ first overlap (e.g., for $n=15$, $F_{\mathrm{IT}}\approx2000$~V/cm).

The calculation of the energy-level structure of non-hydrogenic Rydberg states of atoms and molecules in electric fields can be achieved by constructing the Hamiltonian matrix in a spherical $|n\ell m\rangle$ basis and determining its eigenvalues. This approach has been employed previously to calculate the Stark effect in Rydberg states of alkali metal atoms~\cite{zimmerman79a}. In addition to being suited to treating non-hydrogenic atomic systems, this method has also been extended to Rydberg states of molecules~\cite{fielding91a,qin93a,vrakking96a}. The coefficients of the eigenvectors of the Hamiltonian matrix can then be employed to calculate the spectral intensities of transitions to, and the fluorescence lifetimes of, the resulting eigenstates.

The zero-field matrix is, to a good approximation, diagonal in the $|n\ell m\rangle$ basis and the matrix elements can be calculated using Eq.~(\ref{eq:Ry}). The term in the Hamiltonian in Eq.~(\ref{eq:HStark}) representing the effect of the electric field is
\begin{eqnarray}
H_{\rm S} &=& e\,Fz\nonumber\\
		&=& e\,Fr\cos\theta,\label{eq:HStarkSph}
\end{eqnarray}
in spherical coordinates ($r,\theta,\phi$). This gives rise to matrix elements of the form~\cite{zimmerman79a}
\begin{eqnarray}
\langle n'\,\ell'\,m'|e\,Fr\cos\theta|n\,\ell\,m\rangle &=& e\,F\langle \ell'\,m'|\cos\theta|\ell\,m\rangle\langle n'\,\ell'|r|n\ell\rangle.\label{eq:matStark}
\end{eqnarray}

\noindent Because
\begin{eqnarray}
\cos\theta = \sqrt{\frac{4\pi}{3}}\,Y_{1\,0},
\end{eqnarray}
\noindent the angular components of Eq.~(\ref{eq:matStark}) can be determined by expansion in terms of spherical harmonics, $Y_{\ell\,m}$. Therefore the angular matrix elements are zero unless $m' = m$ and $\ell' = \ell\pm1$ (i.e., $\Delta m = 0$ and $\Delta \ell = \pm1$). Thus the external electric field mixes states with orbital angular momentum differing by one but does not give rise to $m$-mixing. Exploiting the properties of spherical harmonics~\cite{zare88a} permits these non-zero angular integrals to be expressed analytically as~\cite{bethe57a,zimmerman79a}
\begin{eqnarray}
\langle \ell+1\,m|\cos\theta|\ell\,m\rangle &=& \sqrt{\frac{(\ell+1)^2 - m^2}{(2\ell+3)(2\ell+1)}}\label{eq:angint1}\\\nonumber\\
\langle \ell-1\,m|\cos\theta|\ell\,m\rangle &=& \sqrt{\frac{\ell^2 - m^2}{(2\ell+1)(2\ell-1)}}.\label{eq:angint2}
\end{eqnarray}
The radial matrix elements $\langle n'\ell'|r|n\ell\rangle$ can be calculated analytically for the H atom~\cite{bethe57a}. For non-hydrogenic species, they can be determined numerically using the Numerov method~\cite{gallagher94a,zimmerman79a}.

The precision of the Stark energy level structure calculated in this way depends on the accuracy of the quantum defects used and on the range of values of $n$ included in the basis. Tests of convergence must be performed for the particular values of $n$, and the range of field strengths of interest. In weak fields, for which $F\ll F_{\mathrm{IT}}$, contributions from matrix elements coupling states with different values of $n$ are small and basis sets with only a small number of states are therefore often acceptable. However, in fields closer to, and beyond, the Inglis-Teller limit $n$-mixing plays a much more significant role with the result that larger basis sets are necessary. 

\begin{figure}
\begin{center}
\includegraphics[width=0.98\textwidth]{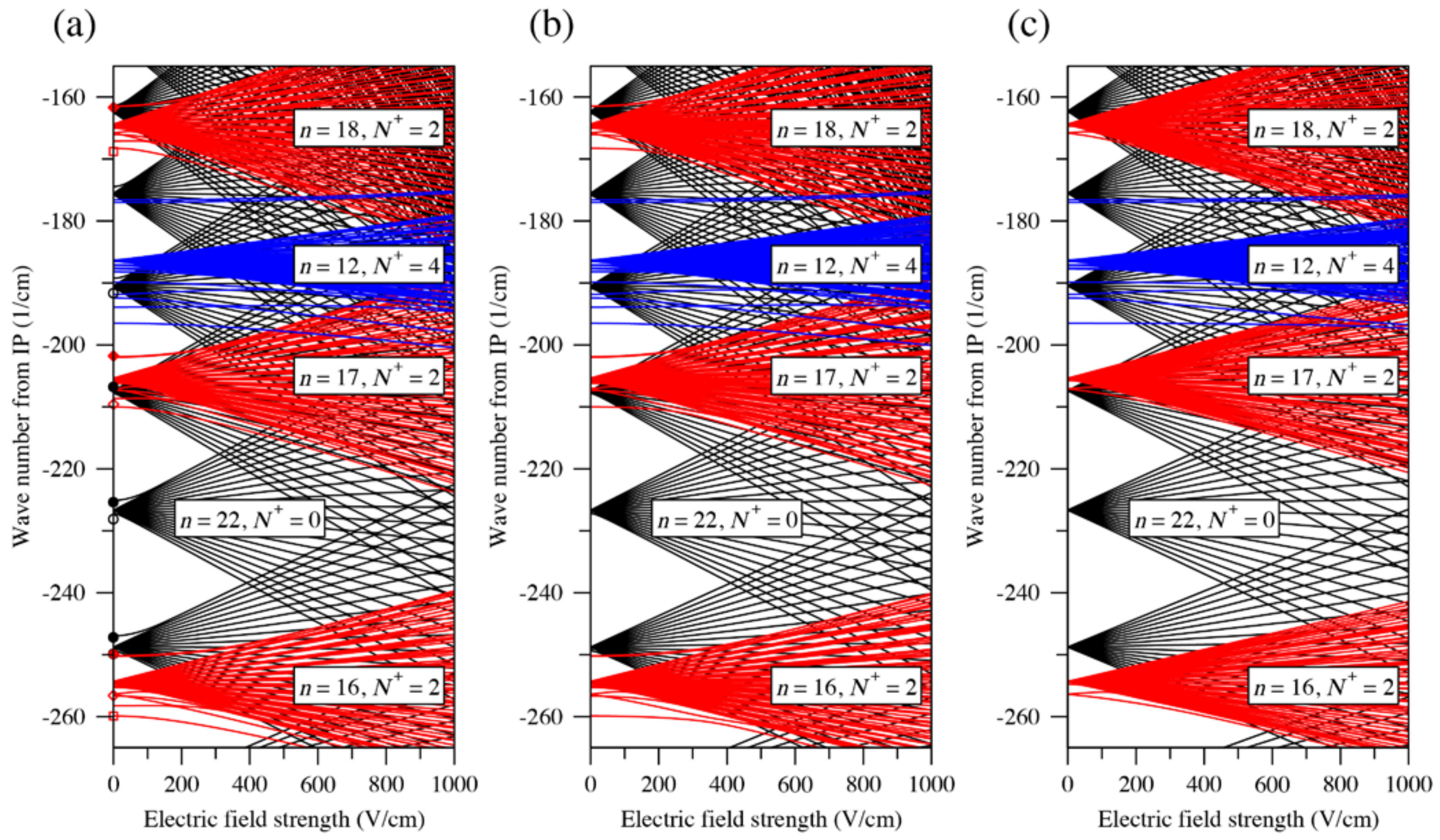}
\caption{\csentence{The Stark effect in Rydberg states of H$_2$.} Calculated Stark maps for para H$_2$. $N^+=0$ states are indicated in black, $N^+=2$ states are indicated red, and $N^+=4$ states are indicated in blue. (a)~$\left|M_J\right|=0$, (b)~$\left|M_J\right|=1$, and (c)~$\left|M_J\right|=3$. From Ref.~\cite{seiler11b}.}
\label{fig:H2starkmaps}
\end{center}
\end{figure}

Similar approaches can be employed in the calculation of the Stark effect in molecular Rydberg states. To account for the vibrational and rotational degrees of freedom of the molecular ion core the basis used must be extended. For example, in the calculation of the Stark effect in Rydberg states of H$_2$, a Hund's-case-(d) zero-field basis $\left|n\ell N^+NM_N\right>$, where $N^+$ is the rotational angular momentum quantum number of the H$_2^+$ ion core, $N$ is the total angular momentum quantum number excluding spin ($\vec{N}=\vec{N}^+ + \vec{\ell}\,$), and $M_N$ is the projection of $\vec{N}$ onto the electric field axis, is appropriate if vibrational channel interactions do not significantly perturb the spectra in the regions of interest~\cite{fielding91a,seiler11b,yamakita04a}. 

Using this approach, and assuming that singlet-triplet mixing can be neglected, the diagonal elements of the electronic Hamiltonian matrix are first determined in a Hund's-case-(b) basis $\left|n\ell \Lambda N\right>$, where $\Lambda$ is the quantum number associated with the projection of the electronic orbital angular momentum vector onto the internuclear axis. This facilitates the inclusion of the appropriate non-zero quantum defects, $\delta_{\ell\,\Lambda}$, for the low-$\ell$ Rydberg series converging to each rotational state of the ion core. In this basis, rotational channel interactions between series with a particular value of $\ell$ (i.e., for which $\Delta\ell=0$, while $\ell\leq3$) can then be included as off-diagonal matrix elements before transforming to the Hund's-case-(d) basis to include the effects of the electric field. In this, Hund's-case-(d) basis the energies of the rotational states of the ion core are then included as diagonal elements. 

Applying the Stark Hamiltonian operator, Eq.~(\ref{eq:HStarkSph}), in the Hund's-case-(d) basis leads to~\cite{fielding91a}
\begin{eqnarray}
\hspace*{-0.7cm}\langle n'\ell' N'N^{+\prime}M_N'\left|eF z\right|n\ell NN^+M_N\rangle
&=& eF\,\left(-1\right)^{N-M_N+N'+N^++\ell+1}\nonumber\\
&&\hspace*{-0.3cm}\times\sqrt{\left(2N+1\right)\left(2N'+1\right)}\nonumber\\
&&\hspace*{-0.3cm}\times\begin{pmatrix}
N' & 1 & N\\
-M_N' & 0 & M_N
\end{pmatrix}
\begin{Bmatrix}
\ell' & N' & N^{+'}\\
N & \ell & 1
\end{Bmatrix}\nonumber\\
&&\hspace*{-0.3cm}\times\langle n'\ell' |r|n\ell\rangle\,\delta_{N^+N^{+\prime}}\,,
\label{eq:H2Stark_hamiltonian}
\end{eqnarray}
which has non-zero values only when $\Delta\ell=\pm1$, $\Delta N=0,\pm1$ ($0 \not\leftrightarrow 0$), and $\Delta N^+=0$. 

Stark energy level diagrams, calculated following this procedure for $|M_N|\equiv|M_J|=0,1$ and~$2$ Rydberg states of para-H$_2$, from $-265$~cm$^{-1}$ to $-155$~cm$^{-1}$ with respect to the adiabatic ionisation limit are displayed in Fig.~\ref{fig:H2starkmaps}. This wavenumber range encompasses $N^+=0$ states for which $n=21\rightarrow27$, $N^+=2$ states for which $n=16\rightarrow18$, and the $N^+=4$ states for which $n=12$~\cite{seiler11b}. 

\subsection*{Lifetimes of molecular Rydberg states}

In general, the fluorescences lifetimes of Rydberg states of molecules are similar to those in atoms, i.e., the fluorescence lifetimes of the pure low-$\ell$ states exceed $1~\mu$s for values of $n>30$, while those of the shortest-lived Rydberg-Stark states are 10 -- 100 times longer. However, molecular Rydberg states can also decay by predissociation. This involves the transfer of energy from the Rydberg electron to the nuclear degrees of freedom of the molecule, typically as a result of the interaction between the potential energy surface of the Rydberg state with that of a repulsive valence state. The outcome of this process is the decay of the molecule into two or more fragments in their ground or excited states\cite{merkt97a,softley04a}. 

Most rapid predissociation occurs on timescales $\ll1~\mu$s. As a result, in general predissociative states are not well suited to Rydberg-Stark deceleration for which lifetimes $\gtrsim10~\mu$s are required. However, the molecular Rydberg states which exhibit the most rapid predissociation are those of core-penetrating low-$\ell$ character, for which the Rydberg electron has a significant charge density in the vicinity of the ion core. For Rydberg electrons in higher $\ell$ states, the centrifugal barrier ensures that they do not significantly perturb the bond in the molecular ion core and predissociation can be inhibited. Because the quantum defects of high Rydberg states depend directly on the penetration of the Rydberg electron into the ion core, molecular Rydberg states with large quantum defects tend to be significantly more susceptible to predissociation than states with quantum defects approaching zero. 

\subsection*{Hydrogenic Rydberg-Stark states of non-hydrogenic atoms and molecules}

For greatest efficiency, Rydberg-Stark deceleration should be implemented using hydrogenic Rydberg-Stark states with linear Stark energy shifts and lifetimes exceeding $\sim10~\mu$s. Ideally the only limitation imposed on the electric fields employed should be that they remain at all times below the ionisation field. In hydrogenic atoms this condition is readily met.  However, in non-hydrogenic species the avoided crossings at and beyond the Inglis-Teller limit [see Fig.~\ref{fig:LiStark}(a) and Fig.~\ref{fig:H2starkmaps}(a)] are often too large to be traversed diabatically during deceleration. The result of this is that for $F>F_{\mathrm{IT}}$ the Stark states lose their electric dipole moments and cannot be efficiently manipulated in these fields. In molecules, predissociation of low-$\ell$ Rydberg states also results in lifetimes which are insufficient for deceleration. 

However, these two challenges associated with decelerating non-hydrogenic atoms and molecules can be circumvented simultaneously by preparing states which do not have core-penetrating low-$\ell$ character. This is achieved by exploiting the selection rules for electric dipole transitions in the photoexcitation process to control the absolute value of the azimuthal quantum number, $|m|$, of the Rydberg states. By carefully controlling the value of $|m|$, $\ell$-mixing induced by the electric fields can be restricted to states for which $\ell\geq |m|$. For example, if Rydberg states with $|m|=1$ are prepared in Li, Fig.~\ref{fig:LiStark}(b), the Stark states do not possess s-character. As a result, the avoided crossings at and beyond the Inglis-Teller limit are reduced (from $\sim1.4$~cm$^{-1}\equiv42$~GHz to $\sim0.1$~cm$^{-1}\equiv3$~GHz). Preparation of states with $|m|=2$, Fig.~\ref{fig:LiStark}(c), removes the contribution from the p-states, the other states with a significant non-zero quantum defect, resulting in a Stark map which is almost identical to that of the hydrogen atom with $m=0$, Fig.~\ref{fig:LiStark}(d). The primary difference between the Stark maps in Fig.~\ref{fig:LiStark}(c) and~(d) is that in the former the outermost Stark states are absent. Photoexcitation of $|m|=2$ Rydberg-Stark states can readily be achieved in Li using a resonance-enhanced two-color two-photon excitation scheme from the 1s$^2$2s\,$^2$S$_{1/2}$ ground state driven using circularly polarised laser radiation with the same helicity for both steps of the excitation process, and propagating parallel to the electric field in the photoexcitation region. Alternatively, non-resonance enhanced single-color two-photon excitation using circularly polarised radiation could also be employed~\cite{wall14a}.

Multiphoton excitation schemes can also be implemented in molecules for the preparation of long-lived hydrogenic Rydberg states. The dependence of the Stark maps of para-H$_2$ on the value of $|M_J|$ can be seen in Fig.~\ref{fig:H2starkmaps}. In this case, if photoexcitation to non--core-penetrating Rydberg states with $\ell\geq3$ is carried out, predissociation is inhibited and the resulting long-lived Rydberg-Stark states with $N^+=0$ exhibit a hydrogenic behaviour. These states have been prepared in deceleration and trapping experiments by resonance-enhanced three-color three-photon excitation using circularly polarised laser radiation following the excitation scheme~\cite{hogan09a,seiler11b}
\begin{eqnarray}
&&\hspace*{-1.0cm}[(1\mathrm{s}\,\sigma_{\mathrm{g}})^1(v^+=0,N^+=0)]\, (n\mathrm{f})^1,(v=0, J=3,|M_J|=3)\nonumber\\
&&\hspace*{0.5cm}\stackrel{\:\,\circlearrowleft}{\longleftarrow}\,(1\mathrm{s}\,\sigma_{\mathrm{g}})^1\, (3\mathrm{d}\,\pi_{\mathrm g})^1\,\mathrm{I}^1\Pi_{\mathrm g}(v'=0,J'=2,|M_{J}'|=2)\nonumber\\
&&\hspace*{1.5cm}\stackrel{\:\,\circlearrowleft}{\longleftarrow}\,(1\mathrm{s}\,\sigma_{\mathrm{g}})^1\, (2\mathrm{p}\,\sigma_{\mathrm u})^1\,\mathrm{B}^1\Sigma^{+}_{\mathrm u}(v''=3,J''=1,|M_{J}''|=1)\nonumber\\
&&\hspace*{2.5cm}\stackrel{\:\,\circlearrowleft}{\longleftarrow}\,(1\mathrm{s}\,\sigma_{\mathrm{g}})^2\,\mathrm{X}^1\Sigma^{+}_{\mathrm g}(v'''=0,J'''=0,|M_\emph{J}'''|=0),
\label{eq:H2excite}
\end{eqnarray}
which raises, after each step, the value of $\ell$, $J$ and $|M_{J}|$ by one. This approach to the preparation of long-lived molecular Rydberg states using carefully chosen multiphoton excitation schemes is quite general and could also be applied to other molecules. Decelerated and trapped molecular samples in hydrogenic Rydberg-Stark states offer the opportunity to observe slow predissociation processes and study their sensitivity to external fields, blackbody radiation and collisions on timescales that are very difficult to achieve in traditional beam experiments. To study collisions and decay processes in states with low-$\ell$ character microwave transitions could be exploited to efficiently change the value of $|M_J|$ after deceleration and trapping.

\subsection*{Effects of blackbody radiation on high Rydberg states}

\begin{figure}[t]
\begin{center}
\includegraphics[width=0.95\textwidth]{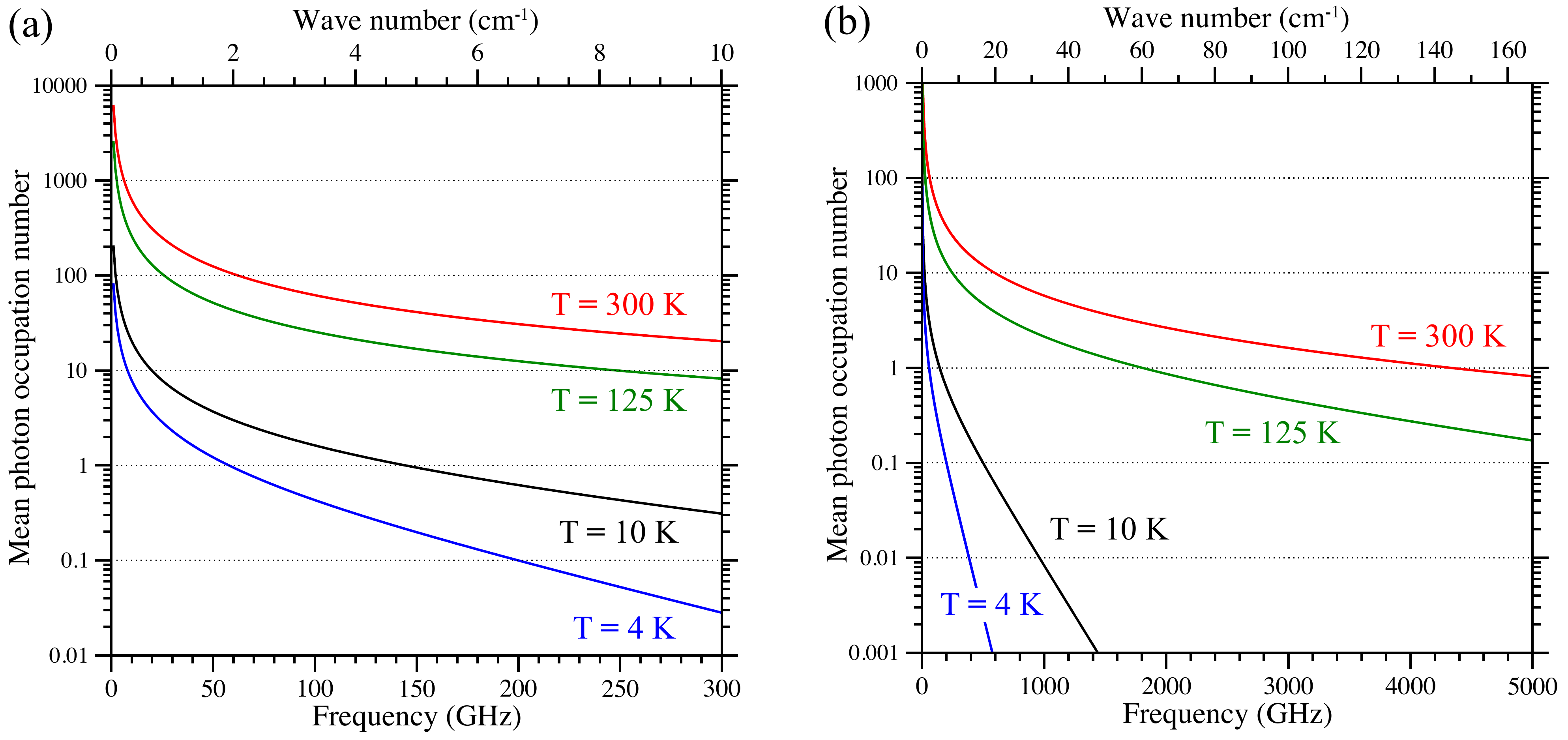}
\caption{\csentence{Thermal photon occupation numbers.} Mean blackbody photon occupation number (a) at frequencies up to 300~GHz (10~cm$^{-1}$) and (b) at at frequencies up to 5000~GHz (160~cm$^{-1}$), for blackbody temperatures of 300~K, 125~K, 10~K and 4~K.}
\label{fig:photonoccupation}
\end{center}
\end{figure}

The small $\Delta n=1$ energy intervals and large electric dipole transition moments of Rydberg states of atoms and molecules make them very sensitive to low-frequency (microwave or millimeter-wave) electromagnetic radiation. In particular, the blackbody radiation field of the environment surrounding the Rydberg atoms or molecules can have a significant effect on the Rydberg state population giving rise to energy level shifts~\cite{hollberg84a}, transitions between Rydberg states~\cite{beiting79a,gallagher79a} and photoionisation~\cite{seiler11a,spencer82a}. The most important aspect of the blackbody radiation field in the treatment of its interaction with a Rydberg atom or molecule is the mean photon occupation number per mode $\overline{n}(\nu)$. $\overline{n}(\nu)$ represents the average number of blackbody photons of one polarisation, with frequency $\nu$, at a temperature $T$, and can be expressed as~\cite{mandel79a}
\begin{eqnarray}
\overline{n}(\nu) &=& \frac{1}{\mathrm{e}^{h\nu/k_{\mathrm{B}}T}-1},
\end{eqnarray}
where $k_{\mathrm{B}}$ is the Boltzmann constant. Mean photon occupation numbers for blackbody temperatures of 300~K, 125~K, 10~K and 4~K are presented in Fig.~\ref{fig:photonoccupation}(a) and~(b) over two different frequency and wavenumber ranges.

In such a thermal radiation field, the transition rate from an initial state $|n\ell m\rangle$ to a state $|n'\ell' m'\rangle$, is given by the product of the mean photon occupation number at the frequency of the transition, and the Einstein A coefficient for the transition. For example, at $n=30$ in the H atom, the wavenumber of the transition from the 30s state to the 31p state is 7.74~cm$^{-1}$ ($\equiv232$~GHz) and the radial integral $|\langle 31\mathrm{p}|r|30\mathrm{s}\rangle|=316\,a_0$. At room temperature (300~K), the mean photon occupation number at this frequency is $\overline{n}(\nu=232~\mathrm{GHz})=26$, and therefore the blackbody transition rate is $2\,436$~s$^{-1}$. After summing the transition rates to all neighbouring Rydberg states, the total blackbody depopulation rate for H atoms in the absence of external fields at $n=30$ can be determined to be $\sim$12\,400~s$^{-1}$ which corresponds to a time constant of $\sim 80~\mu$s. 

In the presence of an electric field, the rates of $n$-changing transitions driven by blackbody radiation are highest between  states with electric dipole moments with a similar orientation with respect to the electric field, and of similar magnitude. Consequently, for an atom or molecule in an outer Rydberg-Stark state with a positive Stark shift confined in an electrostatic trap, the blackbody transitions with the highest rate will be $\Delta n=\pm1$ transitions to other trapped states which also exhibit Stark shifts. As a result, $n$-changing transitions driven by blackbody radiation do not in general lead to an immediate loss of atoms or molecules from an electrostatic trap. However, direct blackbody photoionisation of trapped atoms or molecules does lead to trap loss and plays a significant role on timescales on the order of 100~$\mu$s for states with values of $n$ close to 30 in a room temperature environment~\cite{seiler11a}.

\section*{Rydberg-Stark deceleration}

The first proposals to exploit the large electric dipole moments associated with highly excited Rydberg states for deceleration and electrostatic trapping of atoms and molecules were advanced in the early 1980s. In an article on ``\emph{Electrostatic trapping of neutral atomic particles}'' in 1980, Wing pointed out that ``\emph{At moderate field strengths Rydberg atoms have trap depths comparable to ambient $kT$ ...}''~\cite{wing80a}. While in their article on ``\emph{Stark acceleration of Rydberg atoms in inhomogeneous electric fields}'' in 1981, Breeden and Metcalf wrote that ``\emph{Rydberg atoms exhibit large electric dipole moments suggesting that inhomogeneous fields can exert forces on them}'' and that ``\emph{... the resultant change in kinetic energy is equal to the Stark shift of the Rydberg state}''~\cite{breeden81a}.

\begin{figure}
\begin{center}
\includegraphics[width=0.5\textwidth]{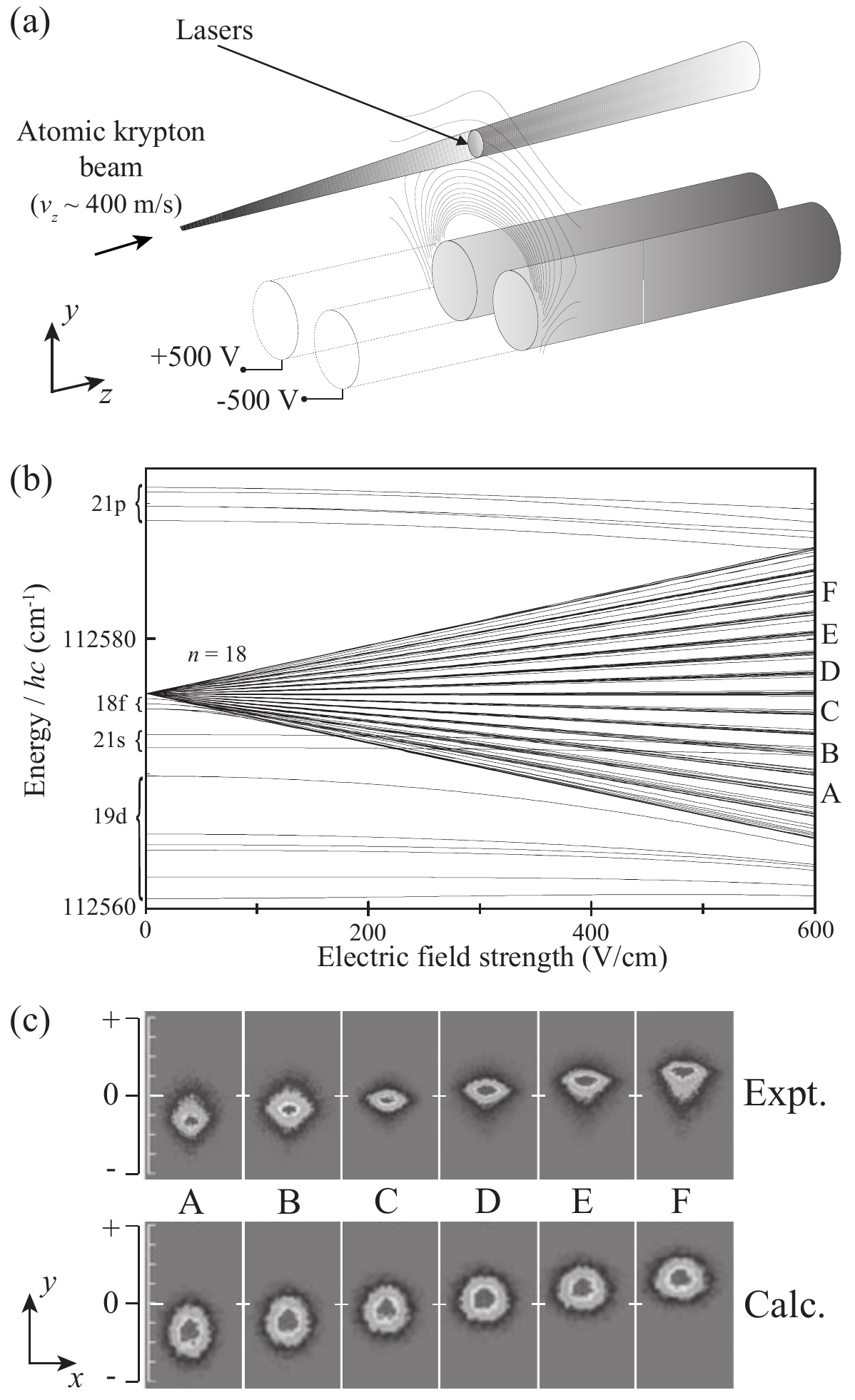}
\caption{\csentence{Transverse deflection of beams of Kr Rydberg atoms.} (a) Schematic diagram of the experimental setup, including the inhomogeneous dipolar electric field distribution above the two cylindrical metallic rods, used to transversely deflect beams of Kr atoms. (b) Stark map for Rydberg states with values of $n$ close to 18 in Kr. The Stark states labelled~A to~F on the righthand side of the figure were selectively excited and subjected to the deflection fields. (c) Experimentally recorded (Expt.) and calculated (Calc.) images of beams of atoms after deflection. From Ref.~\cite{townsend01a} with permission.}
\label{fig:krdef}
\end{center}
\end{figure}

\subsection*{Deceleration in time-independent electric fields}

First experiments in which the interactions of samples in Rydberg states with inhomogeneous electric fields were studied were reported in 2001 by Softley and co-workers~\cite{townsend01a}. This work involved the deflection of a pulsed supersonic beam of Kr atoms in the time-independent dipolar electric-field distribution surrounding a pair of cylindrical electrodes as depicted in Fig.~\ref{fig:krdef}(a). By selectively photoexciting Rydberg-Stark states, each with a different electric dipole moment [see Fig.~\ref{fig:krdef}(b)], directly above the pair of electrodes, deflection in the $y$-dimension, toward or away from the dipole was observed depending on the orientation and magnitude of the dipole moments of the states. This deflection was monitored by imaging the Rydberg atoms when they impinged upon a microchannel plate (MCP) detector~Fig.~\ref{fig:krdef}(c).

\begin{figure}
\begin{center}
\includegraphics[width=0.6\textwidth]{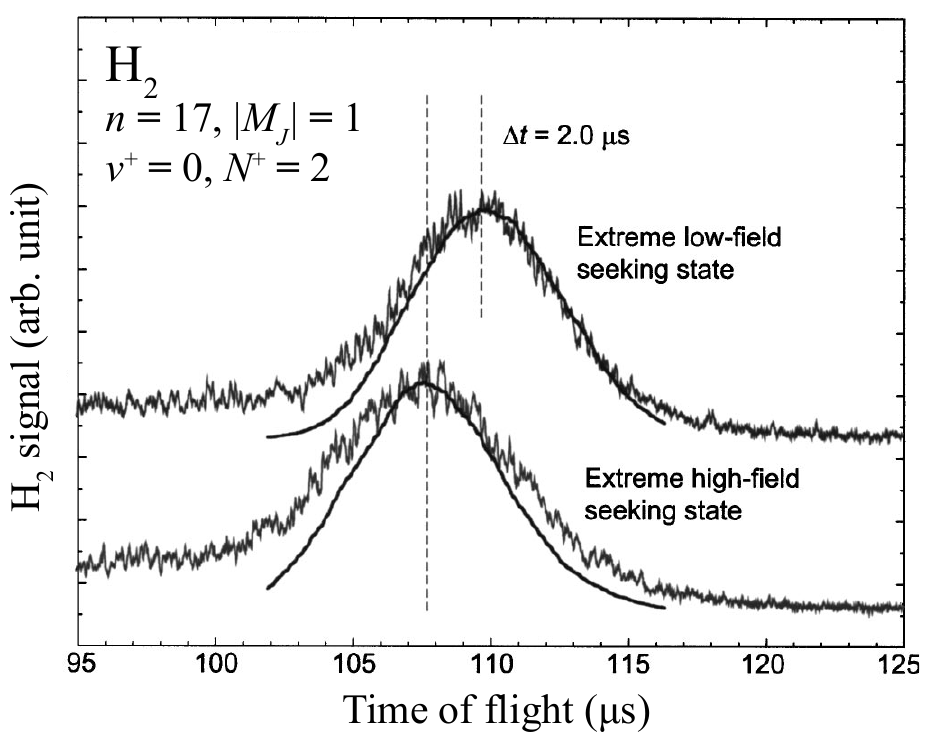}
\caption{\csentence{Acceleration and deceleration of beams of H$_2$.} Time-of-flight distributions of H$_2$ molecules in extreme outer low-field-seeking (upper trace) and high-field-seeking (lower trace) Rydberg-Stark states for which $n=17$, after exiting a time-independent inhomogeneous electric field. From Ref.~\cite{yamakita04a} with permission.}
\label{fig:yamakitaH2}
\end{center}
\end{figure}

These first experiments subsequently led to a proof-of-principle demonstration of the longitudinal acceleration of fast beams of hydrogen molecules in time-independent fields~\cite{yamakita04a,procter03a} (see Fig.~\ref{fig:yamakitaH2}). In this work, a difference in the time of flight of the molecules over a fixed distance was observed for beams of H$_2$ in the extreme low-field-seeking and extreme high-field-seeking $n=17$ Stark states of the $\nu^+=0$, $N^+=2$ Rydberg series when the molecules were decelerated or accelerated in the inhomogeneous electric field of a pair of cylindrical electrodes, respectively. 

Following this, experiments were carried out by Vliegen, Merkt and co-workers~\cite{vliegen04a} using pulsed supersonic beams of Ar atoms. These studies were first performed in the static inhomogeneous electric field distribution of a pair of electrodes in a wedge configuration, generating an electric field gradient along the axis of the atomic beam. The effects of non-hydrogenic low-$\ell$ Rydberg states on the deceleration process in electric fields at and beyond the Inglis-Teller limit were identified in this work. The observation was made that the avoided crossings in these regions of the Stark map were traversed adiabatically under the conditions of the experiments. However, for Rydberg-Stark states of the H atom the opposite behavior was seen, with energy level crossings in fields beyond the Inglis-Teller limit traversed diabatically~\cite{vliegen06b}. This confirmed that the Runge-Lenz vector remains conserved for hydrogenic systems in these deceleration experiments~\cite{englefield72a}. 

\begin{figure}
\begin{center}
\includegraphics[width=0.95\textwidth]{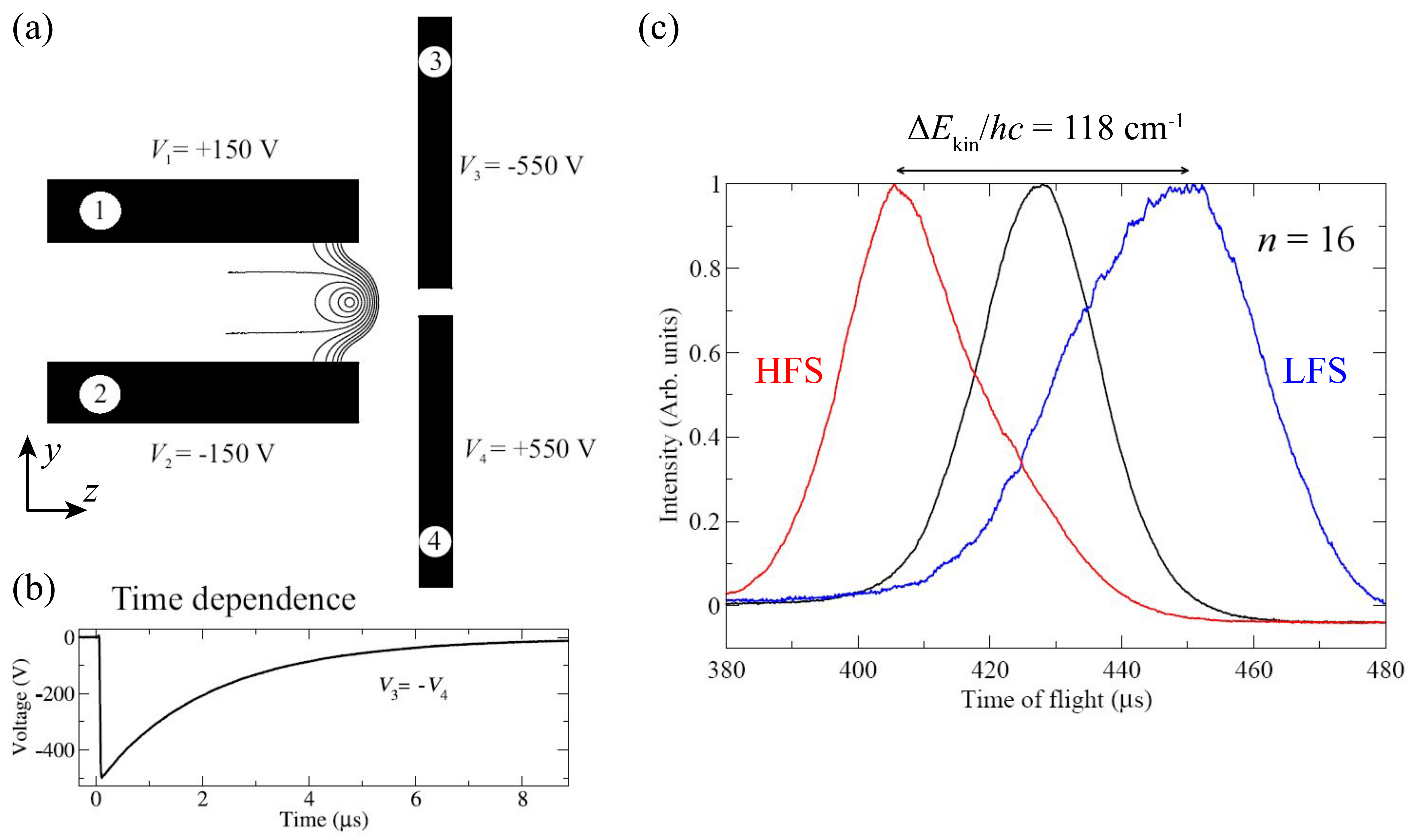}
\caption{\csentence{Acceleration and deceleration of beams of Ar atoms in time-dependent electric fields.} (a) Electrode configuration used in the acceleration/deceleration of Ar Rydberg atoms using time-dependent inhomogeneous electric fields. (b) Time dependence of the potentials applied to electrodes 3 and 4 in (a) for acceleration/deceleration. (c) Experimentally recorded time-of-flight distributions demonstrating the acceleration (left-hand red dataset), deceleration (right-hand blue dataset) of high-field-seeking (HFS) and low-field-seeking (LFS) $n=16$ Rydberg-Stark states, respectively. The central black dataset represents the time-of-flight distribution of the undecelerated Rydberg atom beam. From Ref.~\cite{vliegen05a} with permission.}
\label{fig:vliegenAr}
\end{center}
\end{figure}

\subsection*{Deceleration in time-dependent electric fields}

By introducing time-dependent electric fields the efficiency of the deceleration process could be significantly enhanced, particularly for non-hydrogenic species~\cite{vliegen05a}. Using time-dependent potentials, large, continuously moving, electric field gradients could be generated at the position of the accelerating or decelerating Rydberg atoms while ensuring that the strength of the field was maintained below the Inglis-Teller limit where non-hydrogenic contributions to the deceleration process are most significant~\cite{vliegen04a}. Applying potentials which exponentially decayed in time to a set of four electrodes in a quadrupole configuration, Fig.~\ref{fig:vliegenAr}(a-b), permitted the mean longitudinal kinetic energy of beams of Ar atoms to be modified by up to $\Delta E_{\rm kin}/hc \sim \pm 60$~cm$^{-1}$ [see Fig.~\ref{fig:vliegenAr}(c)]. This change in kinetic energy is 2.7 times the Stark energy in the maximal field experienced by the atoms during acceleration/deceleration. In addition to ensuring that non-hydrogenic samples are not subjected to fields larger than the Inglis-Teller field during deceleration, time-dependent fields can also be exploited to maximise the efficiency with which H atoms, or other atoms or molecules in hydrogenic high-$|m|$ states, can be decelerated while ensuring that they do not experience fields that could result in ionisation during deceleration. 

\begin{figure}
\begin{center}
\includegraphics[width=0.98\textwidth]{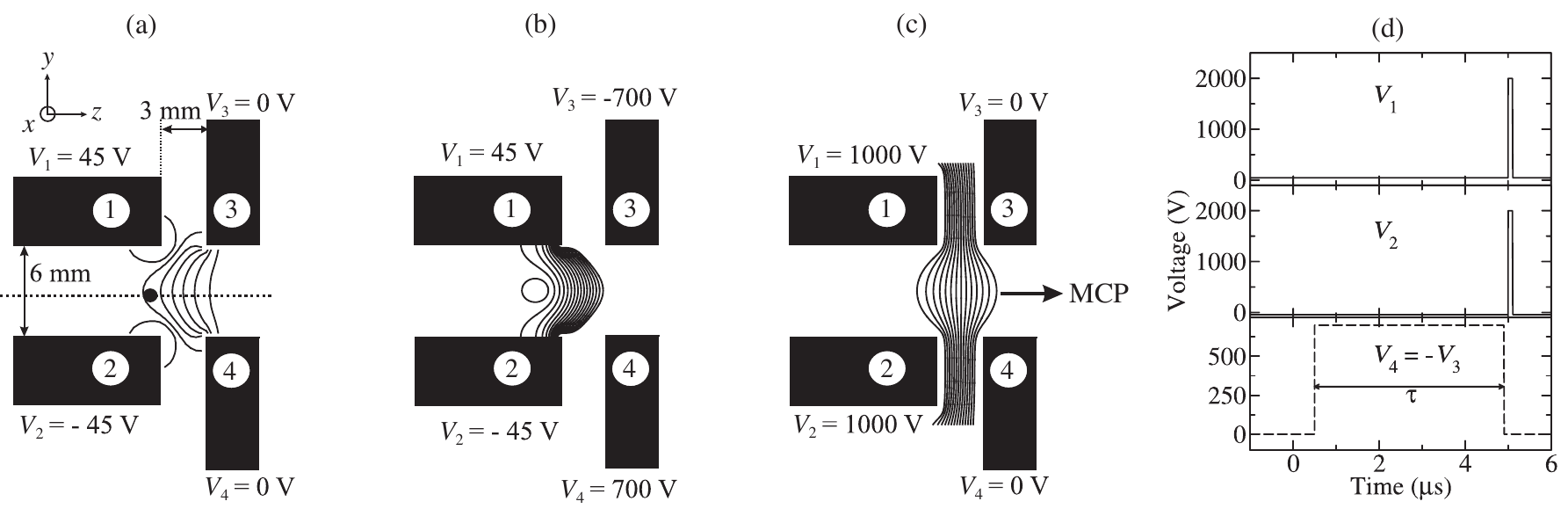}
\caption{\csentence{Electrode configuration of a Rydberg atom mirror.} (a-c) Schematic diagrams of the set of metallic electrodes used to reflect beams of H Rydberg atoms in a normal incidence mirror. The electric potentials and corresponding field distributions at the time of (a) photoexcitation, (b) deceleration/reflection, and (c) detection are displayed. (c) Time-dependence of the electric potentials applied to each mirror electrode. From Ref.~\cite{vliegen06a} with permission.}
\label{fig:mirror_electrodes}
\end{center}
\end{figure}

\subsection*{Rydberg atom mirror}\label{sec:mirror}

The use of time-dependent electric fields subsequently permitted the demonstration of transverse focussing of beams of Ar atoms~\cite{vliegen06a} and the reflection of beams of H atoms in a normal incidence Rydberg atom mirror~\cite{vliegen06c}. The arrangement of metallic electrodes used to realise this mirror are depicted schematically in Fig.~\ref{fig:mirror_electrodes}. In this figure the beam of ground state H atoms propagates in the $z$-dimension between the four mirror electrodes. At the time of photoexcitation, Fig.~\ref{fig:mirror_electrodes}(a), a sufficiently homogeneous electric field was generated at the position between electrodes~1 and~2 where the lasers used for Rydberg state photoexcitation crossed the atomic beam (shaded circle between electrodes~1 and~2) to permit selective excitation of individual $|n,k\rangle=|27,18\rangle$ Rydberg-Stark states. After photoexcitation, pulsed potentials of $\pm700$~V were rapidly applied to electrodes~3 and~4 resulting in a large positive electric field gradient at the position of the excited atoms. This gradient caused atoms in low-field-seeking Rydberg-Stark states to decelerate and was large enough that, if it persisted for a sufficient period of time, the atoms initially travelling at 720~m/s could be decelerated to a standstill and reflected into the negative $z$-dimension. In these experiments, Rydberg atoms located in the region between the four electrodes were detected by pulsed electric field ionisation. This was achieved by applying pulsed potentials of $+1000$~V to electrodes~1 and~2 simultaneously [see Fig.~\ref{fig:mirror_electrodes}(c) and (d)] generating a large field to ionise the excited atoms and accelerate the resulting ions toward a MCP detector located further downstream in the apparatus.

\begin{figure}[h!]
\begin{center}
\includegraphics[width=0.98\textwidth]{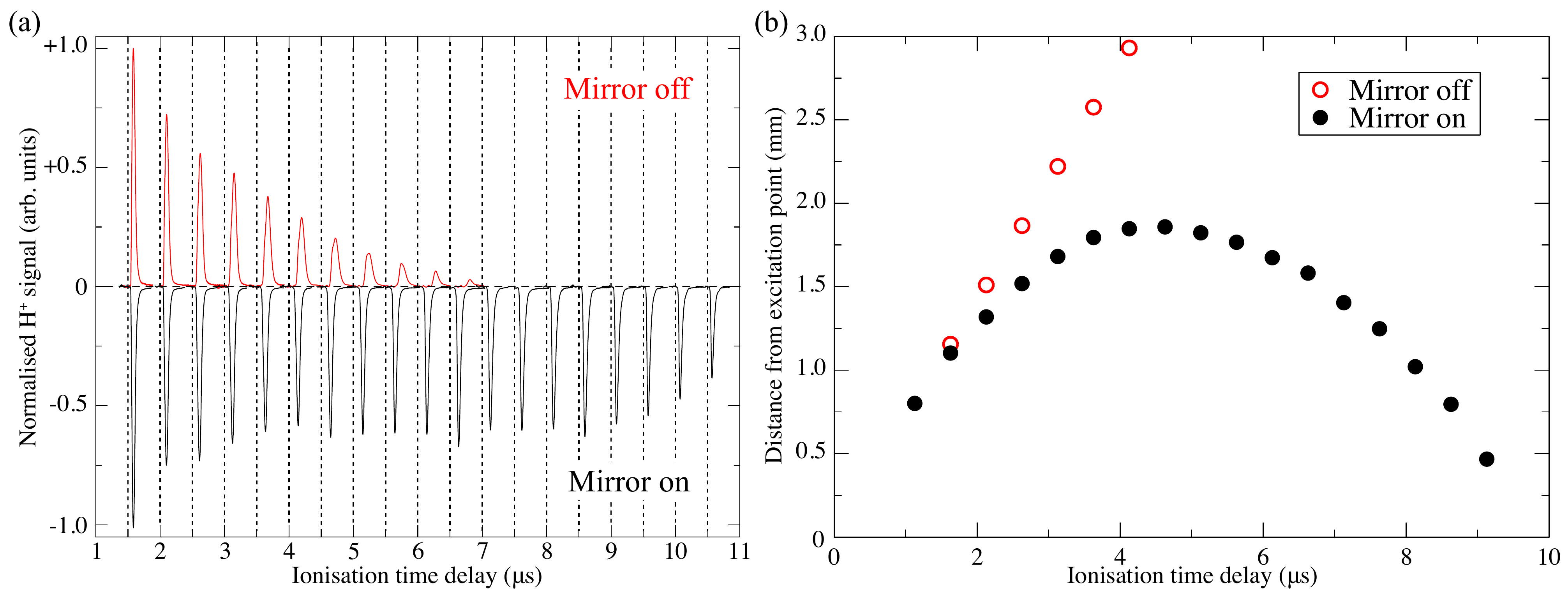}
\caption{\csentence{Measurements of reflected H Rydberg atoms.} (a) Individual H$^+$ time of flight distributions recorded after pulsed electric field ionisation of atoms located between electrodes~1 to~4 in Fig.~\ref{fig:mirror_electrodes}(a) at the times indicated by the dashed vertical lines. Positive-going datasets were recorded with the mirror off, while the negative-going, inverted datasets, were recorded with the mirror on. (b) Dependence of the mean longitudinal position of the cloud of Rydberg atoms, with respect to the position of photoexcitation, on the time delay before pulsed electric field ionisation as extracted from the data in (a) with the mirror off (open red circles), and on (filled black circles). From Ref.~\cite{vliegen06c} and Ref.~\cite{vliegen06d} with permission.}
\label{fig:mirror_data}
\end{center}
\end{figure}

The operation of this Rydberg atom mirror can be most directly seen by comparing the Rydberg atom ionisation signals with the mirror off, and with it active. When off, the Rydberg atoms fly through the region between the four electrodes where they can be ionised by the pulsed electric field, Fig.~\ref{fig:mirror_electrodes}(c), within approximately $6~\mu$s of photoexcitation [Fig.~\ref{fig:mirror_data}(a) positive-going time-of-flight distributions]. On the other hand, if the mirror potentials are activated to decelerated the atoms the H$^+$ ion signal persist for more than $10~\mu$s [Fig.~\ref{fig:mirror_data}(a) inverted negative-going time-of-flight distributions]. This indicates that the electric field gradient associated with the Rydberg atom mirror decelerates the atoms sufficiently that they remain within the detection region for this longer period of time.

However, more detailed information on the longitudinal position of the Rydberg atoms at each ionisation time can be extracted from the ion time-of-flight distributions. In the case of the measurements with the mirror off, it can be seen in Fig.~\ref{fig:mirror_data}(a) that the flight-time of the H$^+$ ions to the MCP detector (i.e., the time interval between each dashed vertical line and the subsequent maximum in the time-of-flight distribution) increases as the delay between excitation and pulsed electric field ionisation increases. This occurs because when the Rydberg atoms move further into the positive $z$ dimension within the mirror electrodes, the H$^+$ ions produced by pulsed electric field ionisation are accelerated through a smaller electric potential difference and therefore travel more slowly to the MCP. If the relation between the flight-time of the ions to the MCP and their position of ionisation is calibrated using a beam with a known longitudinal speed, the position of the atoms at the time of ionisation can be determined. With the mirror off, these positions are indicated by the open red circles in Fig.~\ref{fig:mirror_data}(b). 

With this in mind, it can be seen in the inverted negative-going dataset in Fig.~\ref{fig:mirror_data}(a) recorded with the mirror activated, that although the flight-times of the ions to the MCP gradually increase for early ionisation times, at later times they reduce again. This behaviour is indicative of the atoms first moving forward into the position $z$-dimension and then being reflected backwards. Making the appropriate conversion from the mean time-of-flight of the H$^+$ ions, to the position of ionisation, the trajectory of the ensemble of Rydberg atoms in the $z$-dimension could be reconstructed [Fig.~\ref{fig:mirror_data}(b) filled black circles]. This shows that the atoms were brought from their initial longitudinal speed of 720~m/s, to a stand still $\sim1.75$~mm from their position of photoexcitation in a time of $\sim4.5~\mu$s. In this process they experience an average acceleration of $\sim-1.5\times10^8$~m/s$^2$.

\section*{Electrostatic trapping Rydberg atoms and molecules}

Using time-dependent electric potentials permits deceleration of Rydberg atoms or molecules in a continuously moving electric field gradient. In the electrode configuration used in the experiments described above these gradients form one side of a travelling electric quadrupole trap [see e.g., Fig.~\ref{fig:vliegenAr}(a)]. Such a trap is suitable for confining atoms or molecules in low-field-seeking Rydberg-Stark states. Therefore, if sufficient kinetic energy is removed in the deceleration process and quadrupole electric field distributions are generated with minima at the positions of the decelerated samples, electrostatic trapping can be achieved using only a single deceleration stage~\cite{vliegen07a}. This is the operation principle upon which a set of on-axis and off-axis three-dimensional electrostatic traps for Rydberg atoms and molecules have been developed. 

\begin{figure}
\begin{center}
\includegraphics[width=0.65\textwidth]{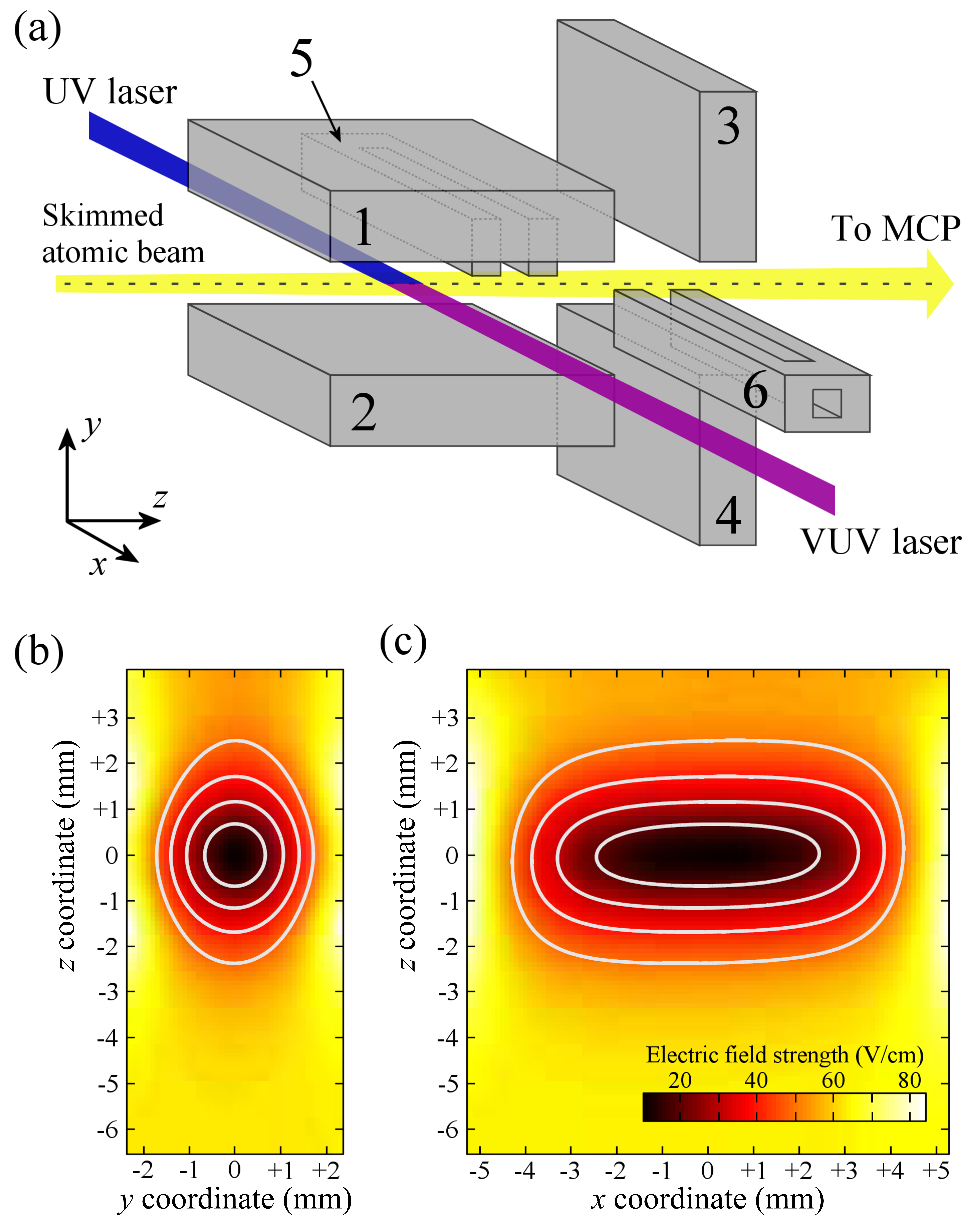}
\caption{\csentence{Three-dimensional electrostatic trap.} (a) Schematic diagram of a single-stage Rydberg-Stark decelerator and three-dimensional electrostatic trap. (b) and (c) electric field distributions in the $yz$~and $xz$~planes at the center of the trap with potentials of $|V_{1,2,3,4}|=20$~V and $|V_{5,6}|=55$~V. The contour lines are spaced by 10~V/cm with the center-most corresponding to a field of 20~V/cm. The color bar indicating the field strength in~(c) also holds for~(b). From Ref.~\cite{hogan08a}.}
\label{fig:3Dtrap}
\end{center}
\end{figure}

\subsection*{Trapping hydrogen and deuterium atoms}

\subsubsection*{On-axis trap}

The first three-dimensional electrostatic trap for atoms in selected Rydberg-Stark states was designed to act as a single-stage decelerator and a trap in which non-zero electric field minima could be generated~\cite{hogan08a}. The field gradients around this minimum gave rise to forces that confined atoms or molecules in low-field-seeking states. The electric potentials used in the first implementation of this device were optimised for H atoms in states for which $|n,k\rangle=|30,25\rangle$. These states possess electric dipole moments of $\sim2900$~D.

\begin{figure}
\begin{center}
\includegraphics[width=0.75\textwidth]{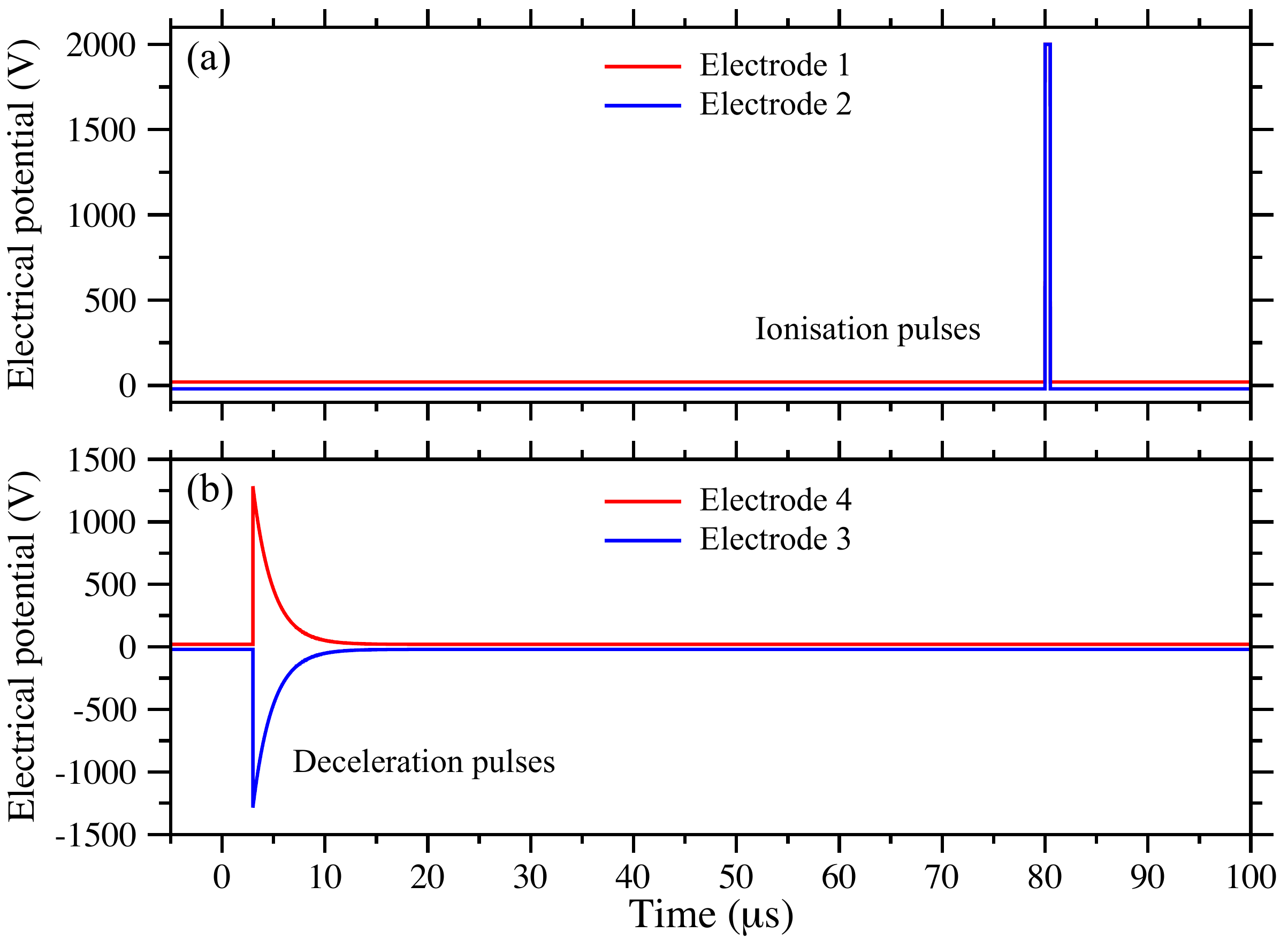}
\caption{\csentence{Time-dependent deceleration and trapping potentials.} Time dependence of the electric potentials applied to electrodes~1--4 in Fig.~\ref{fig:3Dtrap}(a). (a)~Ionisation pulses applied to electrodes~1 and~2 to detect the trapped Rydberg atoms. (b) Exponentially decaying deceleration potentials applied to electrodes~3 and~4. The horizontal axis represents the time after photoexcitation.}
\label{fig:3Dtrap_decel_potentials}
\end{center}
\end{figure}

In these experiments, pulsed supersonic beams of H atoms with a mean longitudinal velocity of 665~m/s were generated by photolysis of NH$_3$ seeded in Ar~\cite{willitsch03b}. After entering the electrode arrangement presented in Fig.~\ref{fig:3Dtrap}(a) the atoms were photoexcited to high Rydberg states using a resonance-enhanced two-colour two-photon excitation scheme via the $2\,^2$P$_{1/2}$ level. The operation of the trap required the application of potentials of $+20$~V ($-20$~V) to electrodes~1 and~4 (electrodes~2 and~3) to form a quadrupole electric field distribution in the $yz$ plane with its minimum located at the mid-point between the four electrodes as in Fig.~\ref{fig:3Dtrap}(b). To close off the trap in the $x$~dimension, and set the minimum electric field to $\sim9$~V/cm, electrodes~5 and~6 were operated at $+55$~V and $-55$~V, respectively [see Fig.~\ref{fig:3Dtrap}(c)]. This ensured that a quantisation axis was maintained throughout the trap volume and atoms would not be lost from the trap through non-adiabatic transitions to untrapped high-field-seeking states. In this configuration the trap had a depth of $E/hc=2.2$~cm$^{-1}$ (or $E/k_{{\rm B}}=3.2$~K) for atoms in $|n,k\rangle=|30,25\rangle$ states.

\begin{figure}
\begin{center}
\includegraphics[width=0.7\textwidth]{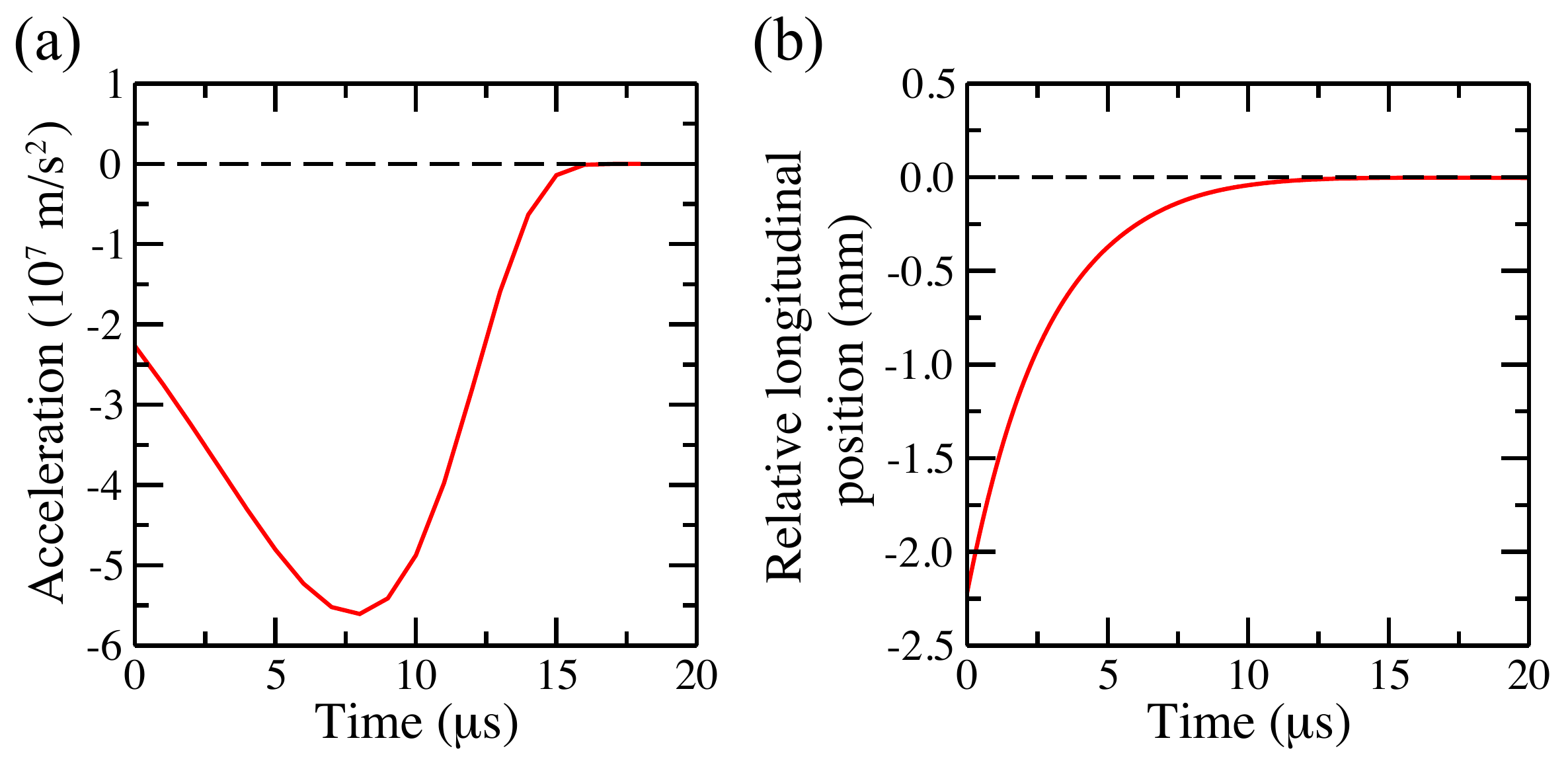}
\caption{\csentence{Acceleration and relative position of atoms during deceleration and trap loading.} (a) Acceleration, and (b) relative longitudinal position in the $z$ dimension with respect to the final position of the trap minimum, of H atoms in the $|n,k\rangle=|30,25\rangle$ state for which the time-dependence of the deceleration potentials was optimised. The origin of the horizontal axis is the activation time of the deceleration potentials.}
\label{fig:3Dtrap_decel_position}
\end{center}
\end{figure}

To decelerate the Rydberg atoms and load them into this trap, shortly after photoexcitation pulsed potentials of $\pm1265$~V were applied to electrodes~3 and~4 (see Fig.~\ref{fig:3Dtrap_decel_potentials}). This gave rise to a large positive electric-field gradient along the $z$ axis in which atoms in low-field-seeking states were decelerated. After being rapidly switched on, these potentials decayed exponentially with a time constant of 1.9~$\mu$s. This exponential decay was optimised so that the decelerating atoms were always subjected to the maximum electric-field gradient that could be generated in the decelerator while never experiencing electric fields large enough to ionise them. The time dependence of the acceleration experienced by the atoms during the deceleration process was determined in numerical calculations of particle trajectories in the decelerator and are presented in Fig.~\ref{fig:3Dtrap_decel_position}(a). The origin of the horizontal axis in this figure represents the time at which the deceleration potentials applied to electrodes~3 and~4 were switched on. The largest acceleration experienced by the $|n,k\rangle=|30,25\rangle$ H Rydberg atoms was $-5.5\times10^7$~m/s$^2$ and they were decelerated to zero velocity within $\sim10~\mu$s of the initial rise of the deceleration potentials, after travelling $\sim2.0$~mm in the $z$ dimension [see Fig.\ref{fig:3Dtrap_decel_position}(b)]. At the end of the deceleration process, the potentials applied to electrodes~3 and~4 returned to their initial values of $\pm20$~V [see Fig.~\ref{fig:3Dtrap_decel_potentials}(b)]. Two unique and essential aspects of this deceleration and trapping procedure are (1) that the Rydberg-atom cloud is stopped exactly at the minimum of the trap with no transverse loss of atoms in the final stages of trap loading, and (2) that the atoms never traverse regions of zero electric field and therefore do not undergo randomisation of $k$ and $m$.

The presence of Rydberg atoms within the trap volume was detected by pulsed electric-field ionisation using potentials of $+2$~kV with rise times of 50~ns and durations of $\sim100$~ns applied to electrodes~1 and~2 [see~Fig.\ref{fig:3Dtrap_decel_potentials}(a)]. The resulting H$^{+}$ ions were then accelerated toward a MCP detector with phosphor screen positioned 20~cm from the trap minimum along the $z$~axis.

\begin{figure}
\begin{center}
\includegraphics[width=0.65\textwidth]{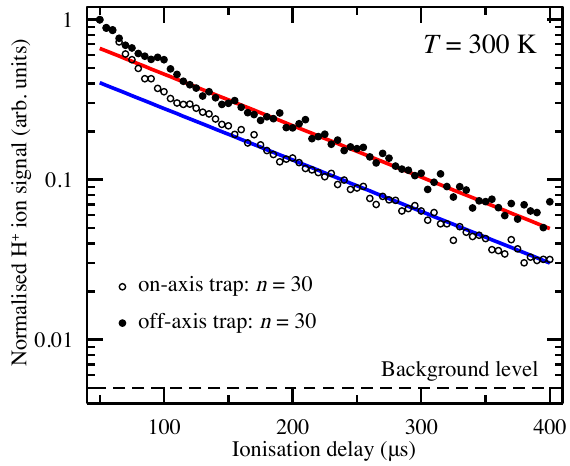}
\caption{\csentence{Decay of H Rydberg atoms from on-axis and off-axis electrostatic traps.} On-axis (open circles) and off-axis (filled circles) decay of H atoms initially prepared in $n=30$ Rydberg-Stark states from traps operated at 300~K. The solid lines are single-exponential functions fitted to the experimental data beyond $200~\mu$s. From Ref.~\cite{seiler11a}.}
\label{fig:300Kdecay}
\end{center}
\end{figure}

The integrated H$^{+}$ signal recorded as a function of the time delay between photoexcitation and field ionisation in this on-axis electrostatic trap is presented in Fig.~\ref{fig:300Kdecay} (open circles). Initially, the atom number density in the trap is $10^{6}-10^{7}$~cm$^{-3}$. Fitting a single exponential function to the data beyond 200~$\mu$s in this figure reveals two time periods in which the trap loss rates are quite different. At early times, up to $\sim150~\mu$s, a rapid loss of atoms from the trap occurs which was seen to be strongly dependent on the density of the atomic beam~\cite{vliegen07a}. This loss of atoms at early trapping times was attributed to collisions with the trailing components of the gas pulse. Beyond 200~$\mu$s the exponential function fitted to the experimental data has a time constant of 135~$\mu$s. This rate of decay of atoms from the trap results from a combination of the fluorescence lifetime of the Rydberg states prepared at photoexcitation, the rate at which transitions driven by the local room-temperature blackbody radiation field occur, and interactions between the trapped atoms. 

\begin{figure}
\begin{center}
\includegraphics[width=0.95\textwidth]{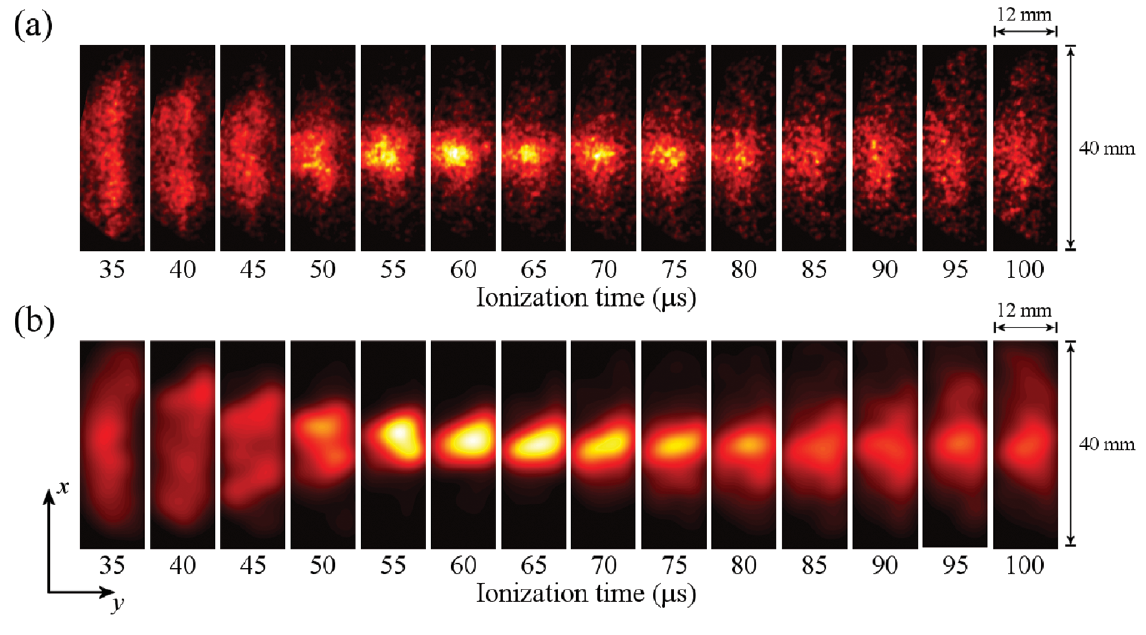}
\caption{\csentence{Imaging the transverse motion of H Rydberg atoms in an on-axis electrostatic trap.} `Breathing' motion of a trapped ensemble of H Rydberg atoms recorded by imaging the spatial distribution of H$^{+}$ ions detected at the MCP for selected times after pulsed electric field ionisation. (a) Experimentally recorded data, and (b) calculated images. From Ref.~\cite{hogan08a}.}
\label{fig:3Dtrap_images}
\end{center}
\end{figure}

In these experiments information on the transverse motion of the ensemble of trapped atoms could be obtained by H$^{+}$ ion imaging. Such data, together with a corresponding set of calculated ion images are displayed in Fig.~\ref{fig:3Dtrap_images}(a) and~(b), respectively. With potentials of $\pm55$~V applied to electrodes~5 and~6, atoms located at the edges of the trap in the $x$~dimension when the trap was initially loaded experienced strong electric-field gradients that forced them toward the centre of the trap, compressing the cloud in this dimension. Under the conditions in which the trap was operated this motion of the ensemble had a period of $\sim100~\mu$s. This can be seen in the experimental data in Fig.~\ref{fig:3Dtrap_images}(a) which were recorded at time intervals of $5~\mu$s. At early times, the cloud of trapped atoms fills the trap in the $x$-dimension. As the atoms are forced toward the centre of the trap, the spatial spread of the ions impinging on the MCP detector reduces until it reaches its minimal size $60-70~\mu$s after excitation. A similar behaviour is seen in the calculated images, validating the interpretation of the experimental data.

In detecting trapped atoms by pulsed electric field ionisation, the ion time-of-flight distributions can also be exploited to provide information on the location and spatial distribution of the Rydberg atoms at the time of ionisation~\cite{vliegen06b}. By switching off the trap and allowing the cloud of Rydberg atoms to expand before the ionising electric field was applied, the rate of expansion in the $z$~dimension could be measured directly. These measurements indicated that the radius of the Rydberg atom cloud increased at a rate of 50~m/s after the trap was switched off. This corresponds to a mean kinetic energy of $\Delta E_{\mathrm{kin}}/k_{{\rm B}}\sim150$~mK of the atoms.

\subsubsection*{Off-axis trap}

\begin{figure}
\begin{center}
\includegraphics[width=0.95\textwidth]{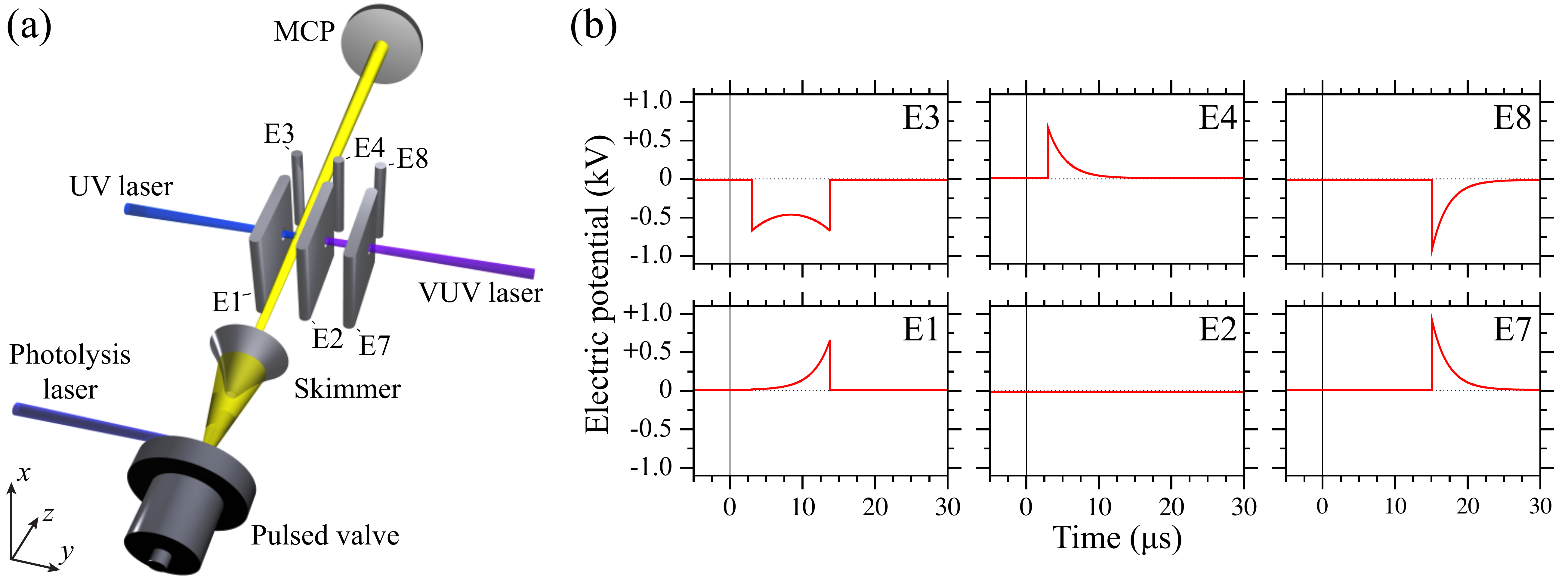}
\caption{\csentence{Off-axis electrostatic trap.} (a) Schematic diagram of the Rydberg-Stark decelerator and off-axis electrostatic trap (not to scale). In this figure, the end-cap electrodes (E5 and E6, and E9 and E10) which close off the on-axis and off-axis quadrupole traps formed between electrodes E1--E4 and E2, E4, E7 and E8 in the $y$ dimension [see Fig.~\ref{fig:3Dtrap}(a)] are omitted for clarity. (b) The sequence of electric potentials applied to the six principal electrodes of the device for deceleration and off-axis trapping. The time on the horizontal axes in (b) is displayed with respect to the time of photoexcitation. From Ref.~\cite{hogan13a}.}
\label{fig:offaxis_trap}
\end{center}
\end{figure}

To isolate and minimise contributions from collisions of trapped atoms with the trailing components of the atomic beams employed in these experiments, and distinguish collisional losses from trap decay driven by blackbody radiation, an off-axis trap which could be cooled to low-temperatures was developed~\cite{seiler11a,hogan13a}. The design of this device is presented in Fig.~\ref{fig:offaxis_trap}(a) as it was initially employed to decelerate beams of H atoms from initial longitudinal speeds of $\sim600$~m/s. In this decelerator and off-axis trap the laser beams required for Rydberg state photoexcitation entered the excitation region through a set of 1.6-mm-diameter holes in electrodes 1, 2, and 7. Following photoexcitation, a series of pulsed electric potentials were applied to decelerate, deflect and load the Rydberg atoms into the off-axis trap located 6~mm away from the initial propagation axis of the atomic beam.

\begin{figure}
\begin{center}
\includegraphics[width=0.7\textwidth]{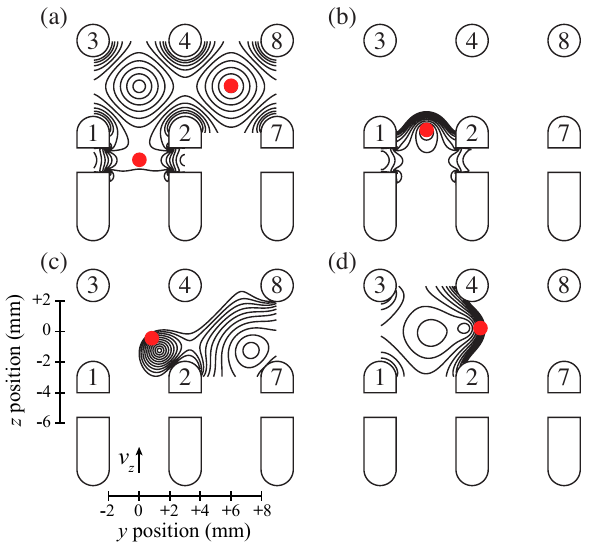}
\caption{\csentence{Electric field distributions in the off-axis decelerator and trap.} Electric field distributions in the $yz$ plane of the off-axis trap (a) at the time of photoexcitation and after completion of the trapping process, (b) in the initial phase of on-axis deceleration, (c) during the 90$^\circ$ deflection process, and (d) in the final deceleration phase. The lines of constant electric field range in (a) from 10 to 100~V/cm in steps of 10~V/cm, and (b--d) from 20 to 200~V/cm in steps of 20~V/cm. The red shaded circles indicate the center of the Rydberg atom cloud at each time. From Ref.~\cite{seiler11a}.}
\label{fig:offaxis_seq}
\end{center}
\end{figure}

The sequence of pulsed potentials used for deceleration and trap loading could be tailored to trap atoms on-axis, between electrodes 1--4, or off-axis between electrodes 2, 4, 7, and 8. The pulsed time-dependent potentials required for off-axis trapping can be seen in Fig.~\ref{fig:offaxis_trap}(b). At the times of photoexcitation and trapping [Fig.~\ref{fig:offaxis_seq}(b)], potentials of $+12$~V (-12~V) were applied to electrodes~1, 4, and~7 (2, 3, and~8) forming two electric quadrupole traps in the $yz$ plane, with minima located at the midpoint between electrodes 1--4 (on-axis trap) and~2, 4, 7, 8 (off-axis trap). End-cap electrodes [not shown in Fig.~\ref{fig:offaxis_trap}(a)], were located above (below) the on-axis and off-axis traps, separated by 10~mm in the $x$ dimension, and operated at potentials of $-22$~V ($+22$~V) to achieve confinement of the Rydberg atom cloud in the $x$ direction and generate a field of 10~V/cm at the trap minimum. This represented the electric field configuration for the first of 5 phases of trap loading. After photoexcitation [$t=0~\mu$s in Fig.~\ref{fig:offaxis_trap}(b)], initial longitudinal deceleration took place ($0<t<5~\mu$s). This was followed by guiding of the atoms off axis ($5~\mu\mathrm{s}< t < 15~\mu\mathrm{s}$), transverse deceleration to the position of the off-axis trap ($15~\mu\mathrm{s}< t < 25~\mu\mathrm{s}$), and off-axis trapping ($t\geq25~\mu$s). The electric-field distributions in the $yz$ plane at the beginning of each of these five phases are displayed in Fig.~\ref{fig:offaxis_seq}(a-d).

The presence of Rydberg atoms in this off-axis trap was measured by pulsed electric field ionisation following the application of pulsed ionisation potentials of $+2$~kV to electrodes~2 and~7 with the resulting H$^+$ ions collected on a MCP detector [see Fig.~\ref{fig:offaxis_trap}(a)]. The results of measurements of this kind for a range of time delays between photoexcitation and pulsed electric field ionisation are presented in Fig.~\ref{fig:300Kdecay} (filled black circles). The data in this figure, all recorded following deceleration and trapping of H atoms initially prepared in low-field-seeing $n=30$ Rydberg-Stark states, permits a direct comparison to be made between the decay of atoms from the on-axis trap with that from the off-axis trap. In these data sets, the decay rates from the two traps are identical at times beyond 170~$\mu$s with the difference at early times evidence for collisional losses of $\sim40\%$ from the on-axis trap induced by the trailing component of atomic beam. This rapid decay of atoms from the trap at early times is suppressed by trapping off-axis. 

Although the rates of decay of electrostatically trapped atoms at later times in Fig.~\ref{fig:300Kdecay} are on the same order of magnitude as those associated with fluorescence from the initially prepared Rydberg-Stark states, no dependence of the decay rate on the value of $n$ was observed in this work. This suggested that decay channels other than fluorescence must play a role in the experiments. Because of the sensitivity of the trapped Rydberg atoms to room temperature blackbody radiation, these additional decay channels were attributed to transitions driven by blackbody radiation. However, contributions from Rydberg-Rydberg interactions could also not be excluded at early times. 

\begin{figure}
\begin{center}
\includegraphics[width=0.98\textwidth]{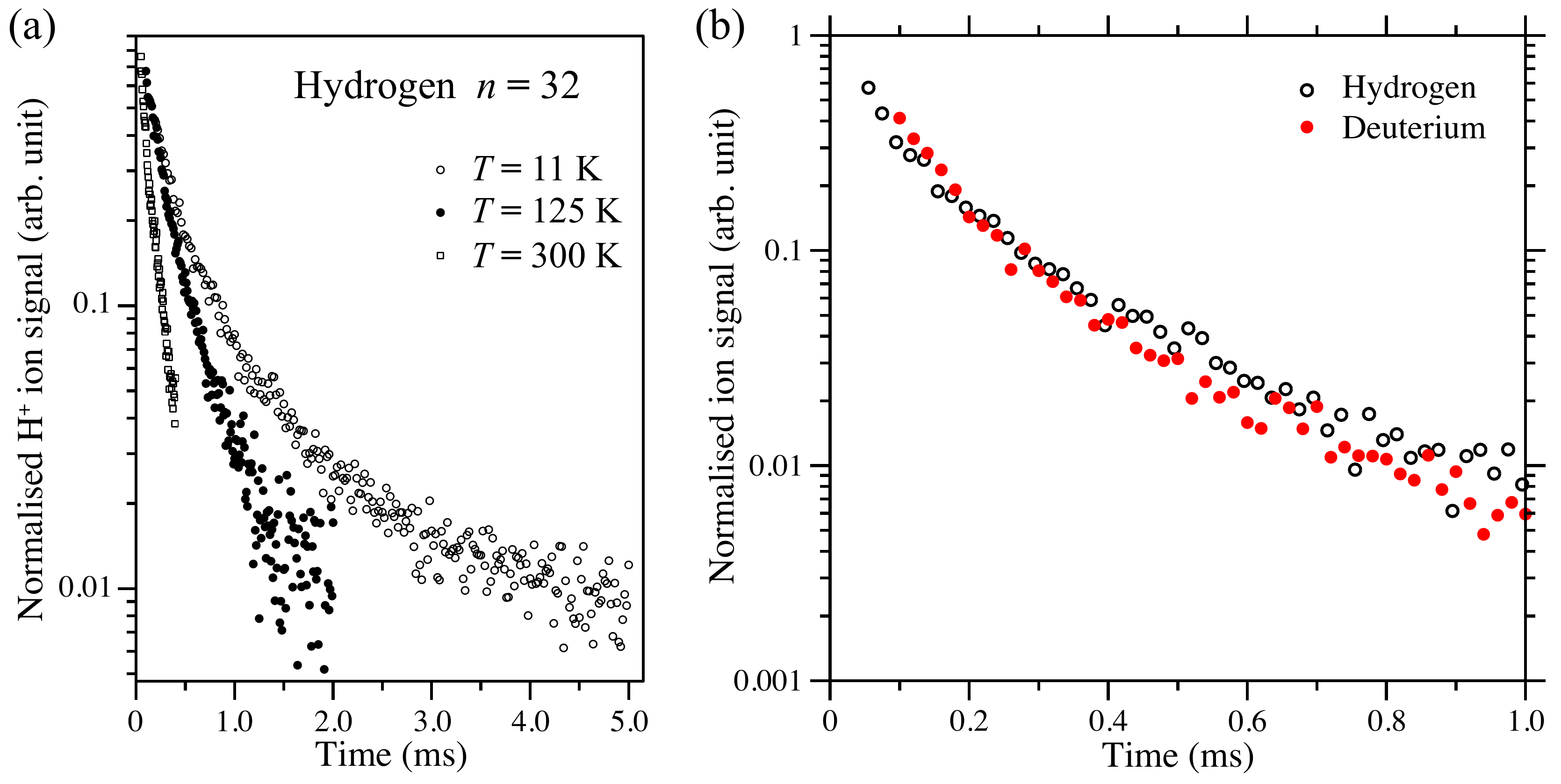}
\caption{\csentence{Blackbody temperature, and isotope dependence of trap decay.} (a) Temperature dependence of the decay of trapped H atoms initially prepared in $n=32$ Rydberg-Stark states. Experiments were performed at 300~K (open squares), 125~K (black filled circles), and 11~K (open circles). From Ref.~\cite{seiler12a}. (b) Decay of H (open circles) and D (filled circles) Rydberg atoms, initially prepared in Stark states for which $n=30$, from an off-axis electrostatic trap operated at 125~K. From Ref.~\cite{hogan13a}.}
\label{fig:trapdecay_temp}
\end{center}
\end{figure}

The effect of the blackbody temperature of the surroundings of the off-axis trap on the rate of decay of H atoms confined within it can be seen in Fig.~\ref{fig:trapdecay_temp}(a). As the blackbody temperature is reduced from room temperature, to 125~K and 11~K, the period of time for which atoms remain in the trap increases. Consequently, the trap decay rate reduces. Since the strongest transitions induced by blackbody radiation between Rydberg states in the presence of an electric field tend be between states with similar electric dipole moments, such transitions do not lead directly to a loss of atoms from the trap. These transitions lead instead to a gradual redistribution of the Rydberg state population among states with a range of values of $n$. At room temperature, this redistribution renders the measured trapping times largely independent of the value of $n$. Therefore the primary blackbody radiation contribution to the loss of trapped atoms at room temperature in Fig.~\ref{fig:trapdecay_temp}(a) is direct blackbody photoionisation. For states with values of $n$ close to~30, photoionisation accounts for approximately $8\%$ of the blackbody-radiation--induced depopulation rate in a 300~K environment~\cite{gallagher94a,spencer82a,beterov09a}. As the blackbody temperature of the environment is reduced the contributions from blackbody photoionisation also reduce. In the data recorded at 125~K and 11~K in Fig.~\ref{fig:trapdecay_temp}(a) the decrease in the decay rate observed at longer trapping times is a result of the slow redistribution of population among states with a range of values of $n$ at these temperatures. As the population is redistributed, atoms de-excited to states with lower values of $n$ decay more rapidly from the trap. Consequently, at later times only atoms in states with the highest values of $n$, which decay more slowly, remain. At 11~K, the initially prepared states are preserved for $\sim500~\mu$s after which time the measured decay rates represent the average rate of decay of the states to which the population has been redistributed following blackbody transitions. 

To confirm that the measured rates of decay of atoms from the off-axis trap are dominated by radiative processes and not affected by the dynamics of the trapped atoms, experiments were also performed with D atoms in states with the same values of $n$~\cite{hogan13a}. A direct comparison between the rate of decay of H and D atoms initially prepared in Rydberg-Stark states for which $n=30$ from an off-axis trap cooled to 125~K can be seen in Fig.~\ref{fig:trapdecay_temp}(b). This indicates that the heavier D atoms can also be efficiently trapped, and the differing dynamics of the ensembles of H and D atoms in the trap, resulting from their slightly different trajectories into the trap, do not significantly affect measurements of their decay.

\begin{figure}
\begin{center}
\includegraphics[width=0.95\textwidth]{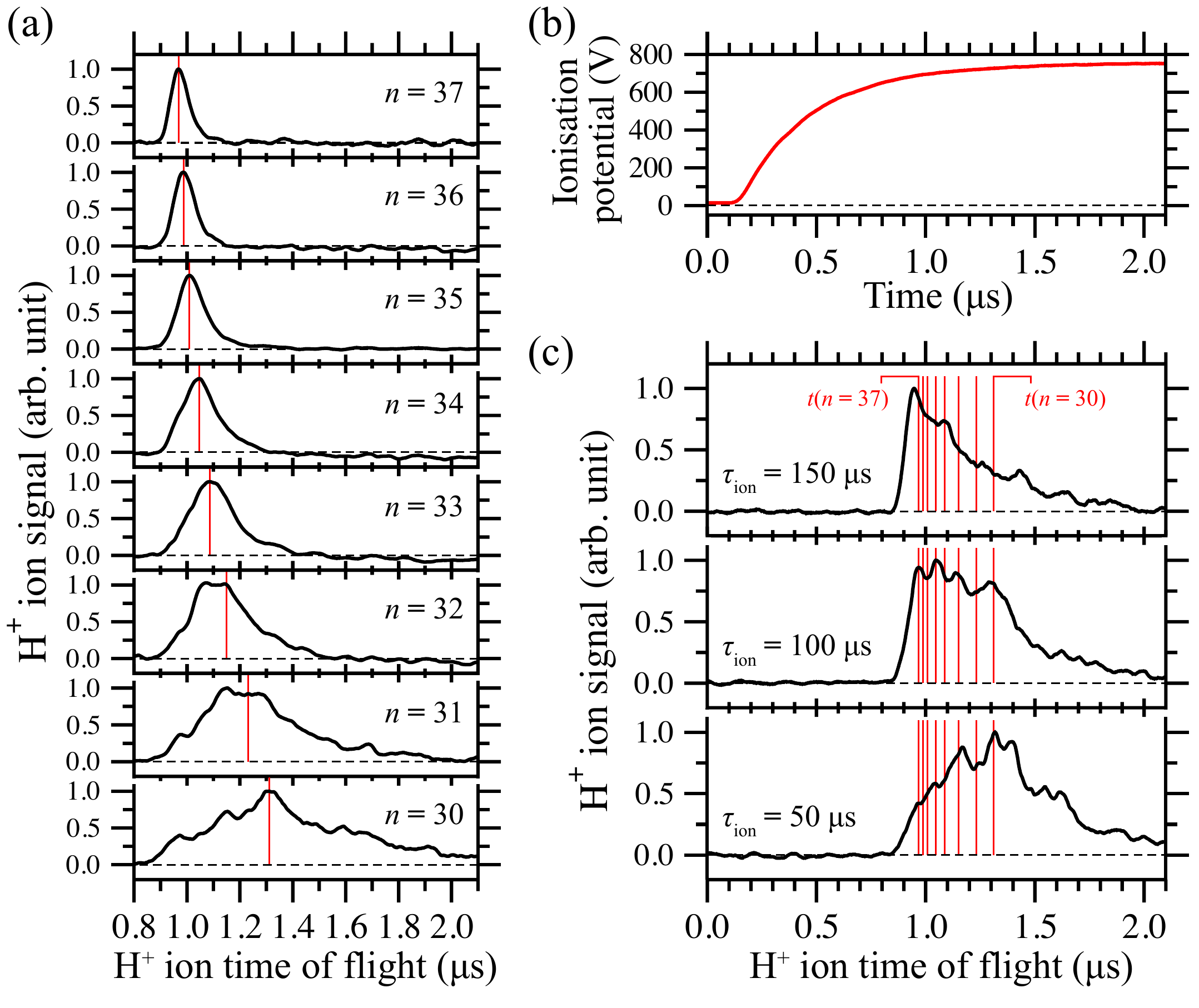}
\caption{\csentence{Evolution of Rydberg state populations in a 125~K environment.} (a) H$^+$ ion time-of-flight distributions recorded following ramped electric field ionisation $50~\mu$s after trapping H Rydberg atoms initially prepared in Stark states with values of $n$ from $30$ to $37$. (b) The time-dependent ionisation potential applied to electrodes~2 and~7 in Fig.~\ref{fig:offaxis_trap}(a) to ionise the trapped atoms. (c) Evolution of the ion time-of-flight distribution for atoms initially prepared in states for which $n=30$ as the trapping time is increased in an environment cooled to 125~K. The vertical red bars indicate the ion arrival times for states with consecutive values of $n$ as determined from (a). From Ref.~\cite{hogan12c}.}
\label{fig:ramped_ionisation}
\end{center}
\end{figure}

Evidence for the redistribution of the Rydberg state population toward higher values of $n$ at longer trapping times when $T=125$~K can be seen in measurements carried out using slowly-rising ionisation electric-field pulses to project the distribution of Rydberg states onto the H$^+$ ion time-of-flight distribution. The results of such measurements, performed with the trap electrodes and surrounding heat shields cooled to 125~K, are presented in Fig.~\ref{fig:ramped_ionisation}. The set of data in Fig.~\ref{fig:ramped_ionisation}(a) are reference measurements recorded after a trapping time of $50~\mu$s for initially excited Rydberg states with values of $n$ ranging from~$30$ to~$37$. The slowly-rising electric potential applied to electrodes~2 and~7 to ionise the atoms from the off-axis trap and extract the resulting ions toward the MCP detector is presented in Fig.~\ref{fig:ramped_ionisation}(b). As the ionisation field rises, high-$n$ states ionise first (at low fields) while lower $n$ states ionise later. The ionisation times are then mapped onto the ion time-of-flight distributions such that ions detected at earlier (later) times correspond to atoms with higher (lower) values of $n$. At 125~K blackbody-radiation--induced transitions do occur during the first $50~\mu$s trapping time. However, although this results in a broadening of the ion time-of-flight distributions, the maximum of each distribution still correspond approximately to the detection time of the initially excited Rydberg state. For increasing values of $n$ the maximum of the corresponding ion time-of-flight distribution therefore shifts to earlier times as indicated by the locations of the red vertical lines in Fig.~\ref{fig:ramped_ionisation}(a). From these measurements a H$^+$ ion flight time can be associated with atoms in each $n$ state present in the trap. This information can then be employed to investigate the time evolution of a single initially excited Rydberg state as shown for $n=30$ in Fig.~\ref{fig:ramped_ionisation}(c). In this figure, the red vertical lines denote the arrival times of the ions produced by pulsed-field-ionisation of Rydberg states from $n=30$ to $n=37$ as determined from Fig.~\ref{fig:ramped_ionisation}(a). After a trapping time of $50~\mu$s, a significant number of ions arrive at the MCP detector with short flight times indicating the presence of Rydberg states with $n>30$ in the trap. However, the larger fraction of the atoms are still in states for which $n=30$. After a trapping time of $100~\mu$s, an increase in the relative intensity of the signal corresponding to Rydberg states for which $n>30$  is observed. A further increase in this signal is seen after increasing the trapping time to $150~\mu$s. At these longer trapping times, the larger fraction of the trapped atoms are in states with $n>33$. As these higher-$n$ states have slower spontaneous emission rates than those of lower $n$, there is a larger probability that they will remain in the trap. This upward shift of the mean value of $n$ in the ensemble of trapped atoms with time gives rise to a reduction in the trap-loss rate by spontaneous emission, and the non-exponential decay of atoms from the trap seen in Fig.~\ref{fig:300Kdecay}.

\subsection*{Trapping hydrogen molecules}

The phase-space acceptance of the on-axis and off-axis traps described above makes them also well suited to deceleration and trapping molecular hydrogen. By preparing pulsed supersonic beams of H$_2$ seeded in Kr, samples with initial kinetic energies equal to those of beams of D atoms can readily be prepared. To decelerate and electrostatically trap these H$_2$ molecules hydrogenic Rydberg-Stark states with $|M_J|=3$ were prepared using the three-photon excitation scheme in Eq.~(\ref{eq:H2excite})~\cite{seiler11b,hogan09a}. This required the generation of coherent, circularly polarised radiation in the vacuum-ultraviolet (vuv) (94093~cm$^{-1}\equiv106.28$~nm), visible (18190~cm$^{-1}\equiv549.75$~nm), and infrared ($\sim1200$~cm$^{-1}\equiv833$~nm) regions of the electromagnetic spectrum, for each step, respectively. In the experiments in which this photoexcitation scheme has been employed, the visible and infrared laser radiation was directly generated using two tuneable nanosecond pulsed dye lasers. The circularly polarised vuv radiation was generated by resonance enhanced sum-frequency mixing in Xe~\cite{seiler11b}. By propagating each of these laser beams through 1.5~mm diameter holes in the electrodes surrounding the photoexcitation region the quantisation axis defined by the laser beams coincided with that defined by the electric field at the position of photoexcitation. The use of this excitation geometry ensured the selective photoexcitation of $n\mathrm{f}$ $|M_J|=3$ Rydberg states converging to the $N^+ = 0$,~2 and~4 rotational states of the $\mathrm{X}\,^2\Sigma_\mathrm{g}^+(v^+=0)$ H$_2^+$ ion core.

The procedure employed to decelerate the H$_2$ molecules into the on-axis electrostatic trap is the same as that described above for the deceleration and electrostatic trapping of H Rydberg atoms. However, because the initial longitudinal velocities of the beams of H$_2$ generated by seeding in Kr were 500~m/s, and lower than those of the beams of H atoms above, the time constant characterising the exponential decay of the deceleration potentials was increased to 3.65~$\mu$s.

\begin{figure}
\begin{center}
\includegraphics[width=0.95\textwidth]{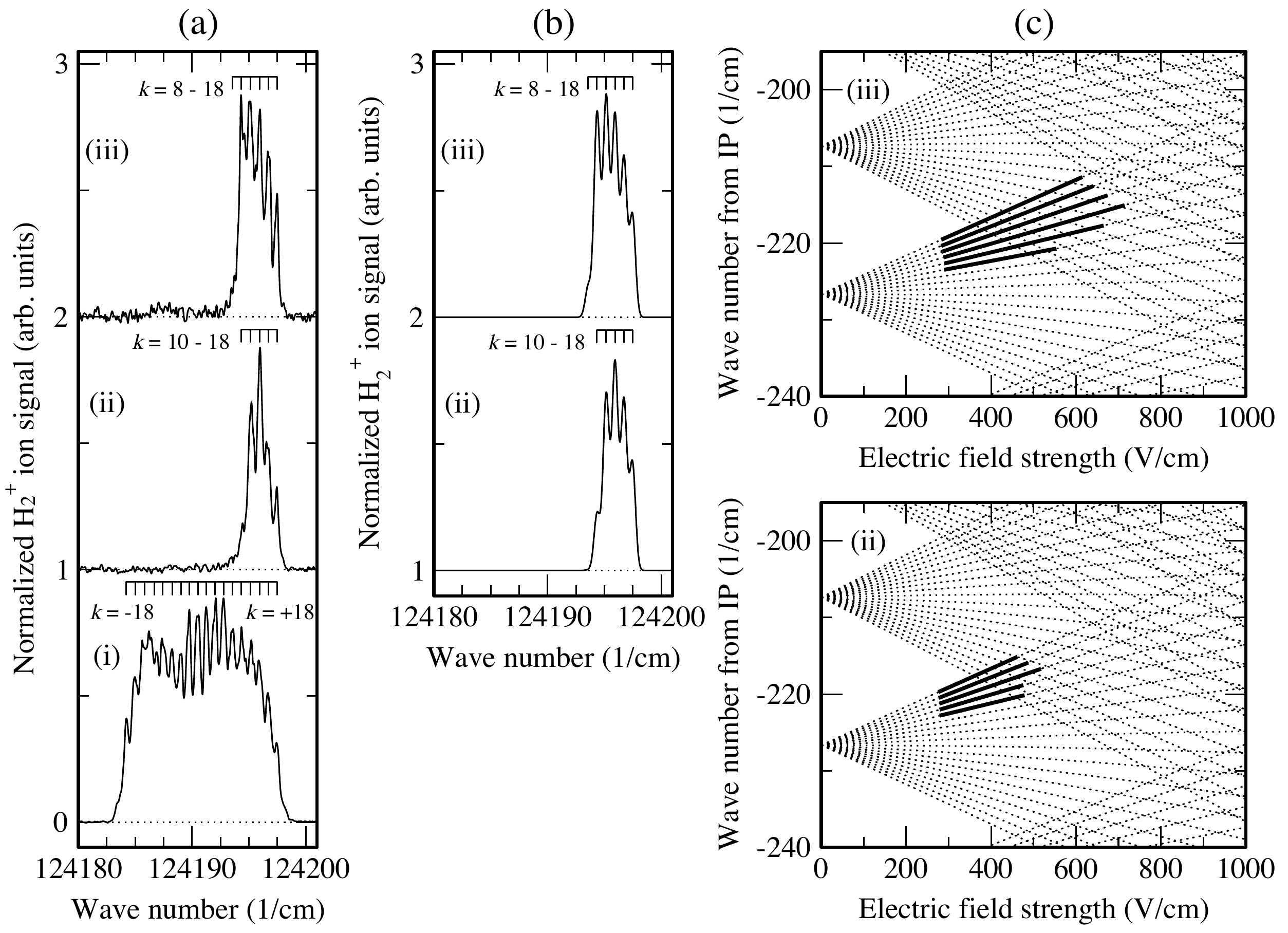}
\caption{\csentence{Trapping H$_2$ molecules in selected Rydberg-Stark states.} (a,i)~$n=22$, $\left|M_J\right|=3$ Stark spectrum of H$_2$ recorded in an electric field of $278\,$V/cm with detection after a time delay of $3~\mu$s. (ii,iii)~Spectra of low-field-seeking $k=10-18$, $n=22$, $\left|M_J\right|=3$ Stark states of H$_2$ detected after a trapping time of $50~\mu$s with deceleration potentials of (ii)~$\pm1.7~$kV and (iii)~$\pm2.3~$kV. (b,ii) and (b,iii) Calculated spectra obtained following numerical simulations of particle trajectories for deceleration potentials of (ii) $\pm1.7~$kV, and (iii) $\pm2.3~$kV. (c)~Calculated $\left|M_J\right|=3$ Stark structure in the vicinity of $n=22$ in H$_2$. The thick lines indicate the range of maximum electric-field strength experienced by molecules during deceleration with potentials of (ii)~$\pm1.7~$kV, and (iii)~$\pm2.3~$kV. From Ref.~\cite{seiler11b}.}
\label{fig:H2_state_selective}
\end{center}
\end{figure}

In these experiments Rydberg states with a wide range of values of $n$ could be decelerated and trapped with any given deceleration pulse sequence. For example, when optimised to decelerate and trap molecules in states for which $n\simeq30$, states with similar electric dipole moments and values of $n$ in the range from $n=21$--$37$ could also be trapped~\cite{hogan09a}. For many spectroscopy and scattering experiments, or studies of excited state decay processes, it is desirable to prepare decelerated samples in selected Stark states. As can be seen in Fig.~\ref{fig:H2_state_selective}, this has been demonstrated in H$_2$ for states with $n=22$. Upon photoexcitation in a field of 278~V/cm and detection after a time delay of $3~\mu$s, all $19$ accessible $n=22$ Stark states for which $\left|M_J\right|=3$, with values of $k$ from $-18$ to $+18$, are resolved [spectrum~(i) of Fig.~\ref{fig:H2_state_selective}(a)]. Spectra (ii) and (iii) in Fig.~\ref{fig:H2_state_selective}(a) were then recorded by monitoring the pulsed-field-ionisation signal of the H$_2$ molecules in the trap $50~\mu$s after photoexcitation. With pulsed deceleration potentials of $\pm1.7~$kV applied to electrodes~3 and~4 (see Fig.~\ref{fig:3Dtrap}), molecules excited to the four outermost low-field-seeking Stark states ($k=12-18$) were trapped efficiently [spectrum~(ii)]. Increasing the deceleration potentials to $\pm2.3\,$kV enabled the generation of larger electric field gradients and therefore also efficient trapping of molecules in Rydberg-Stark states with $k=10$ [spectrum~(iii)], the dipole moment of which was too small for efficient trapping with the lower potentials. The experimentally observed deceleration efficiency for each Stark state is fully accounted for in calculations of particle trajectories in the decelerator and trap, the results of which are displayed in Fig.~\ref{fig:H2_state_selective}(b). Comparison of the results of these calculations with the experimental data leads to the conclusion that the calculations capture all essential aspects of the deceleration and trapping process, including the dynamics at avoided crossings. Indeed, the range of maximal electric fields experiences by molecules within the excited ensemble during deceleration and trap loading, indicated by the think lines in Fig.~\ref{fig:H2_state_selective}(c), show that many molecules are subjected to fields beyond the Inglis-Teller limit. The dynamics at the avoided crossings in these fields were treated using a Landau-Zener model in the calculations~\cite{seiler11b,landau32a,zener32a}. 

\begin{figure}[t]
\begin{center}
\includegraphics[width=0.65\textwidth]{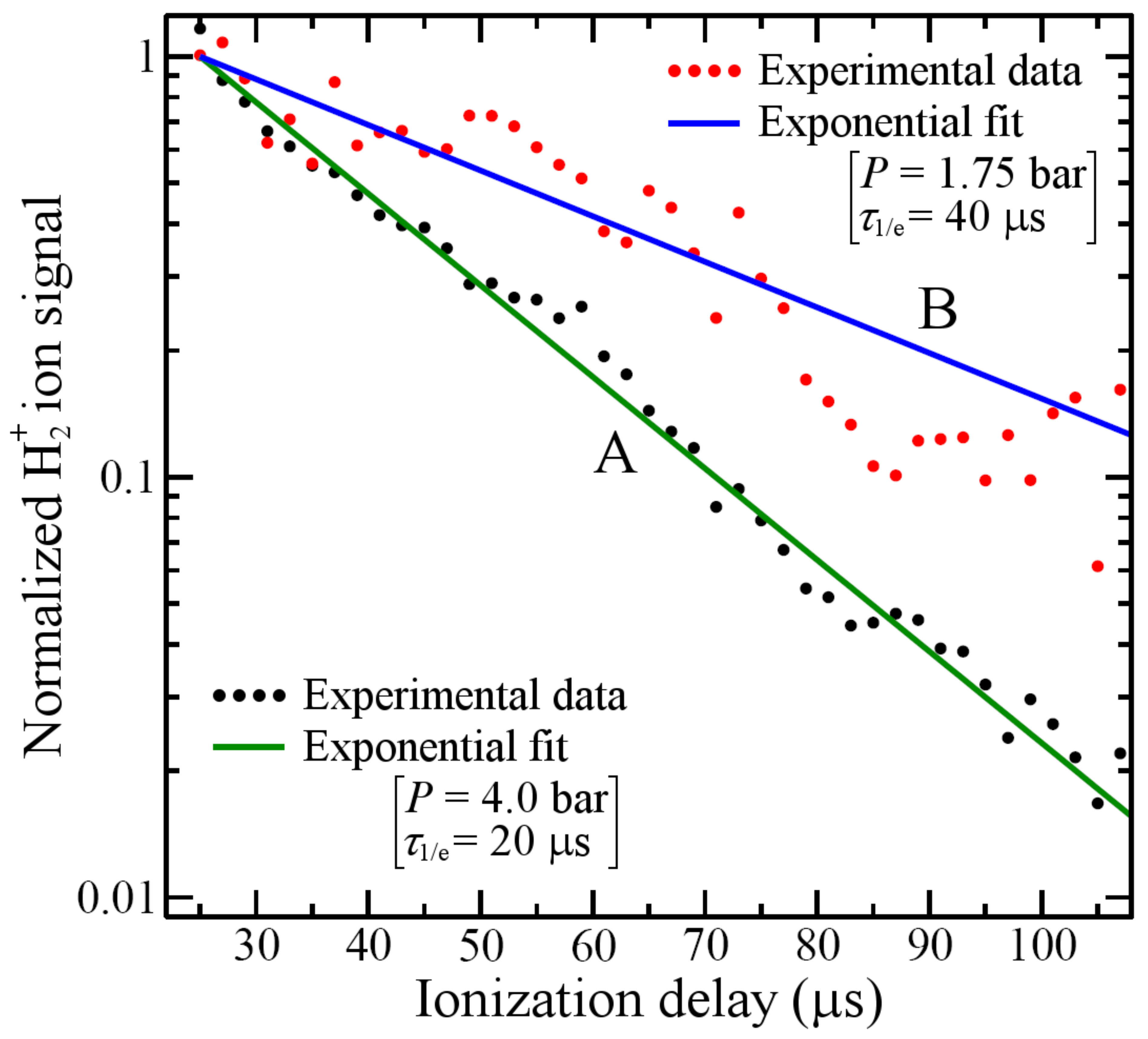}
\caption{\csentence{Decay of trapped H$_2$ molecules.} Measurements of the number of H$^+_2$ ions detected following pulsed electric field ionisation of H$_2$ molecules from an on-axis electrostatic trap in a room temperature environment. Dataset A (B) was recorded with the pulsed valve operated at a stagnation pressure of 4.0~bar (1.75~bar). From Ref.~\cite{hogan09a}.}
\label{fig:H2_trap_decay}
\end{center}
\end{figure}

Measurements of the decay of H$_2$ molecules from the on-axis electrostatic trap following photoexcitation to $|n,k\rangle=|33,23\rangle$ Stark states, deceleration and trapping are presented in Fig.~\ref{fig:H2_trap_decay}~\cite{hogan09a}. These measurements were performed by monitoring the integrated H$^+_2$ ion signal at the MCP detector following pulsed electric field ionisation of the trapped molecules for a range of times after photoexcitation. From the data recorded with the lower density molecular beam (dataset B -- for which the pulsed valve was operated at a stagnation pressure of 4~bar), a trap decay constant of $\tau_{1/\mathrm{e}} = 40~\mu$s was determined. If spontaneous emission to lower $n$ levels, with $\left|M_J\right|=2-4$ and hence $n\geq 3$, were the only decay processes, trapping times exceeding $300~\mu$s would be expected. The significant difference between this time and that measured experimentally indicates that decay by spontaneous emission is not the dominant trap-loss mechanism.

As in the trapping experiments carried out in a room temperature environment with H and D atoms, transitions driven by blackbody radiation can affect the decay of the H$_2$ molecules from this on-axis trap. However, while the overall blackbody depopulation rates of the initially prepared Stark states are expected to be similar to those for H or D, the effects of these transitions on the decay of the H$_2$ molecules from the trap are different. This is because in the hydrogenic Rydberg-Stark states of H$_2$ prepared, blackbody-radiation-induced transitions to $\left|M_J\right|=2$ Rydberg states which can decay by predissociation can occur. Under the assumption that approximately one third of the blackbody-radiation-induced transitions are $\Delta M_J=-1$ transitions one can conclude that $\sim33\%$ of the blackbody-radiation-induced depopulation rate results in trap loss, leading to an expected decay time greater than $100~\mu$s.

Because the trap loss rates in Fig.~\ref{fig:H2_trap_decay} are higher than those expected by the combination of spontaneous emission and blackbody-radiation-induced loss, and a significant dependence of the trap decay rate on the stagnation pressure at which the pulsed valve was operated, and hence the density of the molecular beam, was observed (compare dataset A and dataset B), it must be concluded that collisional losses also play a significant role in these on-axis trapping experiments. In addition, it is also possible that slow predissociation of the $\left|M_J\right|=3$ plays a minor part in the trap decay. Collisional losses can occur either through dipole-dipole interactions between trapped Rydberg molecules or via interactions of the Rydberg molecules, with atoms and molecules in the trailing components of the gas pulse. 

To disentangle these competing decay processes, experiments have recently been performed by Seiler, Merkt and co-workers with H$_2$ molecules trapped in an off-axis electrostatic trap the electrodes of which were cooled to 11~K~\cite{seiler13a}. This work indicates that, as in the H and D atom trapping experiments, under these conditions H$_2$ trapping times exceeding 1~ms are achievable. These experiments now open up a wide range of opportunities for studies of the effects of collisions and blackbody transitions on the decay of long-lived hydrogenic molecular Rydberg states on time-scales that were not previously possible. 

\section*{Chip-based guides, decelerators and traps}

Controlled manipulation of the motional degrees of freedom and internal quantum states of atoms and molecules at vacuum--solid-state interfaces is of importance in several areas of research. Robust and scaleable chip-based electric and magnetic traps and guides have been developed for atomic ions~\cite{chiaverini05a} and neutral ground state atoms~\cite{folman02a}. These devices have been exploited, e.g., in quantum information processing~\cite{home09a} and quantum metrology~\cite{riedel10a}. In addition, the preparation of cold, velocity-controlled samples of polar molecules using chip-based Stark decelerators has been demonstrated as a route ``\emph{towards a gas phase molecular laboratory on a chip}''~\cite{meek08a,meek09a,meek09b}. Approaches directed toward the confinement of Rydberg atoms in the vicinity of surfaces have involved Rydberg photoexcitation in miniature vapor cells~\cite{kubler10a}, and in close proximity to arrays of surface-based permanent-magnet traps~\cite{tauschinsky10a}. Atom chips have also been developed with Rydberg photoexcitation in the strong dipole-blockade regime in mind~\cite{nirrengarten06a,cherry09a}, and for the realisation of sources of single atoms on demand~\cite{saffman02a}. 

\begin{figure}
\begin{center}
\includegraphics[width=0.65\textwidth]{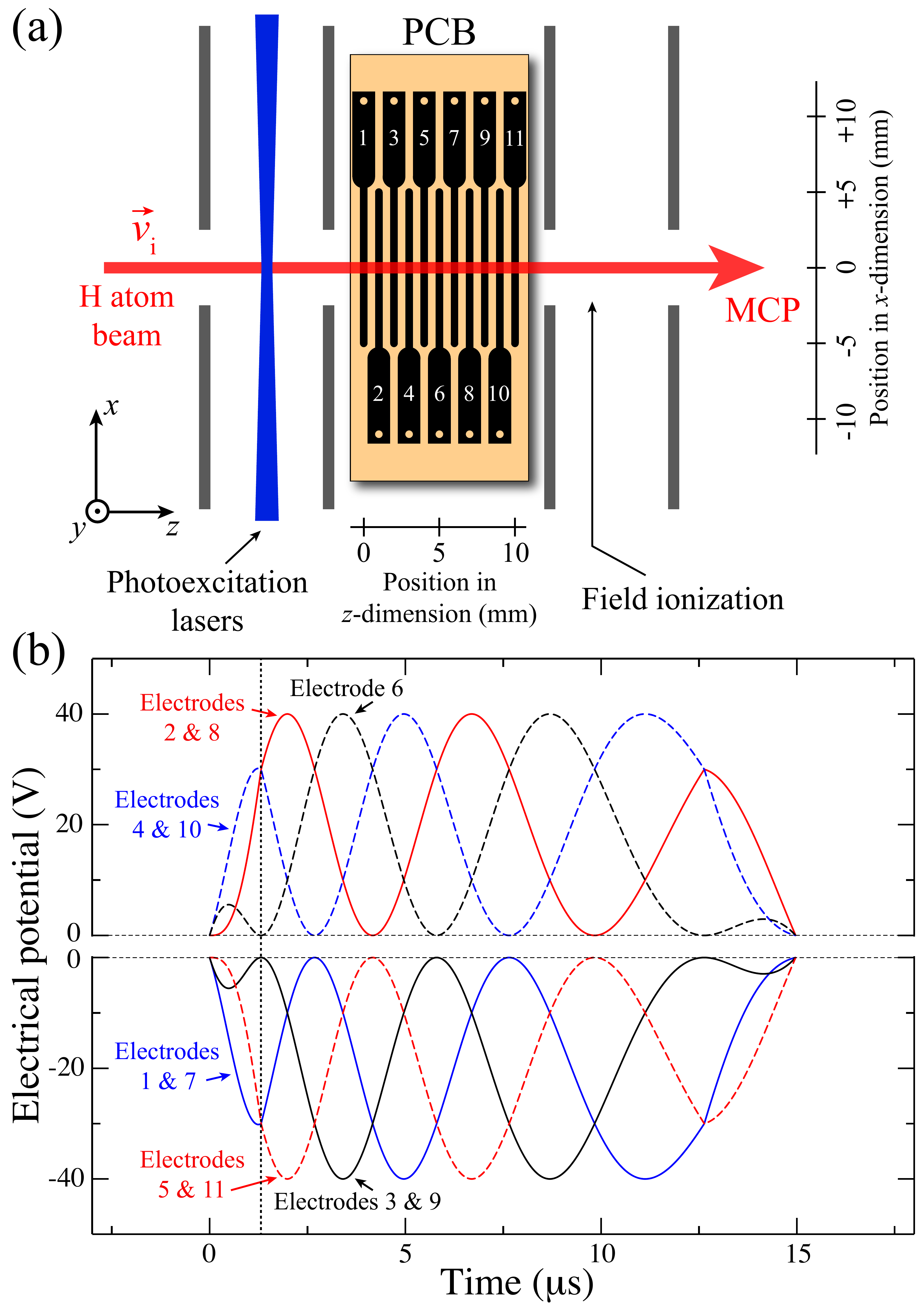}
\caption{\csentence{A surface-electrode Rydberg-Stark decelerator.} (a) Schematic diagram of a surface-electrode-based Rydberg-Stark decelerator and surrounding photoexcitation and electric field ionisation regions. (b) Oscillating potentials applied to the~11 electrodes of the decelerator in (a) for the deceleration of H atoms from $v_{\mathrm{i}}=760$~m/s to $v_{\mathrm{f}}=300$~m/s. From Ref.~\cite{hogan12b}.}
\label{fig:chipdecel_schematic}
\end{center}
\end{figure}

In this context chip-based Rydberg-Stark decelerators have also recently been developed. These devices are composed of arrays of metallic electrodes on electrically insulating substrates and have been used to accelerate, decelerate and trap pulsed beams of H~\cite{hogan12b} and He atoms~\cite{lancuba14a,allmendinger13a}, and H$_2$ molecules~\cite{allmendinger14a} in continuously moving electric traps. Electrostatic trapping at zero mean velocity has also been achieved. These devices are scalable in their construction and therefore well suited to the deceleration of samples with high initial kinetic energies, they can be readily configured to implement complex decelerator or trap geometries, and because the atoms or molecules are always localised about an electric field minimum during deceleration they permit control over the motion of samples in states with a very wide range of principal quantum numbers without losses by electric field ionisation. 

\begin{figure}[t]
\begin{center}
\includegraphics[width=0.6\textwidth]{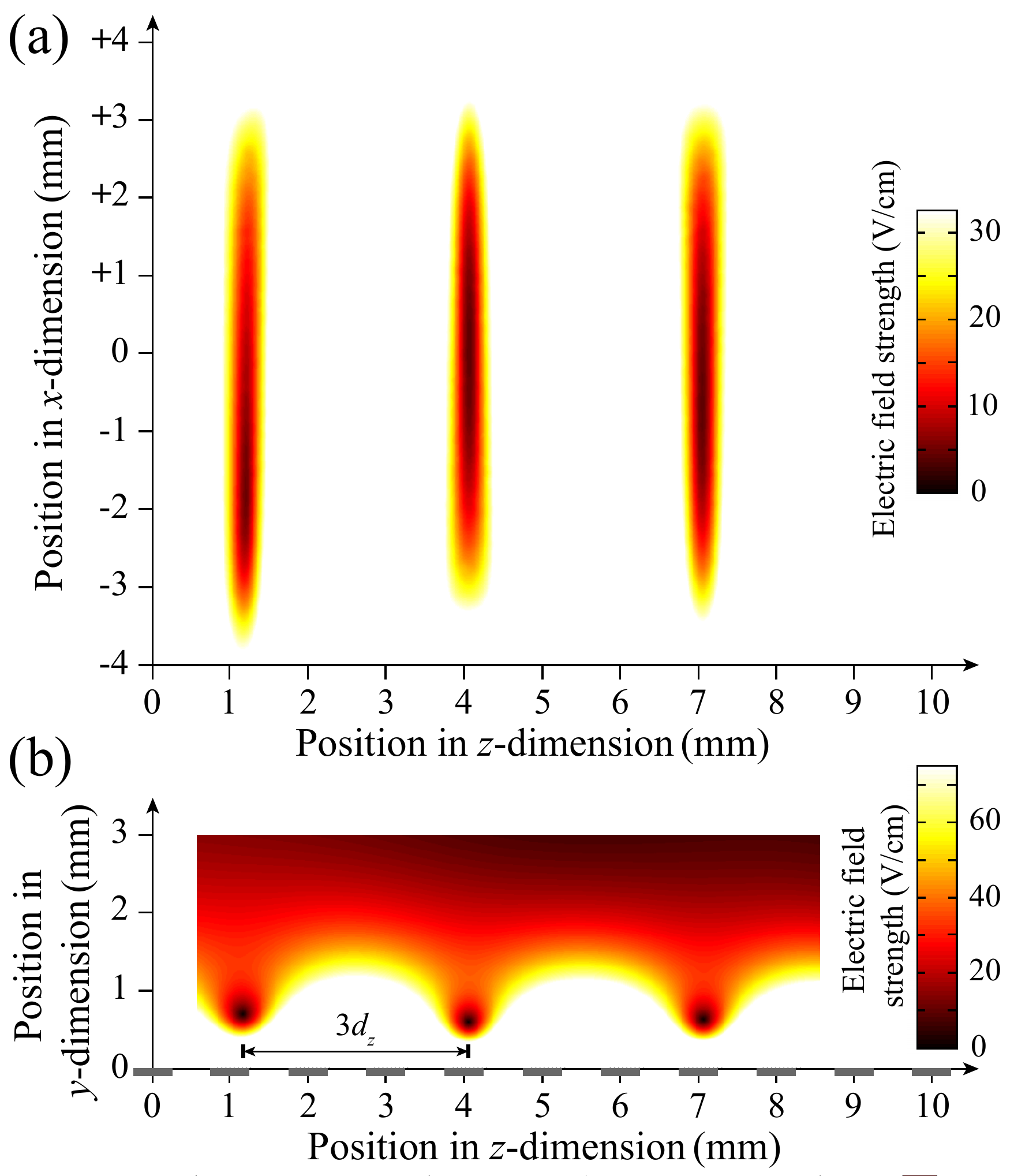}
\caption{\csentence{Electric field distributions in a surface-electrode Rydberg-Stark decelerator.} Electric-field distribution in (a) the $xz$-plane containing the trap minima, and (b) the $x=0$ plane in a surface-electrode Rydberg-Stark decelerator. The positions of the decelerator electrodes are indicated on the horizontal axis in (b). The 0~V plate is located at $y=-0.75$~mm. From Ref.~\cite{hogan12b}.}
\label{fig:chipdecel_fields}
\end{center}
\end{figure}

The design and operation principle of the first chip-based decelerator developed for Rydberg atoms and molecules built on those of chip-based Stark decelerators for polar ground state molecules~\cite{meek08a,meek09b}. However, because the electric dipole moments of the Rydberg states employed in the experiments were three orders of magnitude larger than those of the polar molecules decelerated in these other chip-based Stark decelerators (e.g., CO$^*$~\cite{santambrogio15a}), the electrode dimensions and their spacing could be scaled up, and the amplitudes of the electric potentials applied for deceleration and trapping scaled down, so that entire ensembles of atoms or molecules to be decelerated could be loaded into individual travelling traps of the decelerator. The resulting device, depicted schematically in Fig.~\ref{fig:chipdecel_schematic}(a), was composed of~11 electrodes each with a width of 0.5~mm in the $z$-dimension, and a center-to-center spacing $d_z=1.0$~mm. These dimensions were chosen to match the decelerator acceptance to the phase-space characteristics a beam of H Rydberg atoms. As can be seen in Fig.~\ref{fig:chipdecel_schematic}(a), the ends of the electrodes in the $x$ dimension were enlarged to reduce the oscillatory motion of the minima in this dimension during deceleration. 

The decelerator was operated by applying six oscillating electric potentials, $V_{i}$, to electrodes numbered $i=1-6$ in Fig.~\ref{fig:chipdecel_schematic}(a), and repeating the sequence from the seventh electrode on. These potentials have the general form 
\begin{eqnarray}
V_{i}&=&(-1)^{i}\,V_0[1+\cos(\omega t + \phi_i)],
\end{eqnarray} 
where $2V_0$ is the peak-to-peak potential, $\omega$ is the oscillation angular frequency, and $\phi_i=(1-i)2\pi/3$ is the phase shift from one electrode to the next~\cite{meek08a}. Using this configuration of time-dependent electric potentials, a set of moving electric traps, separated by a distance of $3d_z=3$~mm, were generated above the two-dimensional array of electrodes, as depicted in Fig.~\ref{fig:chipdecel_fields}. The activation time of the decelerator potentials was selected to ensure that all of the excited Rydberg atoms were loaded into a single trap. As the potentials oscillated in time, this trap moved in the positive $z$ dimension with a velocity $v_z=3d_z\omega/(2\pi)$. Acceleration/deceleration, $a_z$, was achieved by applying a linear frequency chirp to the time-dependent potentials such that 
\begin{eqnarray}
\omega(t)=\omega_0+(2\pi/3d_z)a_z t,
\end{eqnarray}
where $\omega_0$ corresponds to the initial velocity $v_{\mathrm{i}}$. An example of a set of potentials tailored for deceleration from $v_{\mathrm{i}}=760$~m/s to a final velocity $v_{\mathrm{f}}=300$~m/s is displayed in Fig.~\ref{fig:chipdecel_schematic}(b).

The potentials used to generate the field distributions depicted in Fig.~\ref{fig:chipdecel_fields} correspond to those at a time of 1.25~$\mu$s in Fig.~\ref{fig:chipdecel_schematic}(b), at the end of the activation phase (dotted vertical line). For the value of $V_0=20$~V used here, the trap minimum into which the atoms to be decelerated were loaded was located 0.6~mm above the surface of the device in the $y$ dimension. The moving trap was chosen to be $\sim5$~mm long in the $x$ dimension and $\sim0.5$~mm wide in the $y$ and $z$ dimensions and the trap depth was $\sim27$~V/cm, corresponding to $E/k_{\mathrm{B}}\simeq2$~K ($\simeq5.5$~K) for the outer low-field-seeking Stark states at $n=30$ ($n=50$).

\begin{figure}[ht!]
\begin{center}
\includegraphics[width=0.98\textwidth]{chipdecel_pseudo_contour_H_Xe.pdf}
\caption{\csentence{Potential energy distributions in the moving frame of reference associated with an electric field minimum in a surface-electrode Rydberg-Stark decelerator.} Potential energy distributions in the $yz$-plane at the mid-point of a surface-electrode Rydberg-Stark decelerator in the $x$-dimension surrounding a moving electric field minimum for accelerations of 0, $-5\times10^{5}$, $-5\times10^{6}$ and $-5\times10^{7}$~m/s$^2$. The contour lines are spaced by $E/k_{\mathrm{B}}=1$~K beginning at $1$~K. Cases for which H atoms in (a-d) $|n,k\rangle=|33,26\rangle$, and (e-h) $|50,40\rangle$ states are displayed. As are those in (i) and (j) for Xe atoms in the $|50,40\rangle$ state.}
\label{fig:chipdecel_pseudo}
\end{center}
\end{figure}

As is the case in the single-stage Rydberg-Stark decelerators described above~\cite{hogan13a}, the acceleration of the continuously moving electric-field minima in these surface-electrode decelerators affects the potential energy distribution in which the trapped Rydberg atoms move. In the surface-electrode decelerator, a constant acceleration is applied as the atoms travel across the device. As a result, the change that this leads to in the potential energy distribution experienced by the atoms in the accelerated frame of reference has a significant effect at all times throughout the acceleration/deceleration process. As in other chip-based Stark decelerators~\cite{meek08a,meek09b} this effect of the acceleration of the moving trap, $\vec{a}_{\mathrm{trap}}$, is most readily seen in the moving frame of reference associated with a single electric field minimum of the decelerator. The transformation from the laboratory-fixed frame of reference to this moving frame is achieved by adding the pseudo potential 
\begin{eqnarray}
V_{\mathrm{pseudo}}=m\,\vec{a}_{\mathrm{trap}}\cdot\vec{s},
\end{eqnarray}
where $m$ is the mass of the atom or molecule being decelerated, and $\vec{s}$ is the displacement from this local field minimum. Potential energy distributions associated with such an electric field minimum in this moving frame of reference of the surface-electrode Rydberg-Stark decelerator are presented in Fig.~\ref{fig:chipdecel_pseudo} for accelerations of 0, $-5\times10^{5}$, $-5\times10^{6}$ and $-5\times10^{7}$~m/s$^2$. In this figure, the contour lines are spaced by $E/k_{\mathrm{B}}=1$~K and begin at $1$~K. For H atoms in $|n,k\rangle=|33,26\rangle$ states for which the decelerator was designed, the shape and size of the moving trap does not significantly change for accelerations from 0 to $-5\times10^{5}$~m/s$^2$ [Fig.~\ref{fig:chipdecel_pseudo}(a) and~(b)]. However, the shape of the trap begins to distort for accelerations of $-5\times10^{6}$~m/s$^2$ and larger [Fig.~\ref{fig:chipdecel_pseudo}(c)] with a significant reduction in size for an acceleration of $-5\times10^{7}$~m/s$^2$. For higher Rydberg states with $|n,k\rangle=|50,40\rangle$, which can also be efficiently decelerated, the moving trap generated using the same electric potentials applied to the electrodes is deeper, and even for an acceleration of $-5\times10^{7}$~m/s$^2$ is only slightly distorted [Fig.~\ref{fig:chipdecel_pseudo}(e-h)]. To assess the suitability of such a decelerator for the manipulation of beams of heavy atoms or molecules, the potential energy distributions for Xe atoms in $|n,k\rangle=|50,40\rangle$ states are presented in Fig.~\ref{fig:chipdecel_pseudo}(i) and~(j) and indicate that accelerations of up to $-5\times10^{5}$~m/s$^2$ are feasible with this device, making it suitable for the deceleration of heavy samples. 

\begin{figure}[t]
\begin{center}
\includegraphics[width=0.85\textwidth]{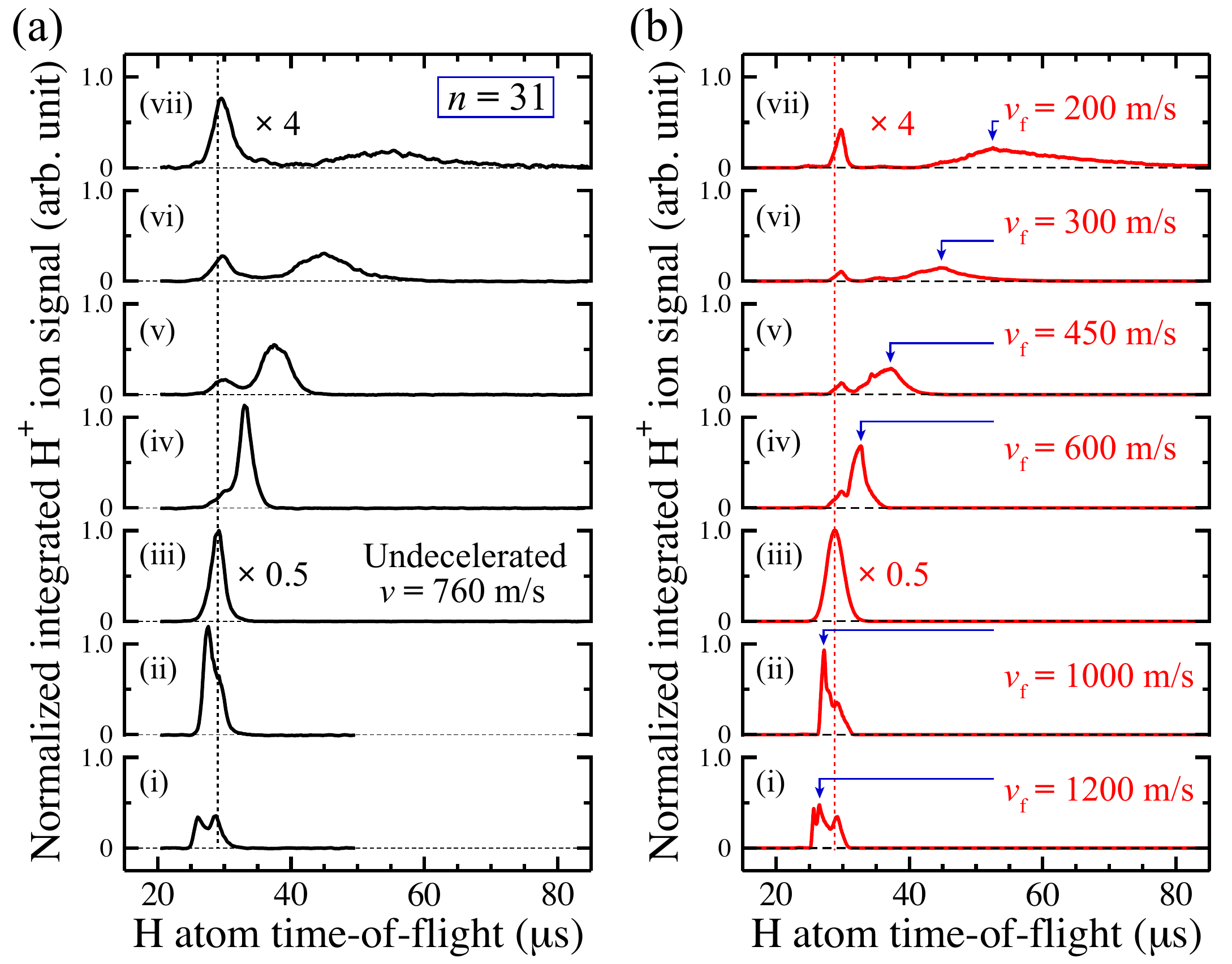}
\caption{\csentence{Acceleration and deceleration of H atoms in a surface-electrode Rydberg-Stark decelerator.} (a) Experimental, and (b) calculated H-atom time-of-flight distributions demonstrating acceleration/deceleration of atoms in states for which $n=31$ from  $v_{\mathrm{i}}=760$~m/s to $v_{\mathrm{f}}=1\,200$, $1\,000$, $600$, $450$, $300$ and $200$~m/s. From Ref.~\cite{hogan12b}.}
\label{fig:chipdecel_tof}
\end{center}
\end{figure}

The operation of this surface-electrode decelerator was investigated experimentally by measuring the times-of-flight of beams of H Rydberg atoms from their position of photoexcitation to that of electric field ionisation (see Fig.~\ref{fig:chipdecel_schematic}). A set of data corresponding to acceleration/deceleration from $v_{\mathrm{i}}=760$~m/s to final velocities between 1\,200~m/s and 200~m/s, is presented in Fig.~\ref{fig:chipdecel_tof}(a). These measurements only differ in the frequency chirp used to achieve the desired final longitudinal speed. The time-of-flight distribution labelled (iii) corresponds to the undecelerated beam of Rydberg atoms, detected after a flight time of $\sim29$~$\mu$s (vertical dotted line). This measurement was made with the decelerator off and its intensity is scaled by a factor of 0.5 with respect to the other measurements. Measurements (i) and (ii) were performed to accelerate the atoms to $v_{\mathrm{f}}=1\,200$ and $1\,000$~m/s, respectively. In these cases the accelerated atoms arrived at the detection region earlier than the undecelerated atoms, at times of $\sim26$ and $\sim27.5$~$\mu$s, respectively. The upper four time-of-flight distributions (iv)--(vii) correspond to deceleration to $v_{\mathrm{f}}=600$, $450$, $300$ and $200$~m/s and indicate progressively later arrival times of the decelerated atoms of 33, 38, 45 and $54$~$\mu$s, respectively. The intensity of the distribution corresponding to deceleration to $v_{\mathrm{f}}=200$~m/s is scaled by a factor of four. 

Calculations of three-dimensional particle-trajectories and the corresponding time-of-flight distributions [Fig.~\ref{fig:chipdecel_tof}(b)] capture the essential features of the experimental data and provide a basis for their interpretation. Comparison of the experimental data with the results of the calculations indicate that the peak observed in all traces at $\sim29$~$\mu$s corresponds to atoms which traverse the decelerator at a position beyond the saddle point of the moving traps in the $y$ dimension (i.e., at $y>1.2$~mm). These atoms follow metastable trajectories across the decelerator and arrive at the field-ionisation point without acceleration or deceleration. The second peak in each time-of-flight distribution corresponds to the accelerated/decelerated atoms. The velocities at each maximum [indicated by the arrows in Fig.~\ref{fig:chipdecel_tof}(b)] exactly match the final velocities for which the deceleration potentials were designed. At low final velocities, the ensemble of decelerated atoms expand in the longitudinal and transverse dimensions as they fly from the end of the device to the detection region. Transverse expansion after deceleration is accompanied by particle loss and leads to a reduction of the overall signal. This behaviour is also reproduced in the calculations.

\begin{figure}[t]
\begin{center}
\includegraphics[width=0.65\textwidth]{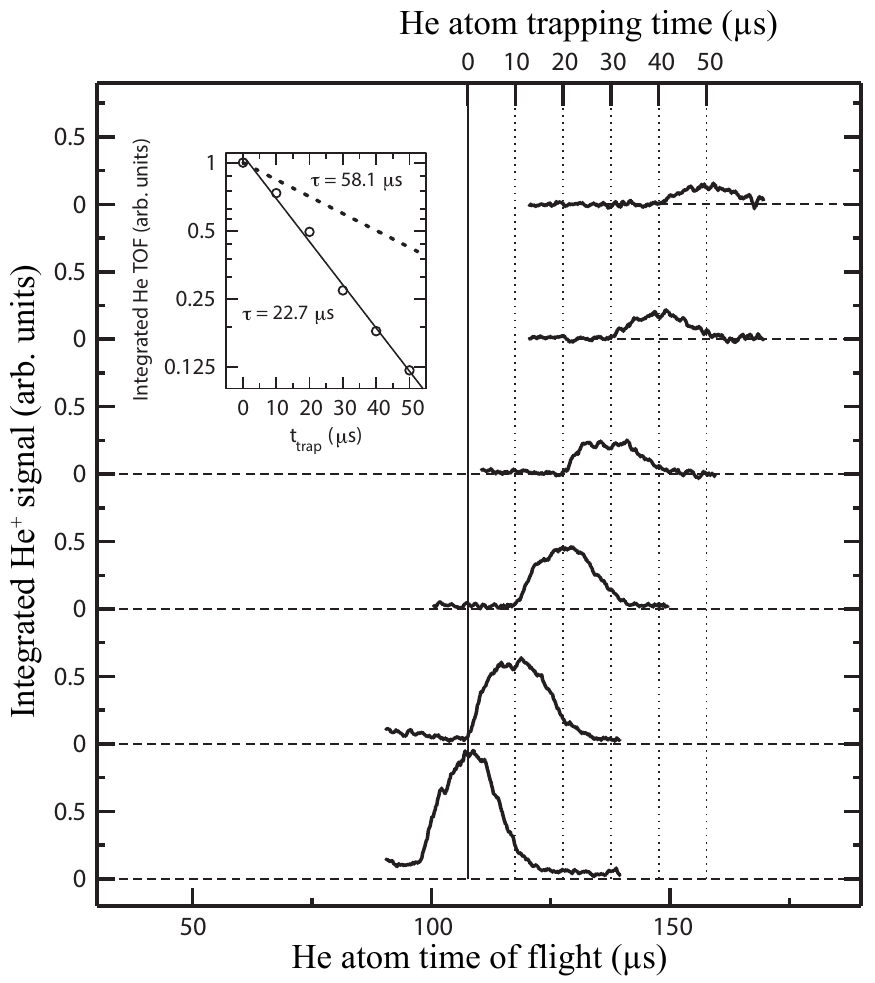}
\caption{\csentence{Trapping stationary samples of He atoms in a surface-electrode Rydberg-Stark decelerator.} He Rydberg atom time-of-flight distributions after deceleration, trapping, and re-acceleration in a surface-electrode decelerator. The states initially prepared at photoexcitation were those for which $|n,k,m\rangle=|30, 23, 0\rangle$. The dashed vertical lines indicate the He atom detection times after trapping stationary samples for times ranging from~0 to 50~$\mu$s. The dependence of the integrated time-of-flight signal on the trapping time is indicated in the inset, together with the decay rate associated fluorescence alone. From Ref.~\cite{allmendinger13a} with permission.}
\label{fig:He_trap}
\end{center}
\end{figure}

Following these initial deceleration and trapping experiments this surface-electrode decelerator was extended to include an array of 44 electrodes~\cite{allmendinger13a}. This longer device permitted the removal of larger amounts of kinetic energy than in previous decelerators and therefore complete deceleration and trapping of fast beams of He atoms in singlet Rydberg states moving at initial speeds of $1200$~m/s. In this work, the deceleration of samples to zero longitudinal velocity in the laboratory-fixed frame of reference, and on-axis trapping was demonstrated in a way that was similar to that employed previously to trap H Rydberg atoms~\cite{hogan12b}. After the traps in which the atoms were decelerated were brought to a stand still, they remained stationary for a selected period of time, the trapping time, $t_{\mathrm{trap}}$, in Fig.~\ref{fig:He_trap}. To detect the trapped atoms by pulsed electric-field ionisation, the traps were subsequently re-accelerated along the axis of the apparatus to 400~m/s and the atoms then released into an electric field ionisation detection region similar to that in Fig.~\ref{fig:chipdecel_schematic}. From the complete time-of-flight distributions of the atoms for a set of selected trapping times, information on the decay of atoms from the stationary traps could be determined. The time constants ($\tau_{1/e}\sim25~\mu$s) associated with the observed trap decay in these experiments were shorter than the calculated lifetimes of the states prepared at photoexcitation ($\tau_{1/e}\sim58~\mu$s). This is a result of the combined effects of collisions with the trailing components of the atomic beam arising from on-axis deceleration and trapping, and transitions driven by the room temperature blackbody radiation field in the environment of the decelerator. 

\begin{figure}
\begin{center}
\includegraphics[width=0.85\textwidth]{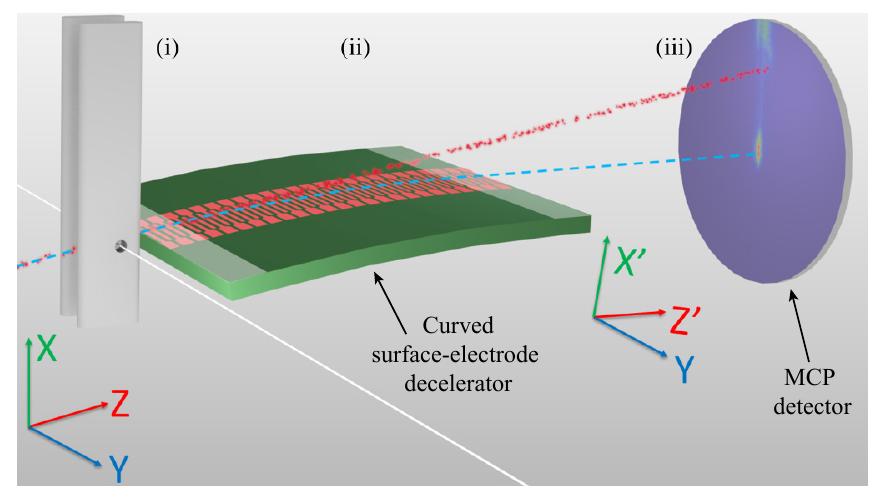}
\caption{\csentence{Surface-electrode decelerator and deflector for Rydberg atoms and molecules.} Schematic diagram of the surface-electrode decelerator and deflector used to manipulate beams of H$_2$ molecules in high Rydberg states. (i) The Rydberg state photoexcitation region is located between two parallel metal plates. (ii) The curved surface-electrode decelerator permitted controlled transport and deflection of the H$_2$ molecules away from their initial axis of propagation. (iii) Deflected and undeflected molecules were directly imaged on a MCP detector. From Ref.~\cite{allmendinger14a} with permission.}
\label{fig:H2_chip}
\end{center}
\end{figure}

With a view to using curved, surface-electrode Rydberg-Stark decelerators to prepare velocity-controlled samples of Rydberg atoms and molecules for merged beam collision experiments (see e.g.,~\cite{osterwalder15a}) devices with curvatures perpendicular to the plane of the electrode surfaces have recently been implemented to transport, decelerate and deflect beams of H$_2$ molecules in high Rydberg states~\cite{allmendinger14a}. A schematic diagram of the experimental apparatus in which such a device was implemented is displayed in Fig.~\ref{fig:H2_chip}. The H$_2$ molecules in these experiments were photoexcited to $n=31$ Rydberg-Stark states between a pair of parallel metallic plates [region (i) in Fig.~\ref{fig:H2_chip}] using the photoexcitation scheme in Eq.~\ref{eq:H2excite}. The excited molecules were then loaded into a single moving electric trap of the surface-electrode decelerator and transported along the device. In this process the molecules were deflected away from the initial propagation axis of the molecular beam. Images of the undeflected and deflected molecules were then recorded at a MCP detector when the decelerator was off, and when it was activated, respectively. The information obtained from these images was then used to characterise the transverse acceptance of the decelerator and determine the phase-space properties of the guided molecules. An example of such an image, recorded with the decelerator active, is included in Fig.~\ref{fig:H2_chip}. Comparisons of the data recorded in these experiments with the results of numerical calculations of the trajectories of the molecules in the device show excellent agreement.  The resulting velocity-controlled beams of cold, state-selected molecules, deflected away from their initial axis of propagation by an angle of 10$^{\circ}$ had translational temperatures of $\sim250$~mK and appear very well suited for use in studies of ion-molecule reactions at low collision energies~\cite{allmendinger14a}. 

\begin{figure}[t]
\begin{center}
\includegraphics[width=0.98\textwidth]{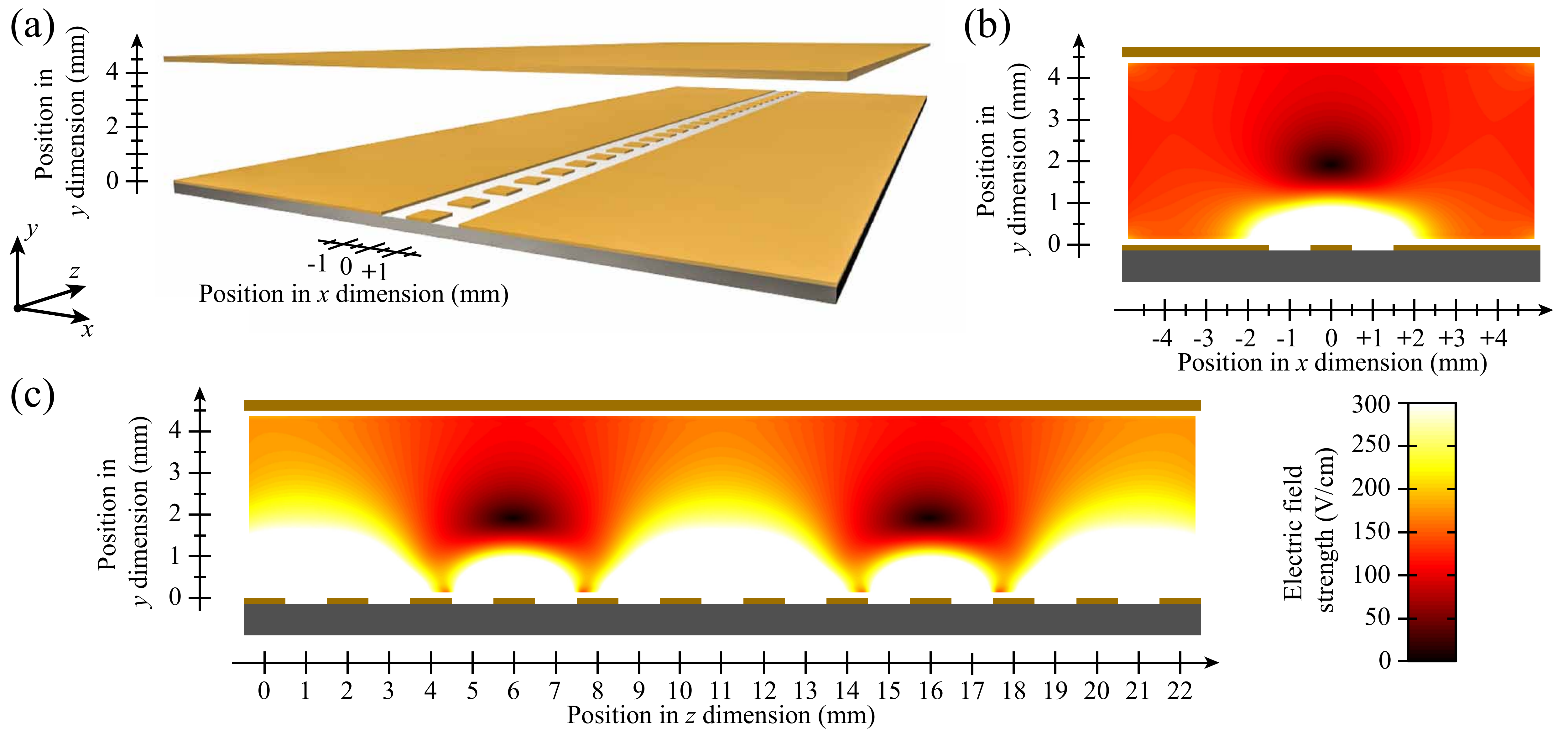}
\caption{\csentence{A transmission-line Rydberg-Stark decelerator.} (a) Schematic diagram of a transmission-line decelerator for Rydberg atoms and molecules. Typical electric field distributions in (b) the transverse $xy$-plane, and (c) the longitudinal $yz$-plane employed for trapping, transport and deceleration are also displayed. From Ref.~\cite{lancuba14a}.}
\label{fig:tlinedecel}
\end{center}
\end{figure}

Most recently, a second kind of chip-based Rydberg-Stark decelerator has been developed. This design of this device, which can be seen in Fig.~\ref{fig:tlinedecel}, is based on the geometry of a two-dimensional electrical transmission line. This design makes it well suited to integration with chip-based microwave circuits (see e.g., Ref.~\cite{hogan12a}). This transmission-line geometry was first implemented in the form of an electrostatic guide for Rydberg atoms or molecules~\cite{lancuba13a}. This device was composed of an electrical transmission line with a continuous center conductor and permitted the transverse positions of beams of Rydberg atoms to be controlled. By then segmenting the center conductor of the transmission line to form an array of square, equally spaced electrodes, a device in which Rydberg atoms or molecules can be transported, accelerated and decelerated while confined in continuously moving electric traps was realised. The resulting transmission-line decelerator operates on a similar principle to the surface-electrode devices described above, but offers the advantages of stronger transverse confinement, resulting in symmetric traps in all three spatial dimensions [compare Fig.~\ref{fig:chipdecel_fields} with Fig.~\ref{fig:tlinedecel}(b) and (c)], and opportunities for the introduction of curvatures in the plane of the two-dimensional electrode arrays from which it is composed. Because atoms or molecules in this decelerator are conveyed in the void between a two-dimensional electrode array and a parallel plane metal plate it lends itself well to cooling to low temperature, directly shielding the samples within it from their surroundings. It also provides a very well defined electromagnetic environment in which to manipulated and trap Rydberg atoms and molecules. This characteristic is a prerequisite for identifying and precisely controlling effects of blackbody radiation on the trapped samples. 

The transmission-line decelerators constructed up to now employ 1~mm square center-conductor segments with center-to-center spacings of $d_{\mathrm{cc}}=2$~mm. The insulating gap between these segments and the ground planes was also selected to be 1~mm, while the upper plane metal plate was positioned 4.5~mm above this electrode array. In this geometry the formation of a three-dimensional electric field minimum can be achieved by applying equal non-zero electric potentials to one single segment of the center conductor and the upper plate while all other electrodes are set to 0~V. However, if a constant potential of $V_{\mathrm{u}} = -V_0/2$ is applied to the upper plate electrode, with potentials of 0~V on the two ground planes, and potentials of $V_i=V_0\cos[-(i-1)\phi]$ on the segments of the center conductor, where $\phi=2\pi/5$ is the phase-shift from one decelerator segment, $i$, to the next, arrays of electric traps spaced by $5d_{\mathrm{cc}}$ are generated within the device.

\begin{figure}[t]
\begin{center}
\includegraphics[width=0.98\textwidth]{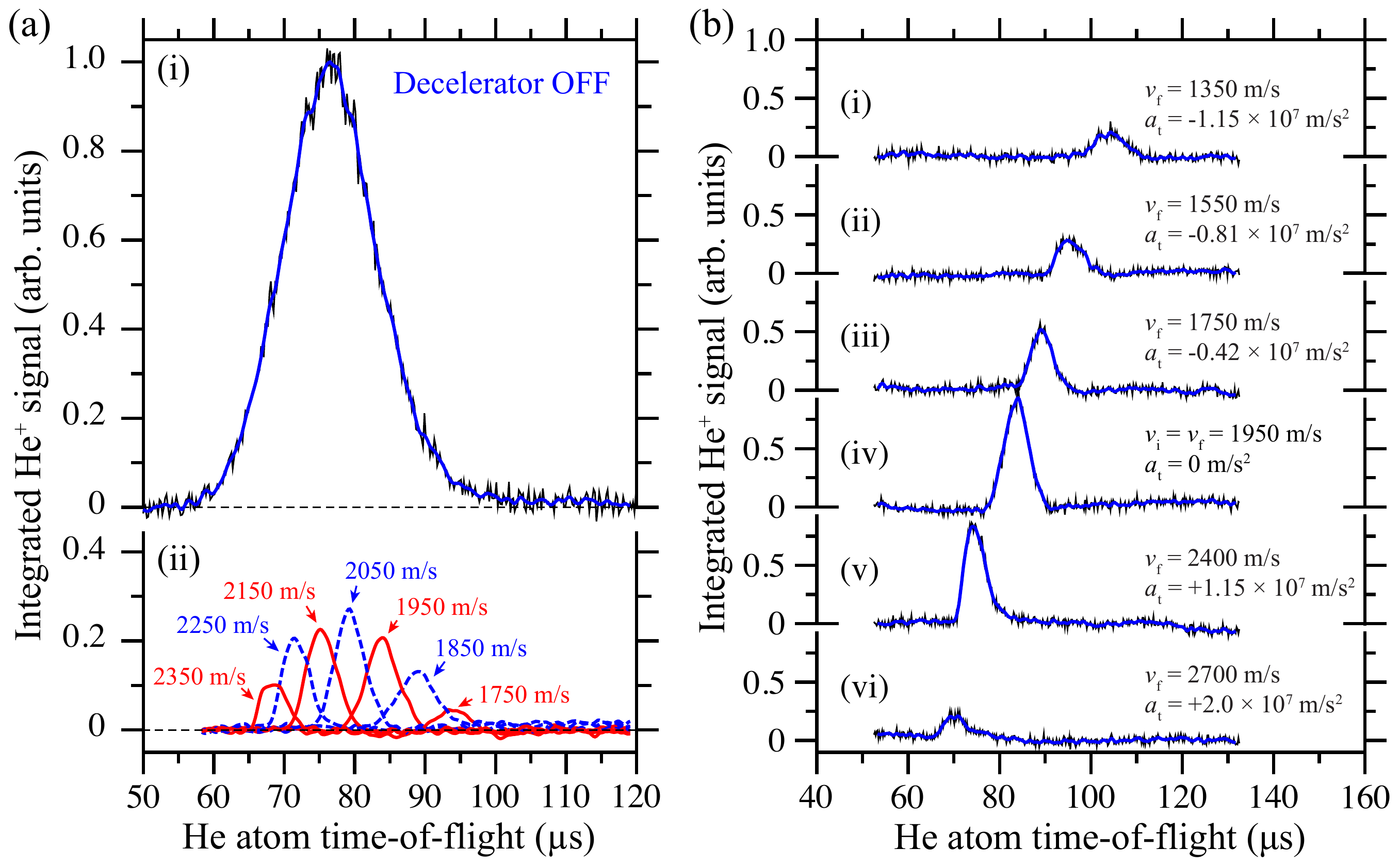}
\caption{\csentence{Guiding, accelerating and decelerating He atoms in a transmission-line decelerator.} (a) He atom time-of-flight distributions recorded after passing through a transmission-line decelerator (i) when the device was off, and (ii) after velocity selection and guiding at a range of longitudinal speeds as indicated. (b) Time-of-flight distributions recorded following acceleration/deceleration of atoms travelling with an initial longitudinal speed of 1950~m/s. From Ref.~\cite{lancuba14a}.}
\label{fig:tlinedata}
\end{center}
\end{figure}

To realise a continuous motion of this array of traps through the decelerator, the electric potentials applied to the segments of the center conductor are set to oscillate in time at an angular frequency $\omega$, such that,
\begin{eqnarray}
V_i(t) = V_0\cos[\omega t - (i-1)\phi].
\end{eqnarray}
In this configuration the speed, $v_{\mathrm{trap}}$, at which the traps move through the decelerator is then 
\begin{eqnarray}
v_{\mathrm{trap}} = 5d_{\mathrm{cc}}/(2\pi).
\end{eqnarray}
To ensure that the traps remain a constant distance above the two-dimensional electrode array as they travel through the decelerator the potentials applied to the ground planes are modulated at a frequency, $\omega_{\mathrm{gp}}=5\omega$, and with an amplitude of $-0.045V_0$.

Using a curved decelerator of this design He Rydberg atoms in $n=52$ Stark states with electric dipole moments of $\sim6900$~D have been velocity selected and guided at constant speed with the results presented in Fig.~\ref{fig:tlinedata}(a). The upper panel, (i), of this figure contains the time-of-flight distribution of the ensemble of initially prepared He Rydberg atoms from their position of photoexcitation to the end of the decelerator with the device off. The results of loading components of this velocity distribution into individual continuously-moving traps of the decelerator, and guiding them through the device at constant speed can be seen in panel (ii). To convey atoms at a constant speed of 1950~m/s through this device required that $\omega=2\pi\times195$~kHz. In the experiments, these oscillating potentials were generated at low voltage using a set of arbitrary waveform generators before amplification by a factor of 50 to amplitudes of up to $V_0=120$~V. In the data presented in Fig.~\ref{fig:tlinedata}(a) effects of the Rydberg atom density on the efficiency with which the ensembles are guided can be seen. This highlights the importance of identifying and minimising the effects of collisions between the trapped atoms in experiments of this kind, particularly at these high values of $n$~\cite{zhelyazkova15a}.

As in the surface-electrode decelerators discussed above, by introducing frequency chirps to the time-dependent electric potentials, the moving electric traps can be accelerated or decelerated as they travel through the device. For a selected initial speed, $v_{\mathrm{trap}}(0)$, and an acceleration, $a_{\mathrm{trap}}$, the time-dependence of the oscillation frequency is
\begin{eqnarray}
\omega(t) = \omega(0) + \frac{\pi a_{\mathrm{trap}}}{5d_{\mathrm{cc}}}t,
\end{eqnarray}
where $\omega(0) = 2\pi v_{\mathrm{trap}}(0)/(5d_{\mathrm{cc}})$ is the initial angular frequency. Data recorded for the acceleration and deceleration of beams of triplet He atoms in $|n,k\rangle=|52,35\rangle$ Stark states initially moving at 1950~m/s are presented in Fig.~\ref{fig:tlinedata}(b). As can be seen from this data, increasing the acceleration of the traps toward $|a_{\mathrm{trap}}|=2\times10^7$~m/s$^2$ leads to a reduction in the intensity of the time-of-flight signals. This is a consequence of the effective reduction in the depths of the moving traps with acceleration as discussed above in the context of the surface-electrode decelerators (see, e.g., Fig.~\ref{fig:chipdecel_pseudo}). In the curved transmission-line decelerator employed in these experiments it was necessary to consider the combined effects of the applied tangential acceleration, and the centripetal accelerations arising from the curvature of the device, on the decelerating atoms. For the Rydberg states prepared in the experiments, the maximum acceleration that could be applied before the traps opened completely was $|a_{\mathrm{trap}}|\simeq2.35\times10^7$~m/s$^2$. 

In these transmission-line decelerators, atomic beams with speeds exceeding 2500~m/s have been manipulated and changes in kinetic energy of up to $E_{\mathrm{kin}}\sim1.3\times10^{-20}$~J  ($\equiv80$~meV or $\equiv650$~cm$^{-1}$) have been achieved. This makes them very well suited to the transport, acceleration and trapping of a wide range of atoms and molecules. At present transmission-line decelerators with \emph{in situ} detection are being developed and tested, and possibilities for the implementation of chip-based storage-rings, and beam-splitters are foreseen. These devices, when combined with co-planar microwave circuitry represent a unique chip-based laboratory with which comprehensive control over the translational motion and internal quantum states of gas-phase Rydberg atoms and molecules can be achieved. 

\section*{Conclusions}

The methods for accelerating, decelerating and trapping Rydberg atoms and molecules initially travelling in pulsed supersonic beams using inhomogeneous electric fields that are reviewed here, open many exciting possibilities for new experimental investigations at the interface between physics and physical chemistry. In the process of developing these techniques measurements have been made of the roles that collisions and transitions driven by blackbody radiation play on the evolution and decay of excited states of atoms and molecules on timescales that were not previously possible. The translationally cold, velocity-controlled beams that can now be prepared by Rydberg-Stark deceleration offer opportunities for investigations of effects of collisions, blackbody radiation and electric and magnetic fields on slow-dissociation processes in long-lived Rydberg states of molecules, and provide an intriguing route to studies of ion-molecule reactions in merged beams~\cite{allmendinger14a}. Tuneable, velocity-controlled beams of atoms and molecules in high Rydberg states provide opportunities for studies of the effects of image states, and the velocity dependence of surface-ionisation processes~\cite{gibbard15a}. In addition, the de-excitation of molecular samples back to their ground states after Rydberg-Stark deceleration represents a route toward to the production of cold samples of ground-state molecules.

In hybrid approaches to quantum information processing involving Rydberg atoms and microwave circuits, the chip-based Rydberg atom guides and decelerators described can be used to transport atoms within cryogenic environments operated a different blackbody temperatures, and with differing cooling capacities. Recent implementations of electrostatic charged-wire guides for atoms or molecules in high-field-seeking Rydberg-Stark states~\cite{ko14a} offer potential opportunities for studies of angular-momentum quantisation in non-commutative space~\cite{baxter95a,zhang96a,zhang04a}. Demonstrations of focussing, decelerating and trapping fast-moving beams of Rydberg atoms highlight the possibility of employing Rydberg-Stark deceleration methods in the manipulation of samples of positronium atoms for precision spectroscopic studies and investigations of the acceleration of particles composed of antimatter in the gravitational field of the Earth~\cite{cassidy14a}. 

In the refinement of the techniques of Rydberg-Stark deceleration and electric trapping for application in each of the above areas, it will be important to study and account for the effects of electric-field noise on the energy level structure of the Rydberg states with their very large electric dipole moments~\cite{zhelyazkova15a,zhelyazkova15b}. In addition, contributions from dipole-dipole interactions~\cite{comparat10a,zhelyazkova15a,reinhard07a,vogt07a} between atoms and molecules during deceleration and trapping must be expected to become increasingly significant particularly on longer experimental timescales.


\begin{backmatter}

\section*{Competing interests}
The author declares no competing interests.

\section*{Acknowledgements}
Parts of the research described in this review were carried out in the context of my habilitation thesis in the Laboratory of Physical Chemistry at ETH Zurich. In relation to this work I am very grateful to Prof. Fr\'ed\'eric Merkt, Dr. Edward Vliegen, Dr. Christian Seiler, Heiner Sa\ss mannshausen, and Pitt Allmendinger for their support and collaboration; and to Ren\'e Gunzinger, Josef Agner, and Hansj\"urg Schmutz for excellent mechanical and electronic support. At University College London I wish to thank Patrick Lancuba for his contributions to the development of the transmission-line Rydberg atom guides and decelerators developed most recently, and the Department of Physics and Astronomy and Faculty of Mathematical and Physical Sciences, together with the Engineering and Physical Sciences Research Council (Grant No. EP/L019620/1) for financial support.


\bibliographystyle{bmc-mathphys} 
\bibliography{hoganbib}      


\begin{thebibliography}{152}
\ifx \bisbn   \undefined \def \bisbn  #1{ISBN #1}\fi
\ifx \binits  \undefined \def \binits#1{#1}\fi
\ifx \bauthor  \undefined \def \bauthor#1{#1}\fi
\ifx \batitle  \undefined \def \batitle#1{#1}\fi
\ifx \bjtitle  \undefined \def \bjtitle#1{#1}\fi
\ifx \bvolume  \undefined \def \bvolume#1{\textbf{#1}}\fi
\ifx \byear  \undefined \def \byear#1{#1}\fi
\ifx \bissue  \undefined \def \bissue#1{#1}\fi
\ifx \bfpage  \undefined \def \bfpage#1{#1}\fi
\ifx \blpage  \undefined \def \blpage #1{#1}\fi
\ifx \burl  \undefined \def \burl#1{\textsf{#1}}\fi
\ifx \doiurl  \undefined \def \doiurl#1{\textsf{#1}}\fi
\ifx \betal  \undefined \def \betal{\textit{et al.}}\fi
\ifx \binstitute  \undefined \def \binstitute#1{#1}\fi
\ifx \binstitutionaled  \undefined \def \binstitutionaled#1{#1}\fi
\ifx \bctitle  \undefined \def \bctitle#1{#1}\fi
\ifx \beditor  \undefined \def \beditor#1{#1}\fi
\ifx \bpublisher  \undefined \def \bpublisher#1{#1}\fi
\ifx \bbtitle  \undefined \def \bbtitle#1{#1}\fi
\ifx \bedition  \undefined \def \bedition#1{#1}\fi
\ifx \bseriesno  \undefined \def \bseriesno#1{#1}\fi
\ifx \blocation  \undefined \def \blocation#1{#1}\fi
\ifx \bsertitle  \undefined \def \bsertitle#1{#1}\fi
\ifx \bsnm \undefined \def \bsnm#1{#1}\fi
\ifx \bsuffix \undefined \def \bsuffix#1{#1}\fi
\ifx \bparticle \undefined \def \bparticle#1{#1}\fi
\ifx \barticle \undefined \def \barticle#1{#1}\fi
\ifx \bconfdate \undefined \def \bconfdate #1{#1}\fi
\ifx \botherref \undefined \def \botherref #1{#1}\fi
\ifx \url \undefined \def \url#1{\textsf{#1}}\fi
\ifx \bchapter \undefined \def \bchapter#1{#1}\fi
\ifx \bbook \undefined \def \bbook#1{#1}\fi
\ifx \bcomment \undefined \def \bcomment#1{#1}\fi
\ifx \oauthor \undefined \def \oauthor#1{#1}\fi
\ifx \citeauthoryear \undefined \def \citeauthoryear#1{#1}\fi
\ifx \endbibitem  \undefined \def \endbibitem {}\fi
\ifx \bconflocation  \undefined \def \bconflocation#1{#1}\fi
\ifx \arxivurl  \undefined \def \arxivurl#1{\textsf{#1}}\fi
\csname PreBibitemsHook\endcsname

\bibitem{rydberg80a}
\begin{barticle}
\bauthor{\bsnm{Rydberg}, \binits{J.R.}}:
\batitle{{\"U}ber den {B}au der {L}inienspektren der chemischen {G}rundstoffe}.
\bjtitle{Z. Phys. Chem.}
\bvolume{5},
\bfpage{227}
(\byear{1890})
\end{barticle}
\endbibitem

\bibitem{merkt97a}
\begin{barticle}
\bauthor{\bsnm{Merkt}, \binits{F.}}:
\batitle{Molecules in high {R}ydberg states}.
\bjtitle{Ann. Rev. Phys. Chem.}
\bvolume{48},
\bfpage{675}
(\byear{1997})
\end{barticle}
\endbibitem

\bibitem{softley04a}
\begin{barticle}
\bauthor{\bsnm{Softley}, \binits{T.P.}}:
\batitle{Applications of molecular {R}ydberg states in chemical dynamics and
  spectroscopy}.
\bjtitle{Int. Rev. Phys. Chem.}
\bvolume{23},
\bfpage{1}
(\byear{2004})
\end{barticle}
\endbibitem

\bibitem{pauli26a}
\begin{barticle}
\bauthor{\bsnm{Pauli}, \binits{W.}}:
\batitle{{\"U}ber das {W}asserstoffspektrum vom {S}tandpunkt der neuen
  {Q}uantunmechanik}.
\bjtitle{Z. Phys.}
\bvolume{36},
\bfpage{336}
(\byear{1926})
\end{barticle}
\endbibitem

\bibitem{damburg83a}
\begin{bchapter}
\bauthor{\bsnm{Damburg}, \binits{R.J.}},
\bauthor{\bsnm{Kolosov}, \binits{V.V.}}:
\bctitle{Theoretical studies of hydrogen {R}ydberg atoms in electric fields}.
In: \beditor{\bsnm{Stebbings}, \binits{R.F.}},
\beditor{\bsnm{Dunning}, \binits{F.B.}} (eds.)
\bbtitle{Rydberg States of Atoms and Molecules},
p. \bfpage{31}.
\bpublisher{Cambridge University Press},
\blocation{Cambridge}
(\byear{1983})
\end{bchapter}
\endbibitem

\bibitem{osterwalder99a}
\begin{barticle}
\bauthor{\bsnm{Osterwalder}, \binits{A.}},
\bauthor{\bsnm{Merkt}, \binits{F.}}:
\batitle{{Using high Rydberg states as electric field sensors}}.
\bjtitle{Phys. Rev. Lett.}
\bvolume{82},
\bfpage{1831}
(\byear{1999})
\end{barticle}
\endbibitem

\bibitem{bennewitz55a}
\begin{barticle}
\bauthor{\bsnm{Bennewitz}, \binits{H.G.}},
\bauthor{\bsnm{Paul}, \binits{W.}},
\bauthor{\bsnm{Schlier}, \binits{{\mbox{{C}h}}.}}:
\batitle{Fokussierung polarer {M}olek\"ule}.
\bjtitle{Z. Phys.}
\bvolume{41},
\bfpage{6}
(\byear{19551})
\end{barticle}
\endbibitem

\bibitem{gordon55a}
\begin{barticle}
\bauthor{\bsnm{Gordon}, \binits{J.P.}},
\bauthor{\bsnm{Zeiger}, \binits{H.J.}},
\bauthor{\bsnm{Townes}, \binits{C.H.}}:
\batitle{The {M}aser -- {N}ew type of microwave amplifier, frequency standard,
  and spectrometer}.
\bjtitle{Phys. Rev.}
\bvolume{99},
\bfpage{1264}
(\byear{1955})
\end{barticle}
\endbibitem

\bibitem{brooks76a}
\begin{barticle}
\bauthor{\bsnm{Brooks}, \binits{P.R.}}:
\batitle{Reactions of oriented molecules}.
\bjtitle{Science}
\bvolume{193},
\bfpage{11}
(\byear{1976})
\end{barticle}
\endbibitem

\bibitem{stolte82a}
\begin{barticle}
\bauthor{\bsnm{Stolte}, \binits{S.}}:
\batitle{Reactive scattering studies on oriented molecules}.
\bjtitle{Ber. Bunsenges. Phys. Chem.}
\bvolume{86},
\bfpage{413}
(\byear{1982})
\end{barticle}
\endbibitem

\bibitem{parker89a}
\begin{barticle}
\bauthor{\bsnm{Parker}, \binits{D.H.}},
\bauthor{\bsnm{Bernstein}, \binits{R.B.}}:
\batitle{Oriented molecule beams via the electrostatic hexapole: {P}reparation,
  characterization, and reactive scattering}.
\bjtitle{Ann. Rev. Phys. Chem.}
\bvolume{40},
\bfpage{561}
(\byear{1989})
\end{barticle}
\endbibitem

\bibitem{bethlem99a}
\begin{barticle}
\bauthor{\bsnm{Bethlem}, \binits{H.L.}},
\bauthor{\bsnm{Berden}, \binits{G.}},
\bauthor{\bsnm{Meijer}, \binits{G.}}:
\batitle{Decelerating neutral dipolar molecules}.
\bjtitle{Phys. Rev. Lett.}
\bvolume{83},
\bfpage{1558}
(\byear{1999})
\end{barticle}
\endbibitem

\bibitem{bethlem00a}
\begin{barticle}
\bauthor{\bsnm{Bethlem}, \binits{H.L.}},
\bauthor{\bsnm{Berden}, \binits{G.}},
\bauthor{\bsnm{Crompvoets}, \binits{F.M.H.}},
\bauthor{\bsnm{Jongma}, \binits{R.T.}},
\bauthor{\bsnm{{\mbox{van~{R}oij}}}, \binits{A.J.A.}},
\bauthor{\bsnm{Meijer}, \binits{G.}}:
\batitle{Electrostatic trapping of ammonia molecules}.
\bjtitle{Nature}
\bvolume{406},
\bfpage{491}
(\byear{2000})
\end{barticle}
\endbibitem

\bibitem{vandemeerakker12a}
\begin{barticle}
\bauthor{\bparticle{van~de} \bsnm{Meerakker}, \binits{S.Y.T.}},
\bauthor{\bsnm{Bethlem}, \binits{H.L.}},
\bauthor{\bsnm{Vanhaecke}, \binits{N.}},
\bauthor{\bsnm{Meijer}, \binits{G.}}:
\batitle{Manipulation and control of molecular beams}.
\bjtitle{Chem. Rev.}
\bvolume{112},
\bfpage{4828}
(\byear{2012})
\end{barticle}
\endbibitem

\bibitem{gerlach21a}
\begin{barticle}
\bauthor{\bsnm{Gerlach}, \binits{W.}},
\bauthor{\bsnm{Stern}, \binits{O.}}:
\batitle{{Der experimentelle Nachweis des magnetischen Moments des
  Silberatoms}}.
\bjtitle{Z. Phys.}
\bvolume{8},
\bfpage{110}
(\byear{1921})
\end{barticle}
\endbibitem

\bibitem{gerlach22a}
\begin{barticle}
\bauthor{\bsnm{Gerlach}, \binits{W.}},
\bauthor{\bsnm{Stern}, \binits{O.}}:
\batitle{Der experimentelle {N}achweis der {R}ichtungsquantelung im
  {M}agnetfeld}.
\bjtitle{Z. Phys.}
\bvolume{9},
\bfpage{349}
(\byear{1922})
\end{barticle}
\endbibitem

\bibitem{gerlach22b}
\begin{barticle}
\bauthor{\bsnm{Gerlach}, \binits{W.}},
\bauthor{\bsnm{Stern}, \binits{O.}}:
\batitle{{Das magnetische Moment des Silberatoms}}.
\bjtitle{Z. Phys.}
\bvolume{9},
\bfpage{353}
(\byear{1922})
\end{barticle}
\endbibitem

\bibitem{gallagher94a}
\begin{bbook}
\bauthor{\bsnm{Gallagher}, \binits{T.F.}}:
\bbtitle{{Rydberg Atoms}}.
\bpublisher{Cambridge University Press},
\blocation{Cambridge}
(\byear{1994})
\end{bbook}
\endbibitem

\bibitem{shuman10a}
\begin{barticle}
\bauthor{\bsnm{Shuman}, \binits{E.S.}},
\bauthor{\bsnm{Barry}, \binits{J.F.}},
\bauthor{\bsnm{DeMille}, \binits{D.}}:
\batitle{Laser cooling of a diatomic molecule}.
\bjtitle{Nature}
\bvolume{467},
\bfpage{820}
(\byear{2010})
\end{barticle}
\endbibitem

\bibitem{zhelyazkova14a}
\begin{barticle}
\bauthor{\bsnm{Zhelyazkova}, \binits{V.}},
\bauthor{\bsnm{Cournol}, \binits{A.}},
\bauthor{\bsnm{Wall}, \binits{T.E.}},
\bauthor{\bsnm{Matsushima}, \binits{A.}},
\bauthor{\bsnm{Hudson}, \binits{J.J.}},
\bauthor{\bsnm{Hinds}, \binits{E.A.}},
\bauthor{\bsnm{Tarbutt}, \binits{M.R.}},
\bauthor{\bsnm{Sauer}, \binits{B.E.}}:
\batitle{Laser cooling and slowing of {CaF} molecules}.
\bjtitle{Phys. Rev. A}
\bvolume{89},
\bfpage{053416}
(\byear{2014})
\end{barticle}
\endbibitem

\bibitem{vanhaecke07a}
\begin{barticle}
\bauthor{\bsnm{Vanhaecke}, \binits{N.}},
\bauthor{\bsnm{Meier}, \binits{U.}},
\bauthor{\bsnm{Andrist}, \binits{M.}},
\bauthor{\bsnm{Meier}, \binits{B.H.}},
\bauthor{\bsnm{Merkt}, \binits{F.}}:
\batitle{Multistage {Z}eeman deceleration of hydrogen atoms}.
\bjtitle{Phys. Rev. A}
\bvolume{75},
\bfpage{031402}
(\byear{2007})
\end{barticle}
\endbibitem

\bibitem{hogan07a}
\begin{barticle}
\bauthor{\bsnm{Hogan}, \binits{S.D.}},
\bauthor{\bsnm{Sprecher}, \binits{D.}},
\bauthor{\bsnm{Andrist}, \binits{M.}},
\bauthor{\bsnm{Vanhaecke}, \binits{N.}},
\bauthor{\bsnm{Merkt}, \binits{F.}}:
\batitle{{Z}eeman deceleration of {H} and {D}}.
\bjtitle{Phys. Rev. A}
\bvolume{76},
\bfpage{023412}
(\byear{2007})
\end{barticle}
\endbibitem

\bibitem{narevicius08a}
\begin{barticle}
\bauthor{\bsnm{Narevicius}, \binits{E.}},
\bauthor{\bsnm{Libson}, \binits{A.}},
\bauthor{\bsnm{Parthey}, \binits{C.G.}},
\bauthor{\bsnm{Chavez}, \binits{I.}},
\bauthor{\bsnm{Narevicius}, \binits{J.}},
\bauthor{\bsnm{Even}, \binits{U.}},
\bauthor{\bsnm{Raizen}, \binits{M.G.}}:
\batitle{Stopping supersonic beams with a series of pulsed electromagnetic
  coils: {A}n atomic coilgun}.
\bjtitle{Phys. Rev. Lett.}
\bvolume{100},
\bfpage{093003}
(\byear{2008})
\end{barticle}
\endbibitem

\bibitem{alramadhan94a}
\begin{barticle}
\bauthor{\bsnm{Al-Ramadhan}, \binits{A.H.}},
\bauthor{\bsnm{Gidley}, \binits{D.W.}}:
\batitle{New precision measurement of the decay rate of singlet positronium}.
\bjtitle{Phys. Rev. Lett.}
\bvolume{72},
\bfpage{1632}
(\byear{1994})
\end{barticle}
\endbibitem

\bibitem{vallery03a}
\begin{barticle}
\bauthor{\bsnm{Vallery}, \binits{R.S.}},
\bauthor{\bsnm{Zitzewitz}, \binits{P.W.}},
\bauthor{\bsnm{Gidley}, \binits{D.W.}}:
\batitle{Resolution of the orthopositronium-lifetime puzzle}.
\bjtitle{Phys. Rev. Lett.}
\bvolume{90},
\bfpage{203402}
(\byear{2003})
\end{barticle}
\endbibitem

\bibitem{bell09a}
\begin{barticle}
\bauthor{\bsnm{Bell}, \binits{M.T.}},
\bauthor{\bsnm{Softley}, \binits{T.P.}}:
\batitle{Ultracold molecules and ultracold chemistry}.
\bjtitle{Mol. Phys.}
\bvolume{107},
\bfpage{99}
(\byear{2009})
\end{barticle}
\endbibitem

\bibitem{carr09a}
\begin{barticle}
\bauthor{\bsnm{Carr}, \binits{L.C.}},
\bauthor{\bsnm{DeMille}, \binits{D.}},
\bauthor{\bsnm{Krems}, \binits{R.V.}},
\bauthor{\bsnm{Ye}, \binits{J.}}:
\batitle{Cold and ultracold molecules: {S}cience, technology and applications}.
\bjtitle{New. J. Phys.}
\bvolume{11},
\bfpage{055049}
(\byear{2009})
\end{barticle}
\endbibitem

\bibitem{fulton04a}
\begin{barticle}
\bauthor{\bsnm{Fulton}, \binits{R.}},
\bauthor{\bsnm{Bishop}, \binits{A.I.}},
\bauthor{\bsnm{Barker}, \binits{P.F.}}:
\batitle{Optical {S}tark decelerator for molecules}.
\bjtitle{Phys. Rev. Lett.}
\bvolume{93},
\bfpage{243004}
(\byear{2004})
\end{barticle}
\endbibitem

\bibitem{messer84a}
\begin{barticle}
\bauthor{\bsnm{Messer}, \binits{J.K.}},
\bauthor{\bsnm{{De~Lucia}}, \binits{F.C.}}:
\batitle{Measurement of pressure-broadening parameters for the {CO}-{H}e system
  at 4~{K}}.
\bjtitle{Phys. Rev. Lett.}
\bvolume{53},
\bfpage{2555}
(\byear{1984})
\end{barticle}
\endbibitem

\bibitem{doyle95a}
\begin{barticle}
\bauthor{\bsnm{Doyle}, \binits{J.M.}},
\bauthor{\bsnm{Friedrich}, \binits{B.}},
\bauthor{\bsnm{Kim}, \binits{J.}},
\bauthor{\bsnm{Patterson}, \binits{D.}}:
\batitle{Buffer-gas loading of atoms and molecules into a magnetic trap}.
\bjtitle{Phys. Rev. A}
\bvolume{52},
\bfpage{2515}
(\byear{1995})
\end{barticle}
\endbibitem

\bibitem{willitsch08a}
\begin{barticle}
\bauthor{\bsnm{Willitsch}, \binits{S.}},
\bauthor{\bsnm{Bell}, \binits{M.T.}},
\bauthor{\bsnm{Gingell}, \binits{A.D.}},
\bauthor{\bsnm{Procter}, \binits{S.R.}},
\bauthor{\bsnm{Softley}, \binits{T.P.}}:
\batitle{Cold reactive collisions between laser-cooled ions and
  velocity-selected neutral molecules}.
\bjtitle{Phys. Rev. Lett.}
\bvolume{100},
\bfpage{043203}
(\byear{2008})
\end{barticle}
\endbibitem

\bibitem{staanum08a}
\begin{barticle}
\bauthor{\bsnm{Staanum}, \binits{P.F.}},
\bauthor{\bsnm{H\o{}jbjerre}, \binits{K.}},
\bauthor{\bsnm{Wester}, \binits{R.}},
\bauthor{\bsnm{Drewsen}, \binits{M.}}:
\batitle{Probing isotope effects in chemical reactions using single ions}.
\bjtitle{Phys. Rev. Lett.}
\bvolume{100},
\bfpage{243003}
(\byear{2008})
\end{barticle}
\endbibitem

\bibitem{choi06a}
\begin{barticle}
\bauthor{\bsnm{Choi}, \binits{J.-H.}},
\bauthor{\bsnm{Guest}, \binits{J.R.}},
\bauthor{\bsnm{Povilus}, \binits{A.P.}},
\bauthor{\bsnm{Hansis}, \binits{E.}},
\bauthor{\bsnm{Raithel}, \binits{G.}}:
\batitle{Magnetic trapping of long-lived cold {R}ydberg atoms}.
\bjtitle{Phys. Rev. Lett.}
\bvolume{95},
\bfpage{243001}
(\byear{2005})
\end{barticle}
\endbibitem

\bibitem{amoretti02a}
\begin{barticle}
\bauthor{\bsnm{Amoretti}, \binits{M.}},
\bauthor{\bsnm{Amsler}, \binits{C.}},
\bauthor{\bsnm{Bonomi}, \binits{G.}},
\bauthor{\bsnm{Bouchta}, \binits{A.}},
\bauthor{\bsnm{Bowe}, \binits{P.}},
\bauthor{\bsnm{Carraro}, \binits{C.}},
\bauthor{\bsnm{Cesar}, \binits{C.L.}},
\bauthor{\bsnm{Charlton}, \binits{M.}},
\bauthor{\bsnm{Collier}, \binits{M.J.T.}},
\bauthor{\bsnm{Doser}, \binits{M.}},
\bauthor{\bsnm{Filippini}, \binits{V.}},
\bauthor{\bsnm{Fine}, \binits{K.S.}},
\bauthor{\bsnm{Fontana}, \binits{A.}},
\bauthor{\bsnm{Fujiwara}, \binits{M.C.}},
\bauthor{\bsnm{Funakoshi}, \binits{R.}},
\bauthor{\bsnm{Genova}, \binits{P.}},
\bauthor{\bsnm{Hangst}, \binits{J.S.}},
\bauthor{\bsnm{Hayano}, \binits{R.S.}},
\bauthor{\bsnm{Holzscheiter}, \binits{M.H.}},
\bauthor{\bsnm{Jorgensen}, \binits{L.V.}},
\bauthor{\bsnm{Lagomarsino}, \binits{V.}},
\bauthor{\bsnm{Landua}, \binits{R.}},
\bauthor{\bsnm{Lindelof}, \binits{D.}},
\bauthor{\bsnm{Rizzini}, \binits{E.L.}},
\bauthor{\bsnm{Macri}, \binits{M.}},
\bauthor{\bsnm{Madsen}, \binits{N.}},
\bauthor{\bsnm{Manuzio}, \binits{G.}},
\bauthor{\bsnm{Marchesotti}, \binits{M.}},
\bauthor{\bsnm{Montagna}, \binits{P.}},
\bauthor{\bsnm{Pruys}, \binits{H.}},
\bauthor{\bsnm{Regenfus}, \binits{C.}},
\bauthor{\bsnm{Riedler}, \binits{P.}},
\bauthor{\bsnm{Rochet}, \binits{J.}},
\bauthor{\bsnm{Rotondi}, \binits{A.}},
\bauthor{\bsnm{Rouleau}, \binits{G.}},
\bauthor{\bsnm{Testera}, \binits{G.}},
\bauthor{\bsnm{Variola}, \binits{A.}},
\bauthor{\bsnm{Watson}, \binits{T.L.}},
\bauthor{\bparticle{van~der} \bsnm{Werf}, \binits{D.P.}}:
\batitle{Production and detection of cold antihydrogen atoms}.
\bjtitle{Nature}
\bvolume{419},
\bfpage{456}
(\byear{2002})
\end{barticle}
\endbibitem

\bibitem{anderson11a}
\begin{barticle}
\bauthor{\bsnm{Anderson}, \binits{S.E.}},
\bauthor{\bsnm{Younge}, \binits{K.C.}},
\bauthor{\bsnm{Raithel}, \binits{G.}}:
\batitle{Trapping {R}ydberg atoms in an optical lattice}.
\bjtitle{Phys. Rev. Lett.}
\bvolume{107},
\bfpage{263001}
(\byear{2011})
\end{barticle}
\endbibitem

\bibitem{stepkin07a}
\begin{barticle}
\bauthor{\bsnm{Stepkin}, \binits{S.V.}},
\bauthor{\bsnm{Konovalenko}, \binits{A.A.}},
\bauthor{\bsnm{Kantharia}, \binits{N.G.}},
\bauthor{\bsnm{{Udaya Shankar}}, \binits{N.}}:
\batitle{Radio recombination lines from the largest bound atoms in space}.
\bjtitle{Mon. Not. R. Astron. Soc.}
\bvolume{374},
\bfpage{852}
(\byear{2007})
\end{barticle}
\endbibitem

\bibitem{wayne93a}
\begin{barticle}
\bauthor{\bsnm{Wayne}, \binits{R.P.}}:
\batitle{Photodissociation dynamics and atmospheric chemistry}.
\bjtitle{J. Geophys. Res.}
\bvolume{98},
\bfpage{13119}
(\byear{1993})
\end{barticle}
\endbibitem

\bibitem{merkt02a}
\begin{barticle}
\bauthor{\bsnm{Merkt}, \binits{F.}},
\bauthor{\bsnm{Osterwalder}, \binits{A.}}:
\batitle{Millimeter wave spectroscopy of high {R}ydberg states}.
\bjtitle{Int. Rev. Phys. Chem.}
\bvolume{21},
\bfpage{385}
(\byear{2002})
\end{barticle}
\endbibitem

\bibitem{hogan12a}
\begin{barticle}
\bauthor{\bsnm{Hogan}, \binits{S.D.}},
\bauthor{\bsnm{Agner}, \binits{J.A.}},
\bauthor{\bsnm{Merkt}, \binits{F.}},
\bauthor{\bsnm{Thiele}, \binits{T.}},
\bauthor{\bsnm{Filipp}, \binits{S.}},
\bauthor{\bsnm{Wallraff}, \binits{A.}}:
\batitle{Driving {R}ydberg-{R}ydberg transitions from a coplanar microwave
  waveguide}.
\bjtitle{Phys. Rev. Lett.}
\bvolume{108},
\bfpage{063004}
(\byear{2012})
\end{barticle}
\endbibitem

\bibitem{thiele14a}
\begin{barticle}
\bauthor{\bsnm{Thiele}, \binits{T.}},
\bauthor{\bsnm{Filipp}, \binits{S.}},
\bauthor{\bsnm{Agner}, \binits{J.A.}},
\bauthor{\bsnm{Schmutz}, \binits{H.}},
\bauthor{\bsnm{Deiglmayr}, \binits{J.}},
\bauthor{\bsnm{Stammeier}, \binits{M.}},
\bauthor{\bsnm{Allmendinger}, \binits{P.}},
\bauthor{\bsnm{Merkt}, \binits{F.}},
\bauthor{\bsnm{Wallraff}, \binits{A.}}:
\batitle{Manipulating {R}ydberg atoms close to surfaces at cryogenic
  temperatures}.
\bjtitle{Phys. Rev. A}
\bvolume{90},
\bfpage{013414}
(\byear{2014})
\end{barticle}
\endbibitem

\bibitem{osterwalder04a}
\begin{barticle}
\bauthor{\bsnm{Osterwalder}, \binits{A.}},
\bauthor{\bsnm{W{\"u}est}, \binits{A.}},
\bauthor{\bsnm{Merkt}, \binits{F.}},
\bauthor{\bsnm{Jungen}, \binits{{\mbox{Ch}}.}}:
\batitle{High-resolution millimeter wave spectroscopy and multichannel quantum
  defect theory of the hyperfine structure in high {R}ydberg states of
  molecular hydrogen {H$_2$}}.
\bjtitle{J. Chem. Phys.}
\bvolume{121},
\bfpage{11810}
(\byear{2004})
\end{barticle}
\endbibitem

\bibitem{liu09b}
\begin{barticle}
\bauthor{\bsnm{Liu}, \binits{J.}},
\bauthor{\bsnm{Salumbides}, \binits{E.J.}},
\bauthor{\bsnm{Hollenstein}, \binits{U.}},
\bauthor{\bsnm{Koelemeij}, \binits{J.C.J.}},
\bauthor{\bsnm{Eikema}, \binits{K.S.E.}},
\bauthor{\bsnm{Ubachs}, \binits{W.}},
\bauthor{\bsnm{Merkt}, \binits{F.}}:
\batitle{Determination of the ionization and dissociation energies of the
  hydrogen molecule}.
\bjtitle{J. Chem. Phys.}
\bvolume{130},
\bfpage{174306}
(\byear{2009})
\end{barticle}
\endbibitem

\bibitem{liu10a}
\begin{barticle}
\bauthor{\bsnm{Liu}, \binits{J.}},
\bauthor{\bsnm{Sprecher}, \binits{D.}},
\bauthor{\bsnm{Jungen}, \binits{{\mbox{Ch}}.}},
\bauthor{\bsnm{Ubachs}, \binits{W.}},
\bauthor{\bsnm{Merkt}, \binits{F.}}:
\batitle{Determination of the ionization and dissociation energies of the
  deuterium molecule {(D$_2$)}}.
\bjtitle{J. Chem. Phys.}
\bvolume{132},
\bfpage{154301}
(\byear{2010})
\end{barticle}
\endbibitem

\bibitem{sprecher10a}
\begin{barticle}
\bauthor{\bsnm{Sprecher}, \binits{D.}},
\bauthor{\bsnm{Liu}, \binits{J.}},
\bauthor{\bsnm{Jungen}, \binits{{\mbox{Ch}}.}},
\bauthor{\bsnm{Ubachs}, \binits{W.}},
\bauthor{\bsnm{Merkt}, \binits{F.}}:
\batitle{The ionization and dissociation energies of {HD}}.
\bjtitle{J. Chem. Phys.}
\bvolume{133},
\bfpage{111102}
(\byear{2010})
\end{barticle}
\endbibitem

\bibitem{ramsey50a}
\begin{barticle}
\bauthor{\bsnm{Ramsey}, \binits{N.F.}}:
\batitle{A molecular beam resonance method with separated oscillating fields}.
\bjtitle{Phys. Rev.}
\bvolume{78},
\bfpage{695}
(\byear{1950})
\end{barticle}
\endbibitem

\bibitem{henson12a}
\begin{barticle}
\bauthor{\bsnm{Henson}, \binits{A.B.}},
\bauthor{\bsnm{Gersten}, \binits{S.}},
\bauthor{\bsnm{Shagam}, \binits{Y.}},
\bauthor{\bsnm{Narevicius}, \binits{J.}},
\bauthor{\bsnm{Narevicius}, \binits{E.}}:
\batitle{Observation of resonances in {P}enning ionization reactions at
  sub-{K}elvin temperatures in merged beams}.
\bjtitle{Science}
\bvolume{338},
\bfpage{234}
(\byear{2012})
\end{barticle}
\endbibitem

\bibitem{shagam13a}
\begin{barticle}
\bauthor{\bsnm{Shagam}, \binits{Y.}},
\bauthor{\bsnm{E.~Narevicius}, \binits{E.}}:
\batitle{Sub-{K}elvin collision temperatures in merged neutral beams by
  correlation in phase-space}.
\bjtitle{J. Phys. Chem. C}
\bvolume{117},
\bfpage{22454}
(\byear{2013})
\end{barticle}
\endbibitem

\bibitem{hinds97a}
\begin{barticle}
\bauthor{\bsnm{Hinds}, \binits{E.A.}},
\bauthor{\bsnm{Lai}, \binits{K.S.}},
\bauthor{\bsnm{Schnell}, \binits{M.}}:
\batitle{Atoms in micron-sized metallic and dielectric waveguides}.
\bjtitle{Phil. Trans. R. Soc. Lond. A}
\bvolume{355},
\bfpage{2353}
(\byear{1997})
\end{barticle}
\endbibitem

\bibitem{kubler10a}
\begin{barticle}
\bauthor{\bsnm{K\"ubler}, \binits{H.}},
\bauthor{\bsnm{Shaffer}, \binits{J.P.}},
\bauthor{\bsnm{Baluktsian}, \binits{T.}},
\bauthor{\bsnm{L\"ow}, \binits{R.}},
\bauthor{\bsnm{Pfau}, \binits{T.}}:
\batitle{Coherent excitation of {R}ydberg atoms in micrometre-sized atomic
  vapour cells}.
\bjtitle{Nature Photon.}
\bvolume{4},
\bfpage{112}
(\byear{2010})
\end{barticle}
\endbibitem

\bibitem{gray88a}
\begin{barticle}
\bauthor{\bsnm{Gray}, \binits{D.F.}},
\bauthor{\bsnm{Zheng}, \binits{Z.}},
\bauthor{\bsnm{Smith}, \binits{K.A.}},
\bauthor{\bsnm{Dunning}, \binits{F.B.}}:
\batitle{Ionization of {K}($n$d) {R}ydberg-state atoms at a surface}.
\bjtitle{Phys. Rev. A}
\bvolume{38},
\bfpage{1601}
(\byear{1988})
\end{barticle}
\endbibitem

\bibitem{anderson88a}
\begin{barticle}
\bauthor{\bsnm{Anderson}, \binits{A.}},
\bauthor{\bsnm{Haroche}, \binits{S.}},
\bauthor{\bsnm{Hinds}, \binits{E.A.}},
\bauthor{\bsnm{Jhe}, \binits{W.}},
\bauthor{\bsnm{Meschede}, \binits{D.}}:
\batitle{Measuring the van~der~{W}aals forces between a {R}ydberg atom and a
  metallic surface}.
\bjtitle{Phys. Rev. A}
\bvolume{37},
\bfpage{3594}
(\byear{1988})
\end{barticle}
\endbibitem

\bibitem{sandoghdar92a}
\begin{barticle}
\bauthor{\bsnm{Sandoghdar}, \binits{V.}},
\bauthor{\bsnm{Sukenik}, \binits{C.I.}},
\bauthor{\bsnm{Hinds}, \binits{E.A.}},
\bauthor{\bsnm{Haroche}, \binits{S.}}:
\batitle{Direct measurement of the van~der~{W}aals interaction between an atom
  and its images in a micron-sized cavity}.
\bjtitle{Phys. Rev. Lett.}
\bvolume{68},
\bfpage{3432}
(\byear{1992})
\end{barticle}
\endbibitem

\bibitem{hill00a}
\begin{barticle}
\bauthor{\bsnm{Hill}, \binits{S.B.}},
\bauthor{\bsnm{Haich}, \binits{C.B.}},
\bauthor{\bsnm{Zhou}, \binits{Z.}},
\bauthor{\bsnm{Nordlander}, \binits{P.}},
\bauthor{\bsnm{Dunning}, \binits{F.B.}}:
\batitle{Ionization of xenon {R}ydberg atoms at a metal surface}.
\bjtitle{Phys. Rev. Lett.}
\bvolume{85},
\bfpage{5444}
(\byear{2000})
\end{barticle}
\endbibitem

\bibitem{lloyd05a}
\begin{barticle}
\bauthor{\bsnm{Lloyd}, \binits{G.R.}},
\bauthor{\bsnm{Procter}, \binits{S.R.}},
\bauthor{\bsnm{Softley}, \binits{T.P.}}:
\batitle{Ionization of hydrogen {R}ydberg molecules at a metal surface}.
\bjtitle{Phys. Rev. Lett.}
\bvolume{95},
\bfpage{133202}
(\byear{2005})
\end{barticle}
\endbibitem

\bibitem{so11a}
\begin{barticle}
\bauthor{\bsnm{So}, \binits{E.}},
\bauthor{\bsnm{Dethlefsen}, \binits{M.}},
\bauthor{\bsnm{Ford}, \binits{M.}},
\bauthor{\bsnm{Softley}, \binits{T.P.}}:
\batitle{Charge transfer of {R}ydberg {H} atoms at a metal surface}.
\bjtitle{Phys. Rev. Lett.}
\bvolume{107},
\bfpage{093201}
(\byear{2011})
\end{barticle}
\endbibitem

\bibitem{gibbard15a}
\begin{barticle}
\bauthor{\bsnm{Gibbard}, \binits{J.A.}},
\bauthor{\bsnm{Dethlefsen}, \binits{M.}},
\bauthor{\bsnm{Kohlhoff}, \binits{M.}},
\bauthor{\bsnm{Rennick}, \binits{C.J.}},
\bauthor{\bsnm{So}, \binits{E.}},
\bauthor{\bsnm{Ford}, \binits{M.}},
\bauthor{\bsnm{Softley}, \binits{T.P.}}:
\batitle{Resonant charge transfer of hydrogen {R}ydberg atoms incident on a
  {C}u(100) projected band-gap surface}.
\bjtitle{Phys. Rev. Lett.}
\bvolume{115},
\bfpage{093201}
(\byear{2015})
\end{barticle}
\endbibitem

\bibitem{bendkowsky09a}
\begin{barticle}
\bauthor{\bsnm{Bendkowsky}, \binits{V.}},
\bauthor{\bsnm{Butscher}, \binits{B.}},
\bauthor{\bsnm{Nipper}, \binits{J.}},
\bauthor{\bsnm{Shaffer}, \binits{J.P.}},
\bauthor{\bsnm{L\"ow}, \binits{R.}},
\bauthor{\bsnm{Pfau}, \binits{T.}}:
\batitle{Observation of ultralong-range {R}ydberg molecules}.
\bjtitle{Nature}
\bvolume{58},
\bfpage{1005}
(\byear{2009})
\end{barticle}
\endbibitem

\bibitem{gallagher08a}
\begin{barticle}
\bauthor{\bsnm{Gallagher}, \binits{T.F.}},
\bauthor{\bsnm{Pillet}, \binits{P.}}:
\batitle{Dipole-dipole interactions of {R}ydberg atoms}.
\bjtitle{Adv. At. Mol. Opt. Phys.}
\bvolume{56},
\bfpage{161}
(\byear{2008})
\end{barticle}
\endbibitem

\bibitem{comparat10a}
\begin{barticle}
\bauthor{\bsnm{Comparat}, \binits{D.}},
\bauthor{\bsnm{P.~Pillet}, \binits{P.}}:
\batitle{Dipole blockade in a cold {R}ydberg atomic sample}.
\bjtitle{J. Opt. Soc. Am. B}
\bvolume{27},
\bfpage{208}
(\byear{2010})
\end{barticle}
\endbibitem

\bibitem{anderson98a}
\begin{barticle}
\bauthor{\bsnm{Anderson}, \binits{W.R.}},
\bauthor{\bsnm{Veale}, \binits{J.R.}},
\bauthor{\bsnm{Gallagher}, \binits{T.F.}}:
\batitle{Resonant dipole-dipole energy transfer in a nearly frozen {R}ydberg
  gas}.
\bjtitle{Phys. Rev. Lett.}
\bvolume{80},
\bfpage{249}
(\byear{1998})
\end{barticle}
\endbibitem

\bibitem{mourachko98a}
\begin{barticle}
\bauthor{\bsnm{Mourachko}, \binits{I.}},
\bauthor{\bsnm{Comparat}, \binits{D.}},
\bauthor{\bparticle{de} \bsnm{Tomasi}, \binits{F.}},
\bauthor{\bsnm{Fioretti}, \binits{A.}},
\bauthor{\bsnm{Nosbaum}, \binits{P.}},
\bauthor{\bsnm{Akulin}, \binits{V.M.}},
\bauthor{\bsnm{Pillet}, \binits{P.}}:
\batitle{Many-body effects in a frozen {R}ydberg gas}.
\bjtitle{Phys. Rev. Lett.}
\bvolume{80},
\bfpage{253}
(\byear{1998})
\end{barticle}
\endbibitem

\bibitem{vogt06a}
\begin{barticle}
\bauthor{\bsnm{Vogt}, \binits{T.}},
\bauthor{\bsnm{Viteau}, \binits{M.}},
\bauthor{\bsnm{Zhao}, \binits{J.}},
\bauthor{\bsnm{Chotia}, \binits{A.}},
\bauthor{\bsnm{Comparat}, \binits{D.}},
\bauthor{\bsnm{Pillet}, \binits{P.}}:
\batitle{Dipole blockade at {F}\"orster resonances in high resolution laser
  excitation of {R}ydberg states of cesium atoms}.
\bjtitle{Phys. Rev. Lett.}
\bvolume{97},
\bfpage{083003}
(\byear{2006})
\end{barticle}
\endbibitem

\bibitem{pritchard10a}
\begin{barticle}
\bauthor{\bsnm{Pritchard}, \binits{J.D.}},
\bauthor{\bsnm{Maxwell}, \binits{D.}},
\bauthor{\bsnm{Gauguet}, \binits{A.}},
\bauthor{\bsnm{Weatherill}, \binits{K.J.}},
\bauthor{\bsnm{Jones}, \binits{M.P.A.}},
\bauthor{\bsnm{Adams}, \binits{C.S.}}:
\batitle{Cooperative atom-light interaction in a blockaded {R}ydberg ensemble}.
\bjtitle{Phys. Rev. Lett.}
\bvolume{105},
\bfpage{193603}
(\byear{2010})
\end{barticle}
\endbibitem

\bibitem{wilk10a}
\begin{barticle}
\bauthor{\bsnm{Wilk}, \binits{T.}},
\bauthor{\bsnm{Ga\"etan}, \binits{A.}},
\bauthor{\bsnm{Evellin}, \binits{C.}},
\bauthor{\bsnm{Wolters}, \binits{J.}},
\bauthor{\bsnm{Miroshnychenko}, \binits{Y.}},
\bauthor{\bsnm{Grangier}, \binits{P.}},
\bauthor{\bsnm{Browaeys}, \binits{A.}}:
\batitle{Entanglement of two individual neutral atoms using {R}ydberg
  blockade}.
\bjtitle{Phys. Rev. Lett.}
\bvolume{104},
\bfpage{010502}
(\byear{2010})
\end{barticle}
\endbibitem

\bibitem{isenhower10a}
\begin{barticle}
\bauthor{\bsnm{Isenhower}, \binits{L.}},
\bauthor{\bsnm{Urban}, \binits{E.}},
\bauthor{\bsnm{Zhang}, \binits{X.L.}},
\bauthor{\bsnm{Gill}, \binits{A.T.}},
\bauthor{\bsnm{Henage}, \binits{T.}},
\bauthor{\bsnm{Johnson}, \binits{T.A.}},
\bauthor{\bsnm{Walker}, \binits{T.G.}},
\bauthor{\bsnm{Saffman}, \binits{M.}}:
\batitle{Demonstration of a neutral atom controlled-{NOT} quantum gate}.
\bjtitle{Phys. Rev. Lett.}
\bvolume{104},
\bfpage{010503}
(\byear{2010})
\end{barticle}
\endbibitem

\bibitem{smith78a}
\begin{barticle}
\bauthor{\bsnm{Smith}, \binits{K.A.}},
\bauthor{\bsnm{Kellert}, \binits{F.G.}},
\bauthor{\bsnm{Rundel}, \binits{R.D.}},
\bauthor{\bsnm{Dunning}, \binits{F.B.}},
\bauthor{\bsnm{Stebbings}, \binits{R.F.}}:
\batitle{Discrete energy transfer in collisions of {Xe}($n\mathrm{f}$)
  {R}ydberg atoms with {NH}$_{3}$ molecules}.
\bjtitle{Phys. Rev. Lett.}
\bvolume{40},
\bfpage{1362}
(\byear{1978})
\end{barticle}
\endbibitem

\bibitem{green00a}
\begin{barticle}
\bauthor{\bsnm{Greene}, \binits{C.H.}},
\bauthor{\bsnm{Dickinson}, \binits{A.S.}},
\bauthor{\bsnm{Sadeghpour}, \binits{H.R.}}:
\batitle{Creation of polar and nonpolar ultra-long-range {R}ydberg molecules}.
\bjtitle{Phys. Rev. Lett.}
\bvolume{85},
\bfpage{2458}
(\byear{2000})
\end{barticle}
\endbibitem

\bibitem{walther06a}
\begin{barticle}
\bauthor{\bsnm{Walther}, \binits{H.}},
\bauthor{\bsnm{Varcoe}, \binits{B.T.H.}},
\bauthor{\bsnm{Englert}, \binits{B.-G.}},
\bauthor{\bsnm{Becker}, \binits{T.}}:
\batitle{Cavity quantum electrodynamics}.
\bjtitle{Rep. Prog. Phys.}
\bvolume{69},
\bfpage{1325}
(\byear{2006})
\end{barticle}
\endbibitem

\bibitem{wallraff04a}
\begin{barticle}
\bauthor{\bsnm{Wallraff}, \binits{A.}},
\bauthor{\bsnm{Schuster}, \binits{D.I.}},
\bauthor{\bsnm{Blais}, \binits{A.}},
\bauthor{\bsnm{Frunzio}, \binits{L.}},
\bauthor{\bsnm{Huang}, \binits{R.-S.}},
\bauthor{\bsnm{Majer}, \binits{J.}},
\bauthor{\bsnm{Kumar}, \binits{S.}},
\bauthor{\bsnm{Girvin}, \binits{S.M.}},
\bauthor{\bsnm{Schoelkopf}, \binits{R.J.}}:
\batitle{Strong coupling of a single photon to a superconducting qubit using
  circuit quantum electrodynamics}.
\bjtitle{Nature}
\bvolume{431},
\bfpage{162}
(\byear{2004})
\end{barticle}
\endbibitem

\bibitem{rabl06a}
\begin{barticle}
\bauthor{\bsnm{Rabl}, \binits{P.}},
\bauthor{\bsnm{DeMille}, \binits{D.}},
\bauthor{\bsnm{Doyle}, \binits{J.M.}},
\bauthor{\bsnm{Lukin}, \binits{M.D.}},
\bauthor{\bsnm{Schoelkopf}, \binits{R.J.}},
\bauthor{\bsnm{Zoller}, \binits{P.}}:
\batitle{Hybrid quantum processors: {M}olecular ensembles as quantum memory for
  solid state circuits}.
\bjtitle{Phys. Rev. Lett.}
\bvolume{97},
\bfpage{033003}
(\byear{2006})
\end{barticle}
\endbibitem

\bibitem{carter12a}
\begin{barticle}
\bauthor{\bsnm{Carter}, \binits{J.D.}},
\bauthor{\bsnm{Cherry}, \binits{O.}},
\bauthor{\bsnm{Martin}, \binits{J.D.D.}}:
\batitle{Electric-field sensing near the surface microstructure of an atom chip
  using cold {R}ydberg atoms}.
\bjtitle{Phys. Rev. A}
\bvolume{86},
\bfpage{053401}
(\byear{2012})
\end{barticle}
\endbibitem

\bibitem{hermann14a}
\begin{barticle}
\bauthor{\bsnm{Hermann-Avigliano}, \binits{C.}},
\bauthor{\bsnm{Teixeira}, \binits{R.C.}},
\bauthor{\bsnm{Nguyen}, \binits{T.L.}},
\bauthor{\bsnm{Cantat-Moltrecht}, \binits{T.}},
\bauthor{\bsnm{Nogues}, \binits{G.}},
\bauthor{\bsnm{Dotsenko}, \binits{I.}},
\bauthor{\bsnm{Gleyzes}, \binits{S.}},
\bauthor{\bsnm{Raimond}, \binits{J.M.}},
\bauthor{\bsnm{Haroche}, \binits{S.}},
\bauthor{\bsnm{Brune}, \binits{M.}}:
\batitle{Long coherence times for {R}ydberg qubits on a superconducting atom
  chip}.
\bjtitle{Phys. Rev. A}
\bvolume{90},
\bfpage{040502}
(\byear{2014})
\end{barticle}
\endbibitem

\bibitem{wall15a}
\begin{barticle}
\bauthor{\bsnm{Wall}, \binits{T.E.}},
\bauthor{\bsnm{Alonso}, \binits{A.M.}},
\bauthor{\bsnm{Cooper}, \binits{B.S.}},
\bauthor{\bsnm{Deller}, \binits{A.}},
\bauthor{\bsnm{Hogan}, \binits{S.D.}},
\bauthor{\bsnm{Cassidy}, \binits{D.B.}}:
\batitle{Selective production of {R}ydberg-{S}tark states of positronium}.
\bjtitle{Phys. Rev. Lett.}
\bvolume{114},
\bfpage{173001}
(\byear{2015})
\end{barticle}
\endbibitem

\bibitem{humberston87a}
\begin{barticle}
\bauthor{\bsnm{Humberston}, \binits{J.W.}},
\bauthor{\bsnm{Charlton}, \binits{M.}},
\bauthor{\bsnm{Jacobson}, \binits{F.M.}},
\bauthor{\bsnm{Deutch}, \binits{B.I.}}:
\batitle{On antihydrogen formation in collisions of antiprotons with
  positronium}.
\bjtitle{J. Phys. B.: At. Mol. Opt. Phys.}
\bvolume{20},
\bfpage{25}
(\byear{1987})
\end{barticle}
\endbibitem

\bibitem{charlton90a}
\begin{barticle}
\bauthor{\bsnm{Charlton}, \binits{M.}}:
\batitle{Antihydrogen production in collisions of antiprotons with excited
  states of positronium}.
\bjtitle{Phys. Lett. A}
\bvolume{143},
\bfpage{143}
(\byear{1990})
\end{barticle}
\endbibitem

\bibitem{storry04a}
\begin{barticle}
\bauthor{\bsnm{Storry}, \binits{C.H.}},
\bauthor{\bsnm{Speck}, \binits{A.}},
\bauthor{\bsnm{Sage}, \binits{D.L.}},
\bauthor{\bsnm{Guise}, \binits{N.}},
\bauthor{\bsnm{Gabrielse}, \binits{G.}},
\bauthor{\bsnm{Grzonka}, \binits{D.}},
\bauthor{\bsnm{Oelert}, \binits{W.}},
\bauthor{\bsnm{Schepers}, \binits{G.}},
\bauthor{\bsnm{Sefzick}, \binits{T.}},
\bauthor{\bsnm{Pittner}, \binits{H.}},
\bauthor{\bsnm{Herrmann}, \binits{M.}},
\bauthor{\bsnm{Walz}, \binits{J.}},
\bauthor{\bsnm{H\"ansch}, \binits{T.W.}},
\bauthor{\bsnm{Comeau}, \binits{D.}},
\bauthor{\bsnm{Hessels}, \binits{E.A.}}:
\batitle{First laser-controlled antihydrogen production}.
\bjtitle{Phys. Rev. Lett.}
\bvolume{93},
\bfpage{263401}
(\byear{2004})
\end{barticle}
\endbibitem

\bibitem{kellerbauer08a}
\begin{barticle}
\bauthor{\bsnm{Kellerbauer}, \binits{A.}},
\bauthor{\bsnm{Amoretti}, \binits{M.}},
\bauthor{\bsnm{Belov}, \binits{A.S.}},
\bauthor{\bsnm{Bonomi}, \binits{G.}},
\bauthor{\bsnm{Boscolo}, \binits{I.}},
\bauthor{\bsnm{Brusa}, \binits{R.S.}},
\bauthor{\bsnm{B{\"u}chner}, \binits{M.}},
\bauthor{\bsnm{Byakov}, \binits{V.M.}},
\bauthor{\bsnm{Cabaret}, \binits{L.}},
\bauthor{\bsnm{Canali}, \binits{C.}},
\bauthor{\bsnm{Carraro}, \binits{C.}},
\bauthor{\bsnm{Castelli}, \binits{F.}},
\bauthor{\bsnm{Cialdi}, \binits{S.}},
\bauthor{\bsnm{{de Combarieu}}, \binits{M.}},
\bauthor{\bsnm{Comparat}, \binits{D.}},
\bauthor{\bsnm{Consolati}, \binits{G.}},
\bauthor{\bsnm{Djourelov}, \binits{N.}},
\bauthor{\bsnm{Doser}, \binits{M.}},
\bauthor{\bsnm{Drobychev}, \binits{G.}},
\bauthor{\bsnm{Dupasquier}, \binits{A.}},
\bauthor{\bsnm{Ferrari}, \binits{G.}},
\bauthor{\bsnm{Forget}, \binits{P.}},
\bauthor{\bsnm{Formaro}, \binits{L.}},
\bauthor{\bsnm{Gervasini}, \binits{A.}},
\bauthor{\bsnm{Giammarchi}, \binits{M.G.}},
\bauthor{\bsnm{Gninenko}, \binits{S.N.}},
\bauthor{\bsnm{Gribakin}, \binits{G.}},
\bauthor{\bsnm{Hogan}, \binits{S.D.}},
\bauthor{\bsnm{Jacquey}, \binits{M.}},
\bauthor{\bsnm{Lagomarsino}, \binits{V.}},
\bauthor{\bsnm{Manuzio}, \binits{G.}},
\bauthor{\bsnm{Mariazzi}, \binits{S.}},
\bauthor{\bsnm{Matveev}, \binits{V.A.}},
\bauthor{\bsnm{Meier}, \binits{J.O.}},
\bauthor{\bsnm{Merkt}, \binits{F.}},
\bauthor{\bsnm{Nedelec}, \binits{P.}},
\bauthor{\bsnm{Oberthaler}, \binits{M.K.}},
\bauthor{\bsnm{Pari}, \binits{P.}},
\bauthor{\bsnm{Prevedelli}, \binits{M.}},
\bauthor{\bsnm{Quasso}, \binits{F.}},
\bauthor{\bsnm{Rotondi}, \binits{A.}},
\bauthor{\bsnm{Sillou}, \binits{D.}},
\bauthor{\bsnm{Stepanov}, \binits{S.V.}},
\bauthor{\bsnm{Stroke}, \binits{H.H.}},
\bauthor{\bsnm{Testera}, \binits{G.}},
\bauthor{\bsnm{Tino}, \binits{G.M.}},
\bauthor{\bsnm{Tr{\'e}nec}, \binits{G.}},
\bauthor{\bsnm{Vairo}, \binits{A.}},
\bauthor{\bsnm{Vigu{\'e}}, \binits{J.}},
\bauthor{\bsnm{Walters}, \binits{H.}},
\bauthor{\bsnm{Warring}, \binits{U.}},
\bauthor{\bsnm{Zavatarelli}, \binits{S.}},
\bauthor{\bsnm{Zvezhinskij}, \binits{D.S.}}:
\batitle{Proposed antimatter gravity measurement with an antihydrogen beam}.
\bjtitle{Nucl. Instr. and Meth. in Phys. Res. B}
\bvolume{266},
\bfpage{351}
(\byear{2008})
\end{barticle}
\endbibitem

\bibitem{alpha10a}
\begin{barticle}
\bauthor{\bsnm{Andresen}, \binits{G.B.}},
\bauthor{\bsnm{Ashkezari}, \binits{M.D.}},
\bauthor{\bsnm{Baquero-Ruiz}, \binits{M.}},
\bauthor{\bsnm{Bertsche}, \binits{W.}},
\bauthor{\bsnm{Bowe}, \binits{P.D.}},
\bauthor{\bsnm{Butler}, \binits{E.}},
\bauthor{\bsnm{Cesar}, \binits{C.L.}},
\bauthor{\bsnm{Chapman}, \binits{S.}},
\bauthor{\bsnm{Charlton}, \binits{M.}},
\bauthor{\bsnm{Deller}, \binits{A.}},
\bauthor{\bsnm{Eriksson}, \binits{S.}},
\bauthor{\bsnm{Fajans}, \binits{J.}},
\bauthor{\bsnm{Friesen}, \binits{T.}},
\bauthor{\bsnm{Fujiwara}, \binits{M.C.}},
\bauthor{\bsnm{Gill}, \binits{D.R.}},
\bauthor{\bsnm{Gutierrez}, \binits{A.}},
\bauthor{\bsnm{Hangst}, \binits{J.S.}},
\bauthor{\bsnm{Hardy}, \binits{W.N.}},
\bauthor{\bsnm{Hayden}, \binits{M.E.}},
\bauthor{\bsnm{Humphries}, \binits{A.J.}},
\bauthor{\bsnm{Hydomako}, \binits{R.}},
\bauthor{\bsnm{Jenkins}, \binits{M.J.}},
\bauthor{\bsnm{Jonsell}, \binits{S.}},
\bauthor{\bsnm{J{\o}rgensen}, \binits{L.V.}},
\bauthor{\bsnm{Kurchaninov}, \binits{L.}},
\bauthor{\bsnm{Madsen}, \binits{N.}},
\bauthor{\bsnm{Menary}, \binits{S.}},
\bauthor{\bsnm{Nolan}, \binits{P.}},
\bauthor{\bsnm{Olchanski}, \binits{K.}},
\bauthor{\bsnm{Olin}, \binits{A.}},
\bauthor{\bsnm{Povilus}, \binits{A.}},
\bauthor{\bsnm{Pusa}, \binits{P.}},
\bauthor{\bsnm{Robicheaux}, \binits{F.}},
\bauthor{\bsnm{Sarid}, \binits{E.}},
\bauthor{\bparticle{Seif~el} \bsnm{{N}asr}, \binits{S.}},
\bauthor{\bsnm{Silveira}, \binits{D.M.}},
\bauthor{\bsnm{So}, \binits{C.}},
\bauthor{\bsnm{Storey}, \binits{J.W.}},
\bauthor{\bsnm{Thompson}, \binits{R.I.}},
\bauthor{\bparticle{van~der} \bsnm{Werf}, \binits{D.P.}},
\bauthor{\bsnm{Wurtele}, \binits{J.S.}},
\bauthor{\bsnm{Yamazaki}, \binits{Y.}}:
\batitle{Trapped antihydrogen}.
\bjtitle{Nature}
\bvolume{468},
\bfpage{673}
(\byear{2010})
\end{barticle}
\endbibitem

\bibitem{alpha11a}
\begin{barticle}
\bauthor{\bsnm{Andresen}, \binits{G.B.}},
\bauthor{\bsnm{Ashkezari}, \binits{M.D.}},
\bauthor{\bsnm{Baquero-Ruiz}, \binits{M.}},
\bauthor{\bsnm{Bertsche}, \binits{W.}},
\bauthor{\bsnm{Bowe}, \binits{P.D.}},
\bauthor{\bsnm{Butler}, \binits{E.}},
\bauthor{\bsnm{Cesar}, \binits{C.L.}},
\bauthor{\bsnm{Charlton}, \binits{M.}},
\bauthor{\bsnm{Deller}, \binits{A.}},
\bauthor{\bsnm{Eriksson}, \binits{S.}},
\bauthor{\bsnm{Fajans}, \binits{J.}},
\bauthor{\bsnm{Friesen}, \binits{T.}},
\bauthor{\bsnm{Fujiwara}, \binits{M.C.}},
\bauthor{\bsnm{Gill}, \binits{D.R.}},
\bauthor{\bsnm{Gutierrez}, \binits{A.}},
\bauthor{\bsnm{Hangst}, \binits{J.S.}},
\bauthor{\bsnm{Hardy}, \binits{W.N.}},
\bauthor{\bsnm{Hayano}, \binits{R.S.}},
\bauthor{\bsnm{Hayden}, \binits{M.E.}},
\bauthor{\bsnm{Humphries}, \binits{A.J.}},
\bauthor{\bsnm{Hydomako}, \binits{R.}},
\bauthor{\bsnm{Jonsell}, \binits{S.}},
\bauthor{\bsnm{Kemp}, \binits{S.L.}},
\bauthor{\bsnm{Kurchaninov}, \binits{L.}},
\bauthor{\bsnm{Madsen}, \binits{N.}},
\bauthor{\bsnm{Menary}, \binits{S.}},
\bauthor{\bsnm{Nolan}, \binits{P.}},
\bauthor{\bsnm{Olchanski}, \binits{K.}},
\bauthor{\bsnm{Olin}, \binits{A.}},
\bauthor{\bsnm{Pusa}, \binits{P.}},
\bauthor{\bsnm{Rasmussen}, \binits{C.{\O}.}},
\bauthor{\bsnm{Robicheaux}, \binits{F.}},
\bauthor{\bsnm{Sarid}, \binits{E.}},
\bauthor{\bsnm{Silveira}, \binits{D.M.}},
\bauthor{\bsnm{So}, \binits{C.}},
\bauthor{\bsnm{Storey}, \binits{J.W.}},
\bauthor{\bsnm{Thompson}, \binits{R.I.}},
\bauthor{\bsnm{{van~der~Werf}}, \binits{D.P.}},
\bauthor{\bsnm{Wurtele}, \binits{J.S.}},
\bauthor{\bsnm{Yamazaki}, \binits{Y.}}:
\batitle{Confinement of antihydrogen for 1000 seconds}.
\bjtitle{Nature Phys.}
\bvolume{7},
\bfpage{558}
(\byear{2011})
\end{barticle}
\endbibitem

\bibitem{mills02a}
\begin{barticle}
\bauthor{\bsnm{{Mills~Jr}}, \binits{A.P.}},
\bauthor{\bsnm{Leventhal}, \binits{M.}}:
\batitle{Can we measure the gravitational free fall of cold {R}ydberg state
  positronium?}
\bjtitle{Nucl. Inst. Meth. Phys. Res. B}
\bvolume{192},
\bfpage{102}
(\byear{2002})
\end{barticle}
\endbibitem

\bibitem{cassidy14a}
\begin{barticle}
\bauthor{\bsnm{Cassidy}, \binits{D.B.}},
\bauthor{\bsnm{Hogan}, \binits{S.D.}}:
\batitle{Atom control and gravity measurements using {R}ydberg positronium}.
\bjtitle{Int. J. Mod. Phys. Conf. Ser.}
\bvolume{30},
\bfpage{1460259}
(\byear{2014})
\end{barticle}
\endbibitem

\bibitem{karshenboim05a}
\begin{barticle}
\bauthor{\bsnm{Karshenboim}, \binits{S.G.}}:
\batitle{Precision physics of simple atoms: {QED} tests, nuclear structure and
  fundamental constants}.
\bjtitle{Phys. Rep.}
\bvolume{422},
\bfpage{1}
(\byear{2005})
\end{barticle}
\endbibitem

\bibitem{ziock90a}
\begin{barticle}
\bauthor{\bsnm{Ziock}, \binits{K.P.}},
\bauthor{\bsnm{Howell}, \binits{R.H.}},
\bauthor{\bsnm{Magnotta}, \binits{F.}},
\bauthor{\bsnm{Failor}, \binits{R.A.}},
\bauthor{\bsnm{Jones}, \binits{K.M.}}:
\batitle{First observation of resonant excitation of high-\textit{n} states in
  positronium}.
\bjtitle{Phys. Rev. Lett.}
\bvolume{64},
\bfpage{2366}
(\byear{1990})
\end{barticle}
\endbibitem

\bibitem{cassidy12a}
\begin{barticle}
\bauthor{\bsnm{Cassidy}, \binits{D.B.}},
\bauthor{\bsnm{Hisakado}, \binits{T.H.}},
\bauthor{\bsnm{Tom}, \binits{H.W.K.}},
\bauthor{\bsnm{Mills}, \binits{A.P.}}:
\batitle{Efficient production of {R}ydberg positronium}.
\bjtitle{Phys. Rev. Lett.}
\bvolume{108},
\bfpage{043401}
(\byear{2012})
\end{barticle}
\endbibitem

\bibitem{hogan13b}
\begin{barticle}
\bauthor{\bsnm{Hogan}, \binits{S.D.}}:
\batitle{Calculated photoexcitation spectra of positronium {R}ydberg states}.
\bjtitle{Phys. Rev. A}
\bvolume{87},
\bfpage{063423}
(\byear{2013})
\end{barticle}
\endbibitem

\bibitem{cassidy10a}
\begin{barticle}
\bauthor{\bsnm{Crivelli}, \binits{P.}},
\bauthor{\bsnm{Gendotti}, \binits{U.}},
\bauthor{\bsnm{Rubbia}, \binits{A.}},
\bauthor{\bsnm{Liszkay}, \binits{L.}},
\bauthor{\bsnm{Perez}, \binits{P.}},
\bauthor{\bsnm{Corbel}, \binits{C.}}:
\batitle{Measurement of the orthopositronium confinement energy in mesoporous
  thin films}.
\bjtitle{Phys. Rev. A}
\bvolume{81},
\bfpage{052703}
(\byear{2010})
\end{barticle}
\endbibitem

\bibitem{lancuba14a}
\begin{barticle}
\bauthor{\bsnm{Lancuba}, \binits{P.}},
\bauthor{\bsnm{Hogan}, \binits{S.D.}}:
\batitle{Transmission-line decelerators for atoms in high {R}ydberg states}.
\bjtitle{Phys. Rev. A}
\bvolume{90},
\bfpage{053420}
(\byear{2014})
\end{barticle}
\endbibitem

\bibitem{kleppner83a}
\begin{bchapter}
\bauthor{\bsnm{Kleppner}, \binits{D.}},
\bauthor{\bsnm{Littman}, \binits{M.G.}},
\bauthor{\bsnm{Zimmerman}, \binits{M.L.}}:
\bctitle{{Rydberg atoms in strong fields}}.
In: \beditor{\bsnm{Stebbings}, \binits{R.F.}},
\beditor{\bsnm{Dunning}, \binits{F.B.}} (eds.)
\bbtitle{Rydberg States of Atoms and Molecules},
p. \bfpage{73}.
\bpublisher{Cambridge University Press},
\blocation{Cambrige}
(\byear{1983})
\end{bchapter}
\endbibitem

\bibitem{bethe57a}
\begin{bbook}
\bauthor{\bsnm{Bethe}, \binits{H.A.}},
\bauthor{\bsnm{Salpeter}, \binits{E.E.}}:
\bbtitle{{Quantum Mechanics of One- and Two-Electron Atoms}}.
\bpublisher{Springer},
\blocation{Berlin}
(\byear{1957})
\end{bbook}
\endbibitem

\bibitem{englefield72a}
\begin{bbook}
\bauthor{\bsnm{Englefield}, \binits{M.J.}}:
\bbtitle{{G}roup Theory and the {C}oulomb Problem}.
\bpublisher{John Wiley \& Sons, Inc},
\blocation{New York}
(\byear{1972})
\end{bbook}
\endbibitem

\bibitem{foot05a}
\begin{bbook}
\bauthor{\bsnm{Foot}, \binits{C.}}:
\bbtitle{Atomic Physics}.
\bpublisher{Oxford University Press},
\blocation{Oxford}
(\byear{2005})
\end{bbook}
\endbibitem

\bibitem{hiskes64a}
\begin{barticle}
\bauthor{\bsnm{Hiskes}, \binits{J.R.}},
\bauthor{\bsnm{Tarter}, \binits{C.B.}},
\bauthor{\bsnm{Moody}, \binits{D.A.}}:
\batitle{Stark lifetimes for the hydrogen atom}.
\bjtitle{Phys. Rev. A}
\bvolume{133},
\bfpage{424}
(\byear{1964})
\end{barticle}
\endbibitem

\bibitem{zimmerman79a}
\begin{barticle}
\bauthor{\bsnm{Zimmerman}, \binits{M.L.}},
\bauthor{\bsnm{Littman}, \binits{M.G.}},
\bauthor{\bsnm{Kash}, \binits{M.M.}},
\bauthor{\bsnm{Kleppner}, \binits{D.}}:
\batitle{{Stark structure of the Rydberg states of alkali-metal atoms}}.
\bjtitle{Phys. Rev. A}
\bvolume{20},
\bfpage{2251}
(\byear{1979})
\end{barticle}
\endbibitem

\bibitem{fielding91a}
\begin{barticle}
\bauthor{\bsnm{Fielding}, \binits{H.H.}},
\bauthor{\bsnm{Softley}, \binits{T.P.}}:
\batitle{Observation of the {S}tark effect in autoionising {R}ydberg states of
  molecular hydrogen}.
\bjtitle{Chem. Phys. Lett.}
\bvolume{185},
\bfpage{199}
(\byear{1991})
\end{barticle}
\endbibitem

\bibitem{qin93a}
\begin{barticle}
\bauthor{\bsnm{Qin}, \binits{K.}},
\bauthor{\bsnm{Bistransin}, \binits{M.}},
\bauthor{\bsnm{Glab}, \binits{W.L.}}:
\batitle{{Stark effect and rotational-series interactions on high Rydberg
  states of molecular hydrogen}}.
\bjtitle{Phys. Rev. A}
\bvolume{47},
\bfpage{4154}
(\byear{1993})
\end{barticle}
\endbibitem

\bibitem{vrakking96a}
\begin{barticle}
\bauthor{\bsnm{Vrakking}, \binits{M.J.}}:
\batitle{{Lifetimes of Rydberg states in ZEKE experiments. III. Calculations of
  the dc electric field dependence of predissociation lifetimes of NO}}.
\bjtitle{J. Chem. Phys.}
\bvolume{105},
\bfpage{7336}
(\byear{1996})
\end{barticle}
\endbibitem

\bibitem{zare88a}
\begin{bbook}
\bauthor{\bsnm{Zare}, \binits{R.N.}}:
\bbtitle{{Angular Momentum}}.
\bpublisher{John Wiley \& Sons},
\blocation{New York}
(\byear{1988})
\end{bbook}
\endbibitem

\bibitem{seiler11b}
\begin{barticle}
\bauthor{\bsnm{Seiler}, \binits{{\mbox{{C}h}}.}},
\bauthor{\bsnm{Hogan}, \binits{S.D.}},
\bauthor{\bsnm{Merkt}, \binits{F.}}:
\batitle{Trapping cold molecular hydrogen}.
\bjtitle{Phys. Chem. Chem. Phys.}
\bvolume{13},
\bfpage{19000}
(\byear{2011})
\end{barticle}
\endbibitem

\bibitem{yamakita04a}
\begin{barticle}
\bauthor{\bsnm{Yamakita}, \binits{Y.}},
\bauthor{\bsnm{Procter}, \binits{S.R.}},
\bauthor{\bsnm{Goodgame}, \binits{A.L.}},
\bauthor{\bsnm{Softley}, \binits{T.P.}},
\bauthor{\bsnm{Merkt}, \binits{F.}}:
\batitle{Deflection and deceleration of hydrogen {R}ydberg molecules in
  inhomogeneous electric fields}.
\bjtitle{J.~Chem. Phys.}
\bvolume{121},
\bfpage{1419}
(\byear{2004})
\end{barticle}
\endbibitem

\bibitem{wall14a}
\begin{barticle}
\bauthor{\bsnm{Wall}, \binits{T.E.}},
\bauthor{\bsnm{Cassidy}, \binits{D.B.}},
\bauthor{\bsnm{Hogan}, \binits{S.D.}}:
\batitle{Single-color two-photon spectroscopy of {R}ydberg states in electric
  fields}.
\bjtitle{Phys. Rev. A}
\bvolume{90},
\bfpage{053430}
(\byear{2014})
\end{barticle}
\endbibitem

\bibitem{hogan09a}
\begin{barticle}
\bauthor{\bsnm{Hogan}, \binits{S.D.}},
\bauthor{\bsnm{Seiler}, \binits{{\mbox{{C}h}}.}},
\bauthor{\bsnm{Merkt}, \binits{F.}}:
\batitle{{R}ydberg-state-enabled deceleration and trapping of cold molecules}.
\bjtitle{Phys. Rev. Lett.}
\bvolume{103},
\bfpage{123001}
(\byear{2009})
\end{barticle}
\endbibitem

\bibitem{hollberg84a}
\begin{barticle}
\bauthor{\bsnm{Hollberg}, \binits{L.}},
\bauthor{\bsnm{Hall}, \binits{J.L.}}:
\batitle{Measurement of the shift of {R}ydberg energy levels induced by
  blackbody radiation}.
\bjtitle{Phys. Rev. Lett.}
\bvolume{53},
\bfpage{230}
(\byear{1984})
\end{barticle}
\endbibitem

\bibitem{beiting79a}
\begin{barticle}
\bauthor{\bsnm{Beiting}, \binits{E.J.}},
\bauthor{\bsnm{Hildebrandt}, \binits{G.F.}},
\bauthor{\bsnm{Kellert}, \binits{F.G.}},
\bauthor{\bsnm{Foltz}, \binits{G.W.}},
\bauthor{\bsnm{Smith}, \binits{K.A.}},
\bauthor{\bsnm{Dunning}, \binits{F.B.}},
\bauthor{\bsnm{Stebbings}, \binits{R.F.}}:
\batitle{The effects of 300~{K} background radiation on {R}ydberg atoms}.
\bjtitle{J. Chem. Phys.}
\bvolume{70},
\bfpage{3551}
(\byear{1979})
\end{barticle}
\endbibitem

\bibitem{gallagher79a}
\begin{barticle}
\bauthor{\bsnm{Gallagher}, \binits{T.F.}},
\bauthor{\bsnm{Cooke}, \binits{W.E.}}:
\batitle{Interactions of blackbody radiation with atoms}.
\bjtitle{Phys. Rev. Lett.}
\bvolume{42},
\bfpage{835}
(\byear{1979})
\end{barticle}
\endbibitem

\bibitem{seiler11a}
\begin{barticle}
\bauthor{\bsnm{Seiler}, \binits{{\mbox{{C}h}}.}},
\bauthor{\bsnm{Hogan}, \binits{S.D.}},
\bauthor{\bsnm{Schmutz}, \binits{H.}},
\bauthor{\bsnm{Agner}, \binits{J.A.}},
\bauthor{\bsnm{Merkt}, \binits{F.}}:
\batitle{Collisional and radiative processes in adiabatic deceleration,
  deflection and off-axis trapping of a {R}ydberg atom beam}.
\bjtitle{Phys. Rev. Lett.}
\bvolume{106},
\bfpage{073003}
(\byear{2011})
\end{barticle}
\endbibitem

\bibitem{spencer82a}
\begin{barticle}
\bauthor{\bsnm{Spencer}, \binits{W.P.}},
\bauthor{\bsnm{Vaidyanathan}, \binits{A.G.}},
\bauthor{\bsnm{Kleppner}, \binits{D.}},
\bauthor{\bsnm{Ducas}, \binits{T.W.}}:
\batitle{Photoionization by blackbody radiation}.
\bjtitle{Phys. Rev. A}
\bvolume{26},
\bfpage{1490}
(\byear{1982})
\end{barticle}
\endbibitem

\bibitem{mandel79a}
\begin{barticle}
\bauthor{\bsnm{Mandel}, \binits{M.}}:
\batitle{Photon occupation numbers in black body radiation}.
\bjtitle{J. Opt. Soc. Am.}
\bvolume{69},
\bfpage{1038}
(\byear{1979})
\end{barticle}
\endbibitem

\bibitem{wing80a}
\begin{barticle}
\bauthor{\bsnm{Wing}, \binits{W.H.}}:
\batitle{{Electrostatic trapping of neutral atomic particles}}.
\bjtitle{Phys. Rev. Lett.}
\bvolume{45},
\bfpage{631}
(\byear{1980})
\end{barticle}
\endbibitem

\bibitem{breeden81a}
\begin{barticle}
\bauthor{\bsnm{Breeden}, \binits{T.}},
\bauthor{\bsnm{Metcalf}, \binits{H.}}:
\batitle{Stark acceleration of {R}ydberg atoms in inhomogeneous electric
  fields}.
\bjtitle{Phys. Rev. Lett.}
\bvolume{47},
\bfpage{1726}
(\byear{1981})
\end{barticle}
\endbibitem

\bibitem{townsend01a}
\begin{barticle}
\bauthor{\bsnm{Townsend}, \binits{D.}},
\bauthor{\bsnm{Goodgame}, \binits{A.L.}},
\bauthor{\bsnm{Procter}, \binits{S.R.}},
\bauthor{\bsnm{Mackenzie}, \binits{S.R.}},
\bauthor{\bsnm{Softley}, \binits{T.P.}}:
\batitle{Deflection of krypton {R}ydberg atoms in the field of an electric
  dipole}.
\bjtitle{J. Phys. B.: At. Mol. Opt. Phys.}
\bvolume{34},
\bfpage{439}
(\byear{2001})
\end{barticle}
\endbibitem

\bibitem{procter03a}
\begin{barticle}
\bauthor{\bsnm{Procter}, \binits{S.R.}},
\bauthor{\bsnm{Yamakita}, \binits{Y.}},
\bauthor{\bsnm{Merkt}, \binits{F.}},
\bauthor{\bsnm{Softley}, \binits{T.P.}}:
\batitle{Controlling the motion of hydrogen molecules}.
\bjtitle{Chem. Phys. Lett.}
\bvolume{374},
\bfpage{667}
(\byear{2003})
\end{barticle}
\endbibitem

\bibitem{vliegen04a}
\begin{barticle}
\bauthor{\bsnm{Vliegen}, \binits{E.}},
\bauthor{\bsnm{W{\"o}rner}, \binits{H.J.}},
\bauthor{\bsnm{Softley}, \binits{T.P.}},
\bauthor{\bsnm{Merkt}, \binits{F.}}:
\batitle{Nonhydrogenic effects in the deceleration of {R}ydberg atoms in
  inhomogeneous electric fields}.
\bjtitle{Phys. Rev. Lett.}
\bvolume{92},
\bfpage{033005}
(\byear{2004})
\end{barticle}
\endbibitem

\bibitem{vliegen06b}
\begin{barticle}
\bauthor{\bsnm{Vliegen}, \binits{E.}},
\bauthor{\bsnm{Merkt}, \binits{F.}}:
\batitle{{S}tark deceleration of hydrogen atoms}.
\bjtitle{J. Phys. B.: At. Mol. Opt. Phys.}
\bvolume{39},
\bfpage{241}
(\byear{2006})
\end{barticle}
\endbibitem

\bibitem{vliegen05a}
\begin{barticle}
\bauthor{\bsnm{Vliegen}, \binits{E.}},
\bauthor{\bsnm{Merkt}, \binits{F.}}:
\batitle{On the electrostatic deceleration of argon atoms in high {R}ydberg
  states by time-dependent inhomogeneous electric fields}.
\bjtitle{J. Phys. B.: At. Mol. Opt. Phys.}
\bvolume{38},
\bfpage{1623}
(\byear{2005})
\end{barticle}
\endbibitem

\bibitem{vliegen06a}
\begin{barticle}
\bauthor{\bsnm{Vliegen}, \binits{E.}},
\bauthor{\bsnm{Limacher}, \binits{P.}},
\bauthor{\bsnm{Merkt}, \binits{F.}}:
\batitle{Measurement of the three-dimensional velocity distribution of
  {S}tark-decelerated {R}ydberg atoms}.
\bjtitle{Eur. Phys. J. D}
\bvolume{40},
\bfpage{73}
(\byear{2006})
\end{barticle}
\endbibitem

\bibitem{vliegen06c}
\begin{barticle}
\bauthor{\bsnm{Vliegen}, \binits{E.}},
\bauthor{\bsnm{Merkt}, \binits{F.}}:
\batitle{Normal-incidence electrostatic {R}ydberg atom mirror}.
\bjtitle{Phys. Rev. Lett.}
\bvolume{97},
\bfpage{033002}
(\byear{2006})
\end{barticle}
\endbibitem

\bibitem{vliegen06d}
\begin{botherref}
\oauthor{\bsnm{Vliegen}, \binits{E.}}:
Rydberg states in atom and molecule optics.
{PhD thesis},
{Eidgen{\"o}ssische Technische Hochschule Z{\"u}rich},
{Z{\"u}rich, Switzerland}
(2006).
{Diss. ETH Nr. 16782}
\end{botherref}
\endbibitem

\bibitem{vliegen07a}
\begin{barticle}
\bauthor{\bsnm{Vliegen}, \binits{E.}},
\bauthor{\bsnm{Hogan}, \binits{S.D.}},
\bauthor{\bsnm{Schmutz}, \binits{H.}},
\bauthor{\bsnm{Merkt}, \binits{F.}}:
\batitle{{S}tark deceleration and trapping of hydrogen {R}ydberg atoms}.
\bjtitle{Phys. Rev. A}
\bvolume{76},
\bfpage{023405}
(\byear{2007})
\end{barticle}
\endbibitem

\bibitem{hogan08a}
\begin{barticle}
\bauthor{\bsnm{Hogan}, \binits{S.D.}},
\bauthor{\bsnm{Merkt}, \binits{F.}}:
\batitle{Demonstration of three-dimensional electrostatic trapping of
  state-selected {R}ydberg atoms}.
\bjtitle{Phys. Rev. Lett.}
\bvolume{100},
\bfpage{043001}
(\byear{2008})
\end{barticle}
\endbibitem

\bibitem{willitsch03b}
\begin{barticle}
\bauthor{\bsnm{Willitsch}, \binits{S.}},
\bauthor{\bsnm{Dyke}, \binits{J.M.}},
\bauthor{\bsnm{Merkt}, \binits{F.}}:
\batitle{Generation and high-resolution photoelectron spectroscopy of small
  organic radicals in cold supersonic expansions}.
\bjtitle{Helv. Chim. Acta}
\bvolume{86},
\bfpage{1152}
(\byear{2003})
\end{barticle}
\endbibitem

\bibitem{hogan13a}
\begin{barticle}
\bauthor{\bsnm{Hogan}, \binits{S.D.}},
\bauthor{\bsnm{Seiler}, \binits{{\mbox{{C}h}}.}},
\bauthor{\bsnm{Merkt}, \binits{F.}}:
\batitle{Motional, isotope and quadratic {S}tark effects in {R}ydberg-{S}tark
  deceleration and off-axis electric trapping of {H} and {D}}.
\bjtitle{J. Phys. B.: At. Mol. Opt. Phys.}
\bvolume{46},
\bfpage{045303}
(\byear{2013})
\end{barticle}
\endbibitem

\bibitem{seiler12a}
\begin{barticle}
\bauthor{\bsnm{Seiler}, \binits{{\mbox{{C}h}}.}},
\bauthor{\bsnm{Hogan}, \binits{S.D.}},
\bauthor{\bsnm{Merkt}, \binits{F.}}:
\batitle{Dynamical processes in {R}ydberg-{S}tark deceleration and trapping of
  atoms and molecules}.
\bjtitle{Chimia}
\bvolume{66},
\bfpage{208}
(\byear{2012})
\end{barticle}
\endbibitem

\bibitem{beterov09a}
\begin{barticle}
\bauthor{\bsnm{Beterov}, \binits{I.I.}},
\bauthor{\bsnm{Tretyakov}, \binits{D.B.}},
\bauthor{\bsnm{Ryabtsev}, \binits{I.I.}},
\bauthor{\bsnm{Entin}, \binits{V.M.}},
\bauthor{\bsnm{Ekers}, \binits{A.}},
\bauthor{\bsnm{Bezuglov}, \binits{N.N.}}:
\batitle{Ionization of {R}ydberg atoms by blackbody radiation}.
\bjtitle{New J. Phys}
\bvolume{11},
\bfpage{013052}
(\byear{2009})
\end{barticle}
\endbibitem

\bibitem{hogan12c}
\begin{botherref}
\oauthor{\bsnm{Hogan}, \binits{S.D.}}:
{Cold atoms and molecules by {Z}eeman deceleration and {R}ydberg-{S}tark
  deceleration}.
{Habilitation thesis},
{Eidgen{\"o}ssische Technische Hochschule Z{\"u}rich},
{Z{\"u}rich, Switzerland}
(2012)
\end{botherref}
\endbibitem

\bibitem{landau32a}
\begin{barticle}
\bauthor{\bsnm{Landau}, \binits{L.D.}}:
\batitle{Zur {T}heorie der {E}nergie\"ubertragung ii}.
\bjtitle{Phys. Z. Sowjetunion}
\bvolume{2},
\bfpage{46}
(\byear{1932})
\end{barticle}
\endbibitem

\bibitem{zener32a}
\begin{barticle}
\bauthor{\bsnm{Zener}, \binits{C.}}:
\batitle{Non-adiabatic crossing of energy levels}.
\bjtitle{Proc. R. Soc. London Ser. A}
\bvolume{137},
\bfpage{696}
(\byear{1932})
\end{barticle}
\endbibitem

\bibitem{seiler13a}
\begin{botherref}
\oauthor{\bsnm{Seiler}, \binits{{\mbox{{C}h}}.}}:
{R}ydberg-{S}tark deceleration and trapping of atoms and molecules.
{PhD thesis},
{Eidgen{\"o}ssische Technische Hochschule Z{\"u}rich},
{Z{\"u}rich, Switzerland}
(2013).
{Diss. ETH Nr. 21340}
\end{botherref}
\endbibitem

\bibitem{chiaverini05a}
\begin{barticle}
\bauthor{\bsnm{Chiaverini}, \binits{J.}},
\bauthor{\bsnm{Blakestad}, \binits{R.B.}},
\bauthor{\bsnm{Britton}, \binits{J.}},
\bauthor{\bsnm{Jost}, \binits{J.D.}},
\bauthor{\bsnm{Langer}, \binits{C.}},
\bauthor{\bsnm{Leibfried}, \binits{D.}},
\bauthor{\bsnm{Ozeri}, \binits{R.}},
\bauthor{\bsnm{Wineland}, \binits{D.J.}}:
\batitle{Surface-electrode architecture for ion-trap quantum information
  processing}.
\bjtitle{Quant. Inf. Comput.}
\bvolume{5},
\bfpage{419}
(\byear{2005})
\end{barticle}
\endbibitem

\bibitem{folman02a}
\begin{barticle}
\bauthor{\bsnm{Folman}, \binits{R.}},
\bauthor{\bsnm{Kr\"uger}, \binits{P.}},
\bauthor{\bsnm{Schmiedmayer}, \binits{J.}},
\bauthor{\bsnm{Denschlag}, \binits{J.}},
\bauthor{\bsnm{Henkel}, \binits{C.}}:
\batitle{Microscopic atom optics: {F}rom wires to an atom chip}.
\bjtitle{Adv. At. Mol. Opt. Phys.}
\bvolume{48},
\bfpage{263}
(\byear{2002})
\end{barticle}
\endbibitem

\bibitem{home09a}
\begin{barticle}
\bauthor{\bsnm{Home}, \binits{J.P.}},
\bauthor{\bsnm{Hanneke}, \binits{D.}},
\bauthor{\bsnm{Jost}, \binits{J.D.}},
\bauthor{\bsnm{Amini}, \binits{J.M.}},
\bauthor{\bsnm{Leibfried}, \binits{D.}},
\bauthor{\bsnm{Wineland}, \binits{D.J.}}:
\batitle{Complete methods set for scalable ion trap quantum information
  processing}.
\bjtitle{Science}
\bvolume{325},
\bfpage{1227}
(\byear{2009})
\end{barticle}
\endbibitem

\bibitem{riedel10a}
\begin{barticle}
\bauthor{\bsnm{Riedel}, \binits{M.F.}},
\bauthor{\bsnm{B\"ohi}, \binits{P.}},
\bauthor{\bsnm{Li}, \binits{Y.}},
\bauthor{\bsnm{H\"ansch}, \binits{T.W.}},
\bauthor{\bsnm{Sinatra}, \binits{A.}},
\bauthor{\bsnm{Treutlein}, \binits{P.}}:
\batitle{Atom-chip-based generation of entanglement for quantum metrology}.
\bjtitle{Nature}
\bvolume{464},
\bfpage{1170}
(\byear{2010})
\end{barticle}
\endbibitem

\bibitem{meek08a}
\begin{barticle}
\bauthor{\bsnm{Meek}, \binits{S.A.}},
\bauthor{\bsnm{Bethlem}, \binits{H.L.}},
\bauthor{\bsnm{Conrad}, \binits{H.}},
\bauthor{\bsnm{Meijer}, \binits{G.}}:
\batitle{Trapping molecules on a chip in traveling potential wells}.
\bjtitle{Phys. Rev. Lett.}
\bvolume{100},
\bfpage{153003}
(\byear{2008})
\end{barticle}
\endbibitem

\bibitem{meek09a}
\begin{barticle}
\bauthor{\bsnm{Meek}, \binits{S.A.}},
\bauthor{\bsnm{Santambrogio}, \binits{G.}},
\bauthor{\bsnm{Conrad}, \binits{H.}},
\bauthor{\bsnm{Meijer}, \binits{G.}}:
\batitle{Taming molecular beams; towards a gas-phase molecular laboratory on a
  chip}.
\bjtitle{J. Phys.: Conf. Ser.}
\bvolume{194},
\bfpage{012063}
(\byear{2009})
\end{barticle}
\endbibitem

\bibitem{meek09b}
\begin{barticle}
\bauthor{\bsnm{Meek}, \binits{S.A.}},
\bauthor{\bsnm{Conrad}, \binits{H.}},
\bauthor{\bsnm{Meijer}, \binits{G.}}:
\batitle{Trapping molecules on a chips}.
\bjtitle{Science}
\bvolume{324},
\bfpage{1699}
(\byear{2009})
\end{barticle}
\endbibitem

\bibitem{tauschinsky10a}
\begin{barticle}
\bauthor{\bsnm{Tauschinsky}, \binits{A.}},
\bauthor{\bsnm{Thijssen}, \binits{R.M.T.}},
\bauthor{\bsnm{Whitlock}, \binits{S.}},
\bauthor{\bparticle{van Linden van~den} \bsnm{Heuvell}, \binits{H.B.}},
\bauthor{\bsnm{Spreeuw}, \binits{R.J.C.}}:
\batitle{Spatially resolved excitation of {R}ydberg atoms and surface effects
  on an atom chip}.
\bjtitle{Phys. Rev. A}
\bvolume{81},
\bfpage{063411}
(\byear{2010})
\end{barticle}
\endbibitem

\bibitem{nirrengarten06a}
\begin{barticle}
\bauthor{\bsnm{Nirrengarten}, \binits{T.}},
\bauthor{\bsnm{Qarry}, \binits{A.}},
\bauthor{\bsnm{Roux}, \binits{C.}},
\bauthor{\bsnm{Emmert}, \binits{A.}},
\bauthor{\bsnm{Nogues}, \binits{G.}},
\bauthor{\bsnm{Brune}, \binits{M.}},
\bauthor{\bsnm{Raimond}, \binits{J.-M.}},
\bauthor{\bsnm{Haroche}, \binits{S.}}:
\batitle{Realization of a superconducting atom chip}.
\bjtitle{Phys. Rev. Lett.}
\bvolume{97},
\bfpage{200405}
(\byear{2006})
\end{barticle}
\endbibitem

\bibitem{cherry09a}
\begin{barticle}
\bauthor{\bsnm{Cherry}, \binits{O.}},
\bauthor{\bsnm{Carter}, \binits{J.D.}},
\bauthor{\bsnm{Martin}, \binits{J.D.D.}}:
\batitle{An atom chip for the manipulation of ultracold atoms}.
\bjtitle{Can. J. Phys.}
\bvolume{87},
\bfpage{633}
(\byear{2009})
\end{barticle}
\endbibitem

\bibitem{saffman02a}
\begin{barticle}
\bauthor{\bsnm{Saffman}, \binits{M.}},
\bauthor{\bsnm{Walker}, \binits{T.G.}}:
\batitle{Creating single-atom and single-photon sources from entangled atomic
  ensembles}.
\bjtitle{Phys. Rev. A}
\bvolume{66},
\bfpage{065403}
(\byear{2002})
\end{barticle}
\endbibitem

\bibitem{hogan12b}
\begin{barticle}
\bauthor{\bsnm{Hogan}, \binits{S.D.}},
\bauthor{\bsnm{Allmendinger}, \binits{P.}},
\bauthor{\bsnm{Sa\ss{}mannshausen}, \binits{H.}},
\bauthor{\bsnm{Schmutz}, \binits{H.}},
\bauthor{\bsnm{Merkt}, \binits{F.}}:
\batitle{Surface-electrode {R}ydberg-{S}tark decelerator}.
\bjtitle{Phys. Rev. Lett.}
\bvolume{108},
\bfpage{063008}
(\byear{2012})
\end{barticle}
\endbibitem

\bibitem{allmendinger13a}
\begin{barticle}
\bauthor{\bsnm{Allmendinger}, \binits{P.}},
\bauthor{\bsnm{Agner}, \binits{J.A.}},
\bauthor{\bsnm{Schmutz}, \binits{H.}},
\bauthor{\bsnm{Merkt}, \binits{F.}}:
\batitle{Deceleration and trapping of a fast supersonic beam of metastable
  helium atoms with a 44-electrode chip decelerator}.
\bjtitle{Phys. Rev. A}
\bvolume{88},
\bfpage{043433}
(\byear{2013})
\end{barticle}
\endbibitem

\bibitem{allmendinger14a}
\begin{barticle}
\bauthor{\bsnm{Allmendinger}, \binits{P.}},
\bauthor{\bsnm{Deiglmayr}, \binits{J.}},
\bauthor{\bsnm{Agner}, \binits{J.A.}},
\bauthor{\bsnm{Schmutz}, \binits{H.}},
\bauthor{\bsnm{Merkt}, \binits{F.}}:
\batitle{Surface-electrode decelerator and deflector for {R}ydberg atoms and
  molecules}.
\bjtitle{Phys. Rev. A}
\bvolume{90},
\bfpage{043403}
(\byear{2014})
\end{barticle}
\endbibitem

\bibitem{santambrogio15a}
\begin{barticle}
\bauthor{\bsnm{Santambrogio}, \binits{G.}}:
\batitle{Trapping molecules on chips}.
\bjtitle{EPJ Tech. Instrum.}
\bvolume{2},
\bfpage{14}
(\byear{2015})
\end{barticle}
\endbibitem

\bibitem{osterwalder15a}
\begin{barticle}
\bauthor{\bsnm{Osterwalder}, \binits{A.}}:
\batitle{Merged neutral beams}.
\bjtitle{EPJ Tech. Instrum.}
\bvolume{2},
\bfpage{10}
(\byear{2015})
\end{barticle}
\endbibitem

\bibitem{lancuba13a}
\begin{barticle}
\bauthor{\bsnm{Lancuba}, \binits{P.}},
\bauthor{\bsnm{Hogan}, \binits{S.D.}}:
\batitle{Guiding {R}ydberg atoms above surface-based transmission lines}.
\bjtitle{Phys. Rev. A}
\bvolume{88},
\bfpage{043427}
(\byear{2013})
\end{barticle}
\endbibitem

\bibitem{zhelyazkova15a}
\begin{barticle}
\bauthor{\bsnm{Zhelyazkova}, \binits{V.}},
\bauthor{\bsnm{Hogan}, \binits{S.D.}}:
\batitle{Probing interactions between {R}ydberg atoms with large electric
  dipole moments in amplitude-modulated electric fields}.
\bjtitle{Phys. Rev. A}
\bvolume{92},
\bfpage{011402}
(\byear{2015})
\end{barticle}
\endbibitem

\bibitem{ko14a}
\begin{barticle}
\bauthor{\bsnm{Ko}, \binits{H.}},
\bauthor{\bsnm{Hogan}, \binits{S.D.}}:
\batitle{High-field-seeking {R}ydberg atoms orbiting a charged wire}.
\bjtitle{Phys. Rev. A}
\bvolume{89},
\bfpage{053410}
(\byear{2014})
\end{barticle}
\endbibitem

\bibitem{baxter95a}
\begin{barticle}
\bauthor{\bsnm{Baxter}, \binits{C.}}:
\batitle{Cold {R}ydberg atoms as realizable analogs of {C}hern-{S}imons
  theory}.
\bjtitle{Phys. Rev. Lett.}
\bvolume{74},
\bfpage{514}
(\byear{1995})
\end{barticle}
\endbibitem

\bibitem{zhang96a}
\begin{barticle}
\bauthor{\bsnm{Zhang}, \binits{J.-Z.}}:
\batitle{Angular momentum of supersymmetric cold {R}ydberg atoms}.
\bjtitle{Phys. Rev. Lett.}
\bvolume{77},
\bfpage{44}
(\byear{1996})
\end{barticle}
\endbibitem

\bibitem{zhang04a}
\begin{barticle}
\bauthor{\bsnm{Zhang}, \binits{J.-Z.}}:
\batitle{Testing spatial noncommutativity via {R}ydberg atoms}.
\bjtitle{Phys. Rev. Lett.}
\bvolume{93},
\bfpage{043002}
(\byear{2004})
\end{barticle}
\endbibitem

\bibitem{zhelyazkova15b}
\begin{barticle}
\bauthor{\bsnm{Zhelyazkova}, \binits{V.}},
\bauthor{\bsnm{Hogan}, \binits{S.D.}}:
\batitle{{Rydberg-Stark states in oscillating electric fields}}.
\bjtitle{Mol. Phys.}
\bvolume{113},
\bfpage{3979}
(\byear{2015})
\end{barticle}
\endbibitem

\bibitem{reinhard07a}
\begin{barticle}
\bauthor{\bsnm{Reinhard}, \binits{A.}},
\bauthor{\bsnm{Liebisch}, \binits{T.C.}},
\bauthor{\bsnm{Knuffman}, \binits{B.}},
\bauthor{\bsnm{Raithel}, \binits{G.}}:
\batitle{Level shifts of rubidium {R}ydberg states due to binary interactions}.
\bjtitle{Phys. Rev. A}
\bvolume{75},
\bfpage{032712}
(\byear{2007})
\end{barticle}
\endbibitem

\bibitem{vogt07a}
\begin{barticle}
\bauthor{\bsnm{Vogt}, \binits{T.}},
\bauthor{\bsnm{Viteau}, \binits{M.}},
\bauthor{\bsnm{Chotia}, \binits{A.}},
\bauthor{\bsnm{Zhao}, \binits{J.}},
\bauthor{\bsnm{Comparat}, \binits{D.}},
\bauthor{\bsnm{Pillet}, \binits{P.}}:
\batitle{Electric-field induced dipole blockade with {R}ydberg atoms}.
\bjtitle{Phys. Rev. Lett.}
\bvolume{99},
\bfpage{073002}
(\byear{2007})
\end{barticle}
\endbibitem

\end{thebibliography}

\newcommand{\BMCxmlcomment}[1]{}

\BMCxmlcomment{

<refgrp>

<bibl id="B1">
  <title><p>{\"U}ber den {B}au der {L}inienspektren der chemischen
  {G}rundstoffe</p></title>
  <aug>
    <au><snm>Rydberg</snm><fnm>J. R.</fnm></au>
  </aug>
  <source>Z. Phys. Chem.</source>
  <pubdate>1890</pubdate>
  <volume>5</volume>
  <fpage>227</fpage>
</bibl>

<bibl id="B2">
  <title><p>Molecules in high {R}ydberg states</p></title>
  <aug>
    <au><snm>Merkt</snm><fnm>F.</fnm></au>
  </aug>
  <source>Ann. Rev. Phys. Chem.</source>
  <pubdate>1997</pubdate>
  <volume>48</volume>
  <fpage>675</fpage>
</bibl>

<bibl id="B3">
  <title><p>Applications of molecular {R}ydberg states in chemical dynamics and
  spectroscopy</p></title>
  <aug>
    <au><snm>Softley</snm><fnm>T. P.</fnm></au>
  </aug>
  <source>Int. Rev. Phys. Chem.</source>
  <publisher>Taylor \& Francis</publisher>
  <pubdate>2004</pubdate>
  <volume>23</volume>
  <fpage>1</fpage>
</bibl>

<bibl id="B4">
  <title><p>{\"U}ber das {W}asserstoffspektrum vom {S}tandpunkt der neuen
  {Q}uantunmechanik</p></title>
  <aug>
    <au><snm>Pauli</snm><fnm>W.</fnm></au>
  </aug>
  <source>Z. Phys.</source>
  <pubdate>1926</pubdate>
  <volume>36</volume>
  <fpage>336</fpage>
</bibl>

<bibl id="B5">
  <title><p>Theoretical studies of hydrogen {R}ydberg atoms in electric
  fields</p></title>
  <aug>
    <au><snm>Damburg</snm><fnm>R. J.</fnm></au>
    <au><snm>Kolosov</snm><fnm>V. V.</fnm></au>
  </aug>
  <source>Rydberg States of Atoms and Molecules</source>
  <publisher>Cambridge: Cambridge University Press</publisher>
  <editor>R. F. Stebbings and F. B. Dunning</editor>
  <pubdate>1983</pubdate>
  <fpage>31</fpage>
</bibl>

<bibl id="B6">
  <title><p>{Using high Rydberg states as electric field sensors}</p></title>
  <aug>
    <au><snm>Osterwalder</snm><fnm>A.</fnm></au>
    <au><snm>Merkt</snm><fnm>F.</fnm></au>
  </aug>
  <source>Phys. Rev. Lett.</source>
  <pubdate>1999</pubdate>
  <volume>82</volume>
  <fpage>1831</fpage>
</bibl>

<bibl id="B7">
  <title><p>Fokussierung polarer {M}olek\"ule</p></title>
  <aug>
    <au><snm>Bennewitz</snm><fnm>H. G.</fnm></au>
    <au><snm>Paul</snm><fnm>W.</fnm></au>
    <au><snm>Schlier</snm><fnm>{\mbox{{C}h}}</fnm></au>
  </aug>
  <source>Z. Phys.</source>
  <pubdate>19551</pubdate>
  <volume>41</volume>
  <fpage>6</fpage>
</bibl>

<bibl id="B8">
  <title><p>The {M}aser -- {N}ew Type of Microwave Amplifier, Frequency
  Standard, and Spectrometer</p></title>
  <aug>
    <au><snm>Gordon</snm><fnm>J. P.</fnm></au>
    <au><snm>Zeiger</snm><fnm>H. J.</fnm></au>
    <au><snm>Townes</snm><fnm>C. H.</fnm></au>
  </aug>
  <source>Phys. Rev.</source>
  <pubdate>1955</pubdate>
  <volume>99</volume>
  <fpage>1264</fpage>
</bibl>

<bibl id="B9">
  <title><p>Reactions of Oriented Molecules</p></title>
  <aug>
    <au><snm>Brooks</snm><fnm>P. R.</fnm></au>
  </aug>
  <source>Science</source>
  <pubdate>1976</pubdate>
  <volume>193</volume>
  <fpage>11</fpage>
</bibl>

<bibl id="B10">
  <title><p>Reactive Scattering Studies on Oriented Molecules</p></title>
  <aug>
    <au><snm>Stolte</snm><fnm>S.</fnm></au>
  </aug>
  <source>Ber. Bunsenges. Phys. Chem.</source>
  <pubdate>1982</pubdate>
  <volume>86</volume>
  <fpage>413</fpage>
</bibl>

<bibl id="B11">
  <title><p>Oriented molecule beams via the electrostatic hexapole:
  {P}reparation, characterization, and reactive scattering</p></title>
  <aug>
    <au><snm>Parker</snm><fnm>D. H.</fnm></au>
    <au><snm>Bernstein</snm><fnm>R. B.</fnm></au>
  </aug>
  <source>Ann. Rev. Phys. Chem.</source>
  <pubdate>1989</pubdate>
  <volume>40</volume>
  <fpage>561</fpage>
</bibl>

<bibl id="B12">
  <title><p>Decelerating neutral dipolar molecules</p></title>
  <aug>
    <au><snm>Bethlem</snm><fnm>HL</fnm></au>
    <au><snm>Berden</snm><fnm>G</fnm></au>
    <au><snm>Meijer</snm><fnm>G</fnm></au>
  </aug>
  <source>Phys. Rev. Lett.</source>
  <pubdate>1999</pubdate>
  <volume>83</volume>
  <fpage>1558</fpage>
</bibl>

<bibl id="B13">
  <title><p>Electrostatic trapping of ammonia molecules</p></title>
  <aug>
    <au><snm>Bethlem</snm><fnm>H. L.</fnm></au>
    <au><snm>Berden</snm><fnm>G.</fnm></au>
    <au><snm>Crompvoets</snm><fnm>F. M. H.</fnm></au>
    <au><snm>Jongma</snm><fnm>R. T.</fnm></au>
    <au><snm>{\mbox{van~{R}oij}}</snm><fnm>A. J. A.</fnm></au>
    <au><snm>Meijer</snm><fnm>G.</fnm></au>
  </aug>
  <source>Nature</source>
  <pubdate>2000</pubdate>
  <volume>406</volume>
  <fpage>491</fpage>
</bibl>

<bibl id="B14">
  <title><p>Manipulation and Control of Molecular Beams</p></title>
  <aug>
    <au><snm>Meerakker</snm><fnm>S. Y. T.</fnm></au>
    <au><snm>Bethlem</snm><fnm>H. L.</fnm></au>
    <au><snm>Vanhaecke</snm><fnm>N.</fnm></au>
    <au><snm>Meijer</snm><fnm>G.</fnm></au>
  </aug>
  <source>Chem. Rev.</source>
  <pubdate>2012</pubdate>
  <volume>112</volume>
  <fpage>4828</fpage>
</bibl>

<bibl id="B15">
  <title><p>{Der experimentelle Nachweis des magnetischen Moments des
  Silberatoms}</p></title>
  <aug>
    <au><snm>Gerlach</snm><fnm>W.</fnm></au>
    <au><snm>Stern</snm><fnm>O.</fnm></au>
  </aug>
  <source>Z. Phys.</source>
  <pubdate>1921</pubdate>
  <volume>8</volume>
  <fpage>110</fpage>
</bibl>

<bibl id="B16">
  <title><p>Der experimentelle {N}achweis der {R}ichtungsquantelung im
  {M}agnetfeld</p></title>
  <aug>
    <au><snm>Gerlach</snm><fnm>W</fnm></au>
    <au><snm>Stern</snm><fnm>O</fnm></au>
  </aug>
  <source>Z. Phys.</source>
  <pubdate>1922</pubdate>
  <volume>9</volume>
  <fpage>349</fpage>
</bibl>

<bibl id="B17">
  <title><p>{Das magnetische Moment des Silberatoms}</p></title>
  <aug>
    <au><snm>Gerlach</snm><fnm>W</fnm></au>
    <au><snm>Stern</snm><fnm>O</fnm></au>
  </aug>
  <source>Z. Phys.</source>
  <pubdate>1922</pubdate>
  <volume>9</volume>
  <fpage>353</fpage>
</bibl>

<bibl id="B18">
  <title><p>{Rydberg Atoms}</p></title>
  <aug>
    <au><snm>Gallagher</snm><fnm>T. F.</fnm></au>
  </aug>
  <publisher>Cambridge: Cambridge University Press</publisher>
  <pubdate>1994</pubdate>
</bibl>

<bibl id="B19">
  <title><p>Laser cooling of a diatomic molecule</p></title>
  <aug>
    <au><snm>Shuman</snm><fnm>E. S.</fnm></au>
    <au><snm>Barry</snm><fnm>J. F.</fnm></au>
    <au><snm>DeMille</snm><fnm>D.</fnm></au>
  </aug>
  <source>Nature</source>
  <pubdate>2010</pubdate>
  <volume>467</volume>
  <fpage>820</fpage>
</bibl>

<bibl id="B20">
  <title><p>Laser cooling and slowing of {CaF} molecules</p></title>
  <aug>
    <au><snm>Zhelyazkova</snm><fnm>V.</fnm></au>
    <au><snm>Cournol</snm><fnm>A.</fnm></au>
    <au><snm>Wall</snm><fnm>T. E.</fnm></au>
    <au><snm>Matsushima</snm><fnm>A.</fnm></au>
    <au><snm>Hudson</snm><fnm>J. J.</fnm></au>
    <au><snm>Hinds</snm><fnm>E. A.</fnm></au>
    <au><snm>Tarbutt</snm><fnm>M. R.</fnm></au>
    <au><snm>Sauer</snm><fnm>B. E.</fnm></au>
  </aug>
  <source>Phys. Rev. A</source>
  <publisher>American Physical Society</publisher>
  <pubdate>2014</pubdate>
  <volume>89</volume>
  <fpage>053416</fpage>
</bibl>

<bibl id="B21">
  <title><p>Multistage {Z}eeman deceleration of hydrogen atoms</p></title>
  <aug>
    <au><snm>Vanhaecke</snm><fnm>N</fnm></au>
    <au><snm>Meier</snm><fnm>U</fnm></au>
    <au><snm>Andrist</snm><fnm>M</fnm></au>
    <au><snm>Meier</snm><fnm>BH</fnm></au>
    <au><snm>Merkt</snm><fnm>F</fnm></au>
  </aug>
  <source>Phys. Rev. A</source>
  <pubdate>2007</pubdate>
  <volume>75</volume>
  <fpage>031402(R)</fpage>
</bibl>

<bibl id="B22">
  <title><p>{Z}eeman deceleration of {H} and {D}</p></title>
  <aug>
    <au><snm>Hogan</snm><fnm>S. D.</fnm></au>
    <au><snm>Sprecher</snm><fnm>D.</fnm></au>
    <au><snm>Andrist</snm><fnm>M.</fnm></au>
    <au><snm>Vanhaecke</snm><fnm>N.</fnm></au>
    <au><snm>Merkt</snm><fnm>F.</fnm></au>
  </aug>
  <source>Phys. Rev. A</source>
  <pubdate>2007</pubdate>
  <volume>76</volume>
  <fpage>023412</fpage>
</bibl>

<bibl id="B23">
  <title><p>Stopping supersonic beams with a series of pulsed electromagnetic
  coils: {A}n atomic coilgun</p></title>
  <aug>
    <au><snm>Narevicius</snm><fnm>E</fnm></au>
    <au><snm>Libson</snm><fnm>A</fnm></au>
    <au><snm>Parthey</snm><fnm>CG</fnm></au>
    <au><snm>Chavez</snm><fnm>I</fnm></au>
    <au><snm>Narevicius</snm><fnm>J</fnm></au>
    <au><snm>Even</snm><fnm>U</fnm></au>
    <au><snm>Raizen</snm><fnm>MG</fnm></au>
  </aug>
  <source>Phys. Rev. Lett.</source>
  <publisher>American Physical Society</publisher>
  <pubdate>2008</pubdate>
  <volume>100</volume>
  <fpage>093003</fpage>
  <url>http://link.aps.org/doi/10.1103/PhysRevLett.100.093003</url>
</bibl>

<bibl id="B24">
  <title><p>New precision measurement of the decay rate of singlet
  positronium</p></title>
  <aug>
    <au><snm>Al Ramadhan</snm><fnm>A. H.</fnm></au>
    <au><snm>Gidley</snm><fnm>D. W.</fnm></au>
  </aug>
  <source>Phys. Rev. Lett.</source>
  <publisher>American Physical Society</publisher>
  <pubdate>1994</pubdate>
  <volume>72</volume>
  <fpage>1632</fpage>
</bibl>

<bibl id="B25">
  <title><p>Resolution of the Orthopositronium-Lifetime Puzzle</p></title>
  <aug>
    <au><snm>Vallery</snm><fnm>R. S.</fnm></au>
    <au><snm>Zitzewitz</snm><fnm>P. W.</fnm></au>
    <au><snm>Gidley</snm><fnm>D. W.</fnm></au>
  </aug>
  <source>Phys. Rev. Lett.</source>
  <publisher>American Physical Society</publisher>
  <pubdate>2003</pubdate>
  <volume>90</volume>
  <fpage>203402</fpage>
</bibl>

<bibl id="B26">
  <title><p>Ultracold molecules and ultracold chemistry</p></title>
  <aug>
    <au><snm>Bell</snm><fnm>M. T.</fnm></au>
    <au><snm>Softley</snm><fnm>T. P.</fnm></au>
  </aug>
  <source>Mol. Phys.</source>
  <pubdate>2009</pubdate>
  <volume>107</volume>
  <fpage>99</fpage>
</bibl>

<bibl id="B27">
  <title><p>Cold and ultracold molecules: {S}cience, technology and
  applications</p></title>
  <aug>
    <au><snm>Carr</snm><fnm>L. C.</fnm></au>
    <au><snm>DeMille</snm><fnm>D.</fnm></au>
    <au><snm>Krems</snm><fnm>R. V.</fnm></au>
    <au><snm>Ye</snm><fnm>J.</fnm></au>
  </aug>
  <source>New. J. Phys.</source>
  <pubdate>2009</pubdate>
  <volume>11</volume>
  <fpage>055049</fpage>
</bibl>

<bibl id="B28">
  <title><p>Optical {S}tark Decelerator for Molecules</p></title>
  <aug>
    <au><snm>Fulton</snm><fnm>R.</fnm></au>
    <au><snm>Bishop</snm><fnm>A. I.</fnm></au>
    <au><snm>Barker</snm><fnm>P. F.</fnm></au>
  </aug>
  <source>Phys. Rev. Lett.</source>
  <pubdate>2004</pubdate>
  <volume>93</volume>
  <fpage>243004</fpage>
</bibl>

<bibl id="B29">
  <title><p>Measurement of pressure-broadening parameters for the {CO}-{H}e
  system at 4~{K}</p></title>
  <aug>
    <au><snm>Messer</snm><fnm>J. K.</fnm></au>
    <au><snm>{De~Lucia}</snm><fnm>FC</fnm></au>
  </aug>
  <source>Phys. Rev. Lett.</source>
  <publisher>American Physical Society</publisher>
  <pubdate>1984</pubdate>
  <volume>53</volume>
  <fpage>2555</fpage>
</bibl>

<bibl id="B30">
  <title><p>Buffer-gas loading of atoms and molecules into a magnetic
  trap</p></title>
  <aug>
    <au><snm>Doyle</snm><fnm>JM</fnm></au>
    <au><snm>Friedrich</snm><fnm>B</fnm></au>
    <au><snm>Kim</snm><fnm>J</fnm></au>
    <au><snm>Patterson</snm><fnm>D</fnm></au>
  </aug>
  <source>Phys. Rev. A</source>
  <pubdate>1995</pubdate>
  <volume>52</volume>
  <fpage>R2515</fpage>
</bibl>

<bibl id="B31">
  <title><p>Cold Reactive Collisions between Laser-Cooled Ions and
  Velocity-Selected Neutral Molecules</p></title>
  <aug>
    <au><snm>Willitsch</snm><fnm>S</fnm></au>
    <au><snm>Bell</snm><fnm>MT</fnm></au>
    <au><snm>Gingell</snm><fnm>AD</fnm></au>
    <au><snm>Procter</snm><fnm>SR</fnm></au>
    <au><snm>Softley</snm><fnm>TP</fnm></au>
  </aug>
  <source>Phys. Rev. Lett.</source>
  <publisher>American Physical Society</publisher>
  <pubdate>2008</pubdate>
  <volume>100</volume>
  <fpage>043203</fpage>
</bibl>

<bibl id="B32">
  <title><p>Probing Isotope Effects in Chemical Reactions Using Single
  Ions</p></title>
  <aug>
    <au><snm>Staanum</snm><fnm>PF</fnm></au>
    <au><snm>H\o{}jbjerre</snm><fnm>K</fnm></au>
    <au><snm>Wester</snm><fnm>R</fnm></au>
    <au><snm>Drewsen</snm><fnm>M</fnm></au>
  </aug>
  <source>Phys. Rev. Lett.</source>
  <publisher>American Physical Society</publisher>
  <pubdate>2008</pubdate>
  <volume>100</volume>
  <fpage>243003</fpage>
</bibl>

<bibl id="B33">
  <title><p>Magnetic trapping of long-lived cold {R}ydberg atoms</p></title>
  <aug>
    <au><snm>Choi</snm><fnm>J. H.</fnm></au>
    <au><snm>Guest</snm><fnm>J. R.</fnm></au>
    <au><snm>Povilus</snm><fnm>A. P.</fnm></au>
    <au><snm>Hansis</snm><fnm>E.</fnm></au>
    <au><snm>Raithel</snm><fnm>G.</fnm></au>
  </aug>
  <source>Phys. Rev. Lett.</source>
  <publisher>American Physical Society</publisher>
  <pubdate>2005</pubdate>
  <volume>95</volume>
  <fpage>243001</fpage>
</bibl>

<bibl id="B34">
  <title><p>Production and detection of cold antihydrogen atoms</p></title>
  <aug>
    <au><snm>Amoretti</snm><fnm>M.</fnm></au>
    <au><snm>Amsler</snm><fnm>C.</fnm></au>
    <au><snm>Bonomi</snm><fnm>G.</fnm></au>
    <au><snm>Bouchta</snm><fnm>A.</fnm></au>
    <au><snm>Bowe</snm><fnm>P.</fnm></au>
    <au><snm>Carraro</snm><fnm>C.</fnm></au>
    <au><snm>Cesar</snm><fnm>C. L.</fnm></au>
    <au><snm>Charlton</snm><fnm>M.</fnm></au>
    <au><snm>Collier</snm><fnm>M. J. T.</fnm></au>
    <au><snm>Doser</snm><fnm>M.</fnm></au>
    <au><snm>Filippini</snm><fnm>V.</fnm></au>
    <au><snm>Fine</snm><fnm>K. S.</fnm></au>
    <au><snm>Fontana</snm><fnm>A.</fnm></au>
    <au><snm>Fujiwara</snm><fnm>M. C.</fnm></au>
    <au><snm>Funakoshi</snm><fnm>R.</fnm></au>
    <au><snm>Genova</snm><fnm>P.</fnm></au>
    <au><snm>Hangst</snm><fnm>J. S.</fnm></au>
    <au><snm>Hayano</snm><fnm>R. S.</fnm></au>
    <au><snm>Holzscheiter</snm><fnm>M. H.</fnm></au>
    <au><snm>Jorgensen</snm><fnm>L. V.</fnm></au>
    <au><snm>Lagomarsino</snm><fnm>V.</fnm></au>
    <au><snm>Landua</snm><fnm>R.</fnm></au>
    <au><snm>Lindelof</snm><fnm>D.</fnm></au>
    <au><snm>Rizzini</snm><fnm>EL</fnm></au>
    <au><snm>Macri</snm><fnm>M.</fnm></au>
    <au><snm>Madsen</snm><fnm>N.</fnm></au>
    <au><snm>Manuzio</snm><fnm>G.</fnm></au>
    <au><snm>Marchesotti</snm><fnm>M.</fnm></au>
    <au><snm>Montagna</snm><fnm>P.</fnm></au>
    <au><snm>Pruys</snm><fnm>H.</fnm></au>
    <au><snm>Regenfus</snm><fnm>C.</fnm></au>
    <au><snm>Riedler</snm><fnm>P.</fnm></au>
    <au><snm>Rochet</snm><fnm>J.</fnm></au>
    <au><snm>Rotondi</snm><fnm>A.</fnm></au>
    <au><snm>Rouleau</snm><fnm>G.</fnm></au>
    <au><snm>Testera</snm><fnm>G.</fnm></au>
    <au><snm>Variola</snm><fnm>A.</fnm></au>
    <au><snm>Watson</snm><fnm>T. L.</fnm></au>
    <au><snm>Werf</snm><fnm>D. P.</fnm></au>
  </aug>
  <source>Nature</source>
  <pubdate>2002</pubdate>
  <volume>419</volume>
  <fpage>456</fpage>
</bibl>

<bibl id="B35">
  <title><p>Trapping {R}ydberg Atoms in an Optical Lattice</p></title>
  <aug>
    <au><snm>Anderson</snm><fnm>S. E.</fnm></au>
    <au><snm>Younge</snm><fnm>K. C.</fnm></au>
    <au><snm>Raithel</snm><fnm>G.</fnm></au>
  </aug>
  <source>Phys. Rev. Lett.</source>
  <publisher>American Physical Society</publisher>
  <pubdate>2011</pubdate>
  <volume>107</volume>
  <fpage>263001</fpage>
  <url>http://link.aps.org/doi/10.1103/PhysRevLett.107.263001</url>
</bibl>

<bibl id="B36">
  <title><p>Radio recombination lines from the largest bound atoms in
  space</p></title>
  <aug>
    <au><snm>Stepkin</snm><fnm>S. V.</fnm></au>
    <au><snm>Konovalenko</snm><fnm>A. A.</fnm></au>
    <au><snm>Kantharia</snm><fnm>N. G.</fnm></au>
    <au><snm>{Udaya Shankar}</snm><fnm>N</fnm></au>
  </aug>
  <source>Mon. Not. R. Astron. Soc.</source>
  <pubdate>2007</pubdate>
  <volume>374</volume>
  <fpage>852</fpage>
</bibl>

<bibl id="B37">
  <title><p>Photodissociation dynamics and atmospheric chemistry</p></title>
  <aug>
    <au><snm>Wayne</snm><fnm>R. P.</fnm></au>
  </aug>
  <source>J. Geophys. Res.</source>
  <pubdate>1993</pubdate>
  <volume>98</volume>
  <fpage>13119</fpage>
</bibl>

<bibl id="B38">
  <title><p>Millimeter wave spectroscopy of high {R}ydberg states</p></title>
  <aug>
    <au><snm>Merkt</snm><fnm>F.</fnm></au>
    <au><snm>Osterwalder</snm><fnm>A.</fnm></au>
  </aug>
  <source>Int. Rev. Phys. Chem.</source>
  <publisher>Taylor \& Francis</publisher>
  <pubdate>2002</pubdate>
  <volume>21</volume>
  <fpage>385</fpage>
</bibl>

<bibl id="B39">
  <title><p>Driving {R}ydberg-{R}ydberg transitions from a coplanar microwave
  waveguide</p></title>
  <aug>
    <au><snm>Hogan</snm><fnm>S. D.</fnm></au>
    <au><snm>Agner</snm><fnm>J. A.</fnm></au>
    <au><snm>Merkt</snm><fnm>F.</fnm></au>
    <au><snm>Thiele</snm><fnm>T.</fnm></au>
    <au><snm>Filipp</snm><fnm>S.</fnm></au>
    <au><snm>Wallraff</snm><fnm>A.</fnm></au>
  </aug>
  <source>Phys. Rev. Lett.</source>
  <publisher>American Physical Society</publisher>
  <pubdate>2012</pubdate>
  <volume>108</volume>
  <fpage>063004</fpage>
</bibl>

<bibl id="B40">
  <title><p>Manipulating {R}ydberg atoms close to surfaces at cryogenic
  temperatures</p></title>
  <aug>
    <au><snm>Thiele</snm><fnm>T.</fnm></au>
    <au><snm>Filipp</snm><fnm>S.</fnm></au>
    <au><snm>Agner</snm><fnm>J. A.</fnm></au>
    <au><snm>Schmutz</snm><fnm>H.</fnm></au>
    <au><snm>Deiglmayr</snm><fnm>J.</fnm></au>
    <au><snm>Stammeier</snm><fnm>M.</fnm></au>
    <au><snm>Allmendinger</snm><fnm>P.</fnm></au>
    <au><snm>Merkt</snm><fnm>F.</fnm></au>
    <au><snm>Wallraff</snm><fnm>A.</fnm></au>
  </aug>
  <source>Phys. Rev. A</source>
  <publisher>American Physical Society</publisher>
  <pubdate>2014</pubdate>
  <volume>90</volume>
  <fpage>013414</fpage>
</bibl>

<bibl id="B41">
  <title><p>High-resolution millimeter wave spectroscopy and multichannel
  quantum defect theory of the hyperfine structure in high {R}ydberg states of
  molecular hydrogen {H$_2$}</p></title>
  <aug>
    <au><snm>Osterwalder</snm><fnm>A.</fnm></au>
    <au><snm>W{\"u}est</snm><fnm>A.</fnm></au>
    <au><snm>Merkt</snm><fnm>F.</fnm></au>
    <au><snm>Jungen</snm><fnm>{\mbox{Ch}}</fnm></au>
  </aug>
  <source>J. Chem. Phys.</source>
  <pubdate>2004</pubdate>
  <volume>121</volume>
  <fpage>11810</fpage>
</bibl>

<bibl id="B42">
  <title><p>Determination of the ionization and dissociation energies of the
  hydrogen molecule</p></title>
  <aug>
    <au><snm>Liu</snm><fnm>J</fnm></au>
    <au><snm>Salumbides</snm><fnm>EJ</fnm></au>
    <au><snm>Hollenstein</snm><fnm>U</fnm></au>
    <au><snm>Koelemeij</snm><fnm>JCJ</fnm></au>
    <au><snm>Eikema</snm><fnm>KSE</fnm></au>
    <au><snm>Ubachs</snm><fnm>W</fnm></au>
    <au><snm>Merkt</snm><fnm>F</fnm></au>
  </aug>
  <source>J. Chem. Phys.</source>
  <pubdate>2009</pubdate>
  <volume>130</volume>
  <fpage>174306</fpage>
</bibl>

<bibl id="B43">
  <title><p>Determination of the ionization and dissociation energies of the
  deuterium molecule {(D$_2$)}</p></title>
  <aug>
    <au><snm>Liu</snm><fnm>J</fnm></au>
    <au><snm>Sprecher</snm><fnm>D</fnm></au>
    <au><snm>Jungen</snm><fnm>{\mbox{Ch}}</fnm></au>
    <au><snm>Ubachs</snm><fnm>W</fnm></au>
    <au><snm>Merkt</snm><fnm>F</fnm></au>
  </aug>
  <source>J. Chem. Phys.</source>
  <pubdate>2010</pubdate>
  <volume>132</volume>
  <fpage>154301</fpage>
</bibl>

<bibl id="B44">
  <title><p>The ionization and dissociation energies of {HD}</p></title>
  <aug>
    <au><snm>Sprecher</snm><fnm>D</fnm></au>
    <au><snm>Liu</snm><fnm>J</fnm></au>
    <au><snm>Jungen</snm><fnm>{\mbox{Ch}}</fnm></au>
    <au><snm>Ubachs</snm><fnm>W</fnm></au>
    <au><snm>Merkt</snm><fnm>F</fnm></au>
  </aug>
  <source>J. Chem. Phys.</source>
  <pubdate>2010</pubdate>
  <volume>133</volume>
  <fpage>111102</fpage>
</bibl>

<bibl id="B45">
  <title><p>A Molecular Beam Resonance Method with Separated Oscillating
  Fields</p></title>
  <aug>
    <au><snm>Ramsey</snm><fnm>NF</fnm></au>
  </aug>
  <source>Phys. Rev.</source>
  <publisher>American Physical Society</publisher>
  <pubdate>1950</pubdate>
  <volume>78</volume>
  <fpage>695</fpage>
</bibl>

<bibl id="B46">
  <title><p>Observation of Resonances in {P}enning Ionization Reactions at
  Sub-{K}elvin Temperatures in Merged Beams</p></title>
  <aug>
    <au><snm>Henson</snm><fnm>A. B.</fnm></au>
    <au><snm>Gersten</snm><fnm>S.</fnm></au>
    <au><snm>Shagam</snm><fnm>Y.</fnm></au>
    <au><snm>Narevicius</snm><fnm>J.</fnm></au>
    <au><snm>Narevicius</snm><fnm>E.</fnm></au>
  </aug>
  <source>Science</source>
  <pubdate>2012</pubdate>
  <volume>338</volume>
  <fpage>234</fpage>
</bibl>

<bibl id="B47">
  <title><p>Sub-{K}elvin Collision Temperatures in Merged Neutral Beams by
  Correlation in Phase-Space</p></title>
  <aug>
    <au><snm>Shagam</snm><fnm>Y.</fnm></au>
    <au><snm>E. Narevicius</snm><fnm>E.</fnm></au>
  </aug>
  <source>J. Phys. Chem. C</source>
  <pubdate>2013</pubdate>
  <volume>117</volume>
  <fpage>22454</fpage>
</bibl>

<bibl id="B48">
  <title><p>Atoms in micron-sized metallic and dielectric
  waveguides</p></title>
  <aug>
    <au><snm>Hinds</snm><fnm>E. A.</fnm></au>
    <au><snm>Lai</snm><fnm>K. S.</fnm></au>
    <au><snm>Schnell</snm><fnm>M.</fnm></au>
  </aug>
  <source>Phil. Trans. R. Soc. Lond. A</source>
  <pubdate>1997</pubdate>
  <volume>355</volume>
  <fpage>2353</fpage>
</bibl>

<bibl id="B49">
  <title><p>Coherent excitation of {R}ydberg atoms in micrometre-sized atomic
  vapour cells</p></title>
  <aug>
    <au><snm>K\"ubler</snm><fnm>H.</fnm></au>
    <au><snm>Shaffer</snm><fnm>J. P.</fnm></au>
    <au><snm>Baluktsian</snm><fnm>T.</fnm></au>
    <au><snm>L\"ow</snm><fnm>R.</fnm></au>
    <au><snm>Pfau</snm><fnm>T.</fnm></au>
  </aug>
  <source>Nature Photon.</source>
  <pubdate>2010</pubdate>
  <volume>4</volume>
  <fpage>112</fpage>
</bibl>

<bibl id="B50">
  <title><p>Ionization of {K}($n$d) {R}ydberg-state atoms at a
  surface</p></title>
  <aug>
    <au><snm>Gray</snm><fnm>D. F.</fnm></au>
    <au><snm>Zheng</snm><fnm>Z.</fnm></au>
    <au><snm>Smith</snm><fnm>K. A.</fnm></au>
    <au><snm>Dunning</snm><fnm>F. B.</fnm></au>
  </aug>
  <source>Phys. Rev. A</source>
  <publisher>American Physical Society</publisher>
  <pubdate>1988</pubdate>
  <volume>38</volume>
  <fpage>1601</fpage>
</bibl>

<bibl id="B51">
  <title><p>Measuring the van~der~{W}aals forces between a {R}ydberg atom and a
  metallic surface</p></title>
  <aug>
    <au><snm>Anderson</snm><fnm>A.</fnm></au>
    <au><snm>Haroche</snm><fnm>S.</fnm></au>
    <au><snm>Hinds</snm><fnm>E. A.</fnm></au>
    <au><snm>Jhe</snm><fnm>W.</fnm></au>
    <au><snm>Meschede</snm><fnm>D.</fnm></au>
  </aug>
  <source>Phys. Rev. A</source>
  <publisher>American Physical Society</publisher>
  <pubdate>1988</pubdate>
  <volume>37</volume>
  <fpage>3594</fpage>
</bibl>

<bibl id="B52">
  <title><p>Direct measurement of the van~der~{W}aals interaction between an
  atom and its images in a micron-sized cavity</p></title>
  <aug>
    <au><snm>Sandoghdar</snm><fnm>V.</fnm></au>
    <au><snm>Sukenik</snm><fnm>C. I.</fnm></au>
    <au><snm>Hinds</snm><fnm>E. A.</fnm></au>
    <au><snm>Haroche</snm><fnm>S</fnm></au>
  </aug>
  <source>Phys. Rev. Lett.</source>
  <publisher>American Physical Society</publisher>
  <pubdate>1992</pubdate>
  <volume>68</volume>
  <fpage>3432</fpage>
</bibl>

<bibl id="B53">
  <title><p>Ionization of xenon {R}ydberg atoms at a metal surface</p></title>
  <aug>
    <au><snm>Hill</snm><fnm>S. B.</fnm></au>
    <au><snm>Haich</snm><fnm>C. B.</fnm></au>
    <au><snm>Zhou</snm><fnm>Z.</fnm></au>
    <au><snm>Nordlander</snm><fnm>P.</fnm></au>
    <au><snm>Dunning</snm><fnm>F. B.</fnm></au>
  </aug>
  <source>Phys. Rev. Lett.</source>
  <publisher>American Physical Society</publisher>
  <pubdate>2000</pubdate>
  <volume>85</volume>
  <fpage>5444</fpage>
</bibl>

<bibl id="B54">
  <title><p>Ionization of hydrogen {R}ydberg molecules at a metal
  surface</p></title>
  <aug>
    <au><snm>Lloyd</snm><fnm>G. R.</fnm></au>
    <au><snm>Procter</snm><fnm>S. R.</fnm></au>
    <au><snm>Softley</snm><fnm>T. P.</fnm></au>
  </aug>
  <source>Phys. Rev. Lett.</source>
  <publisher>American Physical Society</publisher>
  <pubdate>2005</pubdate>
  <volume>95</volume>
  <fpage>133202</fpage>
</bibl>

<bibl id="B55">
  <title><p>Charge transfer of {R}ydberg {H} atoms at a metal
  surface</p></title>
  <aug>
    <au><snm>So</snm><fnm>E.</fnm></au>
    <au><snm>Dethlefsen</snm><fnm>M.</fnm></au>
    <au><snm>Ford</snm><fnm>M.</fnm></au>
    <au><snm>Softley</snm><fnm>T. P.</fnm></au>
  </aug>
  <source>Phys. Rev. Lett.</source>
  <publisher>American Physical Society</publisher>
  <pubdate>2011</pubdate>
  <volume>107</volume>
  <fpage>093201</fpage>
</bibl>

<bibl id="B56">
  <title><p>Resonant Charge Transfer of Hydrogen {R}ydberg Atoms Incident on a
  {C}u(100) Projected Band-Gap Surface</p></title>
  <aug>
    <au><snm>Gibbard</snm><fnm>J. A.</fnm></au>
    <au><snm>Dethlefsen</snm><fnm>M.</fnm></au>
    <au><snm>Kohlhoff</snm><fnm>M.</fnm></au>
    <au><snm>Rennick</snm><fnm>C. J.</fnm></au>
    <au><snm>So</snm><fnm>E.</fnm></au>
    <au><snm>Ford</snm><fnm>M.</fnm></au>
    <au><snm>Softley</snm><fnm>T. P.</fnm></au>
  </aug>
  <source>Phys. Rev. Lett.</source>
  <publisher>American Physical Society</publisher>
  <pubdate>2015</pubdate>
  <volume>115</volume>
  <fpage>093201</fpage>
</bibl>

<bibl id="B57">
  <title><p>Observation of ultralong-range {R}ydberg molecules</p></title>
  <aug>
    <au><snm>Bendkowsky</snm><fnm>V.</fnm></au>
    <au><snm>Butscher</snm><fnm>B.</fnm></au>
    <au><snm>Nipper</snm><fnm>J.</fnm></au>
    <au><snm>Shaffer</snm><fnm>J. P.</fnm></au>
    <au><snm>L\"ow</snm><fnm>R.</fnm></au>
    <au><snm>Pfau</snm><fnm>T.</fnm></au>
  </aug>
  <source>Nature</source>
  <pubdate>2009</pubdate>
  <volume>58</volume>
  <fpage>1005</fpage>
</bibl>

<bibl id="B58">
  <title><p>Dipole-dipole interactions of {R}ydberg atoms</p></title>
  <aug>
    <au><snm>Gallagher</snm><fnm>T. F.</fnm></au>
    <au><snm>Pillet</snm><fnm>P.</fnm></au>
  </aug>
  <source>Adv. At. Mol. Opt. Phys.</source>
  <pubdate>2008</pubdate>
  <volume>56</volume>
  <fpage>161</fpage>
</bibl>

<bibl id="B59">
  <title><p>Dipole blockade in a cold {R}ydberg atomic sample</p></title>
  <aug>
    <au><snm>Comparat</snm><fnm>D.</fnm></au>
    <au><snm>P. Pillet</snm><fnm>P.</fnm></au>
  </aug>
  <source>J. Opt. Soc. Am. B</source>
  <pubdate>2010</pubdate>
  <volume>27</volume>
  <fpage>A208</fpage>
</bibl>

<bibl id="B60">
  <title><p>Resonant Dipole-Dipole Energy Transfer in a Nearly Frozen {R}ydberg
  Gas</p></title>
  <aug>
    <au><snm>Anderson</snm><fnm>W. R.</fnm></au>
    <au><snm>Veale</snm><fnm>J. R.</fnm></au>
    <au><snm>Gallagher</snm><fnm>T. F.</fnm></au>
  </aug>
  <source>Phys. Rev. Lett.</source>
  <pubdate>1998</pubdate>
  <volume>80</volume>
  <fpage>249</fpage>
</bibl>

<bibl id="B61">
  <title><p>Many-Body Effects in a Frozen {R}ydberg Gas</p></title>
  <aug>
    <au><snm>Mourachko</snm><fnm>I.</fnm></au>
    <au><snm>Comparat</snm><fnm>D.</fnm></au>
    <au><snm>Tomasi</snm><fnm>F.</fnm></au>
    <au><snm>Fioretti</snm><fnm>A.</fnm></au>
    <au><snm>Nosbaum</snm><fnm>P.</fnm></au>
    <au><snm>Akulin</snm><fnm>V. M.</fnm></au>
    <au><snm>Pillet</snm><fnm>P.</fnm></au>
  </aug>
  <source>Phys. Rev. Lett.</source>
  <publisher>American Physical Society</publisher>
  <pubdate>1998</pubdate>
  <volume>80</volume>
  <fpage>253</fpage>
</bibl>

<bibl id="B62">
  <title><p>Dipole Blockade at {F}\"orster Resonances in High Resolution Laser
  Excitation of {R}ydberg States of Cesium Atoms</p></title>
  <aug>
    <au><snm>Vogt</snm><fnm>T</fnm></au>
    <au><snm>Viteau</snm><fnm>M</fnm></au>
    <au><snm>Zhao</snm><fnm>J</fnm></au>
    <au><snm>Chotia</snm><fnm>A</fnm></au>
    <au><snm>Comparat</snm><fnm>D</fnm></au>
    <au><snm>Pillet</snm><fnm>P</fnm></au>
  </aug>
  <source>Phys. Rev. Lett.</source>
  <publisher>American Physical Society</publisher>
  <pubdate>2006</pubdate>
  <volume>97</volume>
  <fpage>083003</fpage>
</bibl>

<bibl id="B63">
  <title><p>Cooperative Atom-Light Interaction in a Blockaded {R}ydberg
  Ensemble</p></title>
  <aug>
    <au><snm>Pritchard</snm><fnm>J. D.</fnm></au>
    <au><snm>Maxwell</snm><fnm>D.</fnm></au>
    <au><snm>Gauguet</snm><fnm>A.</fnm></au>
    <au><snm>Weatherill</snm><fnm>K. J.</fnm></au>
    <au><snm>Jones</snm><fnm>M. P. A.</fnm></au>
    <au><snm>Adams</snm><fnm>C. S.</fnm></au>
  </aug>
  <source>Phys. Rev. Lett.</source>
  <pubdate>2010</pubdate>
  <volume>105</volume>
  <fpage>193603</fpage>
</bibl>

<bibl id="B64">
  <title><p>Entanglement of two individual neutral atoms using {R}ydberg
  blockade</p></title>
  <aug>
    <au><snm>Wilk</snm><fnm>T.</fnm></au>
    <au><snm>Ga\"etan</snm><fnm>A.</fnm></au>
    <au><snm>Evellin</snm><fnm>C.</fnm></au>
    <au><snm>Wolters</snm><fnm>J.</fnm></au>
    <au><snm>Miroshnychenko</snm><fnm>Y.</fnm></au>
    <au><snm>Grangier</snm><fnm>P.</fnm></au>
    <au><snm>Browaeys</snm><fnm>A.</fnm></au>
  </aug>
  <source>Phys. Rev. Lett.</source>
  <pubdate>2010</pubdate>
  <volume>104</volume>
  <fpage>010502</fpage>
</bibl>

<bibl id="B65">
  <title><p>Demonstration of a neutral atom controlled-{NOT} quantum
  gate</p></title>
  <aug>
    <au><snm>Isenhower</snm><fnm>L.</fnm></au>
    <au><snm>Urban</snm><fnm>E.</fnm></au>
    <au><snm>Zhang</snm><fnm>X. L.</fnm></au>
    <au><snm>Gill</snm><fnm>A. T.</fnm></au>
    <au><snm>Henage</snm><fnm>T.</fnm></au>
    <au><snm>Johnson</snm><fnm>T. A.</fnm></au>
    <au><snm>Walker</snm><fnm>T. G.</fnm></au>
    <au><snm>Saffman</snm><fnm>M.</fnm></au>
  </aug>
  <source>Phys. Rev. Lett.</source>
  <pubdate>2010</pubdate>
  <volume>104</volume>
  <fpage>010503</fpage>
</bibl>

<bibl id="B66">
  <title><p>Discrete energy transfer in collisions of {Xe}($n\mathrm{f}$)
  {R}ydberg atoms with {NH}$_{3}$ molecules</p></title>
  <aug>
    <au><snm>Smith</snm><fnm>K. A.</fnm></au>
    <au><snm>Kellert</snm><fnm>F. G.</fnm></au>
    <au><snm>Rundel</snm><fnm>R. D.</fnm></au>
    <au><snm>Dunning</snm><fnm>F. B.</fnm></au>
    <au><snm>Stebbings</snm><fnm>R. F.</fnm></au>
  </aug>
  <source>Phys. Rev. Lett.</source>
  <publisher>American Physical Society</publisher>
  <pubdate>1978</pubdate>
  <volume>40</volume>
  <fpage>1362</fpage>
</bibl>

<bibl id="B67">
  <title><p>Creation of polar and nonpolar ultra-long-range {R}ydberg
  molecules</p></title>
  <aug>
    <au><snm>Greene</snm><fnm>CH</fnm></au>
    <au><snm>Dickinson</snm><fnm>A. S.</fnm></au>
    <au><snm>Sadeghpour</snm><fnm>H. R.</fnm></au>
  </aug>
  <source>Phys. Rev. Lett.</source>
  <publisher>American Physical Society</publisher>
  <pubdate>2000</pubdate>
  <volume>85</volume>
  <fpage>2458</fpage>
</bibl>

<bibl id="B68">
  <title><p>Cavity quantum electrodynamics</p></title>
  <aug>
    <au><snm>Walther</snm><fnm>H.</fnm></au>
    <au><snm>Varcoe</snm><fnm>B. T. H.</fnm></au>
    <au><snm>Englert</snm><fnm>B. G.</fnm></au>
    <au><snm>Becker</snm><fnm>T.</fnm></au>
  </aug>
  <source>Rep. Prog. Phys.</source>
  <pubdate>2006</pubdate>
  <volume>69</volume>
  <fpage>1325</fpage>
</bibl>

<bibl id="B69">
  <title><p>Strong coupling of a single photon to a superconducting qubit using
  circuit quantum electrodynamics</p></title>
  <aug>
    <au><snm>Wallraff</snm><fnm>A.</fnm></au>
    <au><snm>Schuster</snm><fnm>D. I.</fnm></au>
    <au><snm>Blais</snm><fnm>A.</fnm></au>
    <au><snm>Frunzio</snm><fnm>L.</fnm></au>
    <au><snm>Huang</snm><fnm>R. S.</fnm></au>
    <au><snm>Majer</snm><fnm>J.</fnm></au>
    <au><snm>Kumar</snm><fnm>S.</fnm></au>
    <au><snm>Girvin</snm><fnm>S. M.</fnm></au>
    <au><snm>Schoelkopf</snm><fnm>R. J.</fnm></au>
  </aug>
  <source>Nature</source>
  <pubdate>2004</pubdate>
  <volume>431</volume>
  <fpage>162</fpage>
</bibl>

<bibl id="B70">
  <title><p>Hybrid Quantum Processors: {M}olecular Ensembles as Quantum Memory
  for Solid State Circuits</p></title>
  <aug>
    <au><snm>Rabl</snm><fnm>P.</fnm></au>
    <au><snm>DeMille</snm><fnm>D.</fnm></au>
    <au><snm>Doyle</snm><fnm>J. M.</fnm></au>
    <au><snm>Lukin</snm><fnm>M. D.</fnm></au>
    <au><snm>Schoelkopf</snm><fnm>R. J.</fnm></au>
    <au><snm>Zoller</snm><fnm>P.</fnm></au>
  </aug>
  <source>Phys. Rev. Lett.</source>
  <publisher>American Physical Society</publisher>
  <pubdate>2006</pubdate>
  <volume>97</volume>
  <fpage>033003</fpage>
</bibl>

<bibl id="B71">
  <title><p>Electric-field sensing near the surface microstructure of an atom
  chip using cold {R}ydberg atoms</p></title>
  <aug>
    <au><snm>Carter</snm><fnm>J. D.</fnm></au>
    <au><snm>Cherry</snm><fnm>O.</fnm></au>
    <au><snm>Martin</snm><fnm>J. D. D.</fnm></au>
  </aug>
  <source>Phys. Rev. A</source>
  <publisher>American Physical Society</publisher>
  <pubdate>2012</pubdate>
  <volume>86</volume>
  <fpage>053401</fpage>
</bibl>

<bibl id="B72">
  <title><p>Long coherence times for {R}ydberg qubits on a superconducting atom
  chip</p></title>
  <aug>
    <au><snm>Hermann Avigliano</snm><fnm>C.</fnm></au>
    <au><snm>Teixeira</snm><fnm>RC</fnm></au>
    <au><snm>Nguyen</snm><fnm>T. L.</fnm></au>
    <au><snm>Cantat Moltrecht</snm><fnm>T.</fnm></au>
    <au><snm>Nogues</snm><fnm>G.</fnm></au>
    <au><snm>Dotsenko</snm><fnm>I.</fnm></au>
    <au><snm>Gleyzes</snm><fnm>S.</fnm></au>
    <au><snm>Raimond</snm><fnm>J. M.</fnm></au>
    <au><snm>Haroche</snm><fnm>S.</fnm></au>
    <au><snm>Brune</snm><fnm>M.</fnm></au>
  </aug>
  <source>Phys. Rev. A</source>
  <publisher>American Physical Society</publisher>
  <pubdate>2014</pubdate>
  <volume>90</volume>
  <fpage>040502</fpage>
</bibl>

<bibl id="B73">
  <title><p>Selective Production of {R}ydberg-{S}tark States of
  Positronium</p></title>
  <aug>
    <au><snm>Wall</snm><fnm>T. E.</fnm></au>
    <au><snm>Alonso</snm><fnm>A. M.</fnm></au>
    <au><snm>Cooper</snm><fnm>B. S.</fnm></au>
    <au><snm>Deller</snm><fnm>A.</fnm></au>
    <au><snm>Hogan</snm><fnm>S. D.</fnm></au>
    <au><snm>Cassidy</snm><fnm>D. B.</fnm></au>
  </aug>
  <source>Phys. Rev. Lett.</source>
  <pubdate>2015</pubdate>
  <volume>114</volume>
  <fpage>173001</fpage>
</bibl>

<bibl id="B74">
  <title><p>On antihydrogen formation in collisions of antiprotons with
  positronium</p></title>
  <aug>
    <au><snm>Humberston</snm><fnm>J. W.</fnm></au>
    <au><snm>Charlton</snm><fnm>M.</fnm></au>
    <au><snm>Jacobson</snm><fnm>F. M.</fnm></au>
    <au><snm>Deutch</snm><fnm>B. I.</fnm></au>
  </aug>
  <source>J. Phys. B.: At. Mol. Opt. Phys.</source>
  <pubdate>1987</pubdate>
  <volume>20</volume>
  <fpage>L25</fpage>
</bibl>

<bibl id="B75">
  <title><p>Antihydrogen production in collisions of antiprotons with excited
  states of positronium</p></title>
  <aug>
    <au><snm>Charlton</snm><fnm>M.</fnm></au>
  </aug>
  <source>Phys. Lett. A</source>
  <pubdate>1990</pubdate>
  <volume>143</volume>
  <fpage>143</fpage>
</bibl>

<bibl id="B76">
  <title><p>First Laser-Controlled Antihydrogen Production</p></title>
  <aug>
    <au><snm>Storry</snm><fnm>C. H.</fnm></au>
    <au><snm>Speck</snm><fnm>A.</fnm></au>
    <au><snm>Sage</snm><fnm>DL</fnm></au>
    <au><snm>Guise</snm><fnm>N.</fnm></au>
    <au><snm>Gabrielse</snm><fnm>G.</fnm></au>
    <au><snm>Grzonka</snm><fnm>D.</fnm></au>
    <au><snm>Oelert</snm><fnm>W.</fnm></au>
    <au><snm>Schepers</snm><fnm>G.</fnm></au>
    <au><snm>Sefzick</snm><fnm>T.</fnm></au>
    <au><snm>Pittner</snm><fnm>H.</fnm></au>
    <au><snm>Herrmann</snm><fnm>M.</fnm></au>
    <au><snm>Walz</snm><fnm>J.</fnm></au>
    <au><snm>H\"ansch</snm><fnm>T. W.</fnm></au>
    <au><snm>Comeau</snm><fnm>D.</fnm></au>
    <au><snm>Hessels</snm><fnm>E. A.</fnm></au>
  </aug>
  <source>Phys. Rev. Lett.</source>
  <publisher>American Physical Society</publisher>
  <pubdate>2004</pubdate>
  <volume>93</volume>
  <fpage>263401</fpage>
</bibl>

<bibl id="B77">
  <title><p>Proposed antimatter gravity measurement with an antihydrogen
  beam</p></title>
  <aug>
    <au><snm>Kellerbauer</snm><fnm>A.</fnm></au>
    <au><snm>Amoretti</snm><fnm>M.</fnm></au>
    <au><snm>Belov</snm><fnm>A.S.</fnm></au>
    <au><snm>Bonomi</snm><fnm>G.</fnm></au>
    <au><snm>Boscolo</snm><fnm>I.</fnm></au>
    <au><snm>Brusa</snm><fnm>R.S.</fnm></au>
    <au><snm>B{\"u}chner</snm><fnm>M.</fnm></au>
    <au><snm>Byakov</snm><fnm>V.M.</fnm></au>
    <au><snm>Cabaret</snm><fnm>L.</fnm></au>
    <au><snm>Canali</snm><fnm>C.</fnm></au>
    <au><snm>Carraro</snm><fnm>C.</fnm></au>
    <au><snm>Castelli</snm><fnm>F.</fnm></au>
    <au><snm>Cialdi</snm><fnm>S.</fnm></au>
    <au><snm>{de Combarieu}</snm><fnm>M.</fnm></au>
    <au><snm>Comparat</snm><fnm>D.</fnm></au>
    <au><snm>Consolati</snm><fnm>G.</fnm></au>
    <au><snm>Djourelov</snm><fnm>N.</fnm></au>
    <au><snm>Doser</snm><fnm>M.</fnm></au>
    <au><snm>Drobychev</snm><fnm>G.</fnm></au>
    <au><snm>Dupasquier</snm><fnm>A.</fnm></au>
    <au><snm>Ferrari</snm><fnm>G.</fnm></au>
    <au><snm>Forget</snm><fnm>P.</fnm></au>
    <au><snm>Formaro</snm><fnm>L.</fnm></au>
    <au><snm>Gervasini</snm><fnm>A.</fnm></au>
    <au><snm>Giammarchi</snm><fnm>M.G.</fnm></au>
    <au><snm>Gninenko</snm><fnm>S.N.</fnm></au>
    <au><snm>Gribakin</snm><fnm>G.</fnm></au>
    <au><snm>Hogan</snm><fnm>S. D.</fnm></au>
    <au><snm>Jacquey</snm><fnm>M.</fnm></au>
    <au><snm>Lagomarsino</snm><fnm>V.</fnm></au>
    <au><snm>Manuzio</snm><fnm>G.</fnm></au>
    <au><snm>Mariazzi</snm><fnm>S.</fnm></au>
    <au><snm>Matveev</snm><fnm>V.A.</fnm></au>
    <au><snm>Meier</snm><fnm>J.O.</fnm></au>
    <au><snm>Merkt</snm><fnm>F.</fnm></au>
    <au><snm>Nedelec</snm><fnm>P.</fnm></au>
    <au><snm>Oberthaler</snm><fnm>M.K.</fnm></au>
    <au><snm>Pari</snm><fnm>P.</fnm></au>
    <au><snm>Prevedelli</snm><fnm>M.</fnm></au>
    <au><snm>Quasso</snm><fnm>F.</fnm></au>
    <au><snm>Rotondi</snm><fnm>A.</fnm></au>
    <au><snm>Sillou</snm><fnm>D.</fnm></au>
    <au><snm>Stepanov</snm><fnm>S.V.</fnm></au>
    <au><snm>Stroke</snm><fnm>H.H.</fnm></au>
    <au><snm>Testera</snm><fnm>G.</fnm></au>
    <au><snm>Tino</snm><fnm>G.M.</fnm></au>
    <au><snm>Tr{\'e}nec</snm><fnm>G.</fnm></au>
    <au><snm>Vairo</snm><fnm>A.</fnm></au>
    <au><snm>Vigu{\'e}</snm><fnm>J.</fnm></au>
    <au><snm>Walters</snm><fnm>H.</fnm></au>
    <au><snm>Warring</snm><fnm>U.</fnm></au>
    <au><snm>Zavatarelli</snm><fnm>S.</fnm></au>
    <au><snm>Zvezhinskij</snm><fnm>D.S.</fnm></au>
  </aug>
  <source>Nucl. Instr. and Meth. in Phys. Res. B</source>
  <pubdate>2008</pubdate>
  <volume>266</volume>
  <fpage>351</fpage>
</bibl>

<bibl id="B78">
  <title><p>Trapped antihydrogen</p></title>
  <aug>
    <au><snm>Andresen</snm><fnm>G. B.</fnm></au>
    <au><snm>Ashkezari</snm><fnm>M. D.</fnm></au>
    <au><snm>Baquero Ruiz</snm><fnm>M.</fnm></au>
    <au><snm>Bertsche</snm><fnm>W.</fnm></au>
    <au><snm>Bowe</snm><fnm>P. D.</fnm></au>
    <au><snm>Butler</snm><fnm>E.</fnm></au>
    <au><snm>Cesar</snm><fnm>C. L.</fnm></au>
    <au><snm>Chapman</snm><fnm>S.</fnm></au>
    <au><snm>Charlton</snm><fnm>M.</fnm></au>
    <au><snm>Deller</snm><fnm>A.</fnm></au>
    <au><snm>Eriksson</snm><fnm>S.</fnm></au>
    <au><snm>Fajans</snm><fnm>J.</fnm></au>
    <au><snm>Friesen</snm><fnm>T.</fnm></au>
    <au><snm>Fujiwara</snm><fnm>M. C.</fnm></au>
    <au><snm>Gill</snm><fnm>D. R.</fnm></au>
    <au><snm>Gutierrez</snm><fnm>A.</fnm></au>
    <au><snm>Hangst</snm><fnm>J. S.</fnm></au>
    <au><snm>Hardy</snm><fnm>W. N.</fnm></au>
    <au><snm>Hayden</snm><fnm>M. E.</fnm></au>
    <au><snm>Humphries</snm><fnm>A. J.</fnm></au>
    <au><snm>Hydomako</snm><fnm>R.</fnm></au>
    <au><snm>Jenkins</snm><fnm>M. J.</fnm></au>
    <au><snm>Jonsell</snm><fnm>S.</fnm></au>
    <au><snm>J{\o}rgensen</snm><fnm>L. V.</fnm></au>
    <au><snm>Kurchaninov</snm><fnm>L.</fnm></au>
    <au><snm>Madsen</snm><fnm>N.</fnm></au>
    <au><snm>Menary</snm><fnm>S.</fnm></au>
    <au><snm>Nolan</snm><fnm>P.</fnm></au>
    <au><snm>Olchanski</snm><fnm>K.</fnm></au>
    <au><snm>Olin</snm><fnm>A.</fnm></au>
    <au><snm>Povilus</snm><fnm>A.</fnm></au>
    <au><snm>Pusa</snm><fnm>P.</fnm></au>
    <au><snm>Robicheaux</snm><fnm>F.</fnm></au>
    <au><snm>Sarid</snm><fnm>E.</fnm></au>
    <au><snm>{N}asr</snm><fnm>S.</fnm></au>
    <au><snm>Silveira</snm><fnm>D. M.</fnm></au>
    <au><snm>So</snm><fnm>C.</fnm></au>
    <au><snm>Storey</snm><fnm>J. W.</fnm></au>
    <au><snm>Thompson</snm><fnm>R. I.</fnm></au>
    <au><snm>Werf</snm><fnm>D. P.</fnm></au>
    <au><snm>Wurtele</snm><fnm>J. S.</fnm></au>
    <au><snm>Yamazaki</snm><fnm>Y.</fnm></au>
  </aug>
  <source>Nature</source>
  <pubdate>2010</pubdate>
  <volume>468</volume>
  <fpage>673</fpage>
</bibl>

<bibl id="B79">
  <title><p>Confinement of antihydrogen for 1000 seconds</p></title>
  <aug>
    <au><snm>Andresen</snm><fnm>G. B.</fnm></au>
    <au><snm>Ashkezari</snm><fnm>M. D.</fnm></au>
    <au><snm>Baquero Ruiz</snm><fnm>M.</fnm></au>
    <au><snm>Bertsche</snm><fnm>W.</fnm></au>
    <au><snm>Bowe</snm><fnm>P. D.</fnm></au>
    <au><snm>Butler</snm><fnm>E.</fnm></au>
    <au><snm>Cesar</snm><fnm>C. L.</fnm></au>
    <au><snm>Charlton</snm><fnm>M.</fnm></au>
    <au><snm>Deller</snm><fnm>A.</fnm></au>
    <au><snm>Eriksson</snm><fnm>S.</fnm></au>
    <au><snm>Fajans</snm><fnm>J.</fnm></au>
    <au><snm>Friesen</snm><fnm>T.</fnm></au>
    <au><snm>Fujiwara</snm><fnm>M. C.</fnm></au>
    <au><snm>Gill</snm><fnm>D. R.</fnm></au>
    <au><snm>Gutierrez</snm><fnm>A.</fnm></au>
    <au><snm>Hangst</snm><fnm>J. S.</fnm></au>
    <au><snm>Hardy</snm><fnm>W. N.</fnm></au>
    <au><snm>Hayano</snm><fnm>R. S.</fnm></au>
    <au><snm>Hayden</snm><fnm>M. E.</fnm></au>
    <au><snm>Humphries</snm><fnm>A. J.</fnm></au>
    <au><snm>Hydomako</snm><fnm>R.</fnm></au>
    <au><snm>Jonsell</snm><fnm>S.</fnm></au>
    <au><snm>Kemp</snm><fnm>S. L.</fnm></au>
    <au><snm>Kurchaninov</snm><fnm>L.</fnm></au>
    <au><snm>Madsen</snm><fnm>N.</fnm></au>
    <au><snm>Menary</snm><fnm>S.</fnm></au>
    <au><snm>Nolan</snm><fnm>P.</fnm></au>
    <au><snm>Olchanski</snm><fnm>K.</fnm></au>
    <au><snm>Olin</snm><fnm>A.</fnm></au>
    <au><snm>Pusa</snm><fnm>P.</fnm></au>
    <au><snm>Rasmussen</snm><fnm>C. {\O}.</fnm></au>
    <au><snm>Robicheaux</snm><fnm>F.</fnm></au>
    <au><snm>Sarid</snm><fnm>E.</fnm></au>
    <au><snm>Silveira</snm><fnm>D. M.</fnm></au>
    <au><snm>So</snm><fnm>C.</fnm></au>
    <au><snm>Storey</snm><fnm>J. W.</fnm></au>
    <au><snm>Thompson</snm><fnm>R. I.</fnm></au>
    <au><snm>{van~der~Werf}</snm><fnm>D. P.</fnm></au>
    <au><snm>Wurtele</snm><fnm>J. S.</fnm></au>
    <au><snm>Yamazaki</snm><fnm>Y.</fnm></au>
  </aug>
  <source>Nature Phys.</source>
  <pubdate>2011</pubdate>
  <volume>7</volume>
  <fpage>558</fpage>
</bibl>

<bibl id="B80">
  <title><p>Can we measure the gravitational free fall of cold {R}ydberg state
  positronium?</p></title>
  <aug>
    <au><snm>{Mills~Jr}</snm><fnm>A. P.</fnm></au>
    <au><snm>Leventhal</snm><fnm>M.</fnm></au>
  </aug>
  <source>Nucl. Inst. Meth. Phys. Res. B</source>
  <pubdate>2002</pubdate>
  <volume>192</volume>
  <fpage>102</fpage>
</bibl>

<bibl id="B81">
  <title><p>Atom control and gravity measurements using {R}ydberg
  positronium</p></title>
  <aug>
    <au><snm>Cassidy</snm><fnm>D. B.</fnm></au>
    <au><snm>Hogan</snm><fnm>S. D.</fnm></au>
  </aug>
  <source>Int. J. Mod. Phys. Conf. Ser.</source>
  <pubdate>2014</pubdate>
  <volume>30</volume>
  <fpage>1460259</fpage>
</bibl>

<bibl id="B82">
  <title><p>Precision physics of simple atoms: {QED} tests, nuclear structure
  and fundamental constants</p></title>
  <aug>
    <au><snm>Karshenboim</snm><fnm>S. G.</fnm></au>
  </aug>
  <source>Phys. Rep.</source>
  <pubdate>2005</pubdate>
  <volume>422</volume>
  <fpage>1</fpage>
</bibl>

<bibl id="B83">
  <title><p>First observation of resonant excitation of high-\textit{n} states
  in positronium</p></title>
  <aug>
    <au><snm>Ziock</snm><fnm>K. P.</fnm></au>
    <au><snm>Howell</snm><fnm>R. H.</fnm></au>
    <au><snm>Magnotta</snm><fnm>F.</fnm></au>
    <au><snm>Failor</snm><fnm>R. A.</fnm></au>
    <au><snm>Jones</snm><fnm>K. M.</fnm></au>
  </aug>
  <source>Phys. Rev. Lett.</source>
  <publisher>American Physical Society</publisher>
  <pubdate>1990</pubdate>
  <volume>64</volume>
  <fpage>2366</fpage>
</bibl>

<bibl id="B84">
  <title><p>Efficient Production of {R}ydberg Positronium</p></title>
  <aug>
    <au><snm>Cassidy</snm><fnm>D. B.</fnm></au>
    <au><snm>Hisakado</snm><fnm>T. H.</fnm></au>
    <au><snm>Tom</snm><fnm>H. W. K.</fnm></au>
    <au><snm>Mills</snm><fnm>A. P.</fnm></au>
  </aug>
  <source>Phys. Rev. Lett.</source>
  <publisher>American Physical Society</publisher>
  <pubdate>2012</pubdate>
  <volume>108</volume>
  <fpage>043401</fpage>
</bibl>

<bibl id="B85">
  <title><p>Calculated photoexcitation spectra of positronium {R}ydberg
  states</p></title>
  <aug>
    <au><snm>Hogan</snm><fnm>S. D.</fnm></au>
  </aug>
  <source>Phys. Rev. A</source>
  <pubdate>2013</pubdate>
  <volume>87</volume>
  <fpage>063423</fpage>
</bibl>

<bibl id="B86">
  <title><p>Measurement of the orthopositronium confinement energy in
  mesoporous thin films</p></title>
  <aug>
    <au><snm>Crivelli</snm><fnm>P</fnm></au>
    <au><snm>Gendotti</snm><fnm>U</fnm></au>
    <au><snm>Rubbia</snm><fnm>A</fnm></au>
    <au><snm>Liszkay</snm><fnm>L</fnm></au>
    <au><snm>Perez</snm><fnm>P</fnm></au>
    <au><snm>Corbel</snm><fnm>C</fnm></au>
  </aug>
  <source>Phys. Rev. A</source>
  <publisher>American Physical Society</publisher>
  <pubdate>2010</pubdate>
  <volume>81</volume>
  <fpage>052703</fpage>
</bibl>

<bibl id="B87">
  <title><p>Transmission-line decelerators for atoms in high {R}ydberg
  states</p></title>
  <aug>
    <au><snm>Lancuba</snm><fnm>P.</fnm></au>
    <au><snm>Hogan</snm><fnm>S. D.</fnm></au>
  </aug>
  <source>Phys. Rev. A</source>
  <pubdate>2014</pubdate>
  <volume>90</volume>
  <fpage>053420</fpage>
</bibl>

<bibl id="B88">
  <title><p>{Rydberg atoms in strong fields}</p></title>
  <aug>
    <au><snm>Kleppner</snm><fnm>D.</fnm></au>
    <au><snm>Littman</snm><fnm>M. G.</fnm></au>
    <au><snm>Zimmerman</snm><fnm>M. L.</fnm></au>
  </aug>
  <source>Rydberg States of Atoms and Molecules</source>
  <publisher>Cambrige: Cambridge University Press</publisher>
  <editor>R. F. Stebbings and F. B. Dunning</editor>
  <pubdate>1983</pubdate>
  <fpage>73</fpage>
</bibl>

<bibl id="B89">
  <title><p>{Quantum Mechanics of One- and Two-Electron Atoms}</p></title>
  <aug>
    <au><snm>Bethe</snm><fnm>HA</fnm></au>
    <au><snm>Salpeter</snm><fnm>EE</fnm></au>
  </aug>
  <publisher>Berlin: Springer</publisher>
  <pubdate>1957</pubdate>
</bibl>

<bibl id="B90">
  <title><p>{G}roup theory and the {C}oulomb Problem</p></title>
  <aug>
    <au><snm>Englefield</snm><fnm>M. J.</fnm></au>
  </aug>
  <publisher>New York: John Wiley \& Sons, Inc</publisher>
  <pubdate>1972</pubdate>
</bibl>

<bibl id="B91">
  <title><p>Atomic Physics</p></title>
  <aug>
    <au><snm>Foot</snm><fnm>C.</fnm></au>
  </aug>
  <publisher>Oxford: Oxford University Press</publisher>
  <pubdate>2005</pubdate>
</bibl>

<bibl id="B92">
  <title><p>Stark lifetimes for the hydrogen atom</p></title>
  <aug>
    <au><snm>Hiskes</snm><fnm>J. R.</fnm></au>
    <au><snm>Tarter</snm><fnm>C. B.</fnm></au>
    <au><snm>Moody</snm><fnm>D. A.</fnm></au>
  </aug>
  <source>Phys. Rev. A</source>
  <pubdate>1964</pubdate>
  <volume>133</volume>
  <fpage>A424</fpage>
</bibl>

<bibl id="B93">
  <title><p>{Stark structure of the Rydberg states of alkali-metal
  atoms}</p></title>
  <aug>
    <au><snm>Zimmerman</snm><fnm>ML</fnm></au>
    <au><snm>Littman</snm><fnm>MG</fnm></au>
    <au><snm>Kash</snm><fnm>MM</fnm></au>
    <au><snm>Kleppner</snm><fnm>D</fnm></au>
  </aug>
  <source>Phys. Rev. A</source>
  <pubdate>1979</pubdate>
  <volume>20</volume>
  <fpage>2251</fpage>
</bibl>

<bibl id="B94">
  <title><p>Observation of the {S}tark effect in autoionising {R}ydberg states
  of molecular hydrogen</p></title>
  <aug>
    <au><snm>Fielding</snm><fnm>H. H.</fnm></au>
    <au><snm>Softley</snm><fnm>T. P.</fnm></au>
  </aug>
  <source>Chem. Phys. Lett.</source>
  <pubdate>1991</pubdate>
  <volume>185</volume>
  <fpage>199</fpage>
</bibl>

<bibl id="B95">
  <title><p>{Stark effect and rotational-series interactions on high Rydberg
  states of molecular hydrogen}</p></title>
  <aug>
    <au><snm>Qin</snm><fnm>K.</fnm></au>
    <au><snm>Bistransin</snm><fnm>M.</fnm></au>
    <au><snm>Glab</snm><fnm>W. L.</fnm></au>
  </aug>
  <source>Phys. Rev. A</source>
  <pubdate>1993</pubdate>
  <volume>47</volume>
  <fpage>4154</fpage>
</bibl>

<bibl id="B96">
  <title><p>{Lifetimes of Rydberg states in ZEKE experiments. III. Calculations
  of the dc electric field dependence of predissociation lifetimes of
  NO}</p></title>
  <aug>
    <au><snm>Vrakking</snm><fnm>MJ</fnm></au>
  </aug>
  <source>J. Chem. Phys.</source>
  <pubdate>1996</pubdate>
  <volume>105</volume>
  <fpage>7336</fpage>
</bibl>

<bibl id="B97">
  <title><p>{Angular Momentum}</p></title>
  <aug>
    <au><snm>Zare</snm><fnm>R. N.</fnm></au>
  </aug>
  <publisher>New York: John Wiley \& Sons</publisher>
  <pubdate>1988</pubdate>
</bibl>

<bibl id="B98">
  <title><p>Trapping cold molecular hydrogen</p></title>
  <aug>
    <au><snm>Seiler</snm><fnm>{\mbox{{C}h}}</fnm></au>
    <au><snm>Hogan</snm><fnm>S. D.</fnm></au>
    <au><snm>Merkt</snm><fnm>F.</fnm></au>
  </aug>
  <source>Phys. Chem. Chem. Phys.</source>
  <pubdate>2011</pubdate>
  <volume>13</volume>
  <fpage>19000</fpage>
</bibl>

<bibl id="B99">
  <title><p>Deflection and deceleration of hydrogen {R}ydberg molecules in
  inhomogeneous electric fields</p></title>
  <aug>
    <au><snm>Yamakita</snm><fnm>Y.</fnm></au>
    <au><snm>Procter</snm><fnm>S. R.</fnm></au>
    <au><snm>Goodgame</snm><fnm>A. L.</fnm></au>
    <au><snm>Softley</snm><fnm>T. P.</fnm></au>
    <au><snm>Merkt</snm><fnm>F.</fnm></au>
  </aug>
  <source>J.~Chem. Phys.</source>
  <pubdate>2004</pubdate>
  <volume>121</volume>
  <fpage>1419</fpage>
</bibl>

<bibl id="B100">
  <title><p>Single-color two-photon spectroscopy of {R}ydberg states in
  electric fields</p></title>
  <aug>
    <au><snm>Wall</snm><fnm>T. E.</fnm></au>
    <au><snm>Cassidy</snm><fnm>D. B.</fnm></au>
    <au><snm>Hogan</snm><fnm>S. D.</fnm></au>
  </aug>
  <source>Phys. Rev. A</source>
  <publisher>American Physical Society</publisher>
  <pubdate>2014</pubdate>
  <volume>90</volume>
  <fpage>053430</fpage>
</bibl>

<bibl id="B101">
  <title><p>{R}ydberg-State-Enabled Deceleration and Trapping of Cold
  Molecules</p></title>
  <aug>
    <au><snm>Hogan</snm><fnm>S. D.</fnm></au>
    <au><snm>Seiler</snm><fnm>{\mbox{{C}h}}</fnm></au>
    <au><snm>Merkt</snm><fnm>F.</fnm></au>
  </aug>
  <source>Phys. Rev. Lett.</source>
  <pubdate>2009</pubdate>
  <volume>103</volume>
  <fpage>123001</fpage>
</bibl>

<bibl id="B102">
  <title><p>Measurement of the shift of {R}ydberg energy levels induced by
  blackbody radiation</p></title>
  <aug>
    <au><snm>Hollberg</snm><fnm>L.</fnm></au>
    <au><snm>Hall</snm><fnm>J. L.</fnm></au>
  </aug>
  <source>Phys. Rev. Lett.</source>
  <publisher>American Physical Society</publisher>
  <pubdate>1984</pubdate>
  <volume>53</volume>
  <fpage>230</fpage>
</bibl>

<bibl id="B103">
  <title><p>The effects of 300~{K} background radiation on {R}ydberg
  atoms</p></title>
  <aug>
    <au><snm>Beiting</snm><fnm>E. J.</fnm></au>
    <au><snm>Hildebrandt</snm><fnm>G. F.</fnm></au>
    <au><snm>Kellert</snm><fnm>F. G.</fnm></au>
    <au><snm>Foltz</snm><fnm>G. W.</fnm></au>
    <au><snm>Smith</snm><fnm>K. A.</fnm></au>
    <au><snm>Dunning</snm><fnm>F. B.</fnm></au>
    <au><snm>Stebbings</snm><fnm>R. F.</fnm></au>
  </aug>
  <source>J. Chem. Phys.</source>
  <publisher>AIP</publisher>
  <pubdate>1979</pubdate>
  <volume>70</volume>
  <fpage>3551</fpage>
</bibl>

<bibl id="B104">
  <title><p>Interactions of blackbody radiation with atoms</p></title>
  <aug>
    <au><snm>Gallagher</snm><fnm>T. F.</fnm></au>
    <au><snm>Cooke</snm><fnm>W. E.</fnm></au>
  </aug>
  <source>Phys. Rev. Lett.</source>
  <publisher>American Physical Society</publisher>
  <pubdate>1979</pubdate>
  <volume>42</volume>
  <fpage>835</fpage>
</bibl>

<bibl id="B105">
  <title><p>Collisional and Radiative Processes in Adiabatic Deceleration,
  Deflection and Off-Axis Trapping of a {R}ydberg Atom Beam</p></title>
  <aug>
    <au><snm>Seiler</snm><fnm>{\mbox{{C}h}}</fnm></au>
    <au><snm>Hogan</snm><fnm>S. D.</fnm></au>
    <au><snm>Schmutz</snm><fnm>H.</fnm></au>
    <au><snm>Agner</snm><fnm>J. A.</fnm></au>
    <au><snm>Merkt</snm><fnm>F.</fnm></au>
  </aug>
  <source>Phys. Rev. Lett.</source>
  <pubdate>2011</pubdate>
  <volume>106</volume>
  <fpage>073003</fpage>
</bibl>

<bibl id="B106">
  <title><p>Photoionization by blackbody radiation</p></title>
  <aug>
    <au><snm>Spencer</snm><fnm>WP</fnm></au>
    <au><snm>Vaidyanathan</snm><fnm>AG</fnm></au>
    <au><snm>Kleppner</snm><fnm>D</fnm></au>
    <au><snm>Ducas</snm><fnm>TW</fnm></au>
  </aug>
  <source>Phys. Rev. A</source>
  <publisher>American Physical Society</publisher>
  <pubdate>1982</pubdate>
  <volume>26</volume>
  <fpage>1490</fpage>
</bibl>

<bibl id="B107">
  <title><p>Photon occupation numbers in black body radiation</p></title>
  <aug>
    <au><snm>Mandel</snm><fnm>M.</fnm></au>
  </aug>
  <source>J. Opt. Soc. Am.</source>
  <pubdate>1979</pubdate>
  <volume>69</volume>
  <fpage>1038</fpage>
</bibl>

<bibl id="B108">
  <title><p>{Electrostatic trapping of neutral atomic particles}</p></title>
  <aug>
    <au><snm>Wing</snm><fnm>WH</fnm></au>
  </aug>
  <source>Phys. Rev. Lett.</source>
  <pubdate>1980</pubdate>
  <volume>45</volume>
  <fpage>631</fpage>
</bibl>

<bibl id="B109">
  <title><p>Stark Acceleration of {R}ydberg Atoms in Inhomogeneous Electric
  Fields</p></title>
  <aug>
    <au><snm>Breeden</snm><fnm>T</fnm></au>
    <au><snm>Metcalf</snm><fnm>H</fnm></au>
  </aug>
  <source>Phys. Rev. Lett.</source>
  <pubdate>1981</pubdate>
  <volume>47</volume>
  <fpage>1726</fpage>
</bibl>

<bibl id="B110">
  <title><p>Deflection of krypton {R}ydberg atoms in the field of an electric
  dipole</p></title>
  <aug>
    <au><snm>Townsend</snm><fnm>D.</fnm></au>
    <au><snm>Goodgame</snm><fnm>A. L.</fnm></au>
    <au><snm>Procter</snm><fnm>S. R.</fnm></au>
    <au><snm>Mackenzie</snm><fnm>S. R.</fnm></au>
    <au><snm>Softley</snm><fnm>T. P.</fnm></au>
  </aug>
  <source>J. Phys. B.: At. Mol. Opt. Phys.</source>
  <pubdate>2001</pubdate>
  <volume>34</volume>
  <fpage>439</fpage>
</bibl>

<bibl id="B111">
  <title><p>Controlling the motion of hydrogen molecules</p></title>
  <aug>
    <au><snm>Procter</snm><fnm>S. R.</fnm></au>
    <au><snm>Yamakita</snm><fnm>Y.</fnm></au>
    <au><snm>Merkt</snm><fnm>F.</fnm></au>
    <au><snm>Softley</snm><fnm>T. P.</fnm></au>
  </aug>
  <source>Chem. Phys. Lett.</source>
  <pubdate>2003</pubdate>
  <volume>374</volume>
  <fpage>667</fpage>
</bibl>

<bibl id="B112">
  <title><p>Nonhydrogenic effects in the deceleration of {R}ydberg atoms in
  inhomogeneous electric fields</p></title>
  <aug>
    <au><snm>Vliegen</snm><fnm>E.</fnm></au>
    <au><snm>W{\"o}rner</snm><fnm>H. J.</fnm></au>
    <au><snm>Softley</snm><fnm>T. P.</fnm></au>
    <au><snm>Merkt</snm><fnm>F.</fnm></au>
  </aug>
  <source>Phys. Rev. Lett.</source>
  <pubdate>2004</pubdate>
  <volume>92</volume>
  <fpage>033005</fpage>
</bibl>

<bibl id="B113">
  <title><p>{S}tark deceleration of hydrogen atoms</p></title>
  <aug>
    <au><snm>Vliegen</snm><fnm>E.</fnm></au>
    <au><snm>Merkt</snm><fnm>F.</fnm></au>
  </aug>
  <source>J. Phys. B.: At. Mol. Opt. Phys.</source>
  <pubdate>2006</pubdate>
  <volume>39</volume>
  <fpage>L241</fpage>
</bibl>

<bibl id="B114">
  <title><p>On the electrostatic deceleration of argon atoms in high {R}ydberg
  states by time-dependent inhomogeneous electric fields</p></title>
  <aug>
    <au><snm>Vliegen</snm><fnm>E.</fnm></au>
    <au><snm>Merkt</snm><fnm>F.</fnm></au>
  </aug>
  <source>J. Phys. B.: At. Mol. Opt. Phys.</source>
  <pubdate>2005</pubdate>
  <volume>38</volume>
  <fpage>1623</fpage>
</bibl>

<bibl id="B115">
  <title><p>Measurement of the three-dimensional velocity distribution of
  {S}tark-decelerated {R}ydberg atoms</p></title>
  <aug>
    <au><snm>Vliegen</snm><fnm>E.</fnm></au>
    <au><snm>Limacher</snm><fnm>P.</fnm></au>
    <au><snm>Merkt</snm><fnm>F.</fnm></au>
  </aug>
  <source>Eur. Phys. J. D</source>
  <pubdate>2006</pubdate>
  <volume>40</volume>
  <fpage>73</fpage>
</bibl>

<bibl id="B116">
  <title><p>Normal-incidence electrostatic {R}ydberg atom mirror</p></title>
  <aug>
    <au><snm>Vliegen</snm><fnm>E.</fnm></au>
    <au><snm>Merkt</snm><fnm>F.</fnm></au>
  </aug>
  <source>Phys. Rev. Lett.</source>
  <pubdate>2006</pubdate>
  <volume>97</volume>
  <fpage>033002</fpage>
</bibl>

<bibl id="B117">
  <title><p>Rydberg states in atom and molecule optics</p></title>
  <aug>
    <au><snm>Vliegen</snm><fnm>E.</fnm></au>
  </aug>
  <source>PhD thesis</source>
  <publisher>{Eidgen{\"o}ssische Technische Hochschule Z{\"u}rich}</publisher>
  <pubdate>2006</pubdate>
  <note>{Diss. ETH Nr. 16782}</note>
</bibl>

<bibl id="B118">
  <title><p>{S}tark deceleration and trapping of hydrogen {R}ydberg
  atoms</p></title>
  <aug>
    <au><snm>Vliegen</snm><fnm>E.</fnm></au>
    <au><snm>Hogan</snm><fnm>S. D.</fnm></au>
    <au><snm>Schmutz</snm><fnm>H.</fnm></au>
    <au><snm>Merkt</snm><fnm>F.</fnm></au>
  </aug>
  <source>Phys. Rev. A</source>
  <pubdate>2007</pubdate>
  <volume>76</volume>
  <fpage>023405</fpage>
</bibl>

<bibl id="B119">
  <title><p>Demonstration of three-dimensional electrostatic trapping of
  state-selected {R}ydberg atoms</p></title>
  <aug>
    <au><snm>Hogan</snm><fnm>S. D.</fnm></au>
    <au><snm>Merkt</snm><fnm>F.</fnm></au>
  </aug>
  <source>Phys. Rev. Lett.</source>
  <pubdate>2008</pubdate>
  <volume>100</volume>
  <fpage>043001</fpage>
</bibl>

<bibl id="B120">
  <title><p>Generation and high-resolution photoelectron spectroscopy of small
  organic radicals in cold supersonic expansions</p></title>
  <aug>
    <au><snm>Willitsch</snm><fnm>S.</fnm></au>
    <au><snm>Dyke</snm><fnm>J. M.</fnm></au>
    <au><snm>Merkt</snm><fnm>F.</fnm></au>
  </aug>
  <source>Helv. Chim. Acta</source>
  <pubdate>2003</pubdate>
  <volume>86</volume>
  <fpage>1152</fpage>
</bibl>

<bibl id="B121">
  <title><p>Motional, isotope and quadratic {S}tark effects in
  {R}ydberg-{S}tark deceleration and off-axis electric trapping of {H} and
  {D}</p></title>
  <aug>
    <au><snm>Hogan</snm><fnm>S. D.</fnm></au>
    <au><snm>Seiler</snm><fnm>{\mbox{{C}h}}</fnm></au>
    <au><snm>Merkt</snm><fnm>F.</fnm></au>
  </aug>
  <source>J. Phys. B.: At. Mol. Opt. Phys.</source>
  <pubdate>2013</pubdate>
  <volume>46</volume>
  <fpage>045303</fpage>
</bibl>

<bibl id="B122">
  <title><p>Dynamical processes in {R}ydberg-{S}tark deceleration and trapping
  of atoms and molecules</p></title>
  <aug>
    <au><snm>Seiler</snm><fnm>{\mbox{{C}h}}</fnm></au>
    <au><snm>Hogan</snm><fnm>S. D.</fnm></au>
    <au><snm>Merkt</snm><fnm>F.</fnm></au>
  </aug>
  <source>Chimia</source>
  <pubdate>2012</pubdate>
  <volume>66</volume>
  <fpage>208</fpage>
</bibl>

<bibl id="B123">
  <title><p>Ionization of {R}ydberg atoms by blackbody radiation</p></title>
  <aug>
    <au><snm>Beterov</snm><fnm>I. I.</fnm></au>
    <au><snm>Tretyakov</snm><fnm>D. B.</fnm></au>
    <au><snm>Ryabtsev</snm><fnm>I. I.</fnm></au>
    <au><snm>Entin</snm><fnm>V. M.</fnm></au>
    <au><snm>Ekers</snm><fnm>A.</fnm></au>
    <au><snm>Bezuglov</snm><fnm>N. N.</fnm></au>
  </aug>
  <source>New J. Phys</source>
  <pubdate>2009</pubdate>
  <volume>11</volume>
  <fpage>013052</fpage>
</bibl>

<bibl id="B124">
  <title><p>{Cold atoms and molecules by {Z}eeman deceleration and
  {R}ydberg-{S}tark deceleration}</p></title>
  <aug>
    <au><snm>Hogan</snm><fnm>S. D.</fnm></au>
  </aug>
  <source>PhD thesis</source>
  <publisher>{Eidgen{\"o}ssische Technische Hochschule Z{\"u}rich}</publisher>
  <pubdate>2012</pubdate>
</bibl>

<bibl id="B125">
  <title><p>Zur {T}heorie der {E}nergie\"ubertragung II</p></title>
  <aug>
    <au><snm>Landau</snm><fnm>L. D.</fnm></au>
  </aug>
  <source>Phys. Z. Sowjetunion</source>
  <pubdate>1932</pubdate>
  <volume>2</volume>
  <fpage>46</fpage>
</bibl>

<bibl id="B126">
  <title><p>Non-adiabatic crossing of energy levels</p></title>
  <aug>
    <au><snm>Zener</snm><fnm>C.</fnm></au>
  </aug>
  <source>Proc. R. Soc. London Ser. A</source>
  <pubdate>1932</pubdate>
  <volume>137</volume>
  <fpage>696</fpage>
</bibl>

<bibl id="B127">
  <title><p>{R}ydberg-{S}tark deceleration and trapping of atoms and
  molecules</p></title>
  <aug>
    <au><snm>Seiler</snm><fnm>{\mbox{{C}h}}</fnm></au>
  </aug>
  <source>PhD thesis</source>
  <publisher>{Eidgen{\"o}ssische Technische Hochschule Z{\"u}rich}</publisher>
  <pubdate>2013</pubdate>
  <note>{Diss. ETH Nr. 21340}</note>
</bibl>

<bibl id="B128">
  <title><p>Surface-electrode architecture for ion-trap quantum information
  processing</p></title>
  <aug>
    <au><snm>Chiaverini</snm><fnm>J.</fnm></au>
    <au><snm>Blakestad</snm><fnm>R. B.</fnm></au>
    <au><snm>Britton</snm><fnm>J.</fnm></au>
    <au><snm>Jost</snm><fnm>J. D.</fnm></au>
    <au><snm>Langer</snm><fnm>C.</fnm></au>
    <au><snm>Leibfried</snm><fnm>D.</fnm></au>
    <au><snm>Ozeri</snm><fnm>R.</fnm></au>
    <au><snm>Wineland</snm><fnm>D. J.</fnm></au>
  </aug>
  <source>Quant. Inf. Comput.</source>
  <pubdate>2005</pubdate>
  <volume>5</volume>
  <fpage>419</fpage>
</bibl>

<bibl id="B129">
  <title><p>Microscopic atom optics: {F}rom wires to an atom chip</p></title>
  <aug>
    <au><snm>Folman</snm><fnm>R.</fnm></au>
    <au><snm>Kr\"uger</snm><fnm>P.</fnm></au>
    <au><snm>Schmiedmayer</snm><fnm>J.</fnm></au>
    <au><snm>Denschlag</snm><fnm>J.</fnm></au>
    <au><snm>Henkel</snm><fnm>C.</fnm></au>
  </aug>
  <source>Adv. At. Mol. Opt. Phys.</source>
  <pubdate>2002</pubdate>
  <volume>48</volume>
  <fpage>263</fpage>
</bibl>

<bibl id="B130">
  <title><p>Complete methods set for scalable ion trap quantum information
  processing</p></title>
  <aug>
    <au><snm>Home</snm><fnm>J. P.</fnm></au>
    <au><snm>Hanneke</snm><fnm>D.</fnm></au>
    <au><snm>Jost</snm><fnm>J. D.</fnm></au>
    <au><snm>Amini</snm><fnm>J. M.</fnm></au>
    <au><snm>Leibfried</snm><fnm>D.</fnm></au>
    <au><snm>Wineland</snm><fnm>D. J.</fnm></au>
  </aug>
  <source>Science</source>
  <pubdate>2009</pubdate>
  <volume>325</volume>
  <fpage>1227</fpage>
</bibl>

<bibl id="B131">
  <title><p>Atom-chip-based generation of entanglement for quantum
  metrology</p></title>
  <aug>
    <au><snm>Riedel</snm><fnm>M. F.</fnm></au>
    <au><snm>B\"ohi</snm><fnm>P.</fnm></au>
    <au><snm>Li</snm><fnm>Y.</fnm></au>
    <au><snm>H\"ansch</snm><fnm>T. W.</fnm></au>
    <au><snm>Sinatra</snm><fnm>A.</fnm></au>
    <au><snm>Treutlein</snm><fnm>P.</fnm></au>
  </aug>
  <source>Nature</source>
  <pubdate>2010</pubdate>
  <volume>464</volume>
  <fpage>1170</fpage>
</bibl>

<bibl id="B132">
  <title><p>Trapping molecules on a chip in traveling potential
  wells</p></title>
  <aug>
    <au><snm>Meek</snm><fnm>S. A.</fnm></au>
    <au><snm>Bethlem</snm><fnm>H. L.</fnm></au>
    <au><snm>Conrad</snm><fnm>H.</fnm></au>
    <au><snm>Meijer</snm><fnm>G.</fnm></au>
  </aug>
  <source>Phys. Rev. Lett.</source>
  <pubdate>2008</pubdate>
  <volume>100</volume>
  <fpage>153003</fpage>
</bibl>

<bibl id="B133">
  <title><p>Taming molecular beams; towards a gas-phase molecular laboratory on
  a chip</p></title>
  <aug>
    <au><snm>Meek</snm><fnm>S. A.</fnm></au>
    <au><snm>Santambrogio</snm><fnm>G.</fnm></au>
    <au><snm>Conrad</snm><fnm>H.</fnm></au>
    <au><snm>Meijer</snm><fnm>G.</fnm></au>
  </aug>
  <source>J. Phys.: Conf. Ser.</source>
  <pubdate>2009</pubdate>
  <volume>194</volume>
  <fpage>012063</fpage>
</bibl>

<bibl id="B134">
  <title><p>Trapping molecules on a chips</p></title>
  <aug>
    <au><snm>Meek</snm><fnm>S. A.</fnm></au>
    <au><snm>Conrad</snm><fnm>H.</fnm></au>
    <au><snm>Meijer</snm><fnm>G.</fnm></au>
  </aug>
  <source>Science</source>
  <pubdate>2009</pubdate>
  <volume>324</volume>
  <fpage>1699</fpage>
</bibl>

<bibl id="B135">
  <title><p>Spatially resolved excitation of {R}ydberg atoms and surface
  effects on an atom chip</p></title>
  <aug>
    <au><snm>Tauschinsky</snm><fnm>A</fnm></au>
    <au><snm>Thijssen</snm><fnm>RMT</fnm></au>
    <au><snm>Whitlock</snm><fnm>S.</fnm></au>
    <au><snm>Heuvell</snm><fnm>H. B.</fnm></au>
    <au><snm>Spreeuw</snm><fnm>R. J. C.</fnm></au>
  </aug>
  <source>Phys. Rev. A</source>
  <publisher>American Physical Society</publisher>
  <pubdate>2010</pubdate>
  <volume>81</volume>
  <fpage>063411</fpage>
</bibl>

<bibl id="B136">
  <title><p>Realization of a superconducting atom chip</p></title>
  <aug>
    <au><snm>Nirrengarten</snm><fnm>T.</fnm></au>
    <au><snm>Qarry</snm><fnm>A.</fnm></au>
    <au><snm>Roux</snm><fnm>C.</fnm></au>
    <au><snm>Emmert</snm><fnm>A.</fnm></au>
    <au><snm>Nogues</snm><fnm>G.</fnm></au>
    <au><snm>Brune</snm><fnm>M.</fnm></au>
    <au><snm>Raimond</snm><fnm>J. M.</fnm></au>
    <au><snm>Haroche</snm><fnm>S.</fnm></au>
  </aug>
  <source>Phys. Rev. Lett.</source>
  <publisher>American Physical Society</publisher>
  <pubdate>2006</pubdate>
  <volume>97</volume>
  <fpage>200405</fpage>
</bibl>

<bibl id="B137">
  <title><p>An atom chip for the manipulation of ultracold atoms</p></title>
  <aug>
    <au><snm>Cherry</snm><fnm>O.</fnm></au>
    <au><snm>Carter</snm><fnm>J. D.</fnm></au>
    <au><snm>Martin</snm><fnm>J. D. D.</fnm></au>
  </aug>
  <source>Can. J. Phys.</source>
  <pubdate>2009</pubdate>
  <volume>87</volume>
  <fpage>633</fpage>
</bibl>

<bibl id="B138">
  <title><p>Creating single-atom and single-photon sources from entangled
  atomic ensembles</p></title>
  <aug>
    <au><snm>Saffman</snm><fnm>M.</fnm></au>
    <au><snm>Walker</snm><fnm>T. G.</fnm></au>
  </aug>
  <source>Phys. Rev. A</source>
  <publisher>American Physical Society</publisher>
  <pubdate>2002</pubdate>
  <volume>66</volume>
  <fpage>065403</fpage>
</bibl>

<bibl id="B139">
  <title><p>Surface-electrode {R}ydberg-{S}tark decelerator</p></title>
  <aug>
    <au><snm>Hogan</snm><fnm>S. D.</fnm></au>
    <au><snm>Allmendinger</snm><fnm>P.</fnm></au>
    <au><snm>Sa\ss{}mannshausen</snm><fnm>H.</fnm></au>
    <au><snm>Schmutz</snm><fnm>H.</fnm></au>
    <au><snm>Merkt</snm><fnm>F.</fnm></au>
  </aug>
  <source>Phys. Rev. Lett.</source>
  <publisher>American Physical Society</publisher>
  <pubdate>2012</pubdate>
  <volume>108</volume>
  <fpage>063008</fpage>
</bibl>

<bibl id="B140">
  <title><p>Deceleration and trapping of a fast supersonic beam of metastable
  helium atoms with a 44-electrode chip decelerator</p></title>
  <aug>
    <au><snm>Allmendinger</snm><fnm>P.</fnm></au>
    <au><snm>Agner</snm><fnm>J. A.</fnm></au>
    <au><snm>Schmutz</snm><fnm>H.</fnm></au>
    <au><snm>Merkt</snm><fnm>F.</fnm></au>
  </aug>
  <source>Phys. Rev. A</source>
  <publisher>American Physical Society</publisher>
  <pubdate>2013</pubdate>
  <volume>88</volume>
  <fpage>043433</fpage>
</bibl>

<bibl id="B141">
  <title><p>Surface-electrode decelerator and deflector for {R}ydberg atoms and
  molecules</p></title>
  <aug>
    <au><snm>Allmendinger</snm><fnm>P.</fnm></au>
    <au><snm>Deiglmayr</snm><fnm>J.</fnm></au>
    <au><snm>Agner</snm><fnm>J. A.</fnm></au>
    <au><snm>Schmutz</snm><fnm>H.</fnm></au>
    <au><snm>Merkt</snm><fnm>F.</fnm></au>
  </aug>
  <source>Phys. Rev. A</source>
  <publisher>American Physical Society</publisher>
  <pubdate>2014</pubdate>
  <volume>90</volume>
  <fpage>043403</fpage>
</bibl>

<bibl id="B142">
  <title><p>Trapping molecules on chips</p></title>
  <aug>
    <au><snm>Santambrogio</snm><fnm>G.</fnm></au>
  </aug>
  <source>EPJ Tech. Instrum.</source>
  <pubdate>2015</pubdate>
  <volume>2</volume>
  <fpage>14</fpage>
</bibl>

<bibl id="B143">
  <title><p>Merged neutral beams</p></title>
  <aug>
    <au><snm>Osterwalder</snm><fnm>A.</fnm></au>
  </aug>
  <source>EPJ Tech. Instrum.</source>
  <pubdate>2015</pubdate>
  <volume>2</volume>
  <fpage>10</fpage>
</bibl>

<bibl id="B144">
  <title><p>Guiding {R}ydberg atoms above surface-based transmission
  lines</p></title>
  <aug>
    <au><snm>Lancuba</snm><fnm>P.</fnm></au>
    <au><snm>Hogan</snm><fnm>S. D.</fnm></au>
  </aug>
  <source>Phys. Rev. A</source>
  <publisher>American Physical Society</publisher>
  <pubdate>2013</pubdate>
  <volume>88</volume>
  <fpage>043427</fpage>
</bibl>

<bibl id="B145">
  <title><p>Probing interactions between {R}ydberg atoms with large electric
  dipole moments in amplitude-modulated electric fields</p></title>
  <aug>
    <au><snm>Zhelyazkova</snm><fnm>V.</fnm></au>
    <au><snm>Hogan</snm><fnm>S. D.</fnm></au>
  </aug>
  <source>Phys. Rev. A</source>
  <publisher>American Physical Society</publisher>
  <pubdate>2015</pubdate>
  <volume>92</volume>
  <fpage>011402</fpage>
</bibl>

<bibl id="B146">
  <title><p>High-field-seeking {R}ydberg atoms orbiting a charged
  wire</p></title>
  <aug>
    <au><snm>Ko</snm><fnm>H.</fnm></au>
    <au><snm>Hogan</snm><fnm>S. D.</fnm></au>
  </aug>
  <source>Phys. Rev. A</source>
  <publisher>American Physical Society</publisher>
  <pubdate>2014</pubdate>
  <volume>89</volume>
  <fpage>053410</fpage>
</bibl>

<bibl id="B147">
  <title><p>Cold {R}ydberg Atoms as Realizable Analogs of {C}hern-{S}imons
  Theory</p></title>
  <aug>
    <au><snm>Baxter</snm><fnm>C.</fnm></au>
  </aug>
  <source>Phys. Rev. Lett.</source>
  <publisher>American Physical Society</publisher>
  <pubdate>1995</pubdate>
  <volume>74</volume>
  <fpage>514</fpage>
</bibl>

<bibl id="B148">
  <title><p>Angular Momentum of Supersymmetric Cold {R}ydberg Atoms</p></title>
  <aug>
    <au><snm>Zhang</snm><fnm>JZ</fnm></au>
  </aug>
  <source>Phys. Rev. Lett.</source>
  <publisher>American Physical Society</publisher>
  <pubdate>1996</pubdate>
  <volume>77</volume>
  <fpage>44</fpage>
</bibl>

<bibl id="B149">
  <title><p>Testing Spatial Noncommutativity via {R}ydberg Atoms</p></title>
  <aug>
    <au><snm>Zhang</snm><fnm>JZ</fnm></au>
  </aug>
  <source>Phys. Rev. Lett.</source>
  <publisher>American Physical Society</publisher>
  <pubdate>2004</pubdate>
  <volume>93</volume>
  <fpage>043002</fpage>
</bibl>

<bibl id="B150">
  <title><p>{Rydberg-Stark states in oscillating electric fields}</p></title>
  <aug>
    <au><snm>Zhelyazkova</snm><fnm>V.</fnm></au>
    <au><snm>Hogan</snm><fnm>S. D.</fnm></au>
  </aug>
  <source>Mol. Phys.</source>
  <pubdate>2015</pubdate>
  <volume>113</volume>
  <fpage>3979</fpage>
</bibl>

<bibl id="B151">
  <title><p>Level shifts of rubidium {R}ydberg states due to binary
  interactions</p></title>
  <aug>
    <au><snm>Reinhard</snm><fnm>A.</fnm></au>
    <au><snm>Liebisch</snm><fnm>TC</fnm></au>
    <au><snm>Knuffman</snm><fnm>B.</fnm></au>
    <au><snm>Raithel</snm><fnm>G.</fnm></au>
  </aug>
  <source>Phys. Rev. A</source>
  <pubdate>2007</pubdate>
  <volume>75</volume>
  <fpage>032712</fpage>
</bibl>

<bibl id="B152">
  <title><p>Electric-Field Induced Dipole Blockade with {R}ydberg
  Atoms</p></title>
  <aug>
    <au><snm>Vogt</snm><fnm>T</fnm></au>
    <au><snm>Viteau</snm><fnm>M</fnm></au>
    <au><snm>Chotia</snm><fnm>A</fnm></au>
    <au><snm>Zhao</snm><fnm>J</fnm></au>
    <au><snm>Comparat</snm><fnm>D</fnm></au>
    <au><snm>Pillet</snm><fnm>P</fnm></au>
  </aug>
  <source>Phys. Rev. Lett.</source>
  <pubdate>2007</pubdate>
  <volume>99</volume>
  <fpage>073002</fpage>
</bibl>

</refgrp>
} 







\end{backmatter}
\end{document}